# $\alpha_{\rm s}(2019)$ – Precision measurements of the QCD coupling

Workshop Proceedings, ECT*, Trento, 11–15 February 2019

*Editors*

David d'Enterria (CERN, Geneva), Stefan Kluth (MPI, München)

*Authors*

S. Alekhin (U. Hamburg), P. A. Baikov (Lomonosov U., Moscow), A. Banfi (U. Sussex, Brighton),
F. Barreiro (UAM, Madrid), A. Bazavov (MSU, Michigan), S. Bethke (MPI, München),
J. Blümlein (DESY, Zeuthen), D. Boito (Univ. São Paulo), N. Brambilla (TU, München),
D. Britzger (MPI, München), S. J. Brodsky (SLAC, Stanford), S. Camarda (CERN, Geneva),
K. G. Chetyrkin (KIT, Karlsruhe), D. d'Enterria (CERN, Geneva),
M. Dalla Brida (U. Milano-Bicocca, INFN), X. Garcia i Tormo (Univ. Bern),
M. Golterman (SF State Univ., San Francisco), R. Horsley (HCTP, Edinburgh),
J. Huston (MSU, Michigan), M. Jamin (IFAE/UAB, Barcelona), A. Kardos (U. Debrecen),
A. Keshavarzi (U. of Mississippi), S. Kluth (T.U. München), J. Kühn (KIT, Karlsruhe),
K. Maltman (York U. Toronto, and U. Adelaide), R. Miravitllas (IFAE/UAB, Barcelona),
S.-O. Moch (U. Hamburg), P. F. Monni (CERN, Geneva), D. Nomura (KEK, Tsukuba),
T. Onogi (Osaka Univ.), R. Pérez-Ramos (IPSA, Paris), S. Peris (UAB, Barcelona),
P. Petreczky (BNL, Upton), J. Pires (CFTP and LIP, Lisbon), A. Poldaru (MPI, München),
K. Rabbertz (KIT, Karlsruhe), F. Ringer (LBNL, Berkeley), S. Sint (Trinity College, Dublin),
R. Sommer (DESY, Zeuthen), G. Somogyi (MTA, Debrecen), J. Soto (Univ. Barcelona),
Z. Szőr (Univ. Mainz), H. Takaura (Kyushu Univ.), T. Teubner (Univ. Liverpool),
Z. Trócsányi (U. Debrecen, and Eötvös Loránd Univ.), Z. Tulipánt (U. Debrecen),
A. Vairo (TU, München), J. H. Weber (MSU, Michigan), X. Weichen (LMU, München),
A. Verbytskyi (MPI, München), G. Zanderighi (MPI, München)

**Abstract**

This document collects a written summary of all contributions presented at the workshop "$\alpha_{\rm s}$(2019): Precision measurements of the strong coupling" held at ECT* (Trento) in Feb. 11–15, 2019. The workshop explored in depth the latest developments on the determination of the QCD coupling $\alpha_{\rm s}$ from the key categories where high precision measurements are available: (i) lattice QCD, (ii) hadronic $\tau$ decays, (iii) deep-inelastic scattering and parton distribution functions, (iv) event shapes, jet cross sections, and other hadronic final-states in $e^+e^-$ collisions, (v) Z boson and W boson hadronic decays, and (vi) hadronic final states in p-p collisions. The status of the current theoretical and experimental uncertainties associated to each extraction method, and future perspectives were thoroughly reviewed. Novel $\alpha_{\rm s}$ determination approaches were discussed, as well as the combination method used to obtain a world-average value of the QCD coupling at the Z mass pole.

# Participants

**S. Alekhin** (U. Hamburg), **F. Barreiro** (UAM, Madrid), **S. Bethke** (MPI, München),
**N. Brambilla** (TU, München), **D. Britzger** (MPI, München), **S. J. Brodsky** (SLAC),
**S. Camarda** (CERN), **D. d'Enterria** (CERN), **M. Dalla Brida** (U. Milano-Bicocca, INFN),
**M. Golterman** (SF State Univ.), **J. Huston** (MSU, Michigan), **S. Kluth** (MPI, München),
**J. Kühn** (KIT, Karlsruhe), **R. Miravitllas** (UAB, Barcelona), **R. Pérez-Ramos** (IPSA, Paris),
**S. Peris** (UAB, Barcelona), **P. Petreczky** (BNL, Upton), **J. Pires** (CFTP and LIP, Lisbon),
**A. Poldaru** (MPI, München), **K. Rabbertz** (KIT, Karlsruhe), **F. Ringer** (LBNL, Berkeley),
**S. Sint** (Trinity College, Dublin), **R. Sommer** (DESY, Zeuthen), **G. Somogyi**
(MTA, Debrecen), **H. Takaura** (Kyushu Univ.), **A. Verbytskyi** (MPI, München),
**G. Zanderighi** (MPI, München)

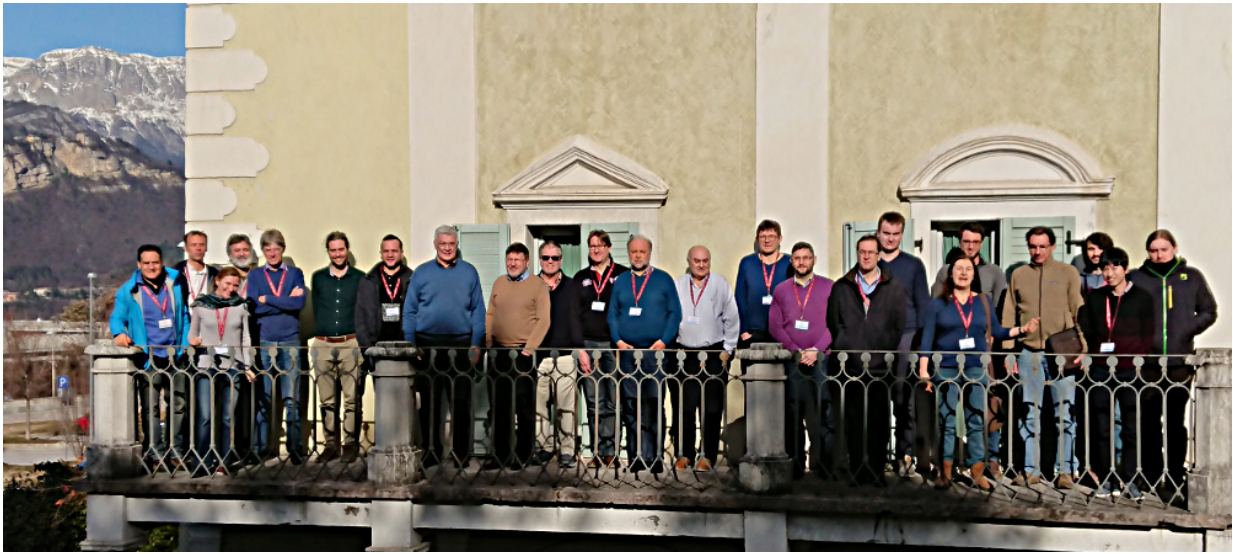

# 1 Introduction

The strong coupling $\alpha_\mathrm{s}$ is one of the fundamental parameters of the Standard Model (SM), setting the scale of the strength of the strong interaction theoretically described by Quantum Chromodynamics (QCD). Its value at the reference Z boson mass scale, in the conventional $\overline{\mathrm{MS}}$ renormalization scheme, amounts today to $\alpha_\mathrm{s}(m_\mathrm{z}) = 0.1181 \pm 0.0011$ with a $\delta\alpha_\mathrm{s}/\alpha_\mathrm{s} \approx 1\%$ uncertainty that is orders of magnitude larger than that of any other fundamental coupling in nature. Improving our knowledge of $\alpha_\mathrm{s}$ is crucial, among other things, to reduce the theoretical uncertainties in the high-precision calculations of all perturbative QCD (pQCD) processes whose cross sections or decay rates depend on higher-order powers of $\alpha_\mathrm{s}$, as is the case for virtually all those measured at the LHC. In the Higgs sector, our imperfect knowledge of $\alpha_\mathrm{s}$ (combined with that of the charm mass) propagates today into total final uncertainties of $\sim$4% for the H $\to gg$ ($c\bar{c}$) partial width(s). In the electroweak sector, the input $\alpha_\mathrm{s}(m_\mathrm{z})$ value is the leading source of uncertainty in the calculation of crucial precision pseudo-observables such as the Z boson width and its Z $\to b\bar{b}$ (and other hadronic) decay widths. The QCD coupling plays also a fundamental role in the calculation of key quantities in top-quark physics, such as the top mass, width, and Yukawa coupling.

The workshop *"$\alpha_\mathrm{s}$(2019) – Precision measurements of the QCD coupling"* was held at ECT*-Trento in February 11–15, 2019 with the aim of bringing together experts from various fields to explore in depth the latest developments on the determination of $\alpha_\mathrm{s}$ from the key categories where high precision measurements and theoretical calculations are currently available. The meeting can be considered as the third one of a "series" that started with the *"Workshop on Precision Measurements of $\alpha_\mathrm{s}$"* (MPI, Munich, February 9–11, 2011; https://arxiv.org/abs/1110.0016), and followed by the *"High-Precision $\alpha_\mathrm{s}$ Measurements from LHC to FCC-ee"* (CERN, Geneva, October 2–13, 2015; https://arxiv.org/abs/1512.05194). The presentations and discussions focused on the following issues:

- What is the current state-of-the-art of each one of the $\alpha_\mathrm{s}$ determination methods, from the theoretical and experimental perspectives?

- What is the status of those $\alpha_\mathrm{s}$ extractions that are not yet included in the world average?

- What is the current size of the theoretical (missing higher pQCD orders, electroweak corrections, power-suppressed corrections, hadronization corrections,...) and experimental uncertainties associated to each measurement?

- Are there improvements to be made in the combination of all $\alpha_\mathrm{s}$ extractions into the world average $\alpha_\mathrm{s}$ of the Particle Data Group report?

One important goal of the workshop was to facilitate the discussion among the different groups, and in particular to give the speakers the opportunity to explain details that one would normally not be able to present at a conference, but which have an important impact on the analyses. About 30 physicists took part in the workshop, and 25 talks were presented. Slides as well as background reference materials are available on the conference website

http://indico.cern.ch/e/alphas2019

The sessions and talks in the workshop program were organized as follows:

- Introduction:
    - "Introduction and goals of the workshop", D. d'Enterria and S. Kluth

- "World Summary of $\alpha_{\rm s}$ before 2019", S. Bethke

- Measurements of $\alpha_{\rm s}$ in the lattice:
    - "$\alpha_{\rm s}$ from the lattice: FLAG 2019 average", R. Sommer
    - "Strong coupling constant from the moments of quarkonium correlators", P. Petreczky
    - "$\alpha_{\rm s}$ from the lattice ALPHA collaboration (part I)", S. Sint
    - "$\alpha_{\rm s}$ from the QCD static energy", N. Brambilla
    - "$\alpha_{\rm s}$ from the lattice ALPHA collaboration (part II)", M. Dalla Brida
    - "$\alpha_{\rm s}$ from the static QCD potential with renormalon subtraction", H. Takaura

- $\alpha_{\rm s}$ and perturbative theory:
    - "The QCD coupling at all scales and the elimination of renormalization scale uncertainties", S. J. Brodsky
    - "The five-loop beta function of QCD", J.H Kühn

- Measurements of $\alpha_{\rm s}$ from e-p collisions and PDF fits:
    - "$\alpha_{\rm s}$, ABM PDFs, and heavy-quark masses", S. Alekhin
    - "$\alpha_{\rm s}$ from H1 jets", D. Britzger
    - "$\alpha_{\rm s}$ from parton densities", J. Huston

- Measurements of $\alpha_{\rm s}$ from $e^+e^-$ final states:
    - "Old and new observables for $\alpha_{\rm s}$ from $e^+e^-$ to hadrons", G. Somogyi
    - "$\alpha_{\rm s}$ from EEC and jet rates in $e^+e^-$", A. Verbytskyi
    - "The strong coupling from low-energy $e^+e^-$ to hadrons", M. Golterman
    - "$\alpha_{\rm s}$ from parton-to-hadron fragmentation", R. Perez-Ramos

- Measurements of $\alpha_{\rm s}$ at the LHC:
    - "$\alpha_{\rm s}$ from jets in pp collisions ", J. Pires
    - "$\alpha_{\rm s}$ jet substructure and a possible determination of the QCD coupling", F. Ringer
    - "Extractions of $\alpha_{\rm s}$ from ATLAS", F. Barreiro
    - "$\alpha_{\rm s}$ determinations from CMS", K. Rabbertz
    - "$\alpha_{\rm s}$ from inclusive W and Z cross sections at the LHC", A. Poldaru
    - "Determination of $\alpha_{\rm s}$ from the Z-boson transverse momentum distribution", S. Camarda

- Measurements of $\alpha_{\rm s}$ from hadronic decays of $\tau$ and electroweak bosons:
    - "$\alpha_{\rm s}$ from hadronic tau decay", S. Peris
    - "QCD coupling: scheme variations and tau decays", R. Miravitllas
    - "$\alpha_{\rm s}$ from hadronic W (and Z) decays", D. d'Enterria

- Discussion and Summary:
    - "$\alpha_{\rm s}$ averaging" discussion, all speakers

These proceedings constitute a collection of few-pages summaries, including relevant bibliographical references, for each one of the presentations, highlighting the most important results and issues of discussion.

ECT*, Trento, winter/spring 2019

# 2 Proceedings Contributions

# Pre-2019 Summaries of $\alpha_\mathrm{s}$

Siegfried Bethke

*Max-Planck-Institut für Physik, Munich*

**Abstract:** Summaries of measurements of $\alpha_\mathrm{s}$ and determinations of world average values of $\alpha_\mathrm{s}(m_\mathrm{z})$ are reviewed, spanning the time from 1989 to the latest update by the Particle Data Group in 2016/2018.

Determinations of $\alpha_\mathrm{s}$, the coupling parameter of the Strong Interaction between quarks and gluons, became available since the early 1980's, based on theoretical predictions of Quantum Chromodynamics (QCD), in next-to-leading or higher order of perturbation theory, and on experimental data at sufficiently large energy scales. Such determinations always were and continue to be challenging, due to the relatively large perturbative and nonperturbative uncertainties which dominate most of the measurements. Determinations of $\alpha_\mathrm{s}$, from different physical processes, energy scales and experiments, therefore do not necessarily agree with each other, within the quoted uncertainties of results. Therefore summaries of $\alpha_\mathrm{s}$ results and the determination of one overall "world average" value became mandatory.

One of the earliest and significant of such summaries and extractions was published by Altarelli in 1989, resulting in $\alpha_\mathrm{s}(m_\mathrm{z}) \approx 0.11 \pm 0.01$, with an overall uncertainty of about 10% [1]. The latest world summary of $\alpha_\mathrm{s}$, in the 2016 and 2018 Reviews of Particle Physics edited by the Particle Data Group (PDG) [2,3], quotes $\alpha_\mathrm{s}(m_\mathrm{z}) = 0.1181 \pm 0.0011$, with an overall uncertainty of just below 1%. The tenfold reduction of the uncertainty of $\alpha_\mathrm{s}$, achieved over the past almost 30 years, is mainly due – in *reverse* order of importance and impact – to

- higher statistics, multitude and quality of data, and improved experimental methods;
- theoretical predictions and calculations at higher perturbative orders (NNLO, N$^3$LO, resummation, ...);
- new theoretical developments in lattice gauge theory.

A (personal) selection of the history of summaries of $\alpha_\mathrm{s}$ is listed and referenced in Table 1 and displayed in Figure 1. Details of the 2016 world summary of $\alpha_\mathrm{s}$ [2] are also presented in Ref. [4]. Note that the overall uncertainty on $\alpha_\mathrm{s}(m_\mathrm{z})$ increased, from its 2014 to the 2016 value, which is mainly due to an adjustment of the procedure to combine systematic uncertainties, as will be discussed below.

In the following, a short recap of procedures used for deriving the most recent world average is given. The first step of summarising results is to define which of (the many) available analyses, measurements and results are to be included:

- the result must be published in a peer-reviewed scientific journal;
- the analysis must be based on at least NLO or higher order QCD perturbation theory (for results of $\alpha_\mathrm{s}(Q^2)$ to be included in the running coupling summary plot);
- results entering the world average determination of $\alpha_\mathrm{s}(m_\mathrm{z})$ must be based on at least NNLO or higher order perturbative QCD;

Table 1: World average values of $\alpha_{\text{s}}(m_{\text{z}})$ over time.

| year | $\alpha_{\text{s}}(m_{\text{z}})$ | $\Delta\alpha_{\text{s}}(m_{\text{z}})$ | comment | ref. |
|------|-----------|-------------|---------|------|
| 1989 | 0.11   | ±0.01   | NLO (pre-LEP) | [1] |
| 1994 | 0.117  | ±0.006  | + LEP + HERA  | [5] |
| 1998 | 0.119  | ±0.004  |               | [6] |
| 2000 | 0.1184 | ±0.0031 | at NNLO       | [7] |
| 2002 | 0.1183 | ±0.0027 |               | [8] |
| 2004 | 0.1182 | ±0.0027 |               | [9] |
| 2006 | 0.1189 | ±0.0010 | + lattice     | [10] |
| 2009 | 0.1184 | ±0.0007 |               | [11] |
| 2012 | 0.1184 | ±0.0007 |               | [12] |
| 2014 | 0.1185 | ±0.0006 |               | [13] |
| 2016 | 0.1181 | ±0.0011 |               | [2] |

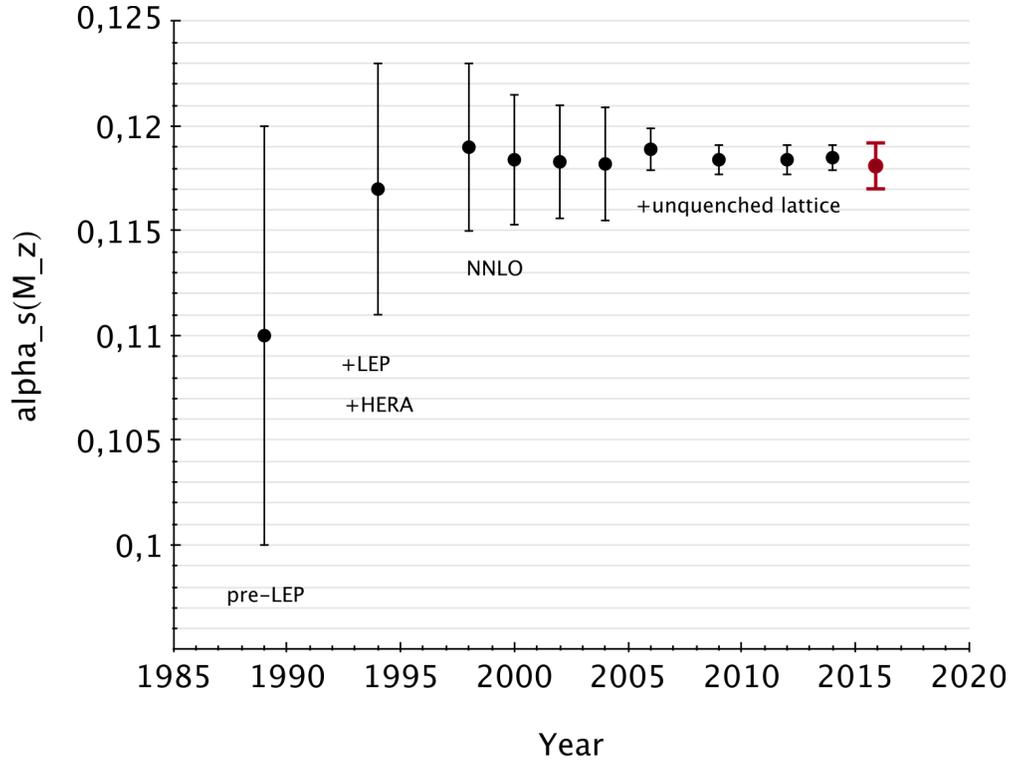

Figure 1: World average values of $\alpha_{\text{s}}(m_{\text{z}})$ over time.

- the analysis must include reliable estimates of experimental, systematic and theoretical uncertainties, based on commonly accepted procedures.

Next, the results are grouped into 6 classes of measurements that are based on similar or identical types of data, calculations or procedures:

- decays of $\tau$-leptons,
- deep inelastic lepton-nucleon scattering (DIS; until recently, only structure functions at NNLO),
- lattice QCD,
- jets and hadronic event shapes in $e^+e^-$ annihilation,
- electro-weak precision fits,
- hadron collider results (so far, only $t\bar{t}$ cross section at NNLO),

and a pre-average is determined for each of these classes. Finally, the world average is then determined from these 6 pre-averages of classes. Pre-averages are determined taking the unweighted mean and average error. This should provide the most unbiased estimator of the average and its uncertainty in case of largely correlated results, with unknown degrees of correlations and unknown "errors on errors".

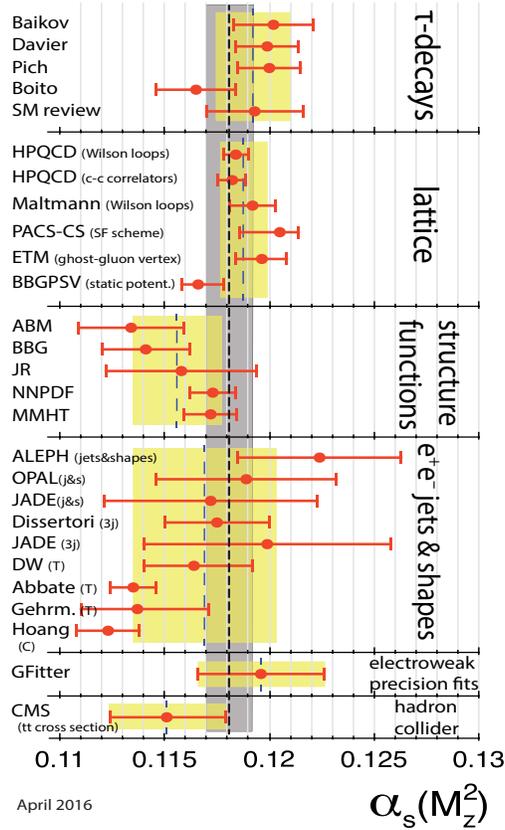

Figure 2: 2016 summary of determinations of $\alpha_s(m_z)$. The light-shaded bands and long-dashed vertical lines indicate the pre-average values; the dark-shaded band and short-dashed line represents the new overall world average of $\alpha_s(m_z)$.

The final world average is then determined as the weighted mean of the class pre-averages, initially treating their uncertainties as being uncorrelated and of Gaussian nature. This determines the final world average value of $\alpha_{\rm s}(m_{\rm z})$. The overall *uncertainty* of the world average is then adjusted according to the following procedure:

If the overall $\chi^2$ is smaller than 1 per degree of freedom (d.o.f.), an overall correlation coefficient is introduced in the error matrix and adjusted such that $\chi^2/{\rm d.o.f.} = 1$. If the overall $\chi^2/{\rm d.o.f.}$ is larger than 1, all uncertainties are enlarged by a common factor such that $\chi^2/{\rm d.o.f.} = 1$. Note that in both cases, adjusting a common correlation factor or enlarging all individual uncertainties, the final uncertainty of the average value increases with respect to the initial, "uncorrelated" starting value!

The results included in the 2016 and 2018 world summary of $\alpha_{\rm s}(m_{\rm z})$, together with the respective pre-averages of classes and the final world average, are displayed in Figure 2. Note that in two of the classes, no pre-averaging has been applied as only one individual result was available in each case, at the time of the analysis (2016).

As shown in Table 1 and Figure 1, the overall quoted uncertainty of $\alpha_{\rm s}(m_{\rm z})$ increased from 0.0006 (in the review of 2014) to 0.0011 (review of 2016). The reason for this increase was mainly procedural: in 2014, pre-averages were not determined by taking the linear average of individual results and their uncertainties, but by a method called "range-averaging". There, pre-averages were determined by taking the central value of the range of input values and half of this range interval, as central value and its uncertainty, respectively. For the lattice results, which were expected to be essentially uncorrelated with each other, the pre-average was determined using the $\chi^2$ method.

Figure 3 summarises the history and values of pre-averages of $\alpha_{\rm s}(m_{\rm z})$ for the different classes of measurements. Note that the change in error determinations predominantly affected the class of

| $\alpha_{\rm s}(M_{\rm Z})$ | 2012 | 2014 | 2016 & 18 |
|---|---|---|---|
| τ | 0.1197 ± 0.0016 | 0.1197 ± 0.0016 | 0.1192 ± 0.0018 |
| Lattice | 0.1185 ± 0.0007 | 0.1185 ± 0.0005 | 0.1188 ± 0.0011, |
| DIS | 0.1151 ± 0.0022 | 0.1154 ± 0.0020 | 0.1156 ± 0.0021, |
| e+e- | 0.1172 ± 0.0037 | 0.1177 ± 0.0046. | 0.1169 ± 0.0034. |
| ew | 0.1193 ± 0.0028 | 0.1193 ± 0.0028 | 0.1196 ± 0.0030 |
| pp | — | — | 0.1151 ± 0.0028 |
| world ave. | 0.1184 ± 0.0007 | 0.1185 ± 0.0006 | 0.1181 ± 0.0011 |

"range" average and error
unweighted average and error
χ² weighted average and error

Figure 3: History and results of pre-averages of $\alpha_{\rm s}(m_{\rm z})$.

lattice results, whose uncertainty thus increased by a factor of two, using the most recent method of taking the unweighted mean and error instead. This, in turn, affected the overall uncertainty of the world average, which was (and still is, albeit to a lesser extent) dominated by the influence of the lattice results.

The current status of determining the world average value of $\alpha_s(m_Z)$ is rather satisfying, showing consistency and agreement within the quoted overall uncertainty of about 1%. The latter is limited by the fact that, within each class of measurements of $\alpha_s$, there are issues which prevail since quite some time, and which could not yet be solved in a convincing manner:

- $\alpha_s$ from $\tau$ decays: uncertainties between different perturbative calculations (FOPT; CIPT) as well as other technical systematics;

- $\alpha_s$ from lattice calculations: size of quoted uncertainties;

- $\alpha_s$ from DIS: unsolved issues between author groups (PDFs);

- $\alpha_s$ from $e^+e^-$ annihilation: analytic vs. classical treatment of (nonperturbative) hadronisation effects;

- $\alpha_s$ from hadron colliders: so far, only one determination in NNLO (more available recently); in NLO analyses: choice of renormalisation/factorisation scales, treatment of top-threshold, non-perturbative/hadronisation corrections;

- $\alpha_s$ from electroweak precision data: correct in strict Standard Model, very sensitive to many beyond-Standard-Model (BSM) effects if present.

Last not least, the methods applied to select and (pre-)average results might have to be revisited and improved.

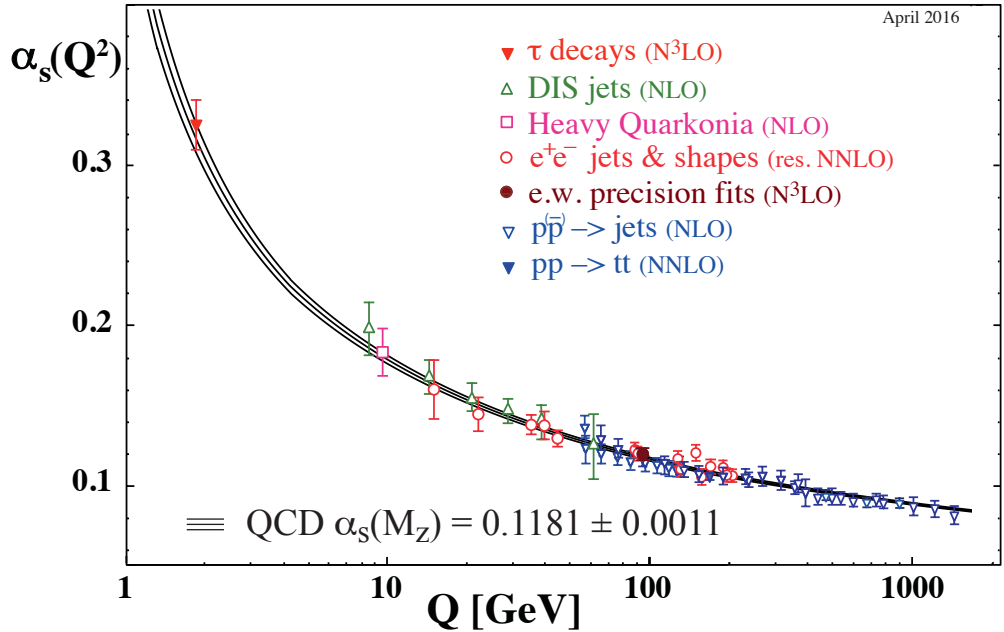

Figure 4: Summary of measurements of $\alpha_s$ as a function of the energy scale Q.

To my personal opinion, significant improvements in the precision of measurements of $\alpha_\mathrm{s}$, below the 1% level, may mainly (only?) be expected from improved lattice calculations, and from high statistics measurements of the $Z^0$ lineshape (also called Giga-Z or Tera-Z), at future high-energy $e^+e^-$ collider projects.

However, and maybe even more important from the viewpoint of testing the fundamental theory of Strong Interactions, the successful and precise confirmation of the concept of Asymptotic Freedom and thus, the experimental "proof" of *the* key feature of QCD, is regarded to be one of the most remarkable achievements of both, theoretical and experimental particle physics, see Figure 4. My personal thanks and respect go to all those who have taken part and actively contributed to the many measurements and results in this field.

# The 2019 lattice FLAG $\alpha_\text{s}$ average

**R. Sommer**[1,2], T. Onogi[3], and R. Horsley[4]

[1] *Institut für Physik, Humboldt-Universität zu Berlin, Newtonstr. 15, 12489 Berlin, Germany*
[2] *John von Neumann Inst. for Computing (NIC), DESY, Platanenallee 6, 15738 Zeuthen, Germany*
[3] *Department of Physics, Osaka University, Toyonaka, Osaka 560-0043, Japan*
[4] *Higgs Centre for Theoretical Physics, School of Physics and Astronomy, University of Edinburgh, Edinburgh EH9 3FD, UK*

**Abstract:** We summarise the recent 2019 average of $\alpha_\text{s}$ by the FLAG collaboration.

## Introduction

Lattice gauge theory is a non-perturbative formulation of QCD, which allows us to evaluate the Euclidean path integral by a Monte Carlo "simulation" for a few suitably chosen values of

$$L/a,\ T/a,\ g_0,\ \{am_i,\ i=1\ldots N_f\}. \tag{1}$$

Here $L/a$ is the number of points of the world in each space dimension, $T$ (often bigger than $L$) is the extent of the time axis, $g_0$ is the bare coupling of the theory, and $am_i$ are the bare quark masses. Once we obtain the relation between the bare parameters and hadronic low-energy quantities, such as $f_\pi, m_\pi, m_K \ldots$, we can in principle predict all physical quantities in QCD, including $\alpha_\text{s}$.

## Methods for the strong coupling.

The general method for extracting $\alpha_{\overline{\text{MS}}}$ with lattice QCD is to consider a short-distance, one-scale, observable with an expansion

$$\mathcal{O}(\mu) = c_1 \alpha_{\overline{\text{MS}}}(\mu) + c_2 \alpha_{\overline{\text{MS}}}(\mu)^2 + \cdots, \tag{2}$$

compute $\mathcal{O}(\mu)$ by lattice QCD and determine $\alpha_{\overline{\text{MS}}}(\mu)$ from Eq. (2). This requires that we are in a region where perturbation theory is valid, i.e. $\alpha_{\overline{\text{MS}}}(\mu)$ is small.

**Advantages.** An important advantage of taking $\mathcal{O}$ from lattice QCD compared to using experimental data is that one is automatically in the Euclidean region where no hadronisation corrections, duality violations etc. are a concern. Furthermore one has a large freedom to design convenient observables.

**Disadvantages.** Determining $\alpha_\text{s}$ is a two stage process, connecting quantities at two disparate scales, high momentum and the hadronic scale – the latter is where lattice QCD naturally resides. Furthermore, lattice QCD simulations are restricted to $N_f = 3$ or $N_f = 4$ quarks at most, because the $b$-quark is simply too heavy. One then relies on perturbative matching across the appropriate quark thresholds to determine $\alpha_\text{s}$ at the $m_Z$ scale where the number of active flavours is $N_f = 5$. Note that this means that many earlier results for $N_f = 2$ cannot be used, as crossing the strange quark threshold needs a non-perturbative procedure. ($N_f = 0$ results being computationally cheap form a useful testbed for checking different methods.)

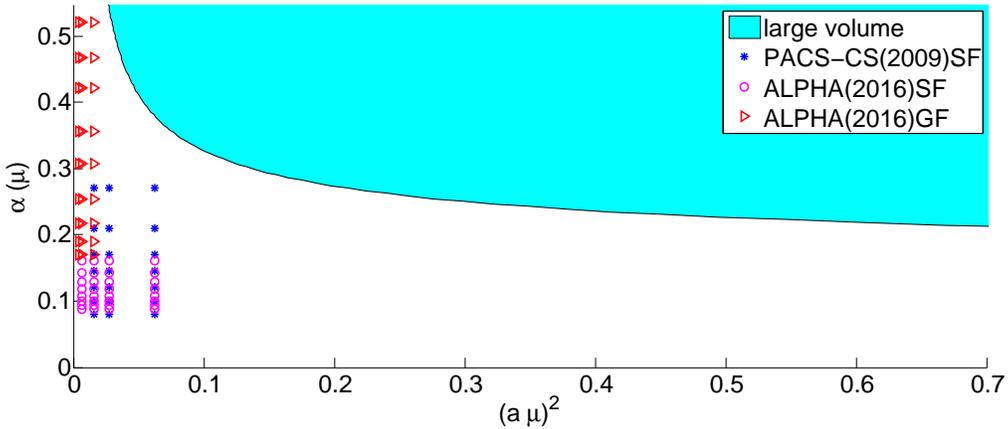

Figure 1: The plane $\alpha^{(3)}_{\overline{\mathrm{MS}}}(\mu)$ for $N_f = 3$ against the scale $\mu$ in lattice units, where $a$ is the lattice spacing and the blue region corresponds to the rough bound $a > 0.04\,\mathrm{fm}$. Note that the continuum limit is approached by extrapolations with $a\mu \ll 1$. The points on the left correspond to actual Monte Carlo simulations in category (III).

## The 2019 FLAG review.

The Flavour Lattice Averaging Group (FLAG) formed a working group (R. Horsley, T. Onogi, R.S.) on $\alpha_\mathrm{s}$ in 2011 and first included determinations of $\alpha_\mathrm{s}(m_\mathrm{Z})$ in its review in 2013 [1]. Updates appeared in 2016 [2] and 2019 [3].

Here we report on this latter work. We first briefly comment on our procedure for determining averages. There are similarities and differences to the PDG approach [4]. The main difference is that FLAG formulates a set of criteria, that computations have to pass in order to enter the average of a given quantity of phenomenological interest [2]. These are based on whether the simulations cover a range of parameters that allow to achieve a satisfactory control of systematic uncertainties (labeled ★) a reasonable attempt can be made at estimating systematic uncertainties ( ○ ), or it is unlikely that systematic uncertainties can be brought under control ( ■ ). The appearance of a ■ even in a single source of systematic error of a given lattice result, disqualifies it from inclusion in the global average.

For the computations of $\alpha_\mathrm{s}$, the usual criteria for chiral and infinite volume extrapolations are somewhat relaxed as they do not play a dominant role. Instead criteria on *perturbative behaviour* and *renormalization scale* try to make sure that the computation is at reasonable high $\mu$, the perturbative knowledge is sufficiently good (i.e. the number of known loops, $n_\mathrm{l}$, is sufficiently high) and $\mu$ could be varied over some range in order to confirm the perturbative $\mu$-dependence. The general idea is that these criteria try to make sure that the available Monte Carlo data have a few points located sufficiently low in the landscape of Fig. 1, while the continuum limit criterion requires us to not be too far on the right. The precise criteria are given in FLAG 19 [3].

In order to arrive at a final average, we first form pre-averages of computations using one and the same method and after combine them to give a final estimate. We now discuss the different methods, following a certain classification (I-III).

**(I) Continuum-limit observables in large volume.** Here $\mathcal{O}$ is a finite observable depending on the scale $\mu$. One can then take the continuum limit

$$\mathcal{O}(\mu) \equiv \lim_{a \to 0} \mathcal{O}_{\text{lat}}(a, \mu) \quad \text{with } \mu \text{ fixed}. \tag{3}$$

One wants $\mu$ to be high such that the expansion Eq. (2) is precise and $a\mu$ small to control the discretization error. However, recall that one is usually in the blue shaded region of Fig. 1 and it is difficult to extrapolate when $\alpha_{\overline{\text{MS}}}$ is small, say $\alpha_{\overline{\text{MS}}} \leq 0.3$.

There are several different methods. They share the necessity for finding a compromise between large $\mu$ and small $a\mu$. In the cases where computations qualify for taking an average (i.e., there is no ■), we perform a weighted average of the different results. According to our judgement the uncertainties are dominantly systematic. They are due to the truncation error of perturbation theory, whether ordinary higher order or non-perturbative effects. We just estimate the perturbative truncation error and take this as the uncertainty of the pre-range, which is usually somewhat more conservative than the uncertainty estimate in the contributing papers.

The individual methods are (we partially have to simplify here):

(1) $Q$-$\overline{Q}$ potential: $\mathcal{O}(\mu) = r^2 F_{\text{static}}(r)$, $\mu = 2/r$, where $F_{\text{static}}(r)$ is the force between static quarks defined by the large-$t$ behaviour of Wilson loops $W(r,t)$. Note that $n_l$ is 3 but $n_l > 3$ terms proportional to $\log \alpha_s$ are also known. Indeed, at fixed order perturbation theory, the basic observable $\mathcal{O}(\mu)$ is infrared divergent. As discussed by N. Brambilla and H. Takaura at this workshop, these divergences can be resummed, leaving terms such as $\alpha_s^4 \log \alpha_s$ in the expansion of $\mathcal{O}(\mu)$.

(2) Vacuum polarization: $\mathcal{O}(\mu) = D(Q^2)$, $\mu^2 = Q^2$, with $D$ the Adler function derived from the V+A two-point function at Euclidean $q$. This method does not yet enter the average.

(3) Two point HH current: moments of heavy-heavy pseudoscalar-current two-point functions. Heavy quarks of masses around the charm and heavier are used. Different discretizations are available that allow also to compare the continuum-limit moments before the extraction of $\alpha_s$. There is quite good agreement.

(4) Gluon-ghost vertex: using gauge fixing, the momentum-space vertex is used. This method does not yet enter the average as the continuum limit criterion is not passed.

(5) Dirac eigenvalues: $\mathcal{O}(\mu) = \partial_\lambda \log(\rho(\lambda))/\partial \log(\lambda)$, $\mu = \lambda$ with $\rho(\lambda)$ the spectral density of the massless Dirac operator. This recently introduced method [5] does not yet pass the continuum-limit criterion.

**(II) Lattice observables at the cutoff.** There is also the possibility to consider lattice observables involving distances of a few lattice spacings, which are not related to a continuum observable. The prominent example is rectangular Wilson loops $W(r,t)$ of extent $r \times t$ with $r = am$ and $t = an$, keeping the integers $n,m$ fixed as one takes the limit $a \to 0$; the loops shrink to size zero in the limit. Such observables have an expansion

$$W(na, ma) \overset{g_0 \to 0}{\sim} \sum_{k \geq 0} c_{m,n}^{(k)} g_0^{2k} \overset{g_0 \to 0}{\sim} \sum_{k \geq 0} \hat{c}_{m,n}^{(k)} g_{\overline{\text{MS}}}^{2k}(1/a), \tag{4}$$

where in the second step use is made of the relation between the bare coupling and a renormalized coupling at the cutoff scale, $g_0^2 = g_{\overline{\text{MS}}}^2(1/a) + \mathcal{O}(g^4)$. The available loop orders are often lower than for continuum perturbation theory. Lattice artefacts can only be separated from perturbative corrections in Eq. (4) by assuming some functional form and fitting to it.

In this category small $(m, n \leq 3)$ Wilson loops $\mathcal{O}(\mu) = W(ma, na)$, $\mu = k/a$, and functions thereof (e.g. $\log(W(a,a))$) are often used. The scale factor $k$ is adjusted to have better apparent convergence of PT. Our estimate of perturbative uncertainties is again somewhat conservative [3].

**(III) Continuum-limit observables in small volume and step scaling.** For finite volume quantities with volume $L^4$ and some technical requirements, Eq. (2) holds but with

$$\mu = 1/L. \tag{5}$$

The advantage is that now $\mu a$ can easily be taken to $a/L = 1/8 \ldots 1/32$ or smaller. However, a number of steps are needed to connect recursively

$$\mu_0 \to s\mu_0 \to s^2\mu_0 \to \ldots \to s^N\mu_0, \tag{6}$$

and in each step a few different lattice spacings $a$ have to be simulated to take the continuum limit. After a few steps, $\mu$ becomes very large so that perturbation theory can be applied with confidence and statistical errors dominate the uncertainty. At this workshop, M. Dalla Brida presented a recent precise three-flavour computation with $\mu_0 \approx 200$ MeV and $s^N \mu_0 \approx 100$ GeV. We perform a straight weighted average for mean and error of the two available results for this method.

**World average from FLAG.** Altogether we have considered 18 computations, of which 9 pass our criteria. These are shown in Fig. 2 and Table 1. For each method, the grey band shows the pre-average as explained above. We are left with the task to combine those pre-averages. Again we take the central value from their weighted average. However, since the errors of the pre-averages are mostly systematic, we feel that the straight error 0.00057 of the weighted average is too optimistic – it would be correct for independent Gaussian distributions. Instead we use the smallest error of the pre-averages. This yields the result

$$\alpha_{\overline{\mathrm{MS}}}^{(5)}(m_Z) = 0.11823(81). \qquad \text{Refs. [6,16,10,18,14,7,15].} \tag{7}$$

## Further progress

Finally, we collect some lessons that we have learned in our forming of a lattice world average of $\alpha_\mathrm{s}$.

The basic problem is simple and has been spelled out often, phrased in varying words. In order to have a precise value with an error that can be estimated by perturbation theory itself, large energy scales $\mu$ have to be reached and theory assumptions have to be kept at a minimum. Further progress will be limited if we include processes where non-perturbative contributions have to be fitted or removed by complicated analyses in order to make lower energies accessible. Dealing with non-perturbative physics is always based on assumptions – if only where the expansion in $1/\mu$ applies and lowest-order terms $(1/\mu)^{N_\mathrm{min}}$ dominate.

We should therefore separate the determination of $\alpha_\mathrm{s}$ at high enough $\mu$, simple theory, from tests of perturbation theory, with resummations, studies of higher-twist contributions, etc.

The concept of criteria introduced by FLAG is very useful in this respect, and we advocate to consider such a procedure for phenomenological determinations. One should at least consider a criterion on minimum values of $\mu$, paired with sufficiently high perturbative order. In FLAG these are the "renormalization scale" / "perturbative behaviour" criteria.

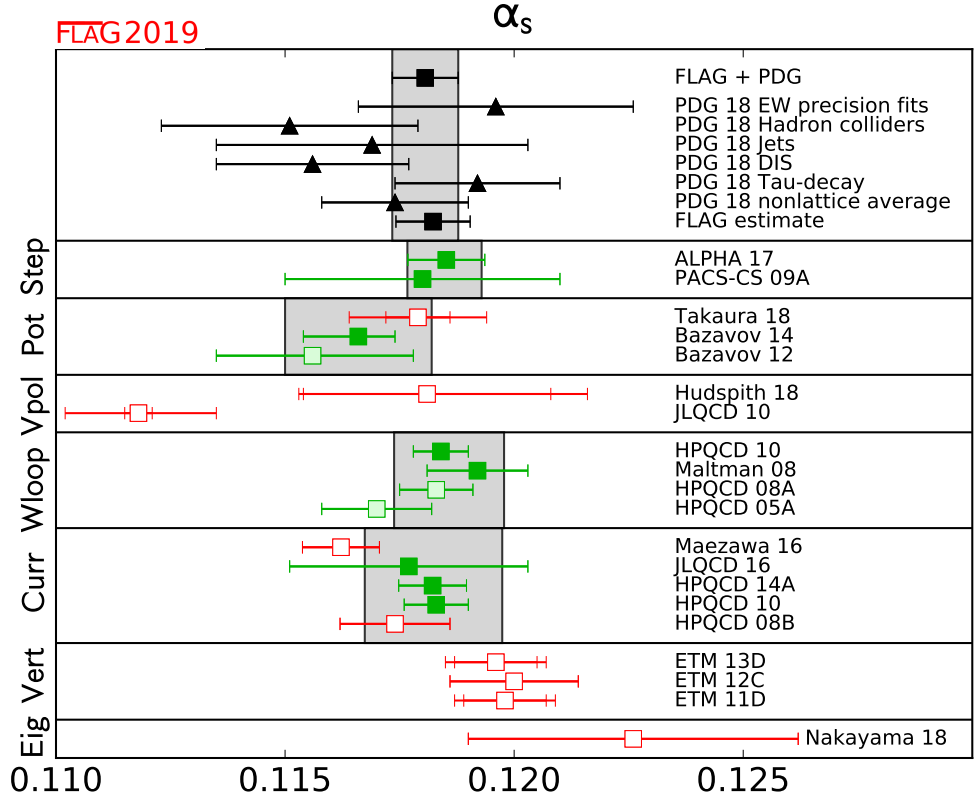

Figure 2: The $\overline{\text{MS}}$ coupling at the $Z$ mass. The PDG 18 [4] entries give the outcome of their analysis from various phenomenological categories including their average. The lattice computations with a filled green box, ■, have no red box, ■, in the previous ratings and therefore qualify for averaging. A □ means the same but the number does not enter an average because it is superseded by a later more complete computation or it was not published at the September 2018 deadline. Computations with □ do not enter the averages because they had at least one ■ before.

We also think that the criteria of FLAG should become more strict as time goes on. This is necessary to avoid situations where complicated procedures, involving e.g. separate estimates of perturbative errors (see above), are needed to arrive at a safe range.

Finally, it seems that the limit of lattice determinations of $\alpha_\text{s}$ is not yet reached; we believe a factor of two reduction in the error is possible with some variation of the developed techniques.

**Acknowledgments.** We thank our colleagues in FLAG for a fruitful collaboration. RS thanks the organizers of the workshop for their initiative and for providing a stimulating atmosphere and the participants of the workshop for interesting discussions.

| Collaboration | Ref. | $N_f$ | publication status | renormalization scale | perturbative behaviour | continuum extrapolation | $\alpha_{\overline{\rm MS}}(m_Z)$ | method | $n_l$ |
|---|---|---|---|---|---|---|---|---|---|
| ALPHA 17 | [6] | 2+1 | A | ★ | ★ | ★ | 0.11852(84) | step scaling | 2 |
| PACS-CS 09A | [7] | 2+1 | A | ★ | ★ | ○ | 0.11800(300) | | 2 |
| pre-range (average) | | | | | | | 0.11848(81) | | |
| Takaura 18 | [8,9] | 2+1 | P | ■ | ○ | ○ | $0.11790(70)(^{+130}_{-120})$ | $Q$-$\bar{Q}$ potential | 3 |
| Bazavov 14 | [10] | 2+1 | A | ○ | ★ | ○ | $0.11660(^{+120}_{-80})$ | | 3 |
| Bazavov 12 | [11] | 2+1 | A | ○ | ○ | ○ | $0.11560(^{+210}_{-220})$ | | 3 |
| pre-range with estimated pert. error | | | | | | | 0.11660(160) | | |
| Hudspith 18 | [12] | 2+1 | P | ○ | ○ | ■ | $0.11810(270)(^{+80}_{-220})$ | vacuum polarization | 3 |
| JLQCD 10 | [13] | 2+1 | A | ■ | ○ | ■ | $0.11180(30)(^{+160}_{-170})$ | | 2 |
| HPQCD 10 | [14] | 2+1 | A | ○ | ★ | ★ | 0.11840(60) | Wilson loops | 2 |
| Maltman 08 | [15] | 2+1 | A | ○ | ○ | ★ | 0.11920(110) | | 2 |
| pre-range with estimated pert. error | | | | | | | 0.11858(120) | | |
| JLQCD 16 | [16] | 2+1 | A | ○ | ○ | ○ | 0.11770(260) | HH current, two points | 2 |
| Maezawa 16 | [17] | 2+1 | A | ○ | ■ | ○ | 0.11622(84) | | 2 |
| HPQCD 14A | [18] | 2+1+1 | A | ○ | ★ | ○ | 0.11822(74) | | 2 |
| HPQCD 10 | [14] | 2+1 | A | ○ | ★ | ○ | 0.11830(70) | | 2 |
| HPQCD 08B | [19] | 2+1 | A | ■ | ■ | ■ | 0.11740(120) | | 2 |
| pre-range with estimated pert. error | | | | | | | 0.11824(150) | | |
| ETM 13D | [20] | 2+1+1 | A | ○ | ○ | ■ | 0.11960(40)(80)(60) | gluon-ghost vertex | 3 |
| ETM 12C | [21] | 2+1+1 | A | ○ | ○ | ■ | 0.12000(140) | | 3 |
| ETM 11D | [22] | 2+1+1 | A | ○ | ○ | ■ | $0.11980(90)(50)(^{+0}_{-50})$ | | 3 |
| Nakayama 18 | [5] | 2+1 | A | ★ | ○ | ■ | 0.12260(360) | Dirac eigenvalues | 2 |

Table 1: Results for $\alpha_{\overline{\rm MS}}(m_Z)$ from simulations that use $2+1$ or $2+1+1$ flavours of quarks. A weighted average of the pre-ranges gives 0.11823(57), using the smallest pre-range gives 0.11823(81) and the average size of ranges as an error gives 0.11823(128).

# Strong coupling constant from moments of quarkonium correlators

Peter Petreczky

*Physics Department, Brookhaven National Laboratory, Upton (NY)*

**Abstract:** I discuss recent progress and challenges in determining $\alpha_\mathrm{s}$ from moments of quarkonium correlators.

The strong coupling constant can be determined using the moments of quarkonium correlators. On the lattice the moments of pseudoscalar quarkonium correlators are the most practical ones, since these have the smallest statistical errors. The moments of the pseudoscalar quarkonium correlator, are defined as

$$G_n = \sum_t t^n G(t), \ G(t) = a^6 \sum_{\mathbf{x}} (am_{h0})^2 \langle j_5(\mathbf{x},t) j_5(0,0) \rangle. \tag{1}$$

Here $j_5 = \bar\psi \gamma_5 \psi$ is the pseudoscalar current, $a$ is the lattice spacing, and $m_{h0}$ is the bare lattice heavy quark mass. The moments $G_n$ are finite for $n \geq 4$ ($n$ even) in the $a \to 0$ limit and do not need renormalization because the explicit factors of the quark mass. The moments can be calculated in perturbation theory in $\overline{\mathrm{MS}}$ scheme

$$G_n = \frac{g_n(\alpha_\mathrm{s}(\mu), \mu/m_h)}{a m_h^{n-4}(\mu_m)}. \tag{2}$$

Here $\mu$ is the $\overline{\mathrm{MS}}$ renormalization scale, and $m_h(\mu_m)$ is the renormalized heavy quark mass in the $\overline{\mathrm{MS}}$ scheme. The scale $\mu_m$ at which the $\overline{\mathrm{MS}}$ heavy quark mass is defined can be different from $\mu$ [1], though most studies assume $\mu_m = \mu$. The coefficient $g_n(\alpha_\mathrm{s}(\mu), \mu/m_h)$ is calculated up to 4-loop, i.e. up to order $\alpha_\mathrm{s}^3$ [2]–[3]. For practical applications it is better to consider the reduced moments

$$R_n = \begin{cases} G_n/G_n^{(0)} & (n=4) \\ \left(G_n/G_n^{(0)}\right)^{1/(n-4)} & (n \geq 6) \end{cases}, \tag{3}$$

where $G_n^{(0)}$ is the moment calculated from the free lattice correlation function, since the leading order lattice artifacts cancel out in this ratio, and thus the cutoff effects in $R_n$ are proportional to $\alpha_\mathrm{s}^m a^{2n}$, $m \geq 1$, $n \geq 1$. It is straightforward to write down the perturbative expansion for $R_n$:

$$R_n = \begin{cases} r_4 & (n=4) \\ r_n \cdot (m_{h0}/m_h(\mu)) & (n \geq 6) \end{cases}, \tag{4}$$

$$r_n = 1 + \sum_{j=1}^{3} r_{nj}(\mu/m_h) \left(\frac{\alpha_\mathrm{s}(\mu)}{\pi}\right)^j. \tag{5}$$

There is also a contribution to the moments of quarkonium correlators from the gluon condensate [4]. From the above equations it is clear that $R_4$ as well as the ratios $R_6/R_8$ and $R_8/R_{10}$ are suitable for the extraction of the strong coupling constant $\alpha_\mathrm{s}(\mu)$. The calculation of $\alpha_\mathrm{s}$ in lattice QCD using the moments of quarkonium correlators was pioneered in Ref. [5] and now is pursued by several groups [5]–[10]. Here I will discussed this approach using the newest lattice results based on the calculations in 3-flavor QCD with Highly Improved Staggered Quark (HISQ) action and several heavy quark masses $m_h = m_c$, $1.5m_c$, $2m_c$ and $3m_c$ with $m_c$ being the charm quark mass [10].

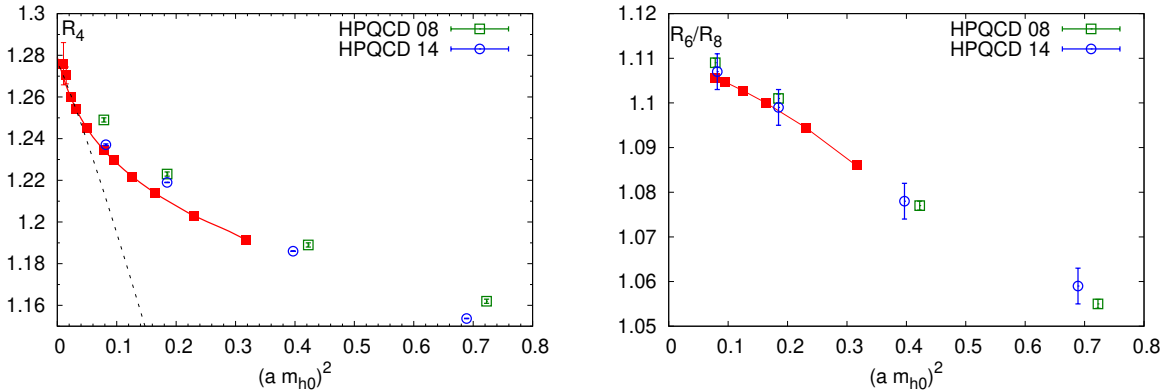

Figure 1: The lattice spacing dependence of $R_4$ and $R_6/R_8$ for $m_h = m_c$. The filled symbols correspond to the lattice results of Ref. [10], while the open symbols correspond to HPQCD results from Refs. [5,7]. The solid line corresponds to polynomial fit, see text. The dashed line corresponds to simple $a^2$ fit. The errors for the HPQCD-14 result for $R_6/R_8$ have been obtained by propagating the errors on $R_6$ and $R_8$.

One of the challenges for accurate determination of the strong coupling constant from the moments of quarkonium correlators is a reliable continuum ($a \to 0$) extrapolation. There is also a window problem. We would like to work with the large value of $m_h$ for perturbation theory to be reliable, at the same time to control the cutoff effects which grow with increasing $m_h$. So, one has to find a window, where $m_h/\Lambda_{QCD} \gg 1$ and $am_h \ll 1$. This problem is not specific to the moments method but is present in all lattice methods of $\alpha_s$ determination, except for the Schrödinger functional method (see discussions in the new FLAG report [11]).

To illustrate the challenge of continuum extrapolation of the moments in Fig. 1, I show the cutoff dependence of $R_4$ and $R_6/R_8$ together with continuum extrapolations. One can see that the cutoff effects is significant and simple $a^2$ extrapolations only work for the smallest three lattice spacings, for details see Ref. [10].

If one has data only at large lattice spacings, the continuum limit for $R_4$ can be easily underestimated, while the continuum limit for $R_6/R_8$ can be easily overestimated. One way to check for correctness of continuum extrapolations is to compare the results obtained for $\alpha_s$ using $R_4$ and $R_6/R_8$. The details of continuum extrapolations are discussed in Ref. [10]. Despite the difficulties of the continuum extrapolations of the moments, the final continuum results obtained in different lattice calculations seem to agree reasonably well, see discussions in Refs. [10]–[11].

From the continuum extrapolated value of $R_4$ or ratios $R_6/R_8$ and $R_8/R_{10}$, the value of $\alpha_s(\mu)$ can be obtained at scales comparable to the heavy quark mass (so that there are no large logarithms). The results for $\alpha_s(\mu = m_h)$ from Ref. [10] are shown in Fig. 2 and Table 1. In Fig. 2, I also compare the results from different lattice determinations. It is clear that performing lattice calculations at different values of the quark mass allows one to map out the running of the coupling constant at relatively low energy scales. It also helps to control the systematic errors of the weak coupling expansion. The running coupling constant extracted from moments of quarkonium correlators in Ref. [10] agrees with the result obtained from the static quark anti-quark energy [12] but is lower than the values of $\alpha_s$ obtained by HPQCD collaboration from the moments of quarkonium correlators. Since the continuum extrapolated lattice results on the moments and their ratios are in a reasonably good agreement with each other the source of this discrepancy must be related to the way comparison of the lattice and weak coupling results is performed. In Refs. [10] $\mu = m_h$, while in HPQCD studies $\mu = 3m_h$.

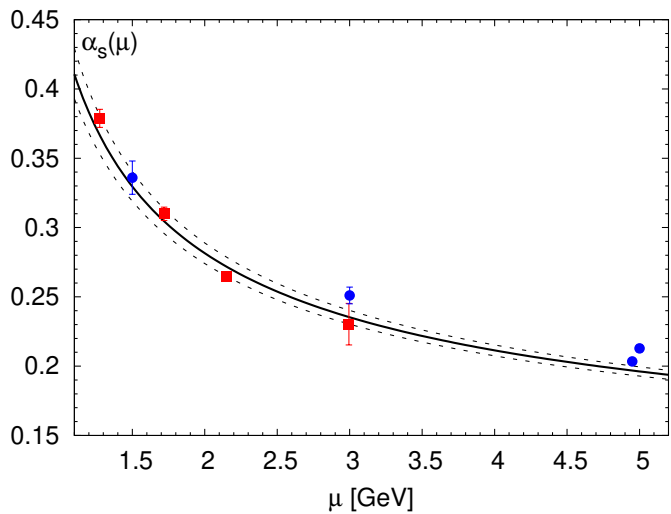

Figure 2: The running coupling in three-flavor QCD constant corresponding to $\Lambda_{\overline{\mathrm{MS}}}^{n_f=3} = 301(16)$ MeV. The solid line corresponds to the central value, while the dashed lines show the error band. The blue circles from left to right correspond to the determination of $\alpha_\mathrm{s}$ for the static quark anti-quark energy [12] and from the moments of quarkonium correlators [5]–[7]. The result of Ref. [6] has been shifted horizontally for better visibility.

Table 1: The values of $\alpha_\mathrm{s}(\mu = m_h)$ for different heavy quark masses, $m_h$, extracted from $R_4$, $R_6/R_8$, and $R_8/R_{10}$. The heavy quark mass is given in units of $m_c$. The first, second, and third errors correspond to the lattice, perturbative truncation, and the error due to the gluon condensate. The fifth column lists the averaged value of $\alpha_\mathrm{s}$. The last column gives the value of $\Lambda_{\overline{\mathrm{MS}}}^{n_f=3}$ in MeV.

| $m_h$ | $R_4$ | $R_6/R_8$ | $R_8/R_{10}$ | average | $\Lambda_{\overline{\mathrm{MS}}}^{n_f=3}$ |
|---|---|---|---|---|---|
| 1.0 | 0.3815(55)(30)(22) | 0.3837(25)(180)(40) | 0.3550(63)(140)(88) | 0.3788(65) | 315(9) |
| 1.5 | 0.3119(28)(4)(4) | 0.3073(42)(63)(7) | 0.2954(75)(60)(17) | 0.3099(48) | 311(10) |
| 2.0 | 0.2651(28)(7)(1) | 0.2689(26)(35)(2) | 0.2587(37)(34)(6) | 0.2649(29) | 285(8) |
| 3.0 | 0.2155(83)(3)(1) | 0.2338(35)(19)(1) | 0.2215(367)(17)(1) | 0.2303(150) | 284(48) |

From the values of $\alpha_\mathrm{s}(\mu = m_h)$ one can extract the 3-flavor $\Lambda$-parameter, $\Lambda_{\overline{\mathrm{MS}}}^{n_f=3}$, which is given in the last column of Table 1. If the perturbative errors are under control, the value of $\Lambda_{\overline{\mathrm{MS}}}^{n_f=3}$ obtained from lattice results at different values of the heavy quark $m_h$ should agree. Table 1, however, shows that there is a tension between $\Lambda_{\overline{\mathrm{MS}}}^{n_f=3}$ obtained for $m_h = 2m_c$ and the values obtained at smaller quark mass. Performing a weighted average of the $\Lambda_{\overline{\mathrm{MS}}}^{n_f=3}$ values in Table 1, I get $\Lambda_{\overline{\mathrm{MS}}}^{n_f=3} = 301 \pm 16$ MeV, where the assigned error reflects the spread of the results in Table 1. This value of the $\Lambda$-parameter corresponds to $\alpha_\mathrm{s}(m_Z, n_f = 5) = 0.1161(12)$, which is about two sigma lower than the most recent result from HPQCD [7], but is in good agreement with the previous determination using the moments of charmonium correlator in 3-flavor QCD [8]. The analysis of Ref. [8] was criticized by the new FLAG report arguing that the perturbative uncertainties have been underestimated and for that reason was given a red symbol for the perturbative behavior [11]. The main argument of this criticism is the fact that $\mu = m_c$ is a low scale and that using higher renormalization scales $\mu = s m_c$, $s > 1$ leads to larger values of $\alpha_s$. While the raised point is certainly valid, the problems with perturbation theory is not specific to the analysis of Ref. [8] and should affect other determinations of $\alpha_s$ from the moments as well. In particular, if $\mu \neq m_h$ other choices of $\mu_m$ need to be considered and varying $\mu$ and $\mu_m$ independently will lead to much larger perturbative error [1].

In summary, the determination of $\alpha_s$ from the moments of quarkonium correlators, while promising also appears to be challenging. One of the challenge is the control of the continuum extrapolations, which requires many calculations at small lattice spacings. So far this requirement is only met in the 3-flavor calculations with HISQ action [10]. Despite this, there seems to be an agreement between the continuum extrapolated lattice results on the moments of the quarkonium correlators from different groups. This implies that differences in the quoted $\alpha_s$ values are not caused by problems in the lattice calculations, but rather the way lattice and perturbative calculations are combined to obtain $\alpha_s$. It should be noted that the moments of the quarkonium correlators can be used to extract also the values of the heavy quark masses, and different lattice results agree quite well, see discussion in Ref. [10].

# $\alpha_\text{s}$ from the ALPHA collaboration (part I)

Stefan Sint

*School of Mathematics and Hamilton Mathematics Institute, Hamilton building, Trinity College Dublin, Dublin 2, Ireland*

**Abstract:** The recent determination of $\alpha_\text{s}(m_\text{z}) = 0.11852(84)$ by the ALPHA collaboration [1] distinguishes itself by the very good control of perturbative truncation and other systematic errors. A variety of tools and methods had to be deployed to enable this result. In this contribution I will give a short account of the step-scaling method and its application to QCD couplings in finite volume renormalization schemes. Tracing the running couplings non-perturbatively between scales $\mu_0$ and $32\mu_0$ (corresponding roughly to the range 4–128 GeV) leads to the intermediate result $\Lambda^{(3)}_{\overline{\text{MS}}}/\mu_0 = 0.0791(19)$ in 3-flavour QCD. By computing this ratio in variety of ways, using perturbation theory in different schemes and at different energy scales at intermediate stages, gives us confidence in the error estimate and also enables a number of useful tests of perturbation theory. The remaining steps required for $\alpha_\text{s}(m_\text{z})$ will be discussed by Mattia Dalla Brida in these proceedings [2].

## Introduction

The recent result for $\alpha_\text{s}$ by the ALPHA collaboration relies on the combination of various tools and techniques that have been developed and improved over the last 20–30 years. A crucial ingredient is the recursive step-scaling method [3] applied to QCD couplings renormalized in a finite Euclidean space-time volume. This allows us to overcome the typical limitation of lattice QCD, whereby large scale differences cannot be resolved on a single lattice without incurring large computational costs [4]. As will become clear in this and in Mattia Dalla Brida's companion contribution [2], we have covered a range of energy scales differing by 2–3 orders of magnitude, thus connecting hadronic scales of $\mathcal{O}(100)$ MeV with electroweak scales of $\mathcal{O}(100)$ GeV. The scale evolution of QCD couplings in so-called Schrödinger functional (SF) schemes is obtained non-perturbatively and in the continuum limit. Given the good perturbative knowledge for the SF schemes one may assess at which scale perturbative behaviour sets in and extract the $\Lambda$ parameter. In this way, the systematic error due to the truncation of the perturbative series can be well-controlled and kept at a level that remains subdominant compared to current statistical errors. This is in contrast to many other lattice determinations of $\alpha_\text{s}$ where perturbative uncertainties arise at much lower energy scales and are thus much harder to quantify.

We remark that all our simulations are carried out for 3-flavour QCD. Therefore the result for $\alpha_\text{s}(m_\text{z})$ in 5-flavour QCD also relies on decoupling relations across charm and bottom quark thresholds; I refer to [2] for references and a discussion. The ALPHA collaboration's strategy involves two different finite volume renormalization schemes for the 3-flavour QCD coupling. At low energies, a coupling based on the gradient flow (GF) has advantageous properties (cf. [2]). The high energy regime is covered using a 1-parameter family of SF couplings, for which the 2-loop matching to the $\overline{\text{MS}}$-coupling and the 3-loop $\beta$-functions are known [5]–[8]. Our strategy then requires a matching between the GF and SF couplings at an intermediate scale, $\mu_0$, which is implicitly defined by the SF coupling and turns out to be around 4 GeV in physical units.

In the following, I will briefly review the step-scaling method and illustrate it with our results for the SF coupling. The exactly known scheme dependence of the $\Lambda$ parameter makes it a useful reference quantity, which enables various tests of perturbation theory. The main intermediate outcome of this first part is $\Lambda_{\overline{\text{MS}}}^{(3)}/\mu_0 = 0.0791(19)$, which defines the starting point for Mattia Dalla Brida's contribution [2].

## Non-perturbatively defined QCD couplings and the $\Lambda$ parameter

Let us assume we have an observable*, $\langle O \rangle$, with a finite continuum limit and also possessing a perturbative expansion starting with $g^2$. We will assume throughout that all three light quark masses are set to zero. If the Euclidean time and space extents are given by $L$ and all dimensionful parameters, such as momenta, distances, or background fields are taken in a fixed proportion to $L$ then the observable depends on a single scale $\mu = 1/L$ and we may define† $\bar{g}^2(L) = \langle O \rangle$. Examples for such finite volume couplings are the GF coupling discussed in [2] and the family of SF couplings introduced in [9]–[11], which derive from the QCD SF [12,13]. For details we refer to [14]. Physically, the SF$\nu$ couplings are response coefficients to the variation of an Abelian colour electric background field. The dependence on the parameter $\nu$ takes the simple form

$$\frac{1}{\bar{g}_\nu^2} = \frac{1}{\bar{g}^2} - \nu \bar{v}, \qquad (1)$$

in terms of two correlation functions $1/\bar{g}^2 = \langle O_1 \rangle$ and $\bar{v} = \langle O_2 \rangle$, measured in a simulation at $\nu = 0$. Given such a coupling, its $\beta$-function $\beta(\bar{g}) = -L \partial \bar{g}/\partial L$ is non-perturbatively defined too. Yet it has the usual weak coupling expansion $\beta(g) = -b_0 g^3 - b_1 g^5 + \ldots$ with the universal coefficients $b_0 = 9/16\pi^2$ and $b_1 = 1/4\pi^4$ (for 3-flavour QCD). Hence also the associated $\Lambda$ parameter, given as an exact solution of the Callan–Symanzik equation, is non-perturbatively defined. Indicating the dependence on the scheme 'x' by a subscript, it takes the form

$$\Lambda_{\text{x}} = L^{-1} \varphi_{\text{x}}(\bar{g}_{\text{x}}(L)), \qquad (2)$$

with

$$\varphi_{\text{x}}(\bar{g}) = (b_0 \bar{g}^2)^{-b_1/(2b_0^2)} e^{-1/(2b_0 \bar{g}^2)} \times \exp\left\{-\int_0^{\bar{g}} dg \left[\frac{1}{\beta_{\text{x}}(g)} + \frac{1}{b_0 g^3} - \frac{b_1}{b_0^2 g}\right]\right\}. \qquad (3)$$

Its behaviour under a change from scheme x to y is *exactly* determined by the one-loop coefficient relating the respective couplings, i.e. if $g_{\text{x}}^2 = g_{\text{y}}^2 + c_{\text{xy}} g_{\text{y}}^4 + \ldots$ then $\Lambda_{\text{x}}/\Lambda_{\text{y}} = \exp(c_{\text{xy}}/2b_0)$. Thus the relations between $\Lambda$ parameters for all SF$_\nu$ schemes and the $\overline{\text{MS}}$ scheme are known. Note that $\Lambda_{\overline{\text{MS}}}$ is thus indirectly defined beyond perturbation theory, even though the $\overline{\text{MS}}$ scheme is otherwise only perturbative. Furthermore, also the 2-loop relations between the respective couplings are known and thus the 3-loop coefficients $b_{\nu,2}$ for SF$_\nu$ schemes can be inferred. Numerical values with parameter $\nu = \text{O}(1)$ seem reasonable from a perturbative viewpoint [14].

---

*In this context, an observable is given as a correlation function of gauge invariant fields defined with the Euclidean (lattice) QCD path integral. These are the quantities estimated in a numerical simulation of lattice QCD.

†To denote the scale dependence we use the convention $\bar{g}^2 = \bar{g}^2(L)$ and $\alpha(\mu = 1/L) = \bar{g}^2(L)/(4\pi)$.

## Step-scaling

Given a QCD coupling in a mass-independent finite volume renormalization scheme, its step-scaling function (SSF) is defined by,

$$\sigma(u) = \bar{g}^2(2L)|_{u=\bar{g}^2(L)}, \tag{4}$$

and thus yields the coupling at $2L$ given the coupling at $L$. In other words it determines the coupling if the scale is changed by a step factor 2 and is related to an integral of the $\beta$-function,

$$\int_{\sqrt{u}}^{\sqrt{\sigma(u)}} \frac{\mathrm{d}g}{\beta(g)} = -\ln 2, \tag{5}$$

For a fixed argument $u$, the SSF can be obtained as the continuum limit of lattice approximants,

$$\sigma(u) = \lim_{a \to 0} \Sigma(u, a/L), \tag{6}$$

where a lattice approximant $\Sigma(u, a/L)$ requires the measurements on pairs of lattices with linear extents $L/a$ and $2L/a$. To keep the lattice spacing $a$ fixed, one uses the same bare lattice coupling, $g_0^2$, for each pair. In principle, keeping $u$ fixed is achieved by tuning the bare coupling $g_0^2$ such that $\bar{g}^2(L) = u$ on an $L/a$-lattice. In practice, however, it is more convenient to produce data for the function $\Sigma(u, a/L)$ at various values of its arguments and then perform a global fit of the form

$$\Sigma(u, a/L) = \sigma(u) + \rho(u) (a/L)^2, \tag{7}$$

where both $\sigma(u)$ and $\rho(u)$ are polynomials in $u$ [14]. A typical parameterization for $\sigma(u)$ is given by

$$\sigma(u) = u + s_0 u^2 + s_1 u^3 + s_2 u^4 + c_1 u^5, \tag{8}$$

where $c_1$ is a fit parameter and $s_{0,1,2}$ are fixed to their perturbative values in terms of $b_{0,1,2}$. The non-perturbatively defined function $\sigma(u)$ is then represented by the fit function for $u$ in some interval $[u_{\min}, u_{\max}]$, cf. Fig. 1.

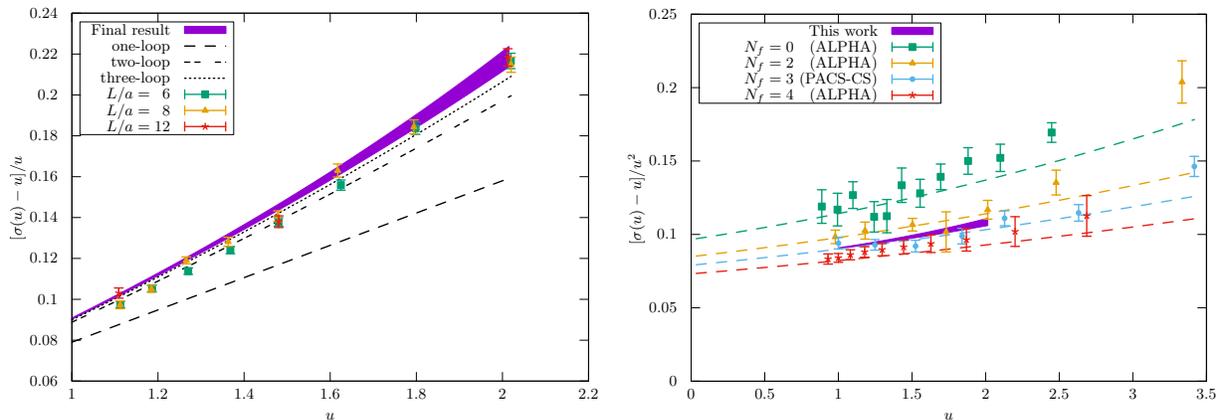

Figure 1: Left: Continuum extrapolation of the lattice data for $\Sigma(u, a/L)$ yielding $\sigma(u)$ for a range of $u$-values with errors indicated by the blue band. Right: Comparison with earlier studies in QCD with $N_\mathrm{f} = 0, 2, 3, 4$ flavours [15]–[18] illustrating the reduced errors in our new $N_\mathrm{f} = 3$ data [14].

Given $\sigma(u)$ one may define the largest coupling $u_{\max} = u_0 = \bar{g}^2(L_0)$ and then recursively step up the energy scale by factors of 2, i.e.

$$u_n = \sigma(u_{n+1}), \qquad u_n = \bar{g}^2(L_n), \qquad L_n = L_0/2^n, \tag{9}$$

until one reaches the smallest coupling still covered by the data[‡]. In our case we set $\bar{g}^2(L_0) = 2.012$ for the SF scheme at our default choice $\nu = 0$, and this implicitly defines the scale $\mu_0 = 1/L_0$. The data shown in Fig. 1 then allows us to make up to $n = 5$ steps from $L_0$, reaching energy scales $\mu_n = 1/L_n$ up to $\mu_5 = 32\mu_0$. In order to do the same steps for any other value of $\nu$ one needs $\bar{v}(L_0) = 0.1199(10)$ to define the start value, $\bar{g}_\nu^2(L_0)$, for the recursion (cf. [14] for details).

## Tests of perturbation theory and extraction of $\Lambda_{\overline{\text{MS}}}$

Taking the $\Lambda$ parameter in the SF scheme with $\nu = 0$ as our reference quantity we can now obtain it in a variety of ways

$$\Lambda L_0 \equiv \Lambda_{\text{SF}_{\nu=0}} L_0 = (\Lambda/\Lambda_{\text{SF}_\nu}) \times 2^n \varphi_{\text{SF}_\nu}\left(\bar{g}_\nu^2(L_n)\right). \tag{10}$$

Obviously, the LHS of this equation must always be the same up to the perturbative approximation to the integral in the exponent of Eq. (3), which reads

$$\int_0^{\bar{g}} dg \left[\frac{1}{\beta_{\text{x}}(g)} + \frac{1}{b_0 g^3} - \frac{b_1}{b_0^2 g}\right] = \frac{b_{\text{x},2} b_0 - b_1^2}{2 b_0^3} \bar{g}^2 + \mathrm{O}(\bar{g}^4). \tag{11}$$

Hence, given that the 3-loop coefficient, $b_{\text{x},2}$, is known for all the SF$_\nu$ schemes, we have a parametric uncertainty of $\mathcal{O}(\alpha^2)$ (with $\alpha = \bar{g}^2/4\pi$) for this integral and thus for $\Lambda$. Obviously, the higher the scale $\mu_n = 1/L_n$, the smaller this uncertainty should become. We test this by evaluating the RHS of Eq. (10) for different values of $\nu$ and $n$, cf. Fig. 2. As expected all points come together as $\alpha$ decreases. We also observe a roughly linear behaviour in $\alpha^2$, as expected from Eq. (11). However, the slope for $\nu = -0.5$ seems rather large, whereas it almost vanishes for $\nu = 0.3$. Our final result, shown as grey band in Fig. 2, is extracted at scales reached after $n = 4$ steps (i.e. around 70 GeV),

$$\Lambda/\mu_0 = 0.0303(7) \quad \Rightarrow \quad \Lambda_{\overline{\text{MS}}}/\mu_0 = 0.0791(19). \tag{12}$$

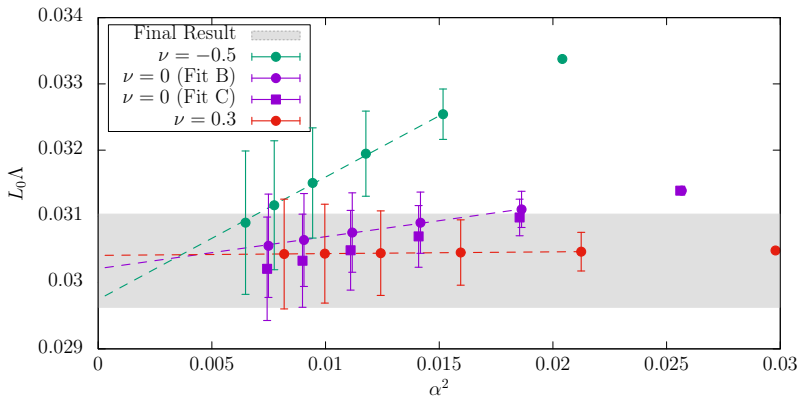

Figure 2: Estimates for $\Lambda L_0$ as a function of the parametric uncertainty $\alpha^2$. "Fit A" and "Fit B" correspond to two different fit functions for $\sigma(u)$, cf. [14] for details.

---

[‡]Note that evolving towards higher energies requires to invert the step-scaling function. This poses no practical problems.

A further test can be performed by first converting the couplings to the $\overline{\text{MS}}$ coupling at 2-loop order, and then extracting the $\Lambda$-parameter within the $\overline{\text{MS}}$ scheme using the $\beta$-function up to 5-loop order [19]–[23]. In the conversion between the couplings we allow for a scale factor, $s$,

$$\begin{aligned}
\Lambda_{\overline{\text{MS}}} L_0 &= s\frac{L_0}{L_n}\varphi_{\overline{\text{MS}}}\left(\bar{g}_{\overline{\text{MS}}}(L_n/s)\right) \\
&= s\, 2^n \varphi_{\overline{\text{MS}}}\left(\sqrt{\bar{g}_\nu^2(L_n) + p_1^\nu(s)\bar{g}_\nu^4(L_n) + p_2^\nu(s)\bar{g}_\nu^6(L_n) + \text{O}\left[\bar{g}_\nu^8(L_n)\right]}\right),
\end{aligned} \quad (13)$$

and the result must be independent of $s$, $\nu$ and $n$. As our best value of $s$ we choose $s = s^*$ such that the one loop coefficient $p_1^\nu(s) \approx 0$, which determines $s^*$ as the ratio of the corresponding $\Lambda$ parameters. We then vary $s$ in the interval $[s^*/2, 2s^*]$, in order to obtain a measure for the uncertainty from neglected higher order terms. This estimate can then be compared with the true deviation from $\Lambda_{\overline{\text{MS}}}/\mu_0$, Eq. (12).

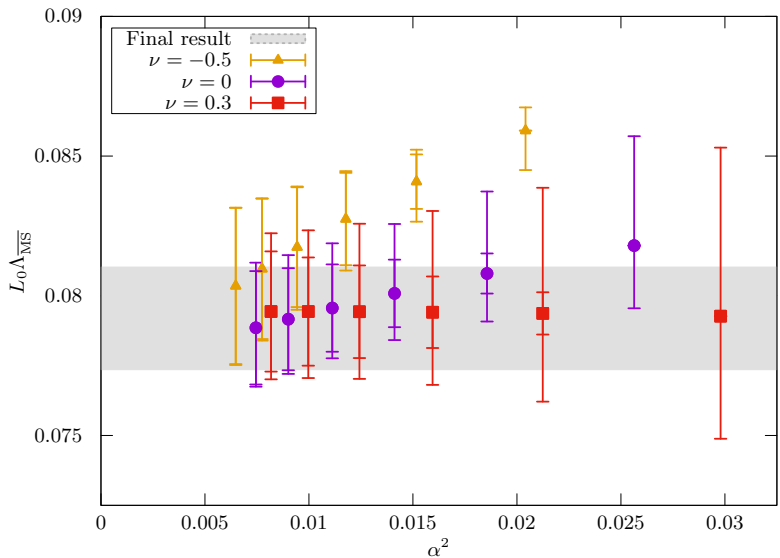

Figure 3: Statistical (interior error band) and total (exterior error band) uncertainties in the determination of $L_0 \Lambda_{\overline{\text{MS}}}$. The total error is the combination in quadrature of the statistical and systematic error, where the latter is obtained by varying $s$ between $s^*/2$ and $2s^*$. The grey band is our final estimate, Eq. (12).

## Conclusion

We have studied the non-perturbative scale evolution of for a 1-parameter family of SF couplings for energies between roughly 4 and 128 GeV. We conclude that one needs to reach $\alpha \approx 0.1$ in order to confidently extract the $\Lambda$ parameter with an error below 3%. In a further consistency check we first converted the SF to the $\overline{\text{MS}}$ coupling and then varied the relative scale within a factor of two either way around a preferred choice. We note that this common recipe may nor may not capture the true perturbative uncertainty. This reinforces the general warning that perturbative truncation errors are easily underestimated.

# $\alpha_\mathrm{s}$ from the ALPHA collaboration (part II)

Mattia Dalla Brida

*Dipartimento di Fisica, Università di Milano-Bicocca, and INFN, Sezione di Milano-Bicocca, 20126 Milan, Italy*

**Abstract:** In this second part we continue the overview of the recent lattice determination of $\alpha_\mathrm{s}$ by the ALPHA collaboration. Starting from the result for $\Lambda_{\overline{\mathrm{MS}}}^{N_\mathrm{f}=3}/\mu_0$ discussed in the first part [1], we first present a precise non-perturbative determination of the $\Lambda$-parameter of $N_\mathrm{f} = 3$ QCD. Using perturbative decoupling to match the $N_\mathrm{f} = 3$ and $N_\mathrm{f} = 5$ theories we then extract a precise value for $\alpha_\mathrm{s}$. The final result: $\alpha_\mathrm{s}(m_\mathrm{Z}) = 0.11852(84)$, reaches subpercent accuracy.

## Introduction

The extraction of $\alpha_\mathrm{s}$ we present is based on the determination of $\Lambda_{\overline{\mathrm{MS}}}^{N_\mathrm{f}=5}$, the $\Lambda$-parameter of $N_\mathrm{f} = 5$ flavour QCD in the $\overline{\mathrm{MS}}$ scheme. The latter is obtained from a non-perturbative determination of $\Lambda_{\overline{\mathrm{MS}}}^{N_\mathrm{f}=3}$, combined with a perturbative estimate for the ratio $\Lambda_{\overline{\mathrm{MS}}}^{N_\mathrm{f}=5}/\Lambda_{\overline{\mathrm{MS}}}^{N_\mathrm{f}=3}$. Our strategy can be summarized into the following equation [2]:

$$\Lambda_{\overline{\mathrm{MS}}}^{N_\mathrm{f}=5} = \left[\frac{\Lambda_{\overline{\mathrm{MS}}}^{N_\mathrm{f}=5}}{\Lambda_{\overline{\mathrm{MS}}}^{N_\mathrm{f}=3}}\right]_\mathrm{PT} \times \Lambda_{\overline{\mathrm{MS}}}^{N_\mathrm{f}=3} \quad \text{where} \quad \Lambda_{\overline{\mathrm{MS}}}^{N_\mathrm{f}=3} = \frac{\Lambda_{\overline{\mathrm{MS}}}^{N_\mathrm{f}=3}}{\mu_0} \times \frac{\mu_0}{\mu_\mathrm{had}} \times \frac{\mu_\mathrm{had}}{f_{\pi K}} \times f_{\pi K}. \qquad (1)$$

In the rest of this contribution, we will briefly review the computation of the different factors entering this expression. For a more complete discussion, we refer the reader to the original reference [2], and to the more extended reviews [3,4,5].

We begin our presentation from the non-perturbative determination of $\Lambda_{\overline{\mathrm{MS}}}^{N_\mathrm{f}=3}$ and the different ratios that compose it. The first ingredient appearing in Eq. (1) is the value of $\Lambda_{\overline{\mathrm{MS}}}^{N_\mathrm{f}=3}$ in units of the technical scale $\mu_0$. This computation is discussed in detail in the first part of this overview [1], which we advise the reader to consult. Here we only quote the final result: $\Lambda_{\overline{\mathrm{MS}}}^{N_\mathrm{f}=3}/\mu_0 = 0.0791(19)$ [6,7], and recall that the scale $\mu_0 \approx 4\,\mathrm{GeV}$ is implicitly defined by the value of the Schrödinger functional (SF) coupling: $\bar{g}_\mathrm{SF}^2(\mu_0) = 2.012$. It is also worth recalling that this ratio has been obtained by studying the non-perturbative running of the SF coupling in the wide energy range $\mu \approx 4-70\,\mathrm{GeV}$. With this result at hand, the value of $\Lambda_{\overline{\mathrm{MS}}}^{N_\mathrm{f}=3}$ in physical units can be obtained by expressing the technical scale $\mu_0$ in terms of some experimentally accessible quantity. We consider a particular combination of the pion and kaon decay constants, $f_\pi$ and $f_K$, given by: $f_{\pi K} = \frac{2}{3}(f_K + \frac{1}{2}f_\pi)$; the reasons for this particular choice will be given later in the text (see Sect. 2). Meson decay constants are typically used to set the physical scale of the lattice theory as they can be accurately determined both phenomenologically and on the lattice.* A direct computation of $\mu_0/f_{\pi K}$, on the other hand, is not really feasible if one wants the systematic uncertainties associated with finite-volume and discretization effects comfortably under control. The large energy separation between $\mu_0$ and $f_{\pi K} = \mathcal{O}(100\,\mathrm{MeV})$ would indeed require us to simulate rather large lattice resolutions, $L/a$, for today's standards; here and in the following we denote by $L$ the physical extent of the lattice in all four space-time directions and by $a$ its spacing. The solution to this problem is to rely, as we did for the determination of $\Lambda_{\overline{\mathrm{MS}}}^{N_\mathrm{f}=3}/\mu_0$, on a *step-scaling strategy* (cf. Ref. [1]). More precisely, by studying the non-perturbative running of a finite-volume coupling, we can relate the scale $\mu_0$ to a lower, finite-volume scale, $\mu_\mathrm{had} = \mathcal{O}(100\,\mathrm{MeV})$, and in a second step connect $\mu_\mathrm{had}$ with $f_{\pi K}$ (cf. Eq. (1)).

---

*A more natural and conceptually clean quantity to consider would be the proton mass. (The masses of the QCD stable mesons are normally used to fix the value of the bare quark masses appearing in the lattice Lagrangian.) The extraction of the decay constants from experimental decay rates is indeed not theoretically straightforward and also relies on the knowledge of CKM matrix elements. Measuring the proton mass precisely on the lattice, however, is at present very challenging.

## The gradient flow coupling and its running to low energy

The obvious strategy we could follow at this point would be to continue the non-perturbative running of the SF coupling started at high-energy down to lower energies. On the other hand, a precise determination of the running of the SF coupling at low energy is impeded by a few technical reasons (see e.g. refs. [6,8]). The main issue is that the statistical variance of the SF coupling as measured in Monte Carlo lattice simulations is such that: $\mathrm{var}(\bar{g}_\mathrm{SF}^2(\mu))/\bar{g}_\mathrm{SF}^4(\mu) = c(a\mu)\,\bar{g}_\mathrm{SF}^4(\mu) + \mathcal{O}(\bar{g}_\mathrm{SF}^6(\mu))$. This implies that it quickly becomes computationally expensive to measure this coupling precisely at low energy where the coupling becomes large. In addition, $\mathrm{var}(\bar{g}_\mathrm{SF}^2)$ is large in general, and increases as the continuum limit of the lattice theory is approached due to: $c(a\mu) \stackrel{a\to 0}{\propto} (a\mu)^{-1}$. For these reasons, it is more convenient to consider a different family of finite-volume couplings for the low-energy end of the running. A particularly compelling family to study is given by couplings defined in terms of the Yang–Mills gradient flow (GF) [9]. The latter is specified by the equations:

$$\partial_t B_\mu(t,x) = D_\nu G_{\nu\mu}(t,x), \qquad G_{\mu\nu} = \partial_\mu B_\nu - \partial_\nu B_\mu + [B_\mu, B_\nu],$$
$$B_\mu(0,x) = A_\mu(x), \qquad D_\mu = \partial_\mu + [B_\mu, \cdot], \qquad (2)$$

where $A_\mu$ is the QCD gauge potential, and $t \geq 0$ is the *flow time* which parametrizes the evolution of the flow field $B_\mu$ along the gradient flow. Gauge invariant fields made out of the flow field $B_\mu$ have the remarkable property of being renormalized once the bare parameters of the theory are [10]. This allows us to define a finite-volume GF coupling as [11,12]:

$$\bar{g}_\mathrm{GF}^2(\mu) = \mathcal{N}^{-1} t^2 \langle E_\mathrm{sp}(t,x)\rangle_\mathrm{SF}\big|_{x_0=L/2}^{\sqrt{8t}=0.3\times L}, \quad E_\mathrm{sp}(t,x) = \frac{1}{4} G_{kl}^a(t,x) G_{kl}^a(t,x), \quad \mu = L^{-1}, \qquad (3)$$

where $\langle \cdot \rangle_\mathrm{SF}$ stands for the (Euclidean) path-integral expectation value in the presence of SF boundary conditions and $\mathcal{N}$ is a constant; we refer the reader to the given references for more details. Here we just note that in order for the GF coupling to depend on a single scale, $L$, we express the flow time $t$ in terms of $L$ through the condition $\sqrt{8t}/L = 0.3$. The nice property of the GF coupling is that $\mathrm{var}(\bar{g}_\mathrm{GF}^2)$ is finite as $a \to 0$, and typically small. In addition, in first approximation, one has that: $\mathrm{var}(\bar{g}_\mathrm{GF}^2)/\bar{g}_\mathrm{GF}^4 \propto \mathrm{const.}$, which, as anticipated, makes this coupling well-suited for low-energy studies.

In order to start computing the running of the GF coupling to low energy, we first need to know its value at the reference scale $\mu_0$. This can be obtained through a non-perturbative matching of the SF and GF couplings. The latter is easily achieved by measuring the two couplings for the very same set of bare lattice parameters for which $\bar{g}_\mathrm{SF}^2(\mu_0) = 2.012$. Combining this matching with a change of scale by a factor of 2, we obtain: $\bar{g}_\mathrm{GF}^2(\mu_0/2) = 2.6723(64)$ [12]. The running to low energy can now proceed in similar fashion to the computation at high energy. In particular, we introduce the step-scaling function (SSF) of the GF coupling and its lattice approximant (cf. Ref. [1]):

$$\sigma(u) = \lim_{a/L \to 0} \Sigma(u, a/L), \qquad \Sigma(u, a/L) = \bar{g}_\mathrm{GF}^2(\mu/2)\big|_{u=\bar{g}_\mathrm{GF}^2(\mu),\,\overline{m}(\mu)=0}, \qquad \mu = L^{-1}. \qquad (4)$$

The SSF encodes the change in the coupling for a finite variation of the energy scale. On the lattice, it is thus a more natural quantity to consider than the $\beta$-function. Once the continuum SSF is known, however, the non-perturbative $\beta$-function can be determined by noticing that:

$$\ln\frac{\mu_2}{\mu_1} = \int_{\bar{g}_{\rm GF}(\mu_1)}^{\bar{g}_{\rm GF}(\mu_2)} \frac{{\rm d}g}{\beta(g)} \quad \Rightarrow \quad \log 2 = -\int_{\sqrt{u}}^{\sqrt{\sigma(u)}} \frac{{\rm d}g}{\beta(g)} \quad \text{where} \quad u = \bar{g}_{\rm GF}^2(\mu). \tag{5}$$

The left panel of Fig. 1 shows the continuum extrapolations of the lattice SSF for values of the GF coupling $\bar{g}_{\rm GF}^2 \approx 2-6.5$, and for the lattice resolutions, $L/a = 8, 12, 16$. As one can see from the figure, discretization errors are significant, particularly so at large values of the coupling (higher sets of points in the plot). Cautious continuum extrapolations are hence needed [12]. Nonetheless, the good statistical precision of the GF coupling allows us to obtain precise continuum results.

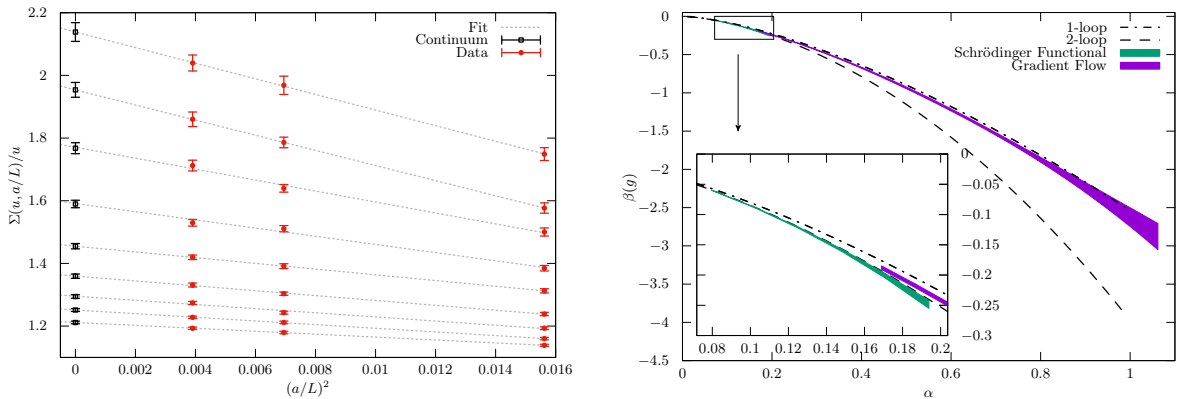

Figure 1: Left: Continuum extrapolations of the lattice SSF of $\bar{g}_{\rm GF}^2$. The lattice data is in red while the black points are the continuum extrapolated results (see Ref. [12] for more details). Right: Non-perturbative $\beta$-function of the GF coupling. For comparison the LO and NLO perturbative results are shown, as well as the results for the non-perturbative $\beta$-function of the SF coupling at high energy [12]. In this plot: $\alpha = g^2/(4\pi)$, with $g^2$ the coupling in the given scheme.

Using these results and Eq. (5) the non-perturbative $\beta$-function of the GF coupling can be computed; this is shown in the right panel of Fig. 1, together with the LO and NLO perturbative predictions, and the non-perturbative $\beta$-function in the SF scheme. It is interesting to observe the peculiar behaviour of the non-perturbative GF $\beta$-function which lies very close to the LO perturbative result even at large values of the coupling, where $\alpha \approx 1$. Note however that the deviation from LO perturbation theory is statistically significant for the most part of the coupling range [12]. Only at values of $\alpha \approx 0.2$ the non-perturbative results start to approach the NLO prediction.

Once the $\beta$-function is known, we can compute the ratio of any two scales associated with two values of the coupling (cf. Eq. (5)). If we define the technical scale $\mu_{\rm had}$ through the relatively large value of the GF coupling: $\bar{g}_{\rm GF}^2(\mu_{\rm had}) = 11.31$, integrating the non-perturbative $\beta$-function we find [12]:

$$\frac{\mu_0}{\mu_{\rm had}} = 21.86(42) \quad \Rightarrow \quad \frac{\Lambda_{\overline{\rm MS}}^{N_{\rm f}=3}}{\mu_{\rm had}} = 1.729(57). \tag{6}$$

# Matching to hadronic physics and $\Lambda_{\overline{\text{MS}}}^{N_{\text{f}}=3}$

Having bridged the gap between the high- and low-energy sectors of QCD, all that is left to do to determine $\Lambda_{\overline{\text{MS}}}^{N_{\text{f}}=3}$ is to relate the technical scale $\mu_{\text{had}}$ with some experimentally accessible quantity. Rather than establishing this relation directly, it is convenient to introduce an intermediate reference scale, $\mu_{\text{ref}}^*$, so that:

$$\Lambda_{\overline{\text{MS}}}^{N_{\text{f}}=3} = \frac{\Lambda_{\overline{\text{MS}}}^{N_{\text{f}}=3}}{\mu_{\text{had}}} \times \frac{\mu_{\text{had}}}{\mu_{\text{ref}}^*} \times \frac{\mu_{\text{ref}}^*}{f_{\pi K}} \times f_{\pi K}. \qquad (7)$$

For the scale $\mu_{\text{ref}}^*$ we must choose a quantity that can be measured very precisely and easily in lattice simulations. The problem of computing $\mu_{\text{had}}/f_{\pi K}$ is thus divided into computing the two ratios $\mu_{\text{ref}}^*/f_{\pi K}$ and $\mu_{\text{had}}/\mu_{\text{ref}}^*$, for which we can consider different strategies in order to achieve the most accurate result. A quantity that satisfies many desirable properties in this respect is given by $\mu_{\text{ref}}^* = 1/\sqrt{8t_0^*}$, where $t_0^*$ is a specific flow time (cf. Eq. (2)), implicitly defined by the equation [9,13,2]:

$$0.3 = (t_0^*)^2 \langle E(t_0^*, x) \rangle |_{m_{u,d,s}=m_{\text{av,phys}}}, \qquad E(t, x) = \frac{1}{4} G_{\mu\nu}^a(t, x) G_{\mu\nu}^a(t, x). \qquad (8)$$

Note that the expectation value appearing in this equation is that of the theory in infinite space-time, i.e., with $L = \infty$. Moreover, it is evaluated at the SU(3) flavour-symmetric point where all quark masses are set equal to the physical average quark mass. As anticipated, $\mu_{\text{ref}}^*$ can be determined very accurately in lattice QCD and with modest computational effort. This is also aided by the fact that it is measured at unphysical values of the quark masses which can be simulated with modest effort, differently from the physical situation which is often reached only through extrapolation. Clearly, $\mu_{\text{ref}}^*$ is not measured in experiments, and its value in physical units must thus be fixed by relating it to some experimentally accessible quantity; in our case $f_{\pi K}$.

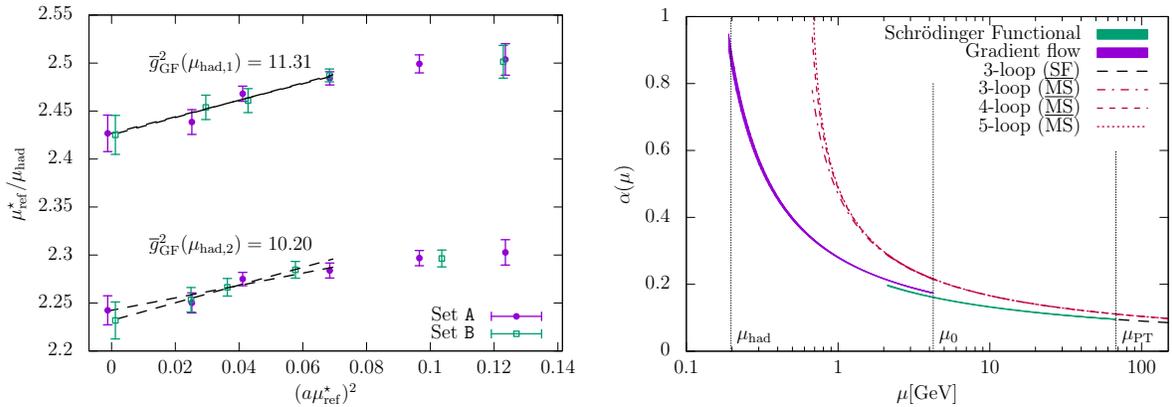

Figure 2: Left: Continuum extrapolations of the ratio $\mu_{\text{ref}}^*/\mu_{\text{had}}$. Note that as a consistency check of our strategy we considered also a second, larger, value for the technical scale $\mu_{\text{had}}$ [2]. The two sets of data, labelled as A,B in the plot, refer to different analysis strategies [2]. Right: Running couplings of $N_{\text{f}} = 3$ QCD obtained from $\Lambda_{\overline{\text{MS}}}^{N_{\text{f}}=3}$ by integrating the non-perturbative $\beta$-functions [2].

The value of $\mu_{\text{ref}}^*$ in physical units was obtained in Ref. [13], to which we refer for any detail. Very briefly, employing an extensive set of state-of-the-art large volume simulations of $N_{\text{f}} = 3$ QCD [14] and a novel strategy for computing the relevant renormalization constants [15,16], the precise continuum result: $\mu_{\text{ref}}^*/f_{\pi K} = 3.24(4)$, was obtained. The particular combination $f_{\pi K} =$ $\frac{2}{3}(f_K + \frac{1}{2}f_\pi)$ was considered as this showed a very mild quark-mass dependence for the chosen set of simulated quark masses. This allowed for robust and precise extrapolations to the physical quark-mass point; the latter identified by computing $\mu^*_{\text{ref}}/m_{\pi,K}$, and taking as inputs the experimental values for the pion and kaon masses, $m_\pi$ and $m_K$ [17]. Using the PDG value for $f_{\pi K}$ [18], one finally arrives at: $\mu^*_{\text{ref}} = 478(7)\,\text{MeV}$.[†] The ratio $\mu^*_{\text{ref}}/\mu_{\text{had}}$ can now easily be evaluated using the results for $a\mu^*_{\text{ref}}$ at several values of the lattice spacing determined in the previous computation [13]. Through a small set of lattice QCD simulations of the SF, $a\mu_{\text{had}}$ can indeed be obtained at matching values of the lattice spacing [2] and the ratio $(a\mu^*_{\text{ref}})/(a\mu_{\text{had}})$ be extrapolated to the continuum. Figure 2 collects these extrapolations, whose final result reads: $\mu^*_{\text{ref}}/\mu_{\text{had}} = 2.428(18)$ [2]. With this last bit of information at our disposal, we can quote (cf. Eq. (7))[2]:

$$\frac{\Lambda^{N_{\text{f}}=3}_{\overline{\text{MS}}}}{\mu^*_{\text{ref}}} = 0.712(24) \quad \Rightarrow \quad \Lambda^{N_{\text{f}}=3}_{\overline{\text{MS}}} = 341(12)\,\text{MeV}. \tag{9}$$

From $\Lambda^{N_{\text{f}}=3}_{\overline{\text{MS}}}$ and the non-perturbative $\beta$-functions of the SF and GF couplings, we can reconstruct the non-perturbative running of the couplings over the whole range of energy we covered, which goes from $\mu_{\text{had}} \approx 200\,\text{MeV}$ up to $\mu_{\text{PT}} = 16\mu_0 \approx 70\,\text{GeV}$. The result is shown in Fig. 2.

## Heavy-quark decoupling and $\alpha_{\text{s}}$

To compute $\alpha_{\text{s}}$ we need $\Lambda^{N_{\text{f}}=5}_{\overline{\text{MS}}}$. How can we obtain this from our result, Eq. (9)? The first issue we address concerns the determination of the scale $\mu^*_{\text{ref}}$, which allows us to express the $\Lambda$-parameter in physical units. As described in the previous section, this determination is based on the computation of several low-energy quantities, $\mathcal{Q} = \mu^*_{\text{ref}}/f_{\pi K}, \mu^*_{\text{ref}}/m_{\pi,K}$, in $N_{\text{f}} = 3$ QCD. Can we consider these results, and hence that for $\mu^*_{\text{ref}}$, valid for the $N_{\text{f}} = 4$ and 5 theories? The decoupling of heavy quarks tells us that for an heavy enough quark we should expect: $\mathcal{Q}_{N_{\text{f}}} = \mathcal{Q}_{N_{\text{f}}-1} + \mathcal{O}(\Lambda^2/M^2)$, where $\mathcal{Q}_{N_{\text{f}}}$ denotes the low-energy quantity computed in the $N_{\text{f}}$ theory where one flavour is much heavier than the others and has renormalization-group invariant mass $M$. $\Lambda$ stands here for a generic low-energy scale of the theory, and clearly the $N_{\text{f}} - 1$ theory is defined only in terms of the lighter quarks (see e.g. Ref. [19]). The $N_{\text{f}} = 3$ results can therefore be considered legitimate for $N_{\text{f}} = 4$ and hence 5, only if the charm mass $M_c$ is actually large enough for the decoupling relation to be valid, and if the leading $\mathcal{O}(\Lambda^2/M_c^2)$ corrections are negligible within the given precision. Dedicated non-perturbative studies show that the typical $\mathcal{O}(\Lambda^2/M_c^2)$ effects in (dimensionless) low-energy quantities are in fact far below the percent level [20]. As the relevant observables are determined to a precision of $\approx 1\%$, we conclude that, within this precision, $\mu^*_{\text{ref}}$ is well-determined from the results of $N_{\text{f}} = 3$ QCD.

The second category of heavy quark effects we must discuss are those affecting the running of the coupling. It is well-known that in a massless renormalization scheme like the $\overline{\text{MS}}$, the decoupling of heavy quarks is not "automatic". Hence, one typically works with the coupling of the relevant effective theory and matches the couplings of the theories with different flavour content according to: $\alpha^{(N_{\text{l}})}_{\overline{\text{MS}}}(\mu) = \xi^2(\alpha^{(N_{\text{f}})}_{\overline{\text{MS}}}, \overline{m}(\mu)/\mu)\,\alpha^{(N_{\text{f}})}_{\overline{\text{MS}}}(\mu)$, where $\overline{m}$ stands for the (renormalized) $N_{\text{f}} - N_{\text{l}}$ heavy quark masses and $\xi$ is a computable function (see e.g. [21]). This allows one to write perturbative expansions that naturally contain only the "active" quarks at the energy scales of the processes of interest and avoids the appearance of large logarithms of the heavy quark masses in the computations. This matching between the two effective theories can equivalently be reformulated in terms of a relation between their $\Lambda$-parameters: $\Lambda_{\overline{\mathrm{MS}}}^{N_{\mathrm{l}}}/\Lambda_{\overline{\mathrm{MS}}}^{N_{\mathrm{f}}} = P_{\mathrm{l,f}}(M/\Lambda_{\overline{\mathrm{MS}}}^{N_{\mathrm{f}}})$. The function $P_{\mathrm{l,f}}$ is expected to be more accurately and reliably determined in perturbation theory the larger the invariant masses $M$ of the decoupling quarks are. Thus, the relevant question in this case is how well does perturbation theory describe the function $P_{3,4}$ for values of $M$ corresponding to the charm mass; for the decoupling of the bottom quark the situation is clearly expected to be better. This issue has been recently investigated in detail and the non-perturbative contributions to $P_{\mathrm{l,f}}$ studied [21]. The conclusions of this work are that perturbation theory describes $P_{3,4}$ at the charm mass with a precision of at least 1.5% – likely much better. As our determination of $\Lambda_{\overline{\mathrm{MS}}}^{N_{\mathrm{f}}=3}$ has a precision of $\approx 3.5\%$ (cf. Eq. (9)), this means that $\Lambda_{\overline{\mathrm{MS}}}^{N_{\mathrm{f}}=5}$ can be safely obtained from $\Lambda_{\overline{\mathrm{MS}}}^{N_{\mathrm{f}}=3}$ using perturbation theory.

---

[†]Note that the hadronic inputs $m_\pi$, $m_K$, and $f_{\pi K}$, used to fix the bare quark masses and to set the physical scale of the lattice theory should be corrected for electromagnetic and $m_u \neq m_d$ effects [13]. This is necessary since our lattice results do not include QED effects and they assume equal up and down quark masses.

We are now in the position of quoting our results for $\alpha_{\mathrm{s}}$. Taking as input our non-perturbatively determined $\Lambda_{\overline{\mathrm{MS}}}^{N_{\mathrm{f}}=3}$, Eq. (9), the values of the charm and bottom masses $\overline{m}_{\overline{\mathrm{MS}}}^{c}$ and $\overline{m}_{\overline{\mathrm{MS}}}^{b}$ from the PDG [18], and the 4- and 5-loop results for the function $\xi$ [22] and the $\beta$-function [23], respectively, perturbative decoupling predicts [2]:

$$\Lambda_{\overline{\mathrm{MS}}}^{N_{\mathrm{f}}=3} \to \Lambda_{\overline{\mathrm{MS}}}^{N_{\mathrm{f}}=5} = 215(10)(3)\,\mathrm{MeV} \quad \Rightarrow \quad \alpha_{\overline{\mathrm{MS}}}^{(N_{\mathrm{f}}=5)}(m_{\mathrm{Z}}) = 0.11852(80)(25). \qquad (10)$$

The second error in $\Lambda_{\overline{\mathrm{MS}}}^{N_{\mathrm{f}}=5}$, then propagated to $\alpha_{\mathrm{s}}$, comes from an estimate within perturbation theory of the truncation errors in the perturbative expansion for $\Lambda_{\overline{\mathrm{MS}}}^{N_{\mathrm{f}}=5}/\Lambda_{\overline{\mathrm{MS}}}^{N_{\mathrm{f}}=3}$ [2]. Our final result for $\alpha_{\mathrm{s}}$ has a precision of $\approx 0.7\%$ and it is well in agreement with the current PDG [18] and FLAG averages [17].

## Conclusions

Lattice QCD offers a very powerful framework for determining $\alpha_{\mathrm{s}}$. By combining finite-volume couplings and a step-scaling strategy, we were able to obtain a subpercent precision determination of $\alpha_{\mathrm{s}}$ where all systematic uncertainties are under control. These include the specific lattice QCD systematics, i.e., discretization and finite-volume effects, as well as the unavoidable uncertainties originating from the use of perturbation theory in extracting $\alpha_{\mathrm{s}}$. Our result for $\alpha_{\mathrm{s}}$ in based on a determination of $\Lambda_{\overline{\mathrm{MS}}}^{N_{\mathrm{f}}=3}$ which relies on perturbation theory only at energy scales of $\mathcal{O}(100\,\mathrm{GeV})$, where we proved it accurate. The strong coupling was then extracted using perturbative decoupling to match the $N_{\mathrm{f}} = 3$ and $N_{\mathrm{f}} = 5$ theories. We argued that non-perturbative corrections to the decoupling relations are not important at our level of precision.

The dominant source of error in our $\alpha_{\mathrm{s}}$ determination comes from $\Lambda_{\overline{\mathrm{MS}}}^{N_{\mathrm{f}}=3}/\mu_0$ (cf. Eq. (1)); in other words from the computation of the non-perturbative running of the SF coupling from about 4 to 70 GeV [2]. This error is predominantly statistical and can therefore be straightforwardly reduced. We want to stress that most other lattice determinations of $\alpha_{\mathrm{s}}$ avoid computing the running of the coupling in this energy range by relying on perturbation theory already at a few GeV (see e.g. refs. [17,24]). In these cases, one ends up dealing with an error which is mostly systematic, and thus much harder to reliably quantify. In the first part of this overview [1], we showed with concrete examples how estimating this sort of error can indeed be very difficult at the level of precision we aim for $\alpha_{\mathrm{s}}$.

In the near future we expect to be able to reduce our error on $\Lambda_{\overline{\mathrm{MS}}}^{N_{\mathrm{f}}=3}$ to about 2%, which would correspond to an error of 0.5% on $\alpha_{\mathrm{s}}$. To further halve this error, on the other hand, requires several issues to be reconsidered. Non-perturbative decoupling effects might not be negligible anymore, and one might need to include electromagnetic and $m_u \neq m_d$ effects in the lattice computations in order to set the physical scale of the theory to a greater level of accuracy.

# $\alpha_\text{s}$ from the QCD static energy

**Nora Brambilla**, Antonio Vairo

*Physik Department, Technische Universität München, D-85748 Garching, Germany*

Alexei Bazavov, Johannes Heinrich Weber

*Department of Computational Mathematics, Science and Engineering and Department of Physics and Astronomy, Michigan State University, East Lansing, MI 48824, USA*

Xavier Garcia i Tormo

*Albert Einstein Center for Fundamental Physics. Institut für Theoretische Physik, Universität Bern, Sidlerstrasse 5, CH-3012 Bern, Switzerland*

Péter Petreczky

*Physics Department, Brookhaven National Laboratory, Upton, NY 11973, USA*

Joan Soto

*Dept. d'Estructura i Constituents de la Matèria and Institut de Ciències del Cosmos, Universitat de Barcelona, 08028 Barcelona, Catalonia, Spain*

**Abstract:** We present our latest determination of the strong coupling constant $\alpha_\text{s}$ from the Quantum Chromodynamics static energy: $\alpha_\text{s}(m_\text{Z}) = 0.1166^{+0.0012}_{-0.0008}$, extracted at three loops with leading ultrasoft log resummation. The determination is based on a combination of lattice data on the static energy at small quark-antiquark distance and perturbative high-order calculations of the static energy for small quark-antiquark distance. We discuss further improvements from an upcoming extraction based on new lattice data, at smaller lattice spacings reaching shorter distances, and on lattice data on the singlet free energy at finite temperature at very small distances.

The QCD static energy $E_0(r)$, i.e. the energy between a static quark and a static antiquark separated by a distance $r$, is a basic object to understand the behavior of strong interactions [1] and constitutes a fundamental ingredient in the description of many physical processes [2]. The short-distance part of $E_0(r)$ has been computed, in the continuum in the $\overline{\text{MS}}$ scheme, using perturbative and effective field theory techniques: it is nowadays known at next-to-next-to-next-to leading-logarithmic (N³LL) accuracy, i.e. including terms up to order $\alpha_\text{s}^{4+n} \ln^n \alpha_\text{s}$ with $n \geq 0$ [3]. The $\ln \alpha_\text{s}$ terms appear due to virtual emissions of ultrasoft gluons, which can change the color state of the quark-antiquark pair [4], and in this context the soft (S) scale is $1/r$ and the ultrasoft (US) scale is $\alpha_\text{s}/r$. $E_0(r)$ is a physical observable (up to an additive constant) and as such it can also be computed on the lattice. It depends only on $\Lambda_\text{QCD}$ and $r$. The comparison between the perturbative and the lattice calculations tests our ability to describe the short-distance regime of QCD, provides information on the region of validity of the perturbative weak-coupling approach and allows for an extraction of $\alpha_\text{s}$. In particular, for distances such that $r\Lambda_\text{QCD} \ll 1$ both the perturbative and the lattice evaluations should agree. Then, one can proceed as follows: fix the scale of the lattice calculation by reproducing a low energy observable*; evaluate $E_0(r)$ for small $r$ perturbatively in the $\overline{\text{MS}}$

---

*Conventionally in these calculations the scale is fixed through the scale parameters $r_0$ or $r_1$ defined by the scheme at the needed order; get $\Lambda_{\overline{\text{MS}}}$ at a given scale by equating the lattice and the perturbative expressions for $E_0$; extract $\alpha_s$ from $\Lambda_{\overline{\text{MS}}}$ and then run it to the $Z$ mass scale. Notice that in such case no lattice-to-$\overline{\text{MS}}$ scheme change is necessary because we deal directly with a physical quantity.

The expression of the static energy in perturbation theory at N³LL is summarized in [9,7]. It contains a residual mass and it depends on $\alpha_s$ at the scale $1/r$ and on the logs of the US scale, that can be resummed at one (N²LL) or two loop accuracy (N³LL) using renormalization group equations in the effective field theory called potential nonrelativistic QCD [5]. A renormalon ambiguity in the series expansion should be appropriately canceled with the residual mass to leave an object well behaved in perturbation theory. When we compare the perturbative curve for the static energy with the lattice data we need to plot $E_0(r) - E_0(r_{ref}) + E_0^{latt}(r_{ref}) = E_0(r) + const$ where $r_{ref}$ is the reference distance where we make the perturbative expression coincide with the lattice data and $E_0^{latt}(r_{ref})$ is the value of the static energy computed on the lattice at that distance.

In Ref. [6] we started a program to extract a precise determination of $\alpha_s$ by using lattice data for the short-distance part of the static energy in $2+1$-flavor QCD [8] and comparing them with the perturbative calculation. This allowed us to determine the strong coupling $\alpha_s$ at three-loop accuracy (including resummation of the leading ultrasoft logarithms), in a way that is largely independent from the other determinations that currently enter in the world average. The natural scale where our determination is performed corresponds to the inverse of the typical distance where we have lattice data, i.e. around 1.5 GeV. Therefore, our analysis provided a determination of $\alpha_s$ at a scale smaller than those entering the world average, and constituted in this way an important ingredient to further test asymptotic freedom in QCD. We obtained $r_0 \Lambda_{\overline{\text{MS}}} = 0.70 \pm 0.07$, which, using $r_0 = 0.468 \pm 0.004$ fm [8] gave $\alpha_s(1.5\text{GeV}, n_f = 3) = 0.326 \pm 0.019$ corresponding to $\alpha_s(m_Z, n_f = 5) = 0.1156^{+0.0021}_{-0.0022}$. The error is dominated by the perturbative uncertainty and could be reduced by using lattice data at shorter quark-antiquark distance.

In our most recent published extraction [9] we therefore used the 2+1 flavor lattice data [15]. The strange-quark mass $m_s$ was fixed to its physical value, while the light-quark masses were chosen to be $m_l = m_s/20$. These correspond to a pion mass of about 160 MeV in the continuum limit, which is very close to the physical value. More precisely, we used lattice QCD data corresponding to the lattice gauge couplings $\beta = 10/g^2 = 7.150, 7.280, 7.373, 7.596$ and $7.825$. The largest gauge coupling, $\beta = 7.825$, corresponds to lattice spacings of $a = 0.041$ fm[†].

Our extraction [9] was improved in several ways and the central value and the error have been scrutinized with a long list of checks that we briefly describe in the following (the details of all this is described in [9]). Lattice artifacts at small distance $r$ may be significant: such artifacts have been removed and the corresponding systematic error has been estimated. The renormalon subtraction has been optimized. We performed fits to the lattice data for the static energy using the perturbative expression at different orders, starting from tree level up to three-loops and we kept only the range of data in which the fit was improving, confirming that we have reached the perturbative window. We repeated the analysis using both the static energy and the force. We performed the analysis with the ultrasoft resummation at N²LL and at N³LL accuracy as well as with N³LO accuracy plus leading US logarithms. In doing so we found that even if all these analyses turn out to be consistent, the size of the leading US logs appeared to be comparable to the three loops correction, which eventually selected the order at which we extracted $\alpha_s$. We varied the analysis considering only some subsets of lattice points and/or varying the reference point $r_{ref}$. We

---

condition: $r^2 \frac{dE_0(r)}{dr}|_{r=r_0} = 1.65$, $r^2 \frac{dE_0(r)}{dr}|_{r=r_1} = 1$. The values of $r_0$ and $r_1$ are extracted from a lattice calculation of a low energy observable.

[†]One may worry about the evolution of the topological charge on such fine lattices, but, as it was shown in Ref. [15], the Monte Carlo evolution of the topological charge is acceptable even for $\beta = 7.825$.

repeated the fits adding $r^3$ and $r^2$ monomials to see if the presence of nonperturbative corrections (nonlocal condensates), not accounted in the previous fits, could distort the analysis: we did not find any evidence of nonperturbative corrections. Lastly, we varied the soft scale and consider the size of the next perturbative correction to estimate the perturbative error.

Our final error comes from the sum in quadrature of the statistical error, the perturbative error and the error on the scale $r_1$. Given all the performed checks, we consider our $\alpha_s$ extraction and the error attached to it pretty solid. We obtained $r_1\Lambda_{\overline{\rm MS}} = 0.495^{+0.028}_{-0.018}$. By converting this result to physical units by using $r_1 = 0.3106 \pm 0.0017$ fm, fixed from the pion decay constant $f_\pi$ [14], we obtained $\Lambda_{\overline{\rm MS}} = 315^{+18}_{-12}$ MeV. This value of $\Lambda_{\overline{\rm MS}}$ gives $\alpha_s(1.5 \text{ GeV}, n_f = 3) = 0.336^{+0.012}_{-0.008}$, corresponding to $\alpha_s(m_Z, n_f = 5) = 0.1166^{+0.0012}_{-0.0008}$. This is an extraction of $\alpha_s$ at three loops plus leading US logs resummation and the number is perfectly compatible, but more accurate, with our previous result given above.

In Fig 1 we show the results one obtains when using larger distance ranges in the fits, up to $r < 0.75 r_1$. The distances $r < 0.6 r_1$ are the ones that passed our $\chi^2$ criteria[‡], and were therefore deemed as suitable for the $\alpha_s$ extraction. The point of showing here the results from larger distance ranges is to illustrate that nothing dramatic happens beyond that point. Figure 1 shows the results for $r_1\Lambda_{\overline{\rm MS}}$ at three-loop accuracy, in all the distance ranges that we have analyzed in Ref. [9].

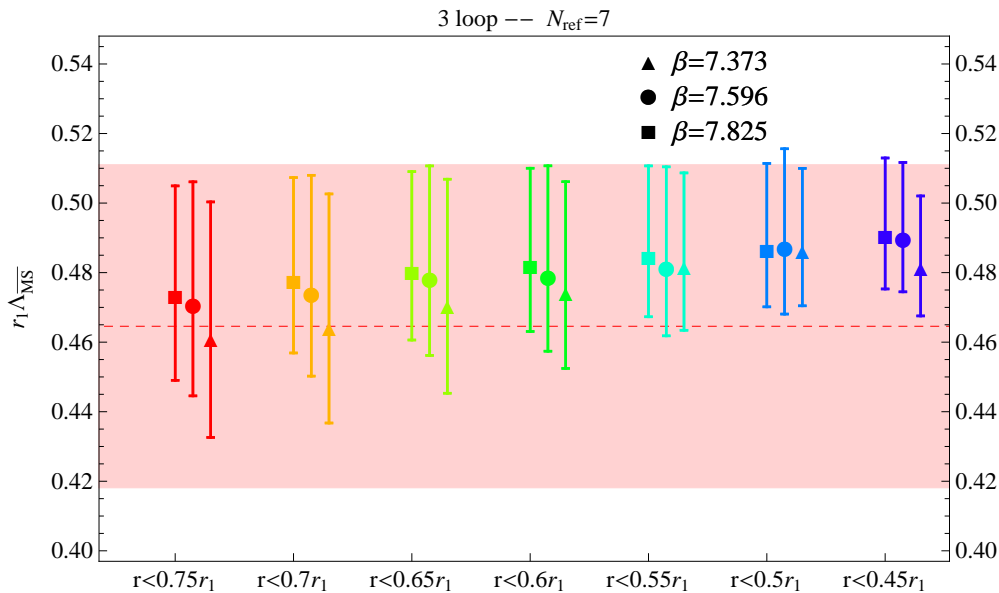

Figure 1: Results for $r_1\Lambda_{\overline{\rm MS}}$ at three-loop accuracy, also showing the outcome of analyses with extended distance ranges. For reference and comparison, the band shows our previous result in Ref. [6]. This figure is taken from [9].

As one can see from the figure, the fits that use distances larger than $0.6 r_1$ give results for $r_1\Lambda_{\overline{\rm MS}}$ that are compatible with those used in our main analysis. The error bars, which come from unknown higher-perturbative orders, are larger in the extended distance ranges. This may be attributed to the fact that those fits involve lower-energy scales and therefore larger values of $\alpha_s$.

In Fig. 2 we put together the data for all the lattice spacings we have, including those used in Ref. [6], i.e. from $\beta = 6.664$ to $\beta = 7.825$, and compare them with the perturbative expressions at different orders of accuracy. The uncertainties due to the normalization of the lattice data to a common scale are now included in the error bars, as it is appropriate when putting together data from different lattice spacings. One can see that the lattice data are perfectly reproduced by perturbation theory and the different perturbative orders converge to the lattice data.

---

[‡]We required that the $\chi^2$ should improve or at least stay constant passing from one perturbative order to the subsequent one and by doing so we selected the perturbative window.

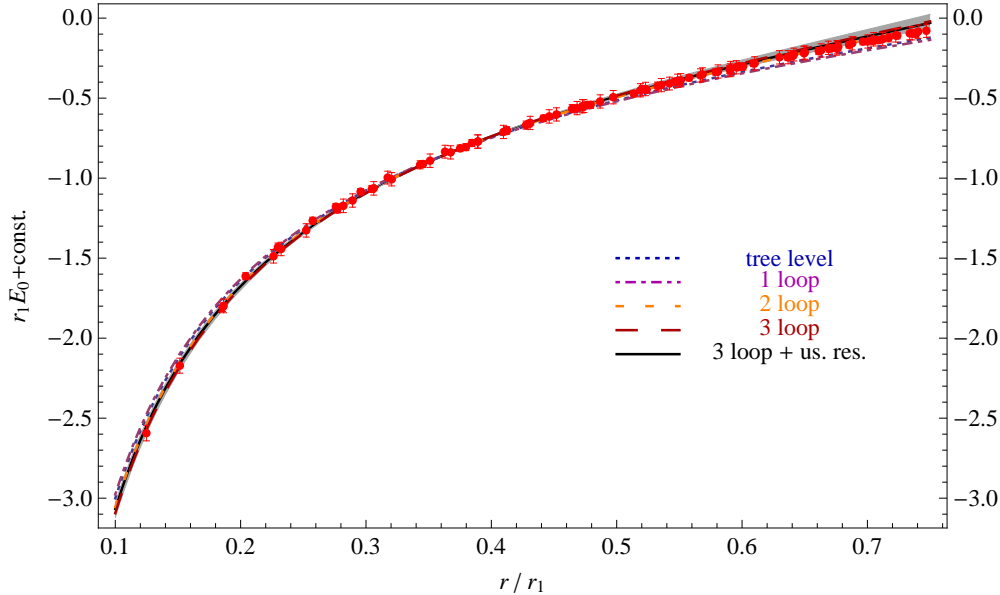

Figure 2: Comparison of the lattice data for the static energy with perturbative expressions at different orders of accuracy. $r_1\Lambda_{\overline{\rm MS}} = 0.495$ is used for all the curves. The grey band corresponds to the variation $r_1\Lambda_{\overline{\rm MS}} = 0.495^{+0.028}_{-0.018}$ for the three-loop plus leading-ultrasoft-resummation accuracy curve. Figure taken from [9].

We would like to further reduce our error: this would entail to get lattice data at smaller spacing and smaller $r$. At the present day, these lattices still pose a major challenge due to critical slowing down, topological freezing, and the need to maintain a sufficiently large volume (in units of the inverse pion mass). In an upcoming paper [10], we use lattices [11] with extraordinarily fine lattice spacing ($a = 0.0246$ fm) to achieve a systematically improved extraction of $\alpha_s$. Additionally, we exploit a new idea. One reason for which it is challenging to reach such fine lattice spacings is that one has to simultaneously maintain the control over finite volume effects from the propagation of the lightest hadronic modes, namely, the Goldstone bosons, at the pion scale. A lattice simulation at high enough temperature avoids this infrared problem, and thus enables reaching much finer lattice spacings using smaller volumes. We use finite temperature lattices with unprecedentedly fine lattice spacing ($a = 0.00848$ fm) [12]. The singlet static free energy is again a function of the static quark-antiquark distance and has been calculated on the lattice [12] and perturbatively using finite temperature effective field theory methods [13]. The comparison between the two offers a novel and independent method to get a precise determination of $\alpha_s$. The results that we are obtaining in these two ways in [10] confirm our 2014 determination of $\alpha_s$ [9] with smaller errors.

# $\alpha_\text{s}$ determination from static QCD potential with renormalon subtraction

Hiromasa Takaura

*Department of Physics, Kyushu University, Fukuoka 819-0395, Japan*

**Abstract:** In the current $\alpha_\text{s}$ determinations based on lattice and perturbation theory, it is generally difficult to take a wide enough matching range. We avoid this problem by improvement of a theoretical calculation: we use the OPE with renormalon subtraction, which is an extended framework of perturbation theory. This allows us to take the matching range widely as $\Lambda_\text{QCD} r \lesssim 0.6$, where relatively low energy scales are included. We obtain $\alpha_\text{s}(m_\text{z}) = 0.1179^{+0.0015}_{-0.0014}$ from a reasonable fit for this wide range.

## Introduction

The strong coupling $\alpha_\text{s}$ is a fundamental parameter in the standard model, and its precision has an impact on various studies of the standard model. This parameter is determined by a matching of a theoretical calculation and an experimental or lattice measurement of a QCD observable. Among determinations from various observables, the determinations using lattice data generally have small errors.

In lattice determinations, however, the so-called window problem has been pointed out: it is difficult to take a wide enough matching range. Accurate lattice simulation can be performed at the scale well below its UV cutoff scale $a^{-1}$, the inverse of the lattice spacing. This lattice result is matched with perturbation theory, where fixed order results are currently accurate at $Q \gtrsim 1$–$2$ GeV. With the typical lattice spacings available today, it is difficult to take the range satisfying $1$–$2$ GeV $\lesssim Q \ll a^{-1}$ widely.

The step-scaling method is known as a solution to the window problem. This method enlarges the validity range of the lattice simulations. (The latest determination has been performed in Ref. [1].) As an alternative approach, we enlarge the validity range of a theoretical calculation to lower energy so that accurate lattice data (due to $Q \ll a^{-1}$) are available. To this end, we use the operator product expansion (OPE), which is an extended framework of perturbation theory.

Perturbation theory suffers from an inevitable uncertainty known as renormalon uncertainty. It is induced from the divergent behavior of perturbative series where perturbative coefficients typically grow as $\sim \beta_0^n n!$ at large orders. For the static QCD potential $V_\text{QCD}(r)$, the leading renormalon uncertainty is $\mathcal{O}(\Lambda_\text{QCD})$, and the next-to-leading one is $\mathcal{O}(\Lambda_\text{QCD}^3 r^2)$. These errors are not negligible at relatively long distances $\Lambda_\text{QCD} r \sim 1$, and give limitations of perturbation theory.

In the OPE, which can be regarded as an extension of perturbation theory, renormalon uncertainties are considered to be eliminated. In the following, we focus on the second renormalon uncertainty rather than the first one (which is $r$-independent and can be eliminated in the QCD force). The OPE of the static QCD potential is performed in the effective field theory, potential non-relativistic QCD (pNRQCD) [2]. It is given in form of multipole expansion as

$$V_\text{QCD}(r) = V_S(r) + \delta E_\text{US}(r) + \cdots, \tag{1}$$

where the singlet potential $V_S(r)$ has a Coulomb-type potential and is the leading behavior at short distances ($V_S(r) \sim 1/r$). A power correction in $r$ is added as $\mathcal{O}(r^2)$ ($\delta E_\text{US}(r) \sim r^2$). Since the second term (and further higher order terms in $r$) are nonperturbative objects*, a perturbative expression of $V_S(r)$ coincides with that of $V_{\rm QCD}(r)$. Therefore, $V_S$ contains the renormalon uncertainty of $\mathcal{O}(\Lambda_{\rm QCD}^3 r^2)$. An advantage of the OPE is that this renormalon is cancelled against that of the second term $\delta E_{\rm US}(r)$. This has been shown explicitly in Ref. [2] at the leading-log (LL) level. Then the OPE prediction has smaller error and has wider validity range than perturbation theory.

However, it is difficult to hold this advantage of the OPE in practical calculations. In particular, with a naive perturbative calculation of $V_S$, one again suffers from the renormalon uncertainty. Consider the case where one adds a power correction term of $Ar^2$ to the perturbative result $V_S$. The fitting parameter $A$ [of $\mathcal{O}(\Lambda_{\rm QCD}^3)$] can be extracted from the $r^2$-term of $V_{\rm QCD}(r) - V_S(r)$. However, since $V_S$ has the error of $\mathcal{O}(\Lambda_{\rm QCD}^3 r^2)$, the nonperturbative effect $A$ has a significant error. (The error is the same size as the nonperturbative effect itself.) Thus, the introduction of the power correction is almost meaningless because its coefficient cannot be determined in practice.

To avoid this feature, we use the OPE while subtracting renormalons in $V_S(r)$. The use of such an OPE allows us to use a wider range as shown below, and thus, it relaxes the window problem. Our fit range is typically taken as $0.6 \text{ GeV} \lesssim r^{-1} \lesssim 4 \text{ GeV}$. This is significantly wider than previous determinations from the static QCD potential, where typically $1 \text{ GeV} \lesssim r^{-1}$ has been used.

## Theoretical framework

We explain how we subtract renormalons and how we use the result in the OPE. First, we consider renormalon subtraction from $V_S$. $V_S(r)$ is given by

$$V_S(r) = -4\pi C_F \int \frac{d^3 \vec{q}}{(2\pi)^3} e^{i\vec{q}\cdot\vec{r}} \frac{\alpha_V(q)}{q^2} \quad (q = |\vec{q}|), \tag{2}$$

where the potential in momentum space $\alpha_V(q)$ is currently known up to $\mathcal{O}(\alpha_s^4)$ [3]. We apply renormalization group (RG) improvement to $\alpha_V(q)$, i.e., we use the next-to-next-to-next-to-LL (N³LL) result $\alpha_V(q)_{\rm N^3LL}$. Then, the above integral becomes just formal because $\alpha_V(q)_{\rm N^3LL}$ has a singularity at $q \sim \Lambda_{\rm QCD}$ due to the running coupling. In other words, the $q$-integration is ambiguous and this corresponds to the renormalon uncertainty. In fact, all the known renormalons of the static QCD potential stem from the $q$ integration of the logarithmic terms in $\alpha_V(q)$. In order to render the integral well-defined, we subtract the IR contribution by an IR cutoff scale $\mu_f$:

$$V_S(r; \mu_f) = -4\pi C_F \int_{q > \mu_f} \frac{d^3 \vec{q}}{(2\pi)^3} e^{i\vec{q}\cdot\vec{r}} \frac{\alpha_V(q)_{\rm N^3LL}}{q^2}, \tag{3}$$

where $\mu_f$ is taken as $\Lambda_{\rm QCD} \ll \mu_f \ll r^{-1}$. The integral is now well-defined. However, it depends on the artificial cutoff scale. This dependence cannot be removed within perturbation theory. (Note that the cutoff $\mu_f$ cannot be sent to zero due to the singularity.) In this sense, this cutoff dependence corresponds to a renormalon uncertainty. On the other hand, a cutoff independent part, which potentially exists, is unambiguously determined within perturbation theory. It corresponds to a renormalon-free part.

To find a renormalon-free part, we separate the cutoff independent part from the cutoff dependent part following Ref. [4]. This is performed by a contour deformation in the complex $q$-plane. We obtain [4]

$$V_S(r; \mu_f) = V_S^{\rm RF}(r) + \mathcal{O}(\mu_f) + \mathcal{O}(\mu_f^3 r^2), \tag{4}$$

---

*Our fit range extends to relatively low energy scale where the ultrasoft scale is not generally perturbative. Hence, the ultrasoft scale is treated as the nonperturbative scale in our analysis.

where $V_S^{\text{RF}}(r)$ is $\mu_f$ independent and renormalon free. It has a Coulomb+linear like form. The cutoff dependence of $\mathcal{O}(\mu_f)$ and $\mathcal{O}(\mu_f^3 r^2)$ correspond to the first and second renormalon uncertainties, respectively.

We list the distinct features of $V_S^{\text{RF}}$. First, it has the N³LL accuracy and is accurate especially at short distances. Secondly, unlike the usual RG improvement, it does not have an unphysical singularity at $r^{-1} \sim \Lambda_{\text{QCD}}$, generally caused by the running coupling. Thirdly, it is free from the renormalon uncertainties of $\mathcal{O}(\mu_f)$ and $\mathcal{O}(\mu_f^3 r^2)$. From the last two features, it is expected that $V_S^{\text{RF}}$ gives a reasonable prediction even at relatively long distances.

The above perturbative result $V_S^{\text{RF}}(r)$ is used as follows in the context of the OPE. Since the IR cutoff scale is introduced to the perturbative calculation, the contribution below $\mu_f$ is represented by nonperturbative objects. Then, we introduce the UV cutoff scale to $\delta E_{\text{US}}(r)$ as $\delta E_{\text{US}}(r; \mu_f)$. In fact, a similar separation of cutoff dependence can be performed for $\delta E_{\text{US}}(r; \mu_f)$, where the opposite cutoff dependence of $\mathcal{O}(\mu_f^3 r^2)$ to $V_S(r; \mu_f)$ is found. That is, the cutoff dependence in $V_S^{\text{RF}}(r; \mu_f)$ cancels[†] that of $\delta E_{\text{US}}^{\text{RF}}(r; \mu_f)$ [5]. Thus, we can perform the OPE in a renormalon-free way:

$$V_{\text{QCD}}(r) = V_S^{\text{RF}}(r) + \delta E_{\text{US}}^{\text{RF}}(r) + \cdots . \tag{5}$$

This is the OPE calculation used in our $\alpha_{\text{s}}$ determination. $V_S^{\text{RF}}(r)$ can be calculated at the N³LL accuracy and has $\Lambda_{\overline{\text{MS}}}$ as the only input parameter. We treat $\delta E_{\text{US}}^{\text{RF}}(r) = A_2 r^2$ where $A_2$ is a fitting parameter. The difference from the naive OPE is that we subtract the renormalons of $V_S(r)$. This prevents a mixing of the renormalon uncertainty and the nonperturbative effect. Also, it serves to reduce higher order uncertainty of $V_S$.

## $\alpha_{\text{s}}$ determination

The above calculation is matched with lattice result to determine $\alpha_{\text{s}}$ [6]. We use the lattice result of $V_{\text{QCD}}(r)$ obtained by the JLQCD collaboration. The simulated lattice sizes are $32^3 \times 64$, $48^3 \times 96$, $64^3 \times 128$, whose lattice spacings are estimated as $a^{-1} = 2.453(4), 3.610(9), 4.496(9)$ GeV, respectively.

In our $\alpha_{\text{s}}$ determination, we perform two analyses. The first one [Analysis (I)] is a step-by-step analysis and the other is a global fit [Analysis (II)]. After examining detailed profiles in Analysis (I), we perform a global fit in Analysis (II), from which our final result is obtained. In this report, we present a consistency check of the OPE, which is a central concern in Analysis (I), and then explain Analysis (II).

We examine consistency of the OPE in Fig. 1, where we compare $V_S^{\text{RF}}(r)$ with the lattice continuum limit. Here, we use the PDG value of $\Lambda_{\overline{\text{MS}}}$ as an input. According to the OPE, the difference between the lattice result and $V_S^{\text{RF}}(r)$, which is shown by the red boxes in the figure, should behave as $\mathcal{O}(r^2)$. Indeed, it is consistent with a quadratic behavior in $r$ up to $\Lambda_{\overline{\text{MS}}} r \lesssim 0.8$. Thus, the validity range of the OPE turns out to be $\Lambda_{\overline{\text{MS}}} r \lesssim 0.8$. This is significantly larger than that of perturbation theory, $\Lambda_{\overline{\text{MS}}} r \lesssim 0.3$.

Now, we explain Analysis (II). This analysis is performed based on the idea that at short distances the OPE prediction should coincide with the lattice result once the discretization errors are removed. Then we assume the continuum limit as

$$V_{\text{latt}}^{\text{cont}}(r) = V_{\text{latt},d,i}(r) - \kappa_{d,i}\left(\frac{1}{r} - \left[\frac{1}{r}\right]_{d,i}\right) + f_d \frac{a_i^2}{r^3} - c_{0,d,i} , \tag{6}$$

---

[†]This is confirmed at the LL level.

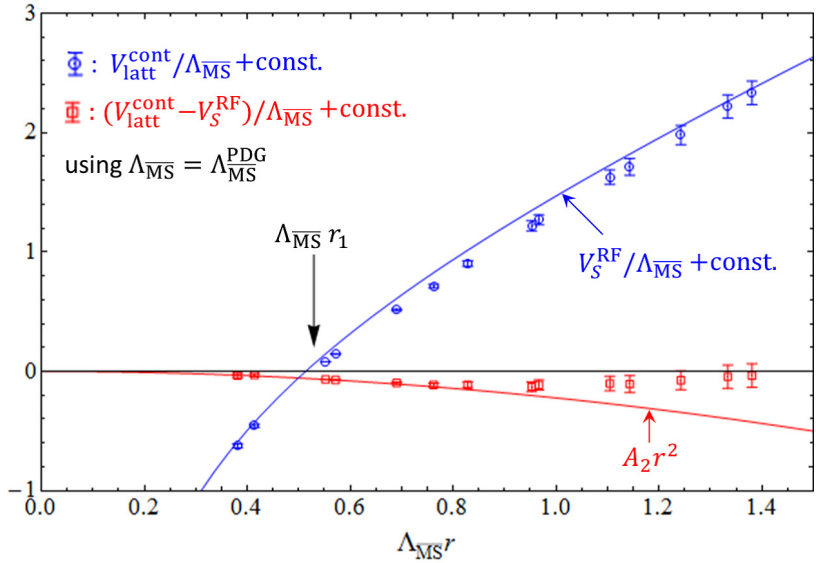

Figure 1: Consistency check of the OPE. The blue date are the lattice continuum limit and the blue line is $V_S^{\rm RF}$. (Both are given in $\Lambda_{\overline{\rm MS}}$ units.) The difference between them are shown by the red data, which are consistent with the quadratic function given by the red line at $\Lambda_{\overline{\rm MS}} r \lesssim 0.8$.

Table 1: Systematic errors in $\alpha_s(m_z)$ in units of $10^{-4}$. See [6] for details.

| finite $a$ | Mass | H.o. | Range | Ultrasoft | Fact. scheme | Latt. spacing |
|---|---|---|---|---|---|---|
| $\pm 2$ | $\pm 0$ | $^{+12}_{-10}$ | $\pm 4$ | $\pm 2$ | $\pm 3$ | $\pm 4$ |

where $V_{{\rm latt},d,i}(r)$ is the original lattice data measured at the $i$-th lattice ($i = 1, 2, 3$) and $d$ denotes the direction of $\vec{r}$;[‡] the second term is a tree-level correction, where $[1/r]$ denotes the LO result of the lattice perturbation theory; the third term removes the remaining error of $\mathcal{O}(\alpha_s^2 a^2)$; the last term adjusts an $r$-independent constant. We give the above lattice result in GeV units. This is matched with the OPE prediction in the same units:

$$V_{\rm OPE}(r) = z[V_S^{\rm RF}/\Lambda_{\overline{\rm MS}}](zr) + A_2 r^2 \,, \qquad (7)$$

where $z = \Lambda_{\overline{\rm MS}}$ GeV. (Note that $V_S(r)$ is originally obtained in $\Lambda_{\overline{\rm MS}}$ units.) In this global fit, we determine 16 parameters in total: $\{z, A_2, \kappa_{d,i}, f_d, c_{0,d,i}\}$. The fit range is $0.07 \leq \Lambda_{\overline{\rm MS}}^{\rm PDG} r < 0.6$, which includes not only short but also relatively long distances.

As a result, we obtain $\alpha_s(m_z) = 0.1179 \pm 0.0007\,({\rm stat})$, where $\chi^2/{\rm d.o.f.} \approx 8.7/14$ shows the validity of this analysis. After including the systematic errors listed in Table 1, we obtain

$$\alpha_s(m_z) = 0.1179 \pm 0.0007\,({\rm stat})^{+0.0014}_{-0.0012}\,({\rm sys}) = 0.1179^{+0.0015}_{-0.0014}\,. \qquad (8)$$

which is consistent with the current world average.

---

[‡] $d = 1$ and $d = 2$ correspond to the spatial directions $(1, 0, 0)$ and $(1, 1, 0)$, respectively.

# Conclusions

Lattice determinations often suffer from the window problem: a matching range cannot be taken sufficiently wide. To avoid this problem, we use the OPE with renormalon subtraction, which is an extended framework of perturbation theory. Such an OPE has the wider validity range than perturbation theory. The fit is performed reasonably for the wide range, which would lead to a reliable value of $\alpha_{\rm s}$. The dominant uncertainty in our determination comes from higher order uncertainty. It can be reduced with finer lattice simulations.

# The QCD coupling $\alpha_s(Q^2)$ at all momentum scales and the elimination of renormalization scale uncertainties

Stanley J. Brodsky

*SLAC National Accelerator Laboratory, Stanford University*

**Abstract:** This contribution discusses two central problems in QCD: (a) the behavior of the QCD running coupling $\alpha_s(Q^2)$ in both the nonperturbative and perturbative domains; and (b) the elimination of perturbative QCD renormalization scale ambiguities.

## Determining the QCD Coupling $\alpha_s(Q^2)$ at all Momentum Scales

The QCD running coupling $\alpha_s(Q^2)$ sets the strength of the interactions of quarks and gluons as a function of the momentum transfer $Q$. The dependence of the coupling at both small and high momenta is needed to describe hadronic interactions at both long and short distances.

As Grunberg has shown [1], the QCD running coupling can be defined at all momentum scales from a perturbatively calculable physical observable. A particularly useful choice is the effective coupling $\alpha_{g_1}^s(Q^2)$, which is defined from Bjorken sum rule and is well measured [2]. At high momentum transfer, such "effective charges" satisfy asymptotic freedom, obey the usual pQCD renormalization group equations, and can be related to each other without scale ambiguity by commensurate scale relations [3].

The "dilaton" soft-wall modification of the AdS$_5$ metric $e^{+\kappa^2 z^2}$, together with LF holography, predicts the functional behavior of the running coupling in the small $Q^2$ nonperturbative domain [4]: $\alpha_{g_1}^s(Q^2) = \pi e^{-Q^2/4\kappa^2}$. Measurements of $\alpha_{g_1}^s(Q^2)$ are remarkably consistent [2] with the Gaussian form predicted by AdSQCD; the best fit gives $\kappa = 0.513 \pm 0.007$ GeV. See Fig. 1.

Deur, de Téramond, and I [4,5,6] have shown how the parameter $\kappa$, which determines the mass scale of hadrons and Regge slopes in the zero quark mass limit [6], can be connected to the mass scale $\Lambda_s$ controlling the evolution of the QCD coupling in the perturbative domain. The high momentum transfer dependence of the coupling $\alpha_{g1}(Q^2)$ is predicted by pQCD. The matching of the high and low momentum transfer regimes of $\alpha_{g1}(Q^2)$ – both its value and its slope – then determines a scale $Q_0 = 0.87 \pm 0.08$ GeV which sets the interface between perturbative and nonperturbative hadron dynamics. This connection can, in fact, be done for any choice of renormalization scheme, such as the $\overline{\text{MS}}$ scheme.

The result of this perturbative/nonperturbative matching is an effective QCD coupling defined at all momenta. The predicted value of $\Lambda_{\overline{\text{MS}}} = 0.339 \pm 0.019$ GeV from this analysis agrees well the measured value [7] $\Lambda_{\overline{\text{MS}}} = 0.332 \pm 0.017$ GeV. These results, combined with the AdS/QCD superconformal predictions [8] for hadron spectroscopy, allow one to compute hadron masses in terms of $\Lambda_{\overline{\text{MS}}}$: $m_p = \sqrt{2}\kappa = 3.21\,\Lambda_{\overline{\text{MS}}}$, $m_\rho = \kappa = 2.2\,\Lambda_{\overline{\text{MS}}}$, and $m_p = \sqrt{2}m_\rho$, meeting a challenge proposed by Zee [9]. The mass scale $\kappa$ underlying confinement and hadron masses can thus be connected to the parameter $\Lambda_{\overline{\text{MS}}}$ in the QCD running coupling by matching the nonperturbative prediction to the perturbative QCD regime.

We have also proposed that the value of $Q_0$, which marks the interface of nonperturbative and perturbative QCD, can be used to set the factorization scale for DGLAP evolution of hadronic structure functions and ERBL evolution of distribution amplitudes [10]. We have also computed the dependence of $Q_0$ on the choice of the effective charge used to define the running coupling and the renormalization scheme used to compute its behavior in the perturbative regime. The use of the scale $Q_0$ to resolve the factorization scale uncertainty in structure functions and fragmentation functions, in combination with the scheme-independent *principle of maximum conformality* (PMC) [11] for setting renormalization scales, can greatly improve the precision of pQCD predictions for collider phenomenology.

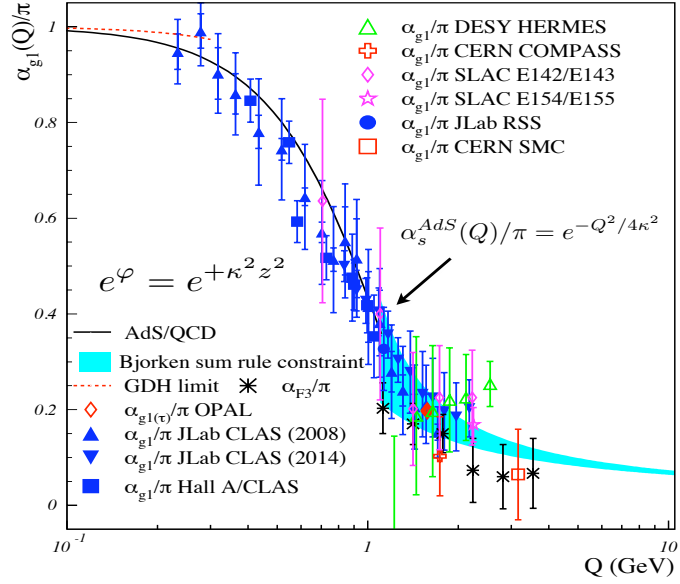
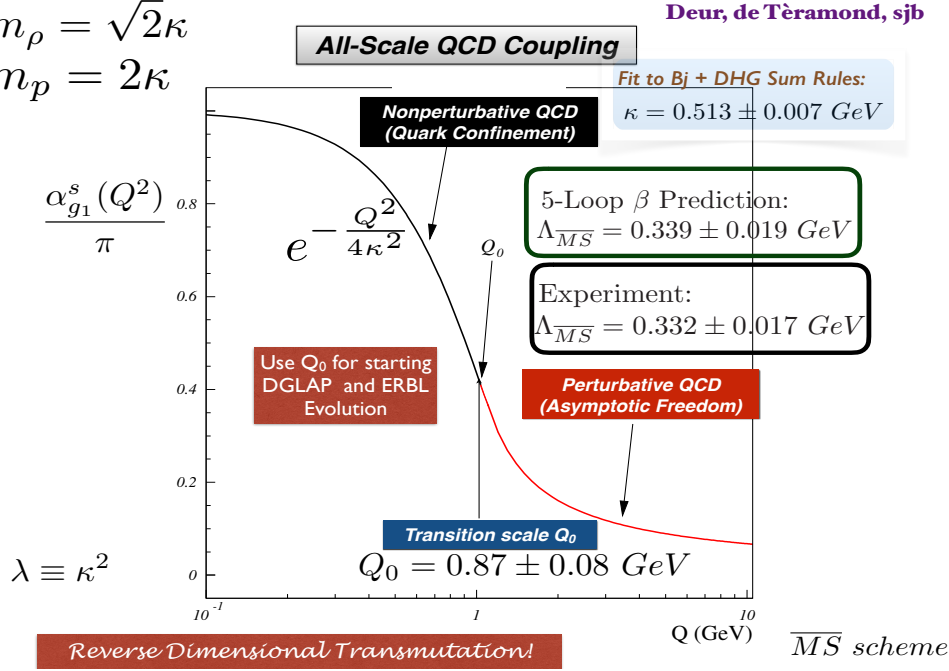

Figure 1: Top: Comparison of the predicted nonperturbative coupling, based on the dilaton $\exp(+\kappa^2 z^2)$ modification of the $AdS_5$ metric, with measurements of the effective charge $\alpha_{g_1}^s(Q^2)$, as defined from the Bjorken sum rule. Bottom: Prediction from LF Holography and pQCD for the QCD running coupling $\alpha_{g_1}^s(Q^2)$ at all scales. The magnitude and derivative of the perturbative and nonperturbative coupling are matched at the scale $Q_0$. This matching connects the perturbative scale $\Lambda_{\overline{MS}}$ in the $\overline{MS}$ scheme to the nonperturbative mass scale $\kappa = \sqrt{\lambda}$, the mass scale which underlies hadronic masses in QCD. See Ref. [6].

The combined approach of light-front holography and superconformal algebra [6] also provides insight into the origin of the QCD mass scale and color confinement. A key observation is the remarkable dAFF principle [12] which shows how a mass scale can appear in the Hamiltonian and the equations of motion while retaining the conformal symmetry of the action. When one applies the dAFF procedure to chiral QCD, a mass scale $\kappa$ appears which determines universal Regge slopes, hadron masses in the absence of the Higgs coupling, and the mass parameter underlying the Gaussian functional form of the nonperturbative QCD running coupling: $\alpha_s(Q^2) \propto \exp{-(Q^2/4\kappa^2)}$. As seen in Fig. 1, this prediction is in remarkable agreement with the effective charge determined from measurements of the Bjorken sum rule.

The potential which underlies color confinement in the effective LF Hamiltonian for the $q\bar{q}$ Fock state of mesons is simply $U(\zeta^2) = \kappa^4 \zeta^2$, a harmonic oscillator potential in the frame-invariant light-front radial variable $\zeta^2 = b_\perp^2 x(1-x)$. This confinement potential also underlies the spectroscopy and structure of baryons and tetraquarks [6]. The parameter $\kappa$ is not determined in absolute units such as MeV; however, the ratios of mass parameters such as $m_p/m_\rho = \sqrt{2}$ are predicted. The same potential can also be derived from the anti–deSitter space representation of the conformal group if the $AdS_5$ is action is modified in the fifth dimension $z$ by the dilaton $e^{+\kappa^2 z^2}$. This correspondence is based on light-front holography [13], the duality between dynamics in physical space-time at fixed LF time and five-dimensional AdS space. The predicted light-front wavefunctions can also be used to model "hadronization at the amplitude level" [14].

# The Thrust Distribution in Electron-Positron Annihilation using the Principle of Maximum Conformality

The Principle of Maximum Conformality (PMC) [11,15,16,17,18] provides a rigorous, systematic way to eliminate renormalization scheme-and-scale ambiguities for perturbative QCD. Since the PMC predictions do not depend on the choice of the renormalization scheme, PMC scale-setting satisfies the principles of renormalization group invariance [19,20].

The PMC provides the underlying principle for extending the Brodsky-Lepage-Mackenzie (BLM) scale-setting method [21] to all orders in pQCD. The essential step is to identify the $\beta$ terms at each order of the pQCD series. The PMC scales are fixed at every order in pQCD by absorbing the $\beta$ terms that govern the behavior of the running coupling via the renormalization group equation (RGE). The divergent renormalon terms disappear, and thus the convergence of the pQCD series is greatly improved. The PMC method also sets the renormalization scales for observables that depend on several invariants. The number of active quark flavors $n_f$ is set at each order, matching the virtuality of the scattering process. The PMC reduces in the Abelian limit, $N_C \to 0$ [22], to the standard Gell-Mann-Low method [23].

The thrust ($T$) variable [24,25] is a frequently studied three-jet event shape observables; it is defined as

$$T = \max_{\vec{n}} \left( \frac{\sum_i |\vec{p}_i \cdot \vec{n}|}{\sum_i |\vec{p}_i|} \right), \tag{1}$$

where the sum runs over all particles in the final state, and $\vec{p}_i$ denotes the three-momentum of particle $i$. The unit vector $\vec{n}$ is varied to define the thrust direction $\vec{n}_T$ by maximizing the sum on the right-hand side.

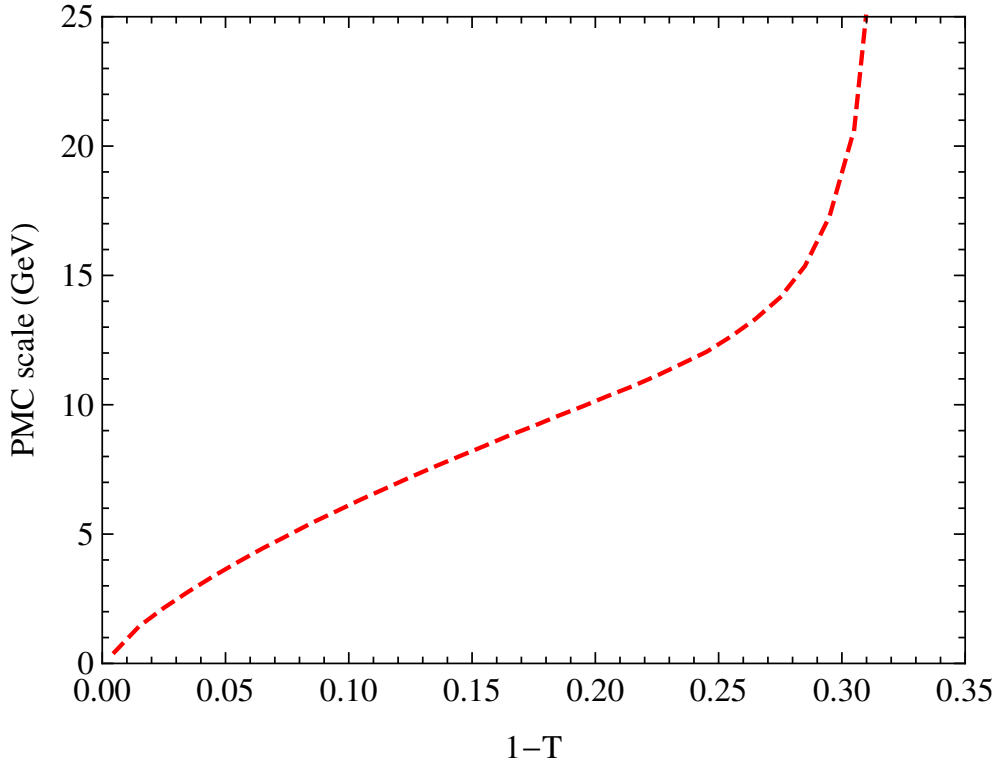

Figure 2: The PMC scales at LL and NLL accuracy for the thrust distribution at $\sqrt{s} = m_Z$.

At the center-of-mass energy $\sqrt{s}$, the differential distribution for thrust variable $\tau = (1 - T)$ for renormalization scale $\mu_r = \sqrt{s} \equiv Q$ can be written as

$$\frac{1}{\sigma_0}\frac{d\sigma}{d\tau} = A(\tau)\, a_s(Q) + B(\tau)\, a_s^2(Q) + \mathcal{O}(a_s^3), \qquad (2)$$

where $a_s(Q) = \alpha_s(Q)/(2\pi)$, $\sigma_0$ is tree-level hadronic cross section. The $A(\tau)$, $B(\tau)$, ... are perturbative coefficients. The experimentally measured thrust distribution is normalized to the total hadronic cross section $\sigma_h$,

$$\frac{1}{\sigma_h}\frac{d\sigma}{d\tau} = \bar{A}(\tau)\, a_s(Q) + \bar{B}(\tau)\, a_s^2(Q) + \mathcal{O}(a_s^3). \qquad (3)$$

The perturbative coefficients $\bar{A}(\tau) = A(\tau)$, and $\bar{B}(\tau) = B(\tau) - 3/2 C_F A(\tau)$, etc., and their general renormalization scale $\mu_r$ dependence $\bar{A}(\tau, \mu_r)$, $\bar{B}(\tau, \mu_r)$ can be restored from the RGE.

In this section, I will review the results of the recent application [26] of PMC scale setting to the thrust distribution by Wang, Wu, de Giustino, and myself. The PMC scale is fixed by absorbing the $\beta_i$-terms into the running coupling; it is itself a perturbative expansion series in $\alpha_s$ and in general shows fast pQCD convergence. A crucial point, as first noted by Gehrmann, Häfliger and Monni [27], is that the pQCD renormalization scale is not a constant; it depends explicitly on the thrust $T$.

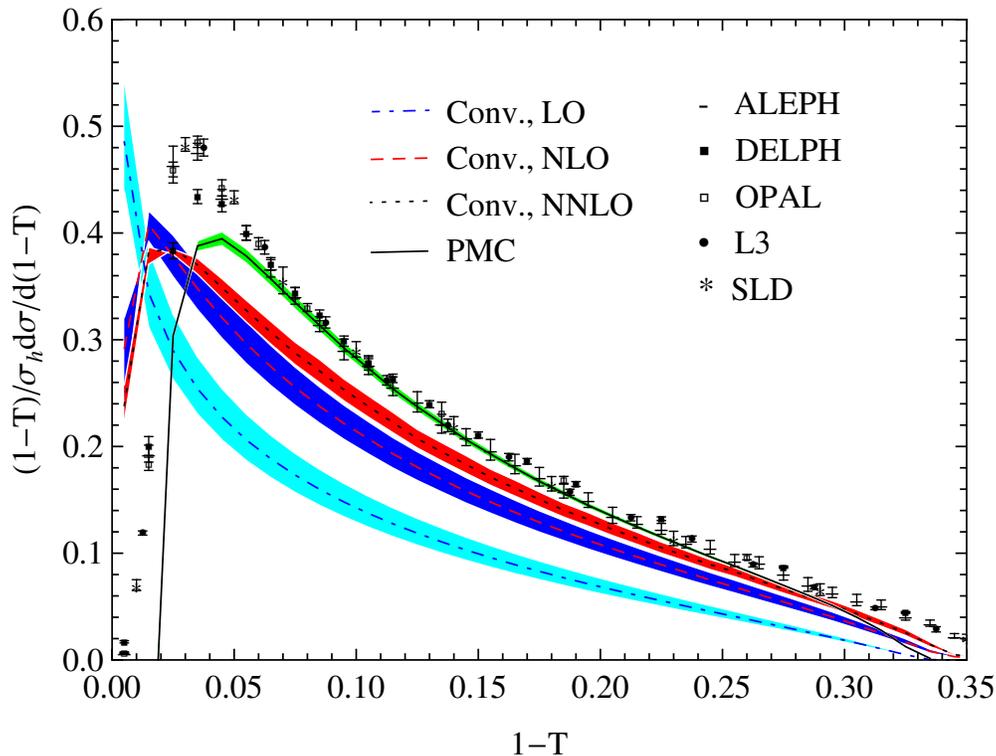

Figure 3: The thrust differential distributions using the conventional (Conv.) and PMC scale settings [26]. The dot-dashed, dashed and dotted lines are the conventional results at LO, NLO and NNLO [28,29], respectively. The solid line is the PMC result [26]. The bands for the conventional theoretical predictions are obtained by varying $\mu_r \in [m_Z/2, 2m_Z]$. The PMC prediction eliminates the renormalization scale $\mu_r$ uncertainty. Its error band is obtained by using $\alpha_s(m_Z) = 0.1181 \pm 0.0011$ [31]. The experimental data are taken from ALEPH [32], DELPHI [33], OPAL [34], L3 [35], and the SLD [36].

In our analysis for the thrust distribution we determine the PMC scale at NLL level by using the pQCD predictions given in Refs. [28,29]. The inclusion of the NNLO correction only slightly changes the PMC scale determined at NLO level. The PMC scale shows fast pQCD convergence, as shown explicitly in Fig. 2.

The renormalization scale using conventional scale-setting is simply set at $\mu_r = m_Z$. The PMC scale, in contrast, is not a single value, but it monotonically increases with $(1-T)$, reflecting the virtuality of the QCD dynamics. The PMC predictions are in excellent agreement with measurements.

The PMC gives the correct physical behavior of the scale and is bounded in the two-jet region. In addition, the number of active flavors $n_f$ changes with $(1-T)$ according to the PMC scale. As the argument of the $\alpha_s$ approaches the two-jet region, the pQCD theory becomes unreliable and non-perturbative effects must be taken into account. One can adopt the predictions from light-front holographic QCD [5] to determine $\alpha_s(Q^2)$ in the low scale domain. The physical behavior of the scale for three-jet processes has also been obtained in Refs. [37,27]. The soft collinear effective theory determines the thrust distribution at different energy scales and also shows that the two-jet region is affected by non-perturbative effects [30].

A remarkable advantage of using the PMC scale setting is that since the PMC scale varies with $(1-T)$, we can extract directly the strong coupling $\alpha_s$ over a wide range of scales using the experimental data at single center-of-mass-energy, $\sqrt{s} = m_{\rm Z}$. In this case we have used the most precise data from ALEPH [32]. We have calculated the thrust differential distribution at each bin corresponding to the bins of the experimental data. We can then extract the $\alpha_{\rm s}$ at different scales bin-by-bin from the comparison of PMC predictions with experimental data. The extracted $\alpha_{\rm s}$ are explicitly presented in Fig. 4. It shows that in the scale range of 3.5 GeV $< Q <$ 16 GeV (corresponding $(1-T)$ range is $0.05 < (1-T) < 0.29$), the extracted $\alpha_{\rm s}$ are in excellent agreement with the world average evaluated from $\alpha_{\rm s}(m_{\rm Z})$ [31]. The pQCD calculation corresponds to a parton-level distribution, while the experimental measurements are the hadron-level. Some previous extractions of $\alpha_{\rm s}$ have applied Monte Carlo generators to correct the effects of hadronization. In our analysis, we have adopted a method similar to [38] in order to extract $\alpha_{\rm s}$.

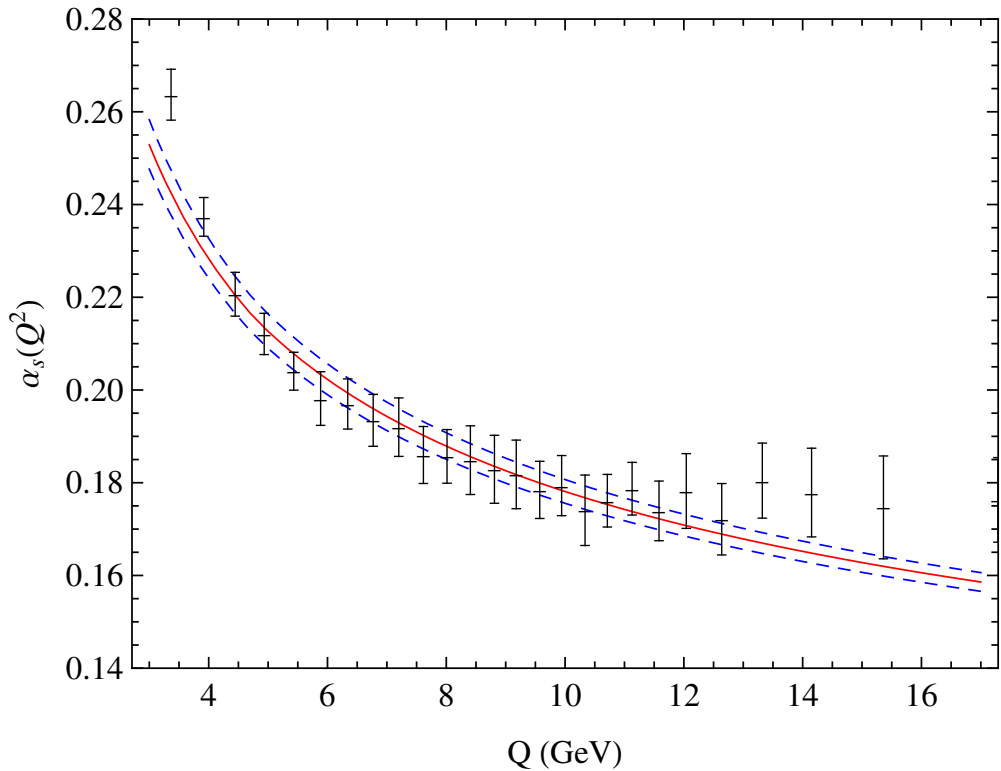

Figure 4: The extracted $\alpha_{\rm s}(Q^2)$ in the $\overline{\rm MS}$ scheme from the comparison of PMC predictions with ALEPH data [32] at $\sqrt{s} = m_{\rm Z}$. The error bars are from the combination of the experimental and theoretical errors. The three lines are the world average evaluated from $\alpha_{\rm s}(m_{\rm Z}) = 0.1181 \pm 0.0011$ [31].

In the case of conventional scale setting, the renormalization scale is simply guessed and set at $\mu_r = \sqrt{s} = m_{\rm Z}$, and thus only one value of $\alpha_{\rm s}$ at scale $m_{\rm Z}$ can be extracted. The resulting predictions does not fit the measured thrust distribution, and it is incorrect for the QED analog. After using the PMC, we obtain a self-consistent determination of $\alpha_{\rm s}$ at different scales over a wide range of the thrust distribution. Moreover, since the PMC predictions eliminate the renormalization scale uncertainty, the extracted values for $\alpha_{\rm s}(Q^2)$ are not plagued by any uncertainty in the choice of $\mu_r$. Thus, remarkably, the PMC provides a new way to determine the running of $\alpha_{\rm s}(Q^2)$ and verify asymptotic freedom from the measurement of jet distributions in $e^+e^-$ annihilation at a single energy of $\sqrt{s}$.

In conclusion, the thrust variable in $e^+e^-$ annihilation provides an ideal platform for testing the QCD. In the case of the conventional scale setting, the predictions are scheme-and-scale dependent, and do not match the experimental results; the extracted coupling constant deviates from the world average. In contrast, after applying PMC scale-setting, we obtain a comprehensive and self-consistent analysis for the thrust measurements, including both the differential distributions and the mean values. The PMC scale reflects the virtuality of the QCD dynamics, and it correctly sets the number of active quark flavors $n_f$ at every order as a function of the thrust. It allows one to determine $\alpha_s(Q^2)$ at different momentum scales by comparing the PMC predictions with the experiment measured at a single center-of-mass-energy $\sqrt{s}$.

This analysis shows the importance of correct renormalization scale-setting. The PMC method rigorously eliminates an unnecessary theoretical uncertainty for all pQCD predictions, and it has general applicability for all precision tests of QCD. A recent review of the PMC is given in Ref. [39].

The work of SJB is supported in part by the Department of Energy under contract DE-AC02-76SF00515.

# Higgs-boson, tau-lepton, and Z-boson decay rates in fourth order and the five-loop $\beta$ function of QCD

P. A. Baikov

*Skobeltsyn Institute of Nuclear Physics, Lomonosov MSU, 1(2), Moscow 119991, Russian Federation*

K.G. Chetyrkin and **J. H. Kühn**

*Institut für Theoretische Teilchenphysik, KIT, 726128 Karlsruhe, Germany*

**Abstract:** The Higgs-boson decay rates into $b\bar{b}$ and into $gg$ have been evaluated in N$^4$LO, corresponding to order $\alpha_s^4$ for $b\bar{b}$ and order $\alpha_s^6$ for $gg$ final states. After inclusion of the four-loop term, nice stabilization of the series is observed. In a similar context the predictions for the $\tau-$ and the Z-decay rate, as well as the $R$-ratio measured in electron-positron annihilation are presented in order $\alpha_s^4$. Similar methods are employed for the evaluation of the beta function which governs the running of the quark-gluon coupling in quantum chromodynamics. The five-loop term of this fundamental quantity has been evaluated and the result has quickly been confirmed and even extended to a general gauge group. This five-loop term leads to a further reduction of the theory uncertainty in $\alpha_s$, evaluated at the Z-boson or Higgs-boson scale, if originally extracted from $\tau$-lepton decays and subsequently evolved to $m_Z$ or $m_Z$.

## Higgs-boson decays

The two dominant decay modes of the Higgs boson are the decay into two gluons and the decay into $b\bar{b}$. With branching ratios of approximately 8% and 65% respectively these are the two most important channels. The decay rate into two gluons is given by [1]

$$\Gamma(H \to gg) = K \frac{G_F m_Z^3}{36\pi\sqrt{2}} \left(\frac{\alpha_s^{(n_l)}(m_Z)}{\pi}\right)^2, \tag{1}$$

$$\begin{aligned} K &= 1 + 17.9167\, a'_s + (156.81 - 5.7083 \ln \frac{m_t^2}{m_Z^2})(a'_s)^2 \\ &\quad + (467.68 - 122.44 \ln \frac{m_t^2}{m_Z^2} + 10.94 \ln^2 \frac{m_t^2}{m_Z^2})\, (a'_s)^3 \\ &= 1 + 0.65038 + 0.20095 + 0.01825, \end{aligned} \tag{2}$$

where $m_t = 175$ GeV, $m_Z = 125$ GeV and $a'_s = \alpha_s^{(5)}(m_Z)/\pi = 0.0363$ has been adopted. The next term, proportional $\alpha_s^6$ and corresponding to N$^4$LO can be found in [2].

The dominant decay channel of the Higgs boson is the one into bottom quarks with a rate given by

$$\Gamma(H \to b\bar{b}) = \frac{G_F m_Z}{4\sqrt{2}\pi} m_b^2 \tilde{R}(s = m_Z^2). \tag{3}$$

Here $\tilde{R}$ stands for the absorptive part of the scalar correlator [3]

$$\begin{aligned}\tilde{R} &= 1 + 5.6667 a_s + 29.147 a_s^2 + 41.758 a_s^3 - 825.7 a_s^4 \\ &= 1 + 0.2041 + 0.0379 + 0.0020 - 0.0014,\end{aligned} \quad (4)$$

where $a_s(m_Z) = \alpha_s(m_Z)/\pi = 0.0360$ and $m_Z = 125$ GeV has been adopted for the numerical evaluation. For the $b$ quark mass we start from the input value

$$m_b(10 \text{ GeV}) = \left(3610 - \frac{\alpha_s - 0.1189}{0.02} 12 \pm 11\right) \text{ MeV}, \quad (5)$$

and evolve to $m_Z = 125$ GeV, arriving [4] at a value

$$m_b(m_Z) = (2771 \pm 8|_{m_b} \pm 15_{\alpha_s}) \text{ MeV}.$$

Last not least there are four-loop corrections to the hadronic decay rate of the Higgs boson which are induced by effective couplings of the Higgs boson to bottom quarks and to gluons and which are mediated by the top quark. These terms have been evaluated to order $\alpha_s^4$ in Ref. [5] and we refer to this paper for details.

## Hadronic $Z-$ and $\tau-$decay rates and the $R$-ratio in order $\alpha_s^4$

Similar methods have been employed for the evaluation of $\mathcal{O}(\alpha_s^4)$ corrections to the ratio $R = \sigma(e^+e^- \to hadrons)/\sigma(e^+e^- \to \mu^+\mu^-)$ at low energies, for the decay rate of the $Z$-boson and for the decay rate of the $\tau$ lepton into hadrons [6,7]. These results have been recently confirmed by an independent calculation [2].

In total, one finds for the QCD corrected decay rate of the $Z$ boson (neglecting for the moment mass suppressed terms of $\mathcal{O}(m_b^2/m_Z^2)$ and electroweak corrections)

$$R^{\text{nc}} = 3\left[\sum_f v_f^2 r_{\text{NS}}^V + \left(\sum_f v_f\right)^2 r_{\text{S}}^V + \sum_f a_f^2 r_{\text{NS}}^A + r_{\text{S;t,b}}^A\right]. \quad (6)$$

The relative importance of the different terms is best seen from the results of the various $r$-ratios introduced above. In numerical form [7]

$$\begin{aligned}r_{\text{NS}} &= 1 + \alpha_s + 1.4092\,\alpha_s^2 - 12.7671\,\alpha_s^3 - 79.9806\,\alpha_s^4 , \\ r_{\text{S}}^V &= -0.4132\,\alpha_s^3 - 4.9841\,\alpha_s^4 , \\ r_{\text{S;t,b}}^A &= (-3.0833 + l_t)\,\alpha_s^2 + (-15.9877 + 3.7222\,l_t + 1.9167\,l_t^2)\,\alpha_s^3 \\ &\quad + (49.0309 - 17.6637\,l_t + 14.6597\,l_t^2 + 3.6736\,l_t^3)\,\alpha_s^4 ,\end{aligned} \quad (7)$$

with $a_s = \alpha_s(m_Z)/\pi$ and $l_t = \ln(m_Z^2/m_t^2)$. Using for the pole mass $m_t$ the value 172 GeV, the axial singlet contribution in numerical form is given by

$$r_{\text{S;t,b}}^A = -4.3524\,\alpha_s^2 - 17.6245\,\alpha_s^3 + 87.5520\,\alpha_s^4 . \quad (8)$$

Let us recall the basic aspects of these results:

- The non-singlet term dominates all different channels. It starts in Born approximation and is identical for $\tau$ decay, for $\sigma(e^+e^- \to hadrons)$ through the vector current (virtual photon) and for $\Gamma(Z \to hadrons)$ through vector and axial current.

- The singlet axial term starts in order $\alpha_s^2$, is present in $Z \to hadrons$ and depends on $\ln(m_Z^2/m_t^2)$. Its origin is the strong imbalance between the masses of top and bottom quarks [8].

- The singlet vector term is present both in $\gamma^* \to hadrons$ and $Z \to hadrons$ and starts in $\mathcal{O}(\alpha_s^3)$.

- All three terms are known up to order $\alpha_s^4$ and the total rate is remarkably stable under scale variations.

The perturbative corrections to the $\tau$ decay rate can be obtained either from fixed order perturbation theory or with "Contour Improvement" [9,10]. Within the two schemes one finds for the perturbative corrections [6]

$$\delta_0^{FO} = a_s + 5.202\, a_s^2 + 26.366\, a_s^3 + 127.079\, a_s^4, \tag{9}$$
$$\delta_0^{CI} = 1.364\, a_s + 2.54\, a_s^2 + 9.71\, a_s^3 + 64.29\, a_s^4. \tag{10}$$

Using the input discussed in [6], one obtains

$$\alpha_s(m_\tau) = 0.332 \pm 0.005|_{exp} \pm 0.015|_{th}. \tag{11}$$

Applying four-loop running and matching this corresponds to

$$\alpha_s(m_Z) = 0.1202 \pm 0.0019 \tag{12}$$

nicely consistent with other determinations.

## Five-Loop Running of the QCD Coupling Constant

Asymptotic freedom, manifested by a decreasing coupling with increasing energy, can be considered as the basic prediction of nonabelian gauge theories [11,12]. The dominant, leading order prediction was quickly followed by the corresponding two-loop [13,14] and three-loop [15,16] results. The next, four-loop calculation was performed almost twenty years later [17] and confirmed in [18]. These results have moved the theory from qualitative agreement with experiment, as observed on the basis of the early results, to precise quantitative predictions, valid over a wide kinematic range, from $\tau$-lepton decays up to LHC results.

There are, of course, a number of phenomenological applications of the five-loop result. On the one hand there is the relation between $Z$-boson and $\tau$-lepton decay rates into hadrons, which involves the strong coupling at two vastly different scales. On the other hand there is the Higgs boson decay rate into bottom quarks and into gluons, which are sensitive to the five-loop running of the QCD coupling.

Let us start with the definition of the beta function

$$\beta(a_s) = \mu^2 \frac{d}{d\mu^2} a_s(\mu) = -\sum_{i \geq 0} \beta_i a_s^{i+2} \tag{13}$$

which describes the running of the quark-gluon coupling $a_s \equiv \alpha_s/\pi$ as a function of the normalization scale $\mu$.

The QCD $\beta$-function in five-loop order reads [19,20,21]

$$\beta_0 = \frac{1}{4}\left\{11 - \frac{2}{3}n_f,\right\}, \quad \beta_1 = \frac{1}{4^2}\left\{102 - \frac{38}{3}n_f\right\}, \quad \beta_2 = \frac{1}{4^3}\left\{\frac{2857}{2} - \frac{5033}{18}n_f + \frac{325}{54}n_f^2\right\},$$

$$\beta_3 = \frac{1}{4^4}\left\{\frac{149753}{6} + 3564\zeta_3 - \left[\frac{1078361}{162} + \frac{6508}{27}\zeta_3\right]n_f + \left[\frac{50065}{162} + \frac{6472}{81}\zeta_3\right]n_f^2 + \frac{1093}{729}n_f^3\right\},$$

$$\begin{aligned}
\beta_4 = \frac{1}{4^5}&\left\{\frac{8157455}{16} + \frac{621885}{2}\zeta_3 - \frac{88209}{2}\zeta_4 - 288090\zeta_5\right. \\
&+ n_f\left[-\frac{336460813}{1944} - \frac{4811164}{81}\zeta_3 + \frac{33935}{6}\zeta_4 + \frac{1358995}{27}\zeta_5\right] \\
&+ n_f^2\left[\frac{25960913}{1944} + \frac{698531}{81}\zeta_3 - \frac{10526}{9}\zeta_4 - \frac{381760}{81}\zeta_5\right] \\
&+ n_f^3\left[-\frac{630559}{5832} - \frac{48722}{243}\zeta_3 + \frac{1618}{27}\zeta_4 + \frac{460}{9}\zeta_5\right] \\
&+ \left.n_f^4\left[\frac{1205}{2916} - \frac{152}{81}\zeta_3\right]\right\},
\end{aligned}$$

where $n_f$ denotes the number of active quark flavors. As expected from the three and four-loop results, the higher transcendentalities $\zeta_6$ and $\zeta_7$ that could be present at five-loop order are actually absent.

The coefficients are surprisingly small. For example, for the particular cases of $n_f = 3, 4, 5,$ and 6 we get:

$$\begin{aligned}
\overline{\beta}(n_f = 3) &= 1 + 1.78\,a_s + 4.47\,a_s^2 + 20.99\,a_s^3 + 56.59\,a_s^4, \\
\overline{\beta}(n_f = 4) &= 1 + 1.54\,a_s + 3.05\,a_s^2 + 15.07\,a_s^3 + 27.33\,a_s^4, \\
\overline{\beta}(n_f = 5) &= 1 + 1.26\,a_s + 1.47\,a_s^2 + 9.83\,a_s^3 + 7.88\,a_s^4, \\
\overline{\beta}(n_f = 6) &= 1 + 0.93\,a_s - 0.29\,a_s^2 + 5.52\,a_s^3 + 0.15\,a_s^4,
\end{aligned}$$

where $\overline{\beta} \equiv \frac{\beta(a_s)}{-\beta_0 a_s^2} = 1 + \sum_{i\geq 1}\overline{\beta}_i a_s^i$.

At this point it may be useful to present the impact of the five-loop term on the running of the strong coupling from low energies, say $\mu = m_\tau$, up to the high energy region $\mu = m_Z$, by comparing the predictions based on three and four versus five-loop results [*]. We start from the scale of $m_\tau$ with $\alpha_s^{(3)}(m_\tau) = 0.33$ (as given in [23]) and evolve the coupling up to 3 GeV. At this point the four-loop matching from 3 to 4 flavours is performed. The strong coupling now runs up to $\mu = 10$ GeV and, at this point, the number of active quark flavours is switched from the 4 to 5. Subsequently, the strong coupling runs again up to $m_Z$ and, finally, up to the Higgs mass $m_Z = 125$ GeV. The relevant values of $\alpha_s$ are listed in Table 1. The combined uncertainty in $\alpha_s^{(5)}(m_Z)$ induced by running and matching can be conservatively estimated by the shift in $\alpha_s^{(5)}(m_Z)$ produced by the use of five-loop running (and, consequently) four-loop matching instead of four-loop running (and three-loop matching). It amounts to a minute $6 \cdot 10^{-5}$ which is by a factor of three less than the similar shift made by the use of four-loop running instead of the three-loop one (see Table 1). Note that the final value of $\alpha_s^{(5)}(m_Z)$ which follows from $\alpha_s^{(3)}(m_\tau)$ is in remarkably good agreement with the fit to electroweak precision data (collected in $Z$ boson decays), namely [24]:

$$\alpha_s^{(5)}(m_Z) = 0.1196 \pm 0.0030. \tag{14}$$

---

[*] For all practical examples in this paper we have used an extended version of the package RunDec [22].

Table 1: Running of $\alpha_s$ from $\mu = m_\tau$ to $\mu = m_Z$. For the threshold values of $c$ and $b$ quarks we have chosen [25,26] $m_c(3 \text{ GeV}) = 0.986$ GeV and $m_b(10 \text{ GeV}) = 3.160$ GeV respectively.

| # of loops | $\alpha_s^{(3)}(m_\tau)$ | $\alpha_s^{(5)}(m_Z)$ | $\alpha_s^{(5)}(m_Z)$ |
|---|---|---|---|
| 3 | $0.33 \pm 0.014$ | $0.1200 \pm 0.0016$ | $0.1145 \pm 0.0014$ |
| 4 | $0.33 \pm 0.014$ | $0.1199 \pm 0.0016$ | $0.1143 \pm 0.0014$ |
| 5 | $0.33 \pm 0.014$ | $0.1198 \pm 0.0016$ | $0.1143 \pm 0.0014$ |

Thus, exact result for the five-loop term of the QCD $\beta$-function allows to relate the strong coupling constant $\alpha_s$, as determined with N³LO accuracy at low energies, say $m_\tau$ with the strong coupling as evaluated at high scales, say $m_Z$ or $m_H$. Including the exact five-loop term has little influence on the central value of the prediction, a consequence of partial cancellations between various contributions from matching and running. However, the five-loop result leads to a considerable further reduction of the theory uncertainty and allows to combine values from low and high energies of appropriate order.

# $\alpha_\text{s}$, ABM PDFs, and heavy-quark masses

**S. Alekhin**[1,2], J. Blümlein[3], and S.-O. Moch[1]

[1] *Institut für Theoretische Physik, Universität Hamburg, D-22761 Hamburg, Germany*
[2] *Institute for High Energy Physics,142281 Protvino, Russia*
[3] *Deutsches Elektronensynchrotron DESY, Platanenallee 6, D15738 Zeuthen, Germany*

**Abstract:** The value of $\alpha_\text{s}$ is extracted from a global QCD analysis of experimental data on inclusive neutral-current (NC) and charged-current (CC) deep-inelastic scattering (DIS), $c$- and $b$-quark production in the NC DIS, $c$-quark production in the CC DIS, and W-, Z-boson, and $t$-quark production in (anti)proton-proton collisions with a simultaneous extraction of parton distribution functions (PDFs). The NNLO value of $\alpha_\text{s}^{(n_f=5)}(m_\text{Z}) = 0.1147 \pm 0.0008\,(\text{exp.}) \pm 0.0022\,(\text{h.o.})$ is obtained with the uncertainty due to missing higher-orders (h.o.) being estimated as one half of the difference between the values of $\alpha_\text{s}$ obtained in the NNLO and NLO variants of this fit. The masses of the heavy-quarks, charm, beauty and top, which are determined in parallel, are employed for cross-check of the theoretical framework consistency of the analysis.

The ABMP16 PDF fit [1] is based on a combination of experimental data on hadronic hard-scattering processes: inclusive neutral-current (NC) and charged-current (CC) deep-inelastic scattering (DIS), $c$- and $b$-quark production in the NC DIS, $c$-quark production in the CC DIS, and W-, Z-boson and $t$-quark production in (anti)proton-proton collisions. A variety of processes provides a complementary set of constraints on the PDFs and the parameters of QCD Lagrangian, which are required for a consistent interpretation of the data, in particular, the heavy-quark masses and $\alpha_\text{s}$. The value of $\alpha_\text{s}$ determined from this fit is predominantly driven by the NC DIS data, which cover a wide range of the momentum transfer squared $Q^2 = 2.5 \div 50000\,\text{GeV}^2$ and can be nicely described by perturbative QCD with the corrections up to next-to-next-to-leading-order (NNLO) taken into account [2] and the heavy flavor corrections [3]. However, at the lowest end of this range the leading-twist (LT) PDF term is accompanied by substantial contributions from the higher-twist (HT) operators [4]. The latter introduce an additional power-like dependence on $Q^2$, which spoils the purely logarithmic behavior of the leading-twist part and, as a result, shifts the fitted value of $\alpha_\text{s}$ upwards [5]. Therefore in order to provide an unbiased determination one has to eliminate the impact of the HT terms either by cutting on the potentially problematic kinematic region or by parameterizing and fitting them in parallel with the LT PDFs. The ABMP16 fit is based on the latter approach, while it has also been checked that the former one provides a consistent value of $\alpha_\text{s}$, cf. Table 1. The HT terms appear at large Bjorken $x$ therefore their isolation can be performed with a cut of $W^2 > 12.5\,\text{GeV}^2$, where $W$ is the invariant mass of the hadronic system. However, such a cut does not affect the small-$x$ part of the HT terms, which manifests itself in the NMC and the HERA data [7]. Thus, in order to allow for a pure leading-twist theoretical treatment of the available DIS data an additional cut of $Q^2 > 10\,\text{GeV}^2$ is also required.

The values of $\alpha_\text{s}$ preferred by four groups of the inclusive DIS data, SLAC, BCDMS, NMC, and HERA, which are used in the ABMP16 fit, are displayed in Fig. 1 in their historical perspective. The earliest experiments, which were performed in SLAC, prefer somewhat larger $\alpha_\text{s}$, while they are also more sensitive to the HT contribution because of the kinematic limitations caused by the relatively low beam energy. The most recent HERA data prefer a smaller value of $\alpha_\text{s}$ with a marginal sensitivity to the HT contribution. It is worth noting that $\alpha_\text{s}$ extracted from the combined Run I+II HERA data is somewhat larger than the one obtained from the earlier Run I sample, which was employed in the earlier version of the ABMP16 PDF fit [8]. Due to the update of the HERA data the value of $\alpha_\mathrm{s}$ moves somewhat up, although still lower than the world average, cf. Fig. 1.

Table 1: The values of $\alpha_s^{(n_f=5)}(m_Z)$ obtained in the NLO and NNLO variants of the ABMP16 fit with various kinematic cuts on the DIS data imposed and different modeling of the higher twist terms. Table from Ref. [6].

| fit ansatz | | $\alpha_s(m_Z)$ | |
| --- | --- | --- | --- |
| higher twist modeling | cuts on DIS data | NLO | NNLO |
| higher twist fitted | $Q^2 > 2.5$ GeV$^2$, $W > 1.8$ GeV | 0.1191(11) | 0.1147(8) |
| higher twist fixed at 0 | $Q^2 > 10$ GeV$^2$, $W^2 > 12.5$ GeV$^2$ | 0.1212(9) | 0.1153(8) |
| | $Q^2 > 15$ GeV$^2$, $W^2 > 12.5$ GeV$^2$ | 0.1201(11) | 0.1141(10) |
| | $Q^2 > 25$ GeV$^2$, $W^2 > 12.5$ GeV$^2$ | 0.1208(13) | 0.1138(11) |

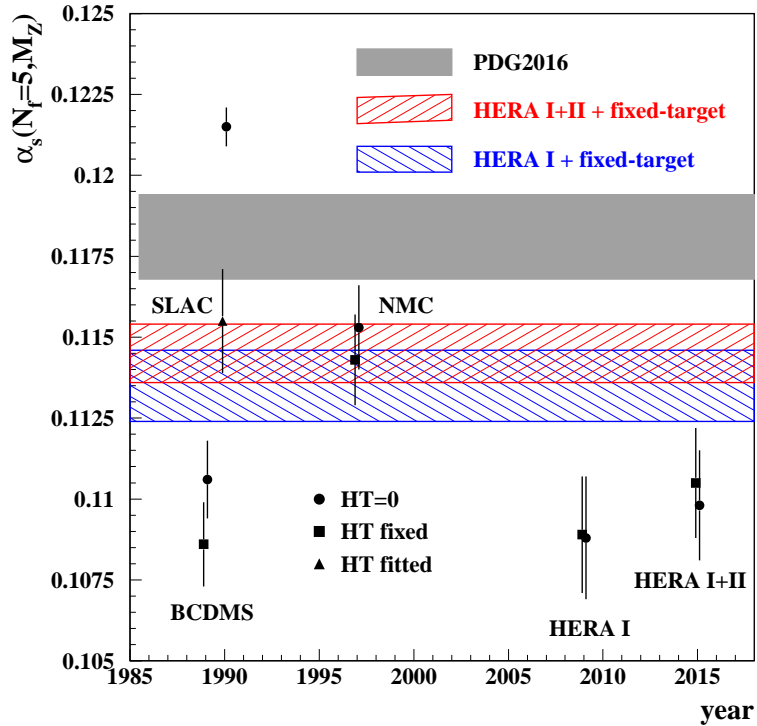

Figure 1: The value of $\alpha_s$ preferred by various DIS data samples employed in the ABMP16 analysis as a function of the year of publication of the data. Three variants of the fit with different treatments of the HT terms are presented: HT set to 0 or to the one obtained in the combined fit (circles and squares, respectively) or fitted to one particular data set (triangles). The $\alpha_s$ bands obtained using combination of the fixed-target SLAC, BCDMS, and NMC samples with the ones from the HERA Run-I (left-tilted hatches) and Run-I+II (right-tilted hatches) as well as the PDG2016 average [9] are given for comparison. Plot from Ref. [1].

An important aspect of the small-$x$ DIS data interpretation is to account for the heavy-quark contribution. In particular final-state configurations including the $c$-quark are responsible for an essential part of the NC inclusive cross section in the region of HERA kinematics. Therefore, an accurate treatment of this term is a necessary ingredient of the related phenomenology [10]. This applies also to the extraction of $\alpha_\mathrm{s}$ from a combination of the DIS data, which include the small-$x$ HERA sample [11]. In this part the ABMP16 fit is based on the fixed-number-flavor (FFN) scheme, which implies only massless partons, gluon and three light quarks, in the initial state, while the heavy-quark contribution is computed within the photon-gluon fusion mechanism including the higher-order QCD corrections up to the NNLO. Furthermore, the $\overline{\mathrm{MS}}$ definition of the heavy-quark mass, which improves the perturbative convergence, is applied in the ABMP16 fit [12]. The relevance of the FFN approach in such a formulation is supported by a good description of the Run I HERA data on the semi-incisive $c$-quark production used in the ABMP16 fit and a good agreement with the more recent Run I+II data [13]. Moreover, the $\overline{\mathrm{MS}}$ value of the $c$-quark mass

$$m_c(m_c) = 1.252 \pm 0.018 \text{ GeV}$$

obtained in the fit simultaneously with $\alpha_\mathrm{s}$ and the PDF parameters is in a good agreement with other determinations [9] that also underpins the consistency of the FFN scheme in the application to the analysis of existing data on $c$-quark DIS production.

The data on hadronic $t$-quark pair production cross sections, which are used in the ABMP16 fit, are also quite sensitive to $\alpha_\mathrm{s}$ since the leading order cross section of this process is proportional to $\alpha_\mathrm{s}^2$. At the same time it is also sensitive to the gluon distribution and value of the $t$-quark mass $m_t$. Therefore, in order to use the potential of these data in the determination of $\alpha_\mathrm{s}$ one has to fix these two ingredients. The gluon distribution at the relevant kinematics is confined by other data employed in the ABMP16 fit, however, none of them are sensitive to $m_t$. At present, the accurate value of $m_t$ also cannot obtained by direct reconstruction in the experiment due to hadronization effects being still not fully under theoretical control [14]. In view of these limitations the value of $m_t$ is fitted simultaneously with $\alpha_\mathrm{s}$ and the PDFs. As a result, impact of the $t$-quark data on $\alpha_\mathrm{s}$ determination is greatly reduced. Indeed, two determinations,

$$\alpha_\mathrm{s}^{(n_f=5)}(m_\mathrm{Z}) = 0.1145 \pm 0.0009$$

and

$$\alpha_\mathrm{s}^{(n_f=5)}(m_\mathrm{Z}) = 0.1147 \pm 0.0008,$$

obtained with and without using $t$-quark data, respectively, are quite similar, both in the central values and uncertainties. An alternative way of illustrating this effect is presented in Fig. 2, which shows a perfect correlation between $\alpha_\mathrm{s}$ and $m_t$ obtained in the ABMP16 fit. However, it is worth mentioning that fitting $m_t$ within the ABMP16 framework allows for its consistent independent determination. Using likewise to the case of heavy-quark DIS production the $\overline{\mathrm{MS}}$ definition we obtain

$$m_t(m_t) = 160.9 \pm 1.1 \text{ GeV},$$

which corresponds to the pole mass value of

$$m_t^\mathrm{pole} = 170.4 \pm 1.2 \text{ GeV}$$

where the relation between these definitions is known to four loops [15]. The value of $m_t^{\text{pole}}$ obtained in this way is smaller than the values of $m_t$, which are directly measured in experiments by $\mathcal{O}(1\,\text{GeV})$. Other data sets, on the W-, Z-boson and single $t$-quark hadronic production, which are used in the ABMP16 fit, demonstrate even less sensitivity to $\alpha_s$ as compared to the $t$-quark pair production cross sections. Therefore the aggregated value of $\alpha_s$ is essentially determined by the DIS data.

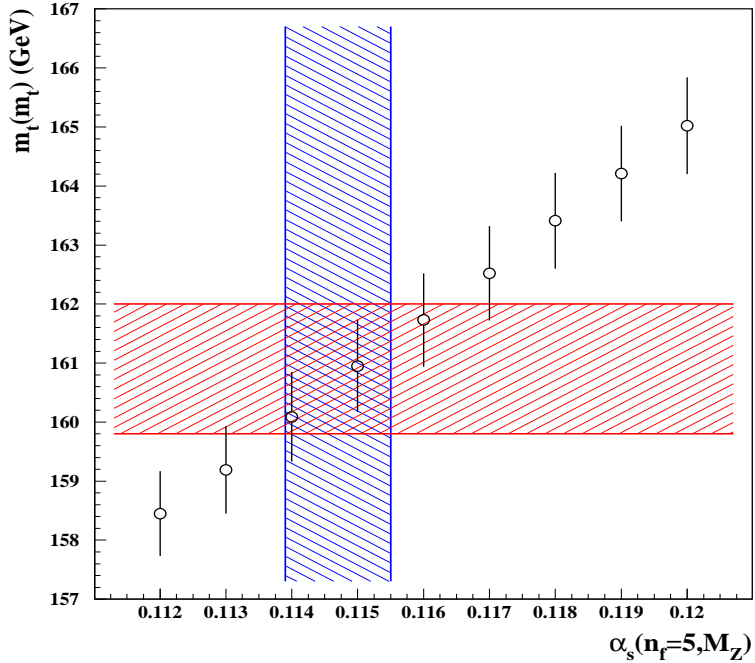

Figure 2: The $\overline{\text{MS}}$ value of the $t$-quark mass $m_t(m_t)$ obtained in the variants of present analysis with the value of $\alpha_s^{(n_f=5)}(m_Z)$ fixed in comparison with the $1\sigma$ bands for $m_t(m_t)$ and $\alpha_s^{(n_f=5)}(m_Z)$ obtained in our nominal fit (left-tilted and right-tilted hatch, respectively). Plot from Ref. [1].

The results of a version of the fit performed with the NLO QCD accuracy [6] can be employed for an estimate of the theoretical uncertainties due to missing higher-order QCD corrections. Taking it as one half of difference between the values obtained in the NNLO and NLO fits we arrive at the following value
$$\alpha_s^{\text{NNLO}}(m_Z) = 0.1147 \pm 0.0008\,(\text{exp.}) \pm 0.0022\,(\text{h.o.}).$$

# $\alpha_\mathrm{s}$ from jet cross section measurements in deep-inelastic $ep$ scattering

Daniel Britzger*

*Max-Planck-Institut für Physik, Föhringer Ring 6, 80805 München, Germany*

**Abstract:** The value of $\alpha_\mathrm{s}(m_\mathrm{Z})$ is determined from inclusive jet and di-jet cross sections in neutral-current deep-inelastic $ep$ scattering (DIS) measured at HERA by the H1 collaboration using next-to-next-to-leading order (NNLO) perturbative QCD predictions. Using inclusive jet and di-jet data together, the strong coupling constant is determined to be $\alpha_\mathrm{s}(m_\mathrm{Z}) = 0.1157(20)_\mathrm{exp}(29)_\mathrm{th}$. Complementary, $\alpha_\mathrm{s}(m_\mathrm{Z})$ is determined together with parton distribution functions of the proton (PDFs) from jet and inclusive DIS data and the value is determined to be $\alpha_\mathrm{s}(m_\mathrm{Z}) = 0.1142(28)_\mathrm{tot}$. Both results are found to be consistent. The running of the strong coupling is tested at different values of the renormalisation scale and the results are found to be in agreement with expectations.

## Introduction

Jet production cross sections in neutral-current deep-inelastic scattering (NC DIS) are measured in the Breit frame, where the virtual photon and the proton collide head on. These measurements are directly sensitive to the value of $\alpha_\mathrm{s}(m_\mathrm{Z})$, since the predictions in perturbative QCD are proportional to $\mathcal{O}(\alpha_s)$ already at leading order. Inclusive jet and di-jet cross sections have been measured by the H1 experiment in $ep$ collisions at HERA in the years 1995–2007 [1,2,3,4,5], at center-of-mass energies of $\sqrt{s} = 300$ GeV and 320 GeV, and for a wide kinematic range in the photon virtuality $Q^2$. For all data, jets are defined using the $k_t$ jet-algorithm with a parameter $R = 1.0$. Inclusive jet cross sections have been measured double-differentially as a function of $Q^2$ and jet transverse momenta, $P_\mathrm{T}^\mathrm{jet}$, with values typically exceeding $P_\mathrm{T}^\mathrm{jet} \gtrsim 5$ GeV, and di-jet cross sections as a function of $Q^2$ and the average $P_\mathrm{T}^\mathrm{jet}$ of the two leading jets, $\langle P_\mathrm{T}^\mathrm{jet} \rangle$. Already in the past, all these data have been used for determinations of $\alpha_\mathrm{s}(m_\mathrm{Z})$ using next-to-leading order pQCD predictions. In the work presented [6], the cross section predictions are performed now for the first time in next-to-next-to-leading order (NNLO) accuracy. These calculations are implemented in the program NNLOJET [7,8], and the coefficients are stored in the fastNLO format [9] to enable a repeated calculation with different values of $\alpha_\mathrm{s}(m_\mathrm{Z})$. Using these new and improved predictions, the strong coupling constant $\alpha_\mathrm{s}(m_\mathrm{Z})$ is determined in two approaches.

## The $\alpha_s$-fit

In the first approach, which is denoted as '$\alpha_s$-fit', the value of $\alpha_\mathrm{s}(m_\mathrm{Z})$ is determined in a fit of NNLO predictions to the inclusive jet and di-jet data, where a statistical goodness-of-fit quantity, $\chi^2$, is minimised. In this fit, both of the $\alpha_s$-dependencies in the predictions, namely in the partonic cross sections and in the PDF, are taken into account. The latter is accounted for by setting the DGLAP-evolution starting-scale to $\mu_0 = 20$ GeV, and thus, the $\alpha_s$-dependence of the evolution kernel can also be considered in the fit. For the central result, the NNPDF3.1 PDF set is used [10]. The renormalisation and factorisation scales are chosen to be $\mu_\mathrm{R}^2 = \mu_\mathrm{F}^2 = Q^2 + P_\mathrm{T}^2$, where $P_\mathrm{T}$ denotes $P_\mathrm{T}^\mathrm{jet}$ in case of inclusive jet cross sections, and $\langle P_\mathrm{T}^\mathrm{jet} \rangle$ in case of di-jets. Subsequently, a representative scale value $\tilde{\mu}$, which is closely related to $\mu_R$ and $\mu_F$, is assigned to each data point and is used for additional cuts, as discussed below.

---

*On behalf of the H1 and NNLOJET collaborations.

In the fits of $\alpha_s(m_Z)$ to each of the nine individual data sets it is found that the results are all consistent, and the data are all found to be well described by the NNLO predictions [6]. The smallest experimental uncertainty ('exp') is then achieved in a fit to all inclusive jet and di-jet cross section data[†], denoted as 'H1 jets', with a value $\alpha_s(m_Z) = 0.1143\,(9)_{\text{exp}}\,(43)_{\text{th}}$. The theoretical uncertainty ('th') comprises multiple uncertainties: various uncertainties of the PDFs (called 'PDF', 'PDF$\alpha_s$', and 'PDFset'), hadronisation uncertainties ('had'), and scale uncertainties. The latter are the dominant source of uncertainty. The main result of this approach is obtained from H1 jet data restricted to $\tilde{\mu} > 28$ GeV. In this fit the value of $\alpha_s(m_Z)$ is determined to

$$\alpha_s(m_Z) = 0.1157\,(20)_{\text{exp}}\,(3)_{\text{PDF}}\,(2)_{\text{PDF}\alpha_s}\,(3)_{\text{PDFset}}\,(6)_{\text{had}}\,(27)_{\text{scale}},$$

while this fit also exhibits a very good agreement of the NNLO predictions with the data, indicated by $\chi^2 = 63.2$ for 91 data points. The cut value of $\tilde{\mu} > 28$ GeV was chosen such that the scale and experimental uncertainty are somewhat balanced: a lower cut on $\tilde{\mu}$ would result in smaller experimental uncertainties, but larger scale uncertainties, and vice-versa.

The fits are repeated with data samples restricted to adjacent small intervals in $\tilde{\mu}$. The resulting values of $\alpha_s(m_Z)$, together with the respective value of $\alpha_s(\mu_R)$, are displayed at a representative value $\tilde{\mu}$ in Fig. 1. This study provides an important consistency test of the data and predictions, and the result nicely illustrates the running of $\alpha_s$. The values are found to be in good agreement with the expectation according to the RGE and with $\alpha_s$-determinations in other reactions at NNLO.

## The PDF+$\alpha_s$-fit

In a complementary approach, which is then denoted as 'PDF+$\alpha_s$' fit, the value of $\alpha_s(m_Z)$ is determined together with the non-perturbative PDFs. For this analysis, normalised inclusive jet and di-jet cross sections [3,4,5], i.e. the jet data are normalised to the inclusive NC DIS cross sections, are exploited, together with all of the H1 inclusive NC and CC DIS cross section data. The latter data samples are equivalent to the one used in the H1PDF2012 PDF fit [11]. The fit ansatz for the determination of the PDF parameters and $\alpha_s(m_Z)$ follows closely the methodology of previous PDF studies, such as HERAPDF2.0 [12] or H1PDF2012. PDFs are parameterised at a starting scale with 12 fit parameters and the DGLAP formalism is employed. All predictions are again performed in NNLO QCD, and the value of $\alpha_s(m_Z)$ is determined in this PDF+$\alpha_s$ fit to

$$\alpha_s(m_Z) = 0.1142\,(11)_{\text{exp,had,PDF}}\,(2)_{\text{mod}}\,(2)_{\text{par}}\,(26)_{\text{scale}},$$

where 'mod' and 'par' denote the model and parameterisation uncertainties, similar to the HERA-PDF2.0 approach. The scale uncertainty is estimated by repeating the fit with scale factors 0.5 and 2 applied to $\mu_R$ and $\mu_F$ for all calculations involved. In this fit, a good agreement of predictions and data is found with a value $\chi^2/n_{\text{dof}} = 1539.7/(1529-13)$. Details of the resulting set of PDFs, denoted as 'H1PDF2017 [NNLO]', are omitted here since this discussion is beyond the scope of the workshop, but these are presented and discussed in greater detail in Ref. [6]. Highlighting one of these results, a simultaneous determination of $\alpha_s$ and the PDFs from HERA inclusive data alone is found not to be reliable, wheras good accuracy on both is reached when including HERA jet data.

---

[†]Some di-jet data are omitted, since their statistical correlations with the respective inclusive jet data have not been determined, and additionally, all data are restricted to twice the $b$-quark mass, $\tilde{\mu} > 2m_b$, since the NNLO predictions are performed with five massless quark flavours.

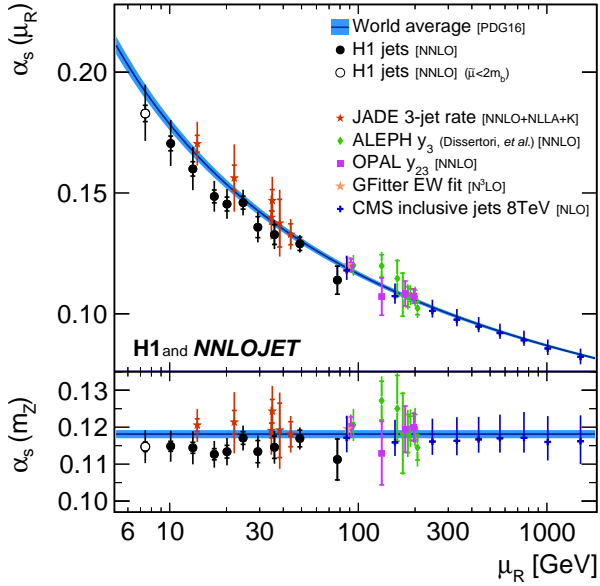
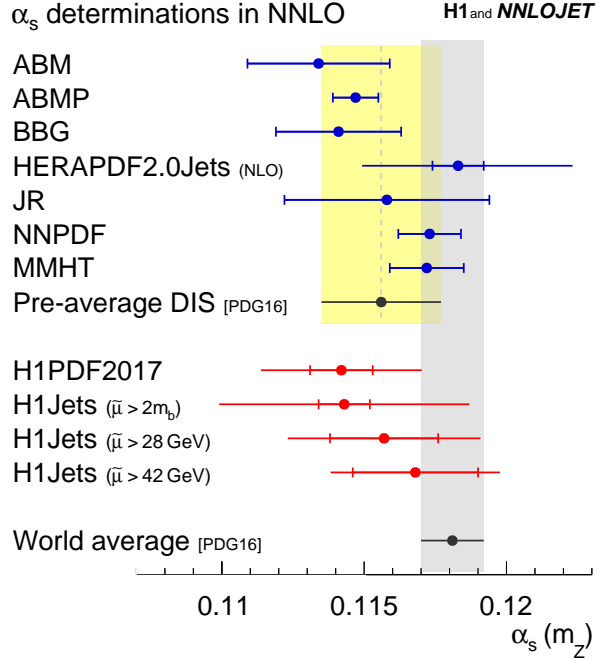

Figure 1: Values of $\alpha_s(m_Z)$ obtained from fits to H1 jet data with similar values of $\mu_R$ (full circles) in comparison to values from other experiments and processes at least in NNLO. The values of $\alpha_s(\mu_R)$ (upper panel) are related to $\alpha_s(m_Z)$ by the solution of the QCD renormalisation group equation as it also enters the predictions. The inner error bars indicate the experimental uncertainties and the outer error barrs the total uncertainties.

Figure 2: Summary of $\alpha_s(m_Z)$ values obtained in the PDF+$\alpha_s$-fit ('H1PDF2017'), and in the alternative $\alpha_s$-fits to H1 jet data ('H1Jets'). The inner error bars indicate the experimental uncertainty and the outer error bars the total uncertainty. The results are compared to the world average and to other determinations from global fits in NNLO and including DIS data.

## Summary

To summarise, the value of $\alpha_s(m_Z)$ has been determined using the new NNLO pQCD predictions for jet production cross section in NC DIS, and analysing the vast majority of inclusive jet and di-jet cross section data published previously by H1. Alternatively, the entirety of inclusive NC and CC DIS cross section data recorded by the H1 experiment, together with normalised jet data, has been considered in an complementary approach, where also the PDFs are determined. The results of the PDF+$\alpha_s$-fit and of the $\alpha_s$-fit are presented in Fig. 2, and compared to the world average value and other analyses [13]. All values are found to be consistent with each other, and the results are found to be consistent with the world average value, although with a tendency to be a bit lower. It was found, that the results yield competitive experimental uncertainties in comparison to the present uncertainty of the world average. The jet data analysed here probe at high experimental and reasonable theoretical precision the region of low scales from about 7 to 90 GeV, where precision data are sparse. Despite of the use of NNLO calculations for this analysis, the dominant uncertainties for most of the results are the scale uncertainty associated to the NNLO predictions. Since the scale uncertainties exceed considerably the experimental uncertainties, improved predictions may yield significantly smaller total uncertainties in the future.

Some improvements of the experimental uncertainties can be possibly achieved by including data from the ZEUS experiment in the analysis. Further improvements are expected from analysing three-jet cross sections, which have an increased sensitivity to $\alpha_\mathrm{s}(m_\mathrm{z})$ since their LO predictions are proportional to $\mathcal{O}(\alpha_s^2)$. Unfortunately, these three-jet predictions are not available yet in NNLO. Event-shape measurements in NC DIS represent alternative observables, which may yield competitive precision in $\alpha_\mathrm{s}(m_\mathrm{z})$. In this case, such data are expected to have large overlap with the jet data analysed here, and also a similar $\alpha_s$ sensitivity. Therefore, most important for future improvements are first of all further reductions of the dominant scale uncertainty, e.g. by adding additional corrections beyond NNLO.

# $\alpha_s$ from parton densities

Joey Huston

*Dept. of Physics and Astronomy, Michigan State University, East Lansing, MI 48824-1116, USA*

**Abstract:** The sensitivity of global parton distribution function (PDF) fits to determine the value of $\alpha_s(m_z)$ is reviewed.

There are over 3500 data points in modern global PDF fits, including data from deep-inelastic scattering (DIS), Drell–Yan (DY) production, inclusive jet and dijet production, and from top production, the latter both singly differential and double-differential in the relevant kinematic variables. The state-of-the art is the production of parton distribution functions (PDFs) at NNLO, although it is common to also produce sets at NLO and LO. The NLO and NNLO PDFs are typically very similar to each other, while the LO PDFs deviate substantially due to the absence of critical higher order corrections. All of the processes in a global PDF fit are sensitive to the value of $\alpha_s(m_z)$, with the power of $\alpha_s(m_z)$ in the prediction depending on the process (and the order). Thus, a global PDF fit can be used in the determination of $\alpha_s(m_z)$.

There are two philosophies in global PDF fitting; either allow $\alpha_s(m_z)$ to be free in the fit, or to to fix its value at some standard, typically the value quoted by the Particle Data Group. The widespread standard is to use a central value of $\alpha_s(m_z)$ of 0.118 (basically an approximation/truncation of the PDG result) and an uncertainty (at the 90% confidence level) of $\pm 0.002$, or $\pm 0.0012$ at the 68% CL. This central value and uncertainty is typically used for both NLO and NNLO global PDF fits. In LO PDF fits, a much larger value is needed, typically in the range 0.130-0.140. (It is difficult to quote an uncertainty for $\alpha_s(m_z)$ for LO PDF fits, just as it is difficult to quote an uncertainty for the PDFs themselves, again due to the deficiencies of the LO matrix elements.) Thus, for example, the PDF4LHC15 PDFs, a combination of the PDFs from the CT, MMHT and NNPDF groups, uses this standard [1].

But, as stated earlier, a global PDF fit can be used for the determination of the $\alpha_s(m_z)$. In fact, previous determinations of $\alpha_s(m_z)$ from the PDG have included the input from PDF fits, though mostly using DIS data [6]. One difficulty, though, is that there is a correlation (or anti-correlation depending on the parton $x$ range) between the value of $\alpha_s(m_z)$ and the strength of the gluon distribution. As the gluon distribution remains with one of the largest uncertainties among the PDFs, this can result in a relatively large uncertainty in the extracted value of $\alpha_s(m_z)$, and thus the philosophy among some groups of instead using the PDG standard value. In any case, it has become the standard among PDF fitting groups to produce PDFs with alternate values of $\alpha_s(m_z)$, in intervals of a multiple of 0.001 above and below the central value, allowing the impact of a different value of $\alpha_s(m_z)$ to be calculated, and thus a determination of the $\alpha_s(m_z)$ uncertainty to go along with the PDF uncertainty.

The gluon distribution at NNLO from MMHT2014 is shown in Fig. 1 for five different values of $\alpha_s(m_z)$, along with the PDF uncertainty for the central value of $\alpha_s(m_z)$ [2]. As stated previously, there is a correlation between the value of $\alpha_s(m_z)$ and the size of the gluon distribution for $x$ values below 0.1, and an anti-correlation for higher $x$.

It is interesting/important that even though the gluon distribution and the value of $\alpha_s(m_z)$ are correlated (or anti-correlated), the uncertainties for those two quantities are un-correlated [3]. Thus, the combined PDF+$\alpha_s(m_z)$ uncertainty can be calculated by computing the one sigma uncertainty with $\alpha_{\rm s}(m_{\rm z})$ fixed at its central value, and adding in quadrature the one-sigma uncertainty in $\alpha_{\rm s}(m_{\rm z})$, and this is the standard that the PDF4LHC working group advocates.

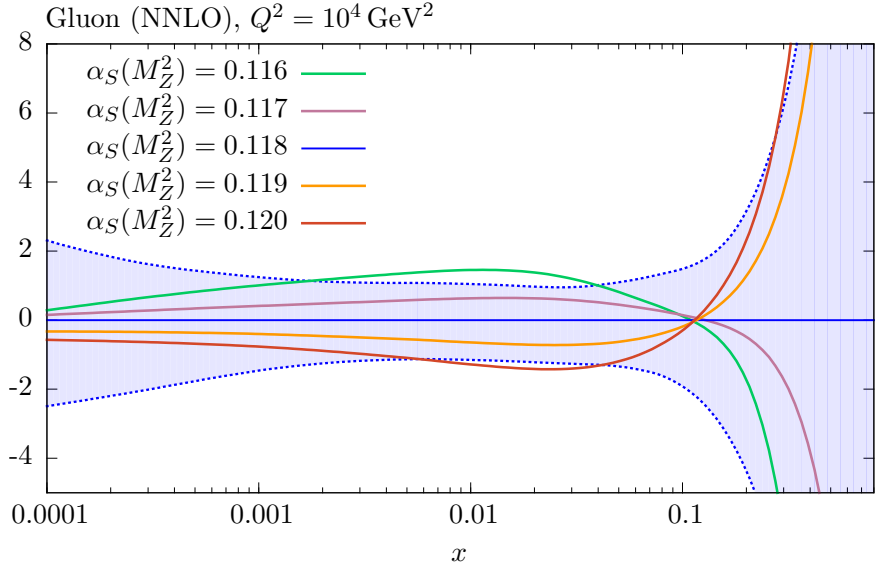

Figure 1: The MMHT NNLO gluon distribution plotted for several values of $\alpha_{\rm s}(m_{\rm z})$, from Ref. [2]. Also shown is the PDF uncertainty for the gluon distribution.

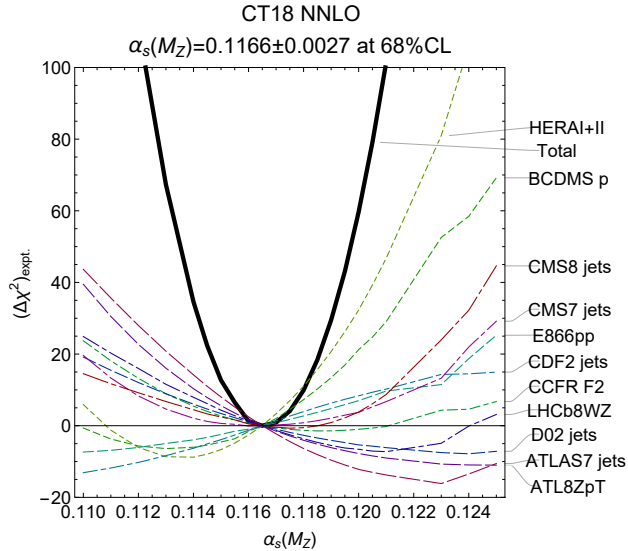

Figure 2: The Lagrange Multiplier study of the sensitivity of the data sets included in CT18 to the value of $\alpha_{\rm s}(m_{\rm z})$.

To make matters more complicated, all of the experiments in a global PDF fit do not speak with one voice, i.e. their preference for the value of $\alpha_{\rm s}(m_{\rm z})$ is not necessarily the same. In Fig. 2 is shown a Lagrange Multiplier study of the preference for the value of $\alpha_{\rm s}(m_{\rm z})$ for some of the data sets included in the CT18 global PDF fit. The $\chi^2$ distribution for the total CT18 data set has a fairly quadratic shape, and a reasonable uncertainty of ±0.0018 (about 1.5 times the uncertainty assumed by the PDF4LHC working group), but the different data sets often prefer a different value, and in some cases the $\chi^2$ distributions are not even quadratic. Some of the strongest constraints come from the LHC jet data and from the HERA Run 1+2 combined data. The former are sensitive to $(\alpha_s(m_z))^2$ at the Born level and have relatively small statistical and systematic errors over a broad kinematic range. The latter have a small sensitivity per point to the value of $\alpha_s(m_z)$ but the large number of points in that data set lead to a significant impact on the value of $\alpha_s(m_z)$ preferred by the CT18 fit. Note that the central value (0.1168) is smaller than the value of 0.118 noted earlier, for both NLO and NNLO. The preference for a smaller value of $\alpha_s(m_z)$ at NNLO compared to NLO has also been observed by MMHT [2].

NNPDF has recently determined a value of $\alpha_s(m_z)$ from their global PDF fit of $0.1185 \pm 0.0012$, or slightly above the world average [5]. See Fig. 3. Their uncertainty is similar to that obtained by the PDG world average. Some care has to be taken, though, for it is difficult to exactly determine a one-sigma error for $\alpha_s(m_z)$ in global PDF fits, due to the issue of tolerance. CT18, for example, has a tolerance in $\Delta\chi^2$ of 100, corresponding to the 90% confidence level, or about 37 at the 68% confidence level. This increased tolerance is motivated by tensions within the data sets (as noted above for the individual experimental sensitivities to the value of $\alpha_s(m_z)$). A decreased tolerance would lead to a decreased uncertainty.

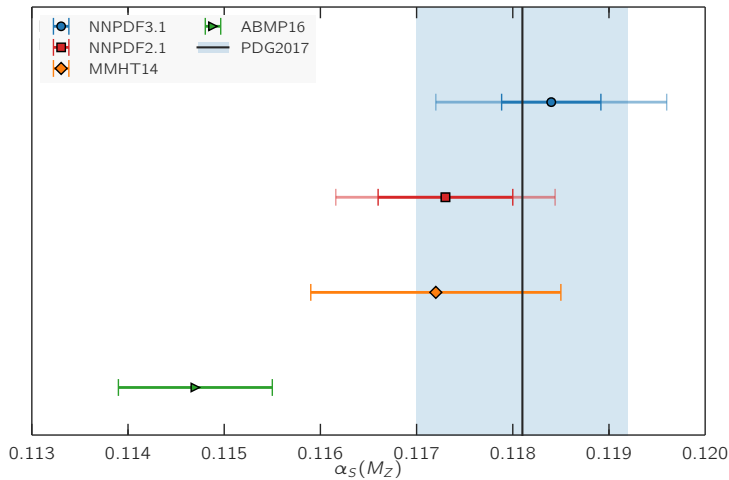

Figure 3: The value of $\alpha_s(m_z)$ and its uncertainty is shown for several PDF fits, and compared with the central value and uncertainty from the world average. From Ref. [5].

To summarize: global PDF fits have a sensitivity to the value of $\alpha_s(m_z)$, somewhat clouded by the remaining sensitivity to the gluon distribution. The deep-inelastic data from HERA will continue to be the most important data set in modern global PDF fits for some time, but the growing number of data sets from the LHC will increase the importance of collider data both for the determination of PDFs and for the determination of $\alpha_s(m_z)$, especially as the full results from the 13 TeV LHC data sets are published in the next few years.

# Old and new observables for $\alpha_\mathrm{s}$ from $e^+e^-$ to hadrons

Adam Kardos[1], Stefan Kluth[2], Zoltán Trócsányi[3,4], **Gábor Somogyi**[3], Zoltan Tulipant[1], and Andrii Verbytskyi[2]

[1]*Institute of Physics, University of Debrecen, 4010 Debrecen, Hungary*
[2]*Max-Planck-Institut für Physik, D-80805 Munich, Germany*
[3]*MTA-DE Particle Physics Research Group, University of Debrecen, 4010 Debrecen, Hungary*
[4]*Institute for Theoretical Physics, Eötvös Loránd University, H-1117 Budapest, Hungary*

**Abstract:** We present a computation of energy-energy correlation in $e^+e^-$ annihilation at next-to-next-to-leading order accuracy in perturbative QCD matched with the next-to-next-to-leading logarithmic resummed calculation for the back-to-back limit. Using these predictions and state-of-the-art Monte Carlo tools to model hadronization corrections, we perform an extraction of the strong coupling from available data sets. We also show next-to-next-to-leading order results for soft-drop thrust, an observable specifically constructed to have reduced hadronization corrections. We study the impact of the soft drop on the convergence of the perturbative prediction and find that generally grooming improves perturbative stability. This improved stability, together with the reduced sensitivity to non-perturbative corrections makes soft-drop thrust a promising observable for precision measurements of the strong coupling at lepton colliders.

## Introduction

Accurate measurements of event shape distributions in $e^+e^-$ annihilation continue to be one of the most precise tools for extracting the strong coupling $\alpha_\mathrm{s}$ value from data [1,2]. Such determinations are typically based on the comparison of differential distributions with perturbative predictions supplemented with hadronization corrections derived either from analytic models or Monte Carlo tools. As new data for $e^+e^-$ annihilation are not foreseen in the near future, progress in such measurements relies solely on improved theoretical understanding of the $e^+e^- \to$ hadrons process.

When discussing the accuracy of theoretical predictions for event shape distributions, two very different sources of uncertainty present themselves. The first of these is simply the uncertainty coming from terms that are not evaluated exactly in perturbation theory. These can be higher-order terms in the coupling that are simply neglected in a fixed-order calculation, or subleading logarithmic terms that are not controlled in an all-order resummation. A second source of uncertainty is that associated with the description of the parton to hadron transition.

These two types of uncertainties have a rather different nature and so their reduction must be addressed in different ways. Clearly, uncertainties associated to the perturbative description of an observable may be reduced, at least in principle, by increasing the perturbative and/or logarithmic order at which the predictions are computed. These days, state-of-the-art computations include exact fixed-order corrections at next-to-next-to-leading order (NNLO) accuracy for three-jet event shapes [3,4,5], as well as next-to-next-to-leading logarithmic (NNLL) (see e.g., Ref. [6] and references therein) and even next-to-next-to-next-to-leading logarithmic (N$^3$LL) resummation [7,8,9] in the two-jet limit. However, it is less obvious how non-perturbative uncertainties could be similarly reduced. In this respect, one idea is to investigate observables that are less sensitive to hadronization corrections. In particular, borrowing ideas from jet grooming, new event shape observables can be defined for which hadronization corrections are much reduced as compared to traditional ones [10].

In this contribution we first present an extraction of the strong coupling $\alpha_\mathrm{s}$ from the energy-energy correlation of particles in $e^+e^-$ collisions, highlighting the role that higher-order perturbative corrections play in reducing the uncertainty of the measurement. Then, we investigate soft-drop thrust, an observable constructed to mitigate the impact of non-perturbative corrections. In particular, we point out that in addition to decreased hadronization corrections, this new observable also exhibits an increased perturbative stability, making it an appealing candidate for a precise determination of the strong coupling.

## An old observable: energy-energy correlation

Energy-energy correlation (EEC) was one of the first infrared- and collinear-safe event shapes to be considered in the literature [11]. It is defined as the normalized energy-weighted distribution with respect to angles $\chi$ between the three-momenta of particles in an event,

$$\frac{1}{\sigma_\mathrm{t}} \frac{\mathrm{d}\Sigma(\chi)}{\mathrm{d}\cos\chi} = \frac{1}{\sigma_\mathrm{t}} \int \sum_{i,j} \frac{E_i E_j}{Q^2} \mathrm{d}\sigma_{e^+e^- \to ij+X} \delta(\cos\chi - \cos\theta_{ij}), \qquad (1)$$

where $E_i$ and $E_j$ are particle energies, $Q$ is the total center-of-mass energy, $\theta_{ij}$ is the angle between the three-momenta of particles $i$ and $j^*$ and $\sigma_\mathrm{t}$ is the total hadronic cross section.

The fixed-order prediction for EEC in perturbative QCD has been known numerically at NLO accuracy for some time (see e.g., Ref. [14] and references therein), while the NNLO correction has been computed more recently [15] using the CoLoRFulNNLO method [5,16,17]. At the renormalization scale $\mu$ the fixed-order result can be written as

$$\left[\frac{1}{\sigma_\mathrm{t}} \frac{\mathrm{d}\Sigma(\chi,\mu)}{\mathrm{d}\cos\chi}\right]_{(\mathrm{f.o.})} = \frac{\alpha_\mathrm{s}(\mu)}{2\pi} \frac{\mathrm{d}\bar{A}(\chi,\mu)}{\mathrm{d}\cos\chi} + \left(\frac{\alpha_\mathrm{s}(\mu)}{2\pi}\right)^2 \frac{\mathrm{d}\bar{B}(\chi,\mu)}{\mathrm{d}\cos\chi} + \left(\frac{\alpha_\mathrm{s}(\mu)}{2\pi}\right)^3 \frac{\mathrm{d}\bar{C}(\chi,\mu)}{\mathrm{d}\cos\chi} + \mathcal{O}(\alpha_\mathrm{s}^4), \quad (2)$$

where the perturbative coefficients at LO, NLO and NNLO, $\bar{A}$, $\bar{B}$ and $\bar{C}$, have been normalized to the total hadronic cross section. On the left panel of Fig. 1 we show the physical predictions for EEC in fixed-order perturbation theory up to NNLO accuracy together with data measured by the OPAL collaboration [18]. The bands in the plot represent the effect of varying the renormalization scale by a factor of two around its central value of $\mu = Q$.

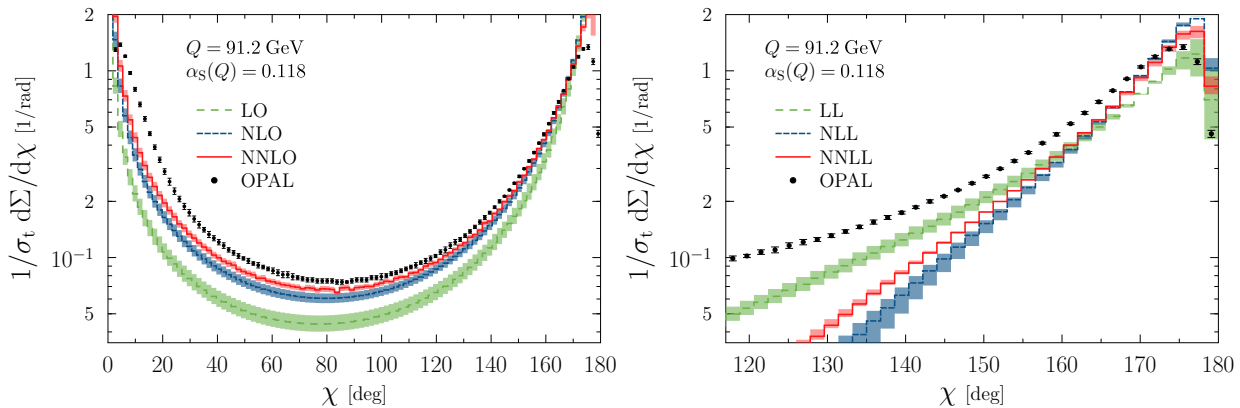

Figure 1: Fixed-order (left) and resumed (right) predictions for EEC.

---
*Refs. [12,13] use the opposite $\chi = 180° - \theta_{ij}$ convention such that the back-to-back region corresponds to $\chi \to 0$.

Clearly the inclusion of higher-order corrections improves the agreement of the prediction and the data, although there are pronounced differences around the forward ($\chi \to 0°$) and back-to-back ($\chi \to 180°$) regions, where in fact the fixed-order predictions diverge. This is due to the presence of large logarithmic corrections of infrared origin in these phase space regions that must be resummed to all orders to obtain a physically valid description of EEC around the endpoints of the distribution. The resummation of large logarithms of

$$y = \cos^2 \frac{\chi}{2} \qquad (3)$$

in the back-to-back region around $\chi = 180°$ has been known for some time at NNLL accuracy [12]. The resummed prediction can be written as

$$\left[\frac{1}{\sigma_\text{t}} \frac{\text{d}\Sigma(\chi,\mu)}{\text{d}\cos\chi}\right]_{(\text{res.})} = \frac{Q^2}{8} H(\alpha_\text{s}(\mu)) \int_0^\infty \text{d}b\, J_0(bQ\sqrt{y}) S(Q,b)\,, \qquad (4)$$

where the logarithmically enhanced terms are collected in the Sudakov form factor

$$S(Q,b) = \exp\left\{-\int_{b_0^2/b^2}^{Q^2} \frac{\text{d}q^2}{q^2} \left[A(\alpha_\text{s}(q^2)) \ln \frac{Q^2}{q^2} + B(\alpha_\text{s}(q^2))\right]\right\}\,. \qquad (5)$$

The functions $A$, $B$ and $H$ are free of logarithmic corrections and can be computed in perturbation theory. Their explicit expressions can be found in Refs. [12,13]. On the right panel of Fig. 1 we present purely resummed predictions in the back-to-back limit for EEC up to NNLL accuracy together with OPAL data. The resummed calculation is finite and captures the trends of the data correctly for angles $\chi$ close to $180°$, but does not do a good job of describing the measurement away from the back-to-back region.

From Fig. 1 it is evident that data is best described by fixed-order or resummed results over different angular ranges. In particular, fixed-order predictions are reliable for moderate to large $y$ ($\alpha_\text{s} \ln^2 y \ll 1$), while the resummed calculation applies to small $y$ ($y \ll 1$). Predictions that are valid over a wide kinematical range can be obtained by combining the fixed-order and resummed results. The matched predictions are obtained in the log-$R$ scheme. The details of this procedure are presented in Ref. [13]. Note that the description of the EEC distribution over the full angular range would require resummation also in the forward limit.

In order to extract the strong coupling $\alpha_\text{s}$ from measurements of EEC, the theoretical prediction described above must be combined with hadronization corrections. We modeled these non-perturbative effects using the state-of-the-art particle-level Monte Carlo generators `SHERPA` [19] and `Herwig 7` [20]. The exact Monte Carlo generation setups employed are discussed in Ref. [21] as well as by the contribution of A. Verbytskyi in these proceedings. Hadronization corrections were derived on a bin-by-bin basis as ratios of the EEC distribution at hadron and parton level in the simulated samples.

The perturbative results, corrected for hadronization effects as described above, were confronted with available data sets from the SLD, OPAL, L3, DELPHI, TOPAZ, TASSO, PLUTO, JADE, CELLO, MARKII, and MAC experiments. The details of data selection are described in Ref. [21]. The optimal value of $\alpha_\text{s}$ was determined by a chi-squared minimization procedure employing the `MINUIT 2` program [22], see Ref. [21] and A. Verbytskyi's contribution in these proceedings for details. In Fig. 2 we show representative results of fits to data obtained with theoretical predictions at NNLO+NNLL as well as NLO+NNLL accuracy. Our best fit value for $\alpha_\text{s}$ at NNLO+NNLL accuracy is

$$\alpha_\text{s}(m_\text{z}) = 0.11750 \pm 0.00287\,(\text{comb.})\,, \qquad (6)$$

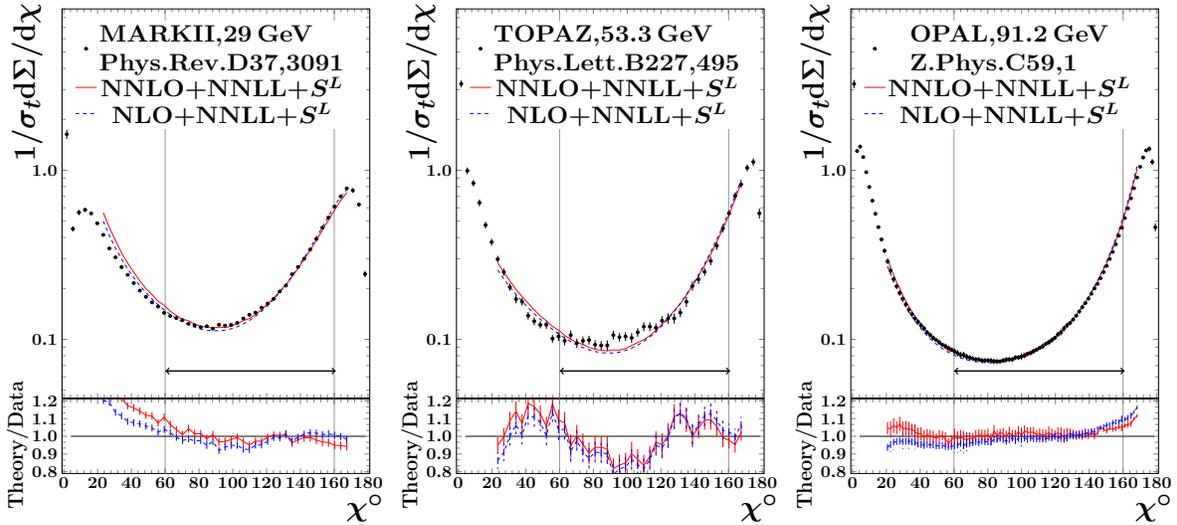

Figure 2: Selected results of fits to data at NNLO+NNLL and NLO+NNLL accuracy.

in agreement with the world average as of 2017 [23]. The quoted combined error takes into account uncertainties associated with the variation of the renormalization and resummation scales in the perturbative calculation, the choice of hadronization model employed (Lund string fragmentation or cluster model), as well as fit uncertainty (obtained with the $\chi^2 + 1$ criterion as implemented in MINUIT2). A detailed description of the estimation of the various uncertainties is given in Ref. [21] (see also A. Verbytskyi's contribution in these proceedings).

In order to highlight the impact of NNLO corrections on the determination, the fit was repeated with theoretical predictions computed at NLO+NNLL accuracy. The corresponding best fit value for $\alpha_s$ is $\alpha_s(m_Z) = 0.12200 \pm 0.00535$ (comb.). We see that the inclusion of the NNLO correction has a moderate but non-negligible effect on the extracted value of $\alpha_s$, while the uncertainty of the determination is reduced substantially, by a factor of two.

## New observables: soft-drop event shapes

We now turn to the issue of how the uncertainty associated with the estimation of hadronization corrections might be mitigated in measurements of $\alpha_s$. As mentioned in the introduction, one possible approach is to construct observables with reduced sensitivity to non-perturbative effects. The idea is simple: if the overall size of the hadronization correction is small, then even a sizable relative uncertainty on this contribution will correspond to a small overall uncertainty on $\alpha_s$. Thus the limited precision of the hadronization correction becomes less of an issue.

Soft-drop event shapes constitute a generic class of observables that are constructed to have reduced hadronization uncertainties. Indeed, soft drop is a kind of grooming procedure, designed to remove soft and wide-angle radiation from jets that are defined in an event. For Cambridge–Aachen jets of radius $R$, soft-drop grooming is defined as follows [24]:

1. Undo the last step of clustering for jet $J$ and split it into two subjets.

2. Check if the subjets pass the soft-drop condition, which for $e^+e^-$ collisions reads

$$\frac{\min\{E_i, E_j\}}{E_i + E_j} > z_{\text{cut}} \left(\frac{1 - \cos\theta_{ij}}{1 - \cos R}\right)^{\beta/2} \quad \text{or} \quad \frac{\min\{E_i, E_j\}}{E_i + E_j} > z_{\text{cut}} (1 - \cos\theta_{ij})^{\beta/2} \quad (7)$$

for jets of radius $R$ or hemisphere jets, respectively.

3. If the splitting fails this condition, the softer subjet is discarded and the groomer continues to the next step in the clustering.

4. If the splitting passes, the procedure ends and $J$ is the soft-drop jet.

The grooming parameter $z_{\text{cut}}$ sets an energy threshold for discarding soft radiation ($z_{\text{cut}} \to 0$ corresponds to no grooming), while $\beta$ controls how strongly wide-angle emissions are rejected ($\beta \to \infty$ corresponds to no grooming).

With the soft-drop procedure, one can define event shapes by first performing a special kind of grooming of the event, and then computing the value of the event shape from the groomed event. As an example, consider soft-drop thrust (more specifically $T'_{\text{SD}}$), which was defined for $e^+e^-$ collisions in Ref. [10]:

(a) Compute the thrust axis, $\vec{n}_T$, and divide the event into two hemispheres.

(b) Apply soft-drop grooming to each hemisphere.

(c') The set of particles left in the two hemispheres after the soft-drop constitute the soft-drop hemispheres $\mathcal{H}^L_{\text{SD}}$ and $\mathcal{H}^R_{\text{SD}}$, on which the soft-drop thrust $T'_{\text{SD}}$ is defined as

$$T'_{\text{SD}} = \frac{\sum_{i \in \mathcal{H}^L_{\text{SD}}} |\vec{n}_L \cdot \vec{p}_i|}{\sum_{i \in \mathcal{E}_{\text{SD}}} |\vec{p}_i|} + \frac{\sum_{i \in \mathcal{H}^R_{\text{SD}}} |\vec{n}_R \cdot \vec{p}_i|}{\sum_{i \in \mathcal{E}_{\text{SD}}} |\vec{p}_i|}, \qquad (8)$$

where $\vec{n}_L$ and $\vec{n}_R$ are the jet axes of the original left and right hemispheres and $\mathcal{E}_{\text{SD}}$ is the soft-drop event, $\mathcal{E}_{\text{SD}} = \mathcal{H}^L_{\text{SD}} \cup \mathcal{H}^R_{\text{SD}}$.

Hadronization corrections to soft-drop thrust were studied in Ref. [10]. There it was demonstrated that non-perturbative corrections are indeed much reduced over a wide range of the event shape, with the precise magnitude of the reduction depending on the choice of grooming parameters $z_{\text{cut}}$ and $\beta$. This property makes soft-drop event shapes attractive candidates for extractions of $\alpha_s$, however, it should be noted that grooming also reduces the cross section, hence the soft-drop parameters must be chosen carefully to avoid the loss of too much data.

Furthermore, the precision of potential $\alpha_s$ measurements based on soft-drop event shapes is also influenced by the perturbative stability of the observables. Hence, it is important to investigate how grooming affects the convergence of perturbative predictions. In order to assess this, in Ref. [25] we computed the QCD corrections to soft-drop thrust ($T'_{\text{SD}}$), hemisphere jet mass ($e_2^{(2)}$), and narrow jet mass ($\rho$). (The precise definitions of $e_2^{(2)}$ and $\rho$ are given in Ref. [10].) We quantify the convergence of the perturbative results with $K$-factors, defined as the ratios of distributions at subsequent orders in perturbation theory,

$$K_{\text{NLO}}(\mu) = \frac{\text{d}\sigma_{\text{NLO}}(\mu)}{\text{d}O} \bigg/ \frac{\text{d}\sigma_{\text{LO}}(Q)}{\text{d}O} \qquad \text{and} \qquad K_{\text{NNLO}}(\mu) = \frac{\text{d}\sigma_{\text{NNLO}}(\mu)}{\text{d}O} \bigg/ \frac{\text{d}\sigma_{\text{NLO}}(Q)}{\text{d}O}. \qquad (9)$$

Clearly the less the $K$-factors deviate from unity, the better the convergence of the perturbative prediction.

We present our results for soft-drop thrust in Fig. 3, where the left panel shows the distribution of $\tau'_{\text{SD}} \equiv 1 - T'_{\text{SD}}$ at LO, NLO, and NNLO accuracy for grooming parameters $z_{\text{cut}} = 0.1$ and $\beta = 0$. The bands represent the effects of varying the renormalization scale by a factor of two around the central value $\mu = Q$. The dependence of the $K$-factors on grooming parameters is studied on the right panel of Fig. 3, where $K_{\text{NLO}}(\mu)$ and $K_{\text{NNLO}}(\mu)$ are plotted in dashed blue and solid red. We observe that stronger grooming (larger $z_{\text{cut}}$ and smaller $\beta$) leads to $K$-factors closer to unity and hence a more stable perturbative prediction. This conclusion is of course not unexpected. For $\tau'_{\text{SD}} \gtrsim 10^{-2}$, i.e., in the range where the bulk of the cross section is generated, we see that the perturbative result is the most stable for $z_{\text{cut}} = 0.1$ and $\beta = 0$.

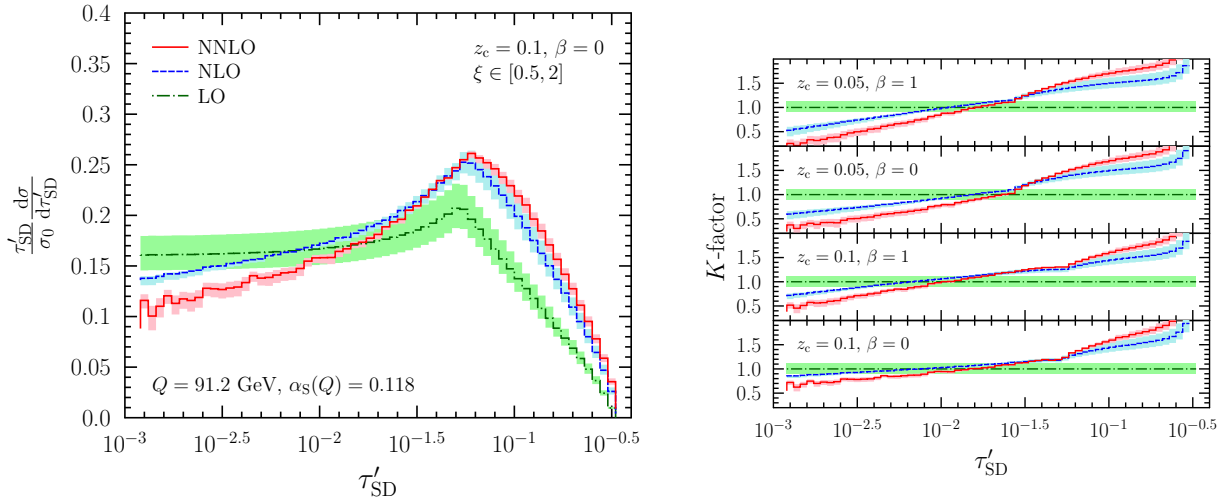

Figure 3: The soft-drop thrust distribution at LO, NLO and NNLO accuracy with $z_{\text{cut}} = 0.1$ and $\beta = 0$ (left), and $K$-factors as defined in Eq. (9) for different choices of grooming parameters (right).

## Summary

In this contribution we examined potential ways of increasing the precision of $\alpha_s$ measurements from $e^+e^-$ annihilation into hadrons. On the one hand, we stressed the important role that higher-order perturbative corrections play in reducing the uncertainty of the determination. We highlighted this by presenting an extraction of $\alpha_s$ from measurements of energy-energy correlation, based on theoretical predictions with NNLO+NNLL accuracy and hadronization corrections derived using modern Monte Carlo tools. We find that the inclusion of NNLO corrections has a dramatic effect on the uncertainty of the measurement, which is reduced by a factor of two as compared to the result at only NLO+NNLL accuracy. Our analysis provides a determination of $\alpha_s(m_z)$ with the highest numerical and theoretical precision obtained from this observable to date,

$$\alpha_s(m_z) = 0.11750 \pm 0.00287 \,(\text{comb.})\,.$$

On the other hand, we pointed out that a possible strategy for reducing uncertainties in measurements of $\alpha_s$ associated with the modeling of hadronization corrections is to employ observables for which these corrections are small. In particular, we examined soft-drop event shapes, observables specifically tailored so as to show less sensitivity to non-perturbative effects. Through the example of soft-drop thrust, we demonstrated that in addition to reducing hadronization corrections, soft-drop grooming also enhances the perturbative stability of the theoretical predictions. These features make soft-drop event shapes promising candidates for precision measurements of the strong coupling at lepton colliders.

# $\alpha_\text{s}$ from energy-energy correlations and jet rates in $e^+e^-$ collisions

Andrea Banfi[1], Adam Kardos[2], Stefan Kluth[3], Pier Francesco Monni[4], Gábor Somogyi[5], Zoltán Szőr[6], Zoltán Trócsányi[5,7], Zoltán Tulipánt[5], **Andrii Verbytskyi**[3], and Giulia Zanderighi[3]

[1]*University of Sussex, Brighton, BN1 9RH United Kingdom*
[2]*University of Debrecen, 4010 Debrecen, PO Box 105, Hungary*
[3]*Max-Planck-Institut für Physik, D-80805 Munich, Germany*
[4]*CERN, Theory Department, CH-1211 Geneva 23, Switzerland*
[5]*MTA-DE Particle Physics Research Group, University of Debrecen, 4010 Debrecen, Hungary*
[6]*PRISMA Cluster of Excellence, Institut für Physik, Universität Mainz, D-55099 Mainz, Germany*
[7]*Institute for Theoretical Physics, Eötvös Loránd University, H-1117 Budapest, Hungary*

**Abstract:** We present a comparison of the computation of energy-energy correlations and Durham algorithm jet rates in $e^+e^-$ collisions at next-to-next-to-leading logarithmic accuracy matched with the $\mathcal{O}(\alpha_\text{s}^3)$ perturbative prediction to LEP, PEP, PETRA, SLC, and TRISTAN data. With these predictions we perform extractions of the strong coupling constant taking into account non-perturbative effects modelled with modern Monte Carlo event generators that simulate NLO QCD corrections.

## Introduction

The strong interaction in the Standard Model (SM) is described by Quantum Chromodynamics (QCD), see Ref. [1] for a review. The theory successfully describes the interactions between quarks and gluons and is a source of numerous predictions. One of the precise QCD predictions that depends strongly on the only theory parameter, the coupling constant of the strong interaction $\alpha_\text{s}$, is the topology of the $e^+e^- \to$ hadrons events. In these events at high energies, hadrons predominantly appear in collimated bunches, called *jets*. The topologies of $e^+e^- \to$ partons events can be predicted with high precision in perturbation theory and the observables of the final hadronic state observed in the experiments are closely related to them.

The state of the art predictions for QCD for such observables currently includes exact fixed-order next-to-next-to-leading order (NNLO) corrections for the three-jet event shapes and jet rates. The specialized numerical matrix element integration codes allow a straightforward computation of any suitable, i.e. collinear and infrared safe, event shape or jet observable.

In this paper we describe two analyses that utilise the NNLO predictions matched to next-to-next-leading-log (NNLL) resummed calculations for the region with $e^+e^- \to$ 2-partons topology.

The first analysis considers the energy-energy correlation (EEC). EEC is the normalised energy-weighted cross section defined in terms of the angle between two particles $i$ and $j$ in an event [2]:

$$\frac{1}{\sigma_\text{t}} \frac{\text{d}\Sigma(\chi)}{\text{d}\cos\chi} \equiv \frac{1}{\sigma_\text{t}} \int \sum_{i,j} \frac{E_i E_j}{Q^2} \text{d}\sigma_{e^+e^- \to ij+X} \delta(\cos\chi - \cos\theta_{ij}),$$

where $E_i$ and $E_j$ are the particle energies, $Q$ is the centre-of-mass energy, $\theta_{ij} = \chi$ is the angle between the two particles, and $\sigma_{tot}$ is the total hadronic cross section. EEC was the first event shape for which a complete NNLL resummation was performed [3] while the fixed-order NNLO corrections to this observable were computed only recently [4].

The second analysis considers the 2- and 3-jet rates obtained with the Durham jet algorithm [5]. The algorithm is described in detail elsewhere [5], only a brief description is given below. As every jet clustering algorithm, the Durham jet algorithm combines the energy and the momenta of particles (partons or hadrons) into jet objects. This is done using a measure in phase space between pairs of particles $i$ and $j$ with corresponding energies $E_i$ and $E_j$ as

$$d_{ij} = 2\min(E_i^2, E_j^2)(1 - \cos\theta_{ij}),$$

where $\theta_{ij}$ is the angle between the momenta of particles. At a given stage of the combination procedure a pair of objects $i$ and $j$ with minimal $d_{ij}$ is found. The object $i$ is merged (e.g. by adding 4-vectors) with object $j$. Therefore, at every given stage, the number of objects (jets) can be related to the parameter $y = \min\{d_{ij}\}/Q^2$. Consequently, the jet rates are defined as $R_n(y) = \frac{\sigma_{n\text{-jets}}(y)}{\sigma_{tot}}$, where $\sigma_{n\text{-jet}}(y)$ is the cross-section of $n$-jet events. In this analysis we used the implementations of the algorithm from the `FastJet3.1` [6] package. The NNLL resummation for the 2-jet rates is described in Ref. [7].

## Extraction procedure

The $\alpha_\text{s}$ extraction procedure is based on the comparison of data to the perturbative QCD prediction combined with non-perturbative (hadronization) corrections, and contains ingredients described below.

### Fixed-order and resummed calculations

In NNLO perturbative QCD at the default renormalization scale of $\mu = Q$, the fixed-order predictions for observable $O$, vanishing in the 2-jet limit, reads

$$O_\text{f.o.} = \frac{\alpha_\text{s}(Q)}{2\pi}A + \left(\frac{\alpha_\text{s}(Q)}{2\pi}\right)^2 B + \left(\frac{\alpha_\text{s}(Q)}{2\pi}\right)^3 C + \mathcal{O}(\alpha_\text{s}^4),$$

where $A$, $B$ and $C$ are the perturbative coefficients at LO, NLO and NNLO, normalised to the LO cross section for $e^+e^- \to$ hadrons, $\sigma_0$. In the presented analyses the coefficients $A$, $B$, $C$ were calculated using the CoLoRFulNNLO method [8,4] as function of angle $\chi$ (for EEC) or $y$ (for jet rates). The NNLL resummed predictions and matching procedures were used as described in Ref. [9] and Ref. [10] (for EEC) and in Ref. [7] (for 2-jet rates). For the three jet rate $R_3$ the resummed prediction has a much lower logarithmic accuracy [5] and does not guarantee a good theoretical control in the region where logarithms are large. Therefore, for the three jet rate $R_3$ in this analysis only fixed order predictions were used.

### Finite $b$-quark mass corrections

The theoretical predictions described above are computed in massless QCD. In order to take into account finite $b$-quark mass effects, we subtract the fraction of $b$-quark events, $r_b(Q)$ from the massless result and add back the corresponding massive contribution. Hence, we include mass effects directly at the level of matched distributions of corresponding observables $O$,

$$O = (1 - r_b(Q))O_\text{massless} + r_b(Q)O_\text{massive}^{\text{NNLO}^*}.$$

Here $O_{\text{massless}}$ is the matched distribution, computed in massless QCD as outlined above, while $O_{\text{massive}}^{\text{NNLO}^*}$ is the fixed-order massive distribution. The complete massive NNLO corrections are currently unknown, so we model them by supplementing the massive NLO prediction of the parton level Monte Carlo generator `Zbb4` [11], with the NNLO coefficient of the massless fixed-order result. The fraction of $b$-quark events $r_b(Q)$ is defined as

$$r_b(Q) \equiv \sigma_{\text{massive}}(e^+e^- \to b\bar{b})/\sigma_{\text{massive}}(e^+e^- \to \text{hadrons}),$$

where all quantities are calculated up to $\mathcal{O}(\alpha_s^3)$.

### Data sets

To extract the strong coupling the predictions described above were confronted with the available data sets. The criteria to include the data were high precision measurements obtained with charged and neutral final state particles, presence of corrections for detector effects, correction for initial state photon radiation and sufficient amount of supplementary information. Namely, for the EEC analysis the data obtained in SLD, L3, DELPHI, OPAL, TOPAZ, TASSO, JADE, MAC, MARKII, CELLO, and PLUTO experiments were included, see details in Ref. [10]. The corresponding centre-of-mass energy range is $\sqrt{s} = 14 - 91.2\,\text{GeV}$. For the jet rates analysis, the data obtained in the OPAL, JADE, DELPHI, L3, and ALEPH experiments were included, see details in Ref. [12]. The corresponding centre-of-mass energy range is $\sqrt{s} = 35 - 207\,\text{GeV}$.

### Monte Carlo generation setup

In both analyses, the non-perturbative effects in the $e^+e^- \to$ hadrons process are modelled using state-of-the-art particle-level Monte Carlo (MC) generators `SHERPA` [13] and `Herwig 7` [14]. The MC generated event samples describe the data relatively well, see Fig. 1.

The full description of the MC event generator setups is given in Ref. [12] and [10], only a brief overview is given below. The `SHERPA` samples were generated using the matrix element generators `AMEGIC` and `COMIX`. The `Herwig 7` samples were generated using the matrix element generator `MadGraph5`. To simulate one-loop QCD correction the `GoSam` one-loop library for the EEC analysis and the `OpenLoops` one-loop library for the jet rates analysis are employed. In all cases the 2-parton final state processes had NLO accuracy in perturbative QCD and the matrix elements were calculated assuming massive $b$-quarks.

To test the fragmentation and hadronization model dependence, the parton level events were hadronized with different hadronization setups. Here and below the results of the $\alpha_s$ extraction are labelled according to these hadronization setups.

### MC event samples for EEC analysis

The events generated by `SHERPA` were hadronized with a native implementation of the cluster model (label $S^C$) and the Lund string fragmentation model as implemented in `Pythia 6` (label $S^L$). The events generated by the `Herwig 7` were hadronized by the native implementation of the cluster model (label $H^M$). $S^L$ was chosen to be the default setup. The hadronization corrections were used multiplicatively, i.e. $\text{EEC}^{(\text{hadrons})}(\chi) = k(\chi,s) \times \text{EEC}^{(\text{partons})}(\chi)$, where the coefficients $k(\chi,s)$ are extracted for every bin from the MC simulated samples. Before the extraction, the MC simulated samples were re-weighted on an event-by-event basis so the energy-energy correlation distributions on hadron level coincide with data, see Ref. [10] for details.

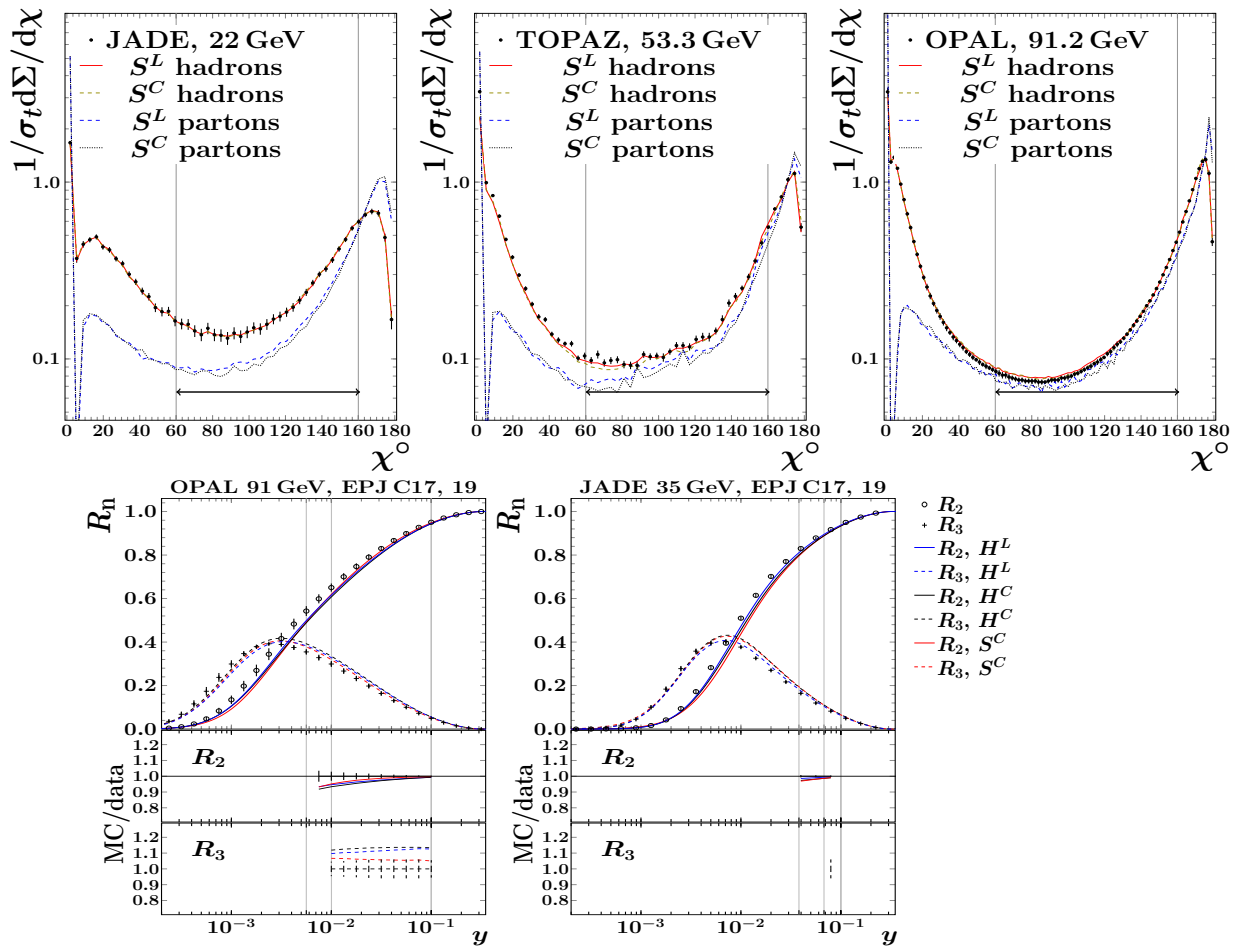

Figure 1: Selected data and predictions with different Monte Carlo setups for EEC (top) and jet rates (bottom) analyses. For the EEC analysis the hadron level distributions are accompanied with corresponding parton level distributions.

**MC event samples for jet rates analysis**

The events generated by `Herwig` were hadronized with the native implementation of the cluster model (label $H^C$) and the Lund string fragmentation model as implemented in `Pythia 8` (label $H^L$). The events generated by `SHERPA` were hadronized by the native implementation of the cluster model (label $S^C$). $H^L$ was chosen to be the default setup. The hadronization correction procedure is designed to take into account that the jet rates add up to unity, see Ref. [12] for details.

**Fit procedure and estimation of uncertainties**

The perturbative part of the predictions was calculated for every data point as described in previous sections. To find the optimal value of $\alpha_s$, the `MINUIT2` program was used to minimise the value of

$$\chi^2(\alpha_s) = \sum_{\text{data sets}} \chi^2(\alpha_s)_{\text{data set}},$$

where $\chi^2(\alpha_s)$ was calculated for each data set as

$$\chi^2(\alpha_s) = (\vec{D} - \vec{P}(\alpha_s))V^{-1}(\vec{D} - \vec{P}(\alpha_s))^T,$$

with $\vec{D}$ standing for the vector of data points, $\vec{P}(alphas)$ for the vector of calculated predictions and $V$ for the covariance matrix for $\vec{D}$. The default scale used in the fit procedure was $\mu = Q = \sqrt{s}$. The fit ranges were chosen to avoid regions where resummed predictions or hadronization correction calculations are not reliable. The uncertainty on the fit result ('exp.') was estimated with the $\chi^2 + 1$ criterion as implemented in the MINUIT2 program. For both analyses the fits were performed taking into account the correlations between measurements within each data set, that were estimated from Monte Carlo simulations. The distributions obtained in the reference fits are shown in Fig. 2.

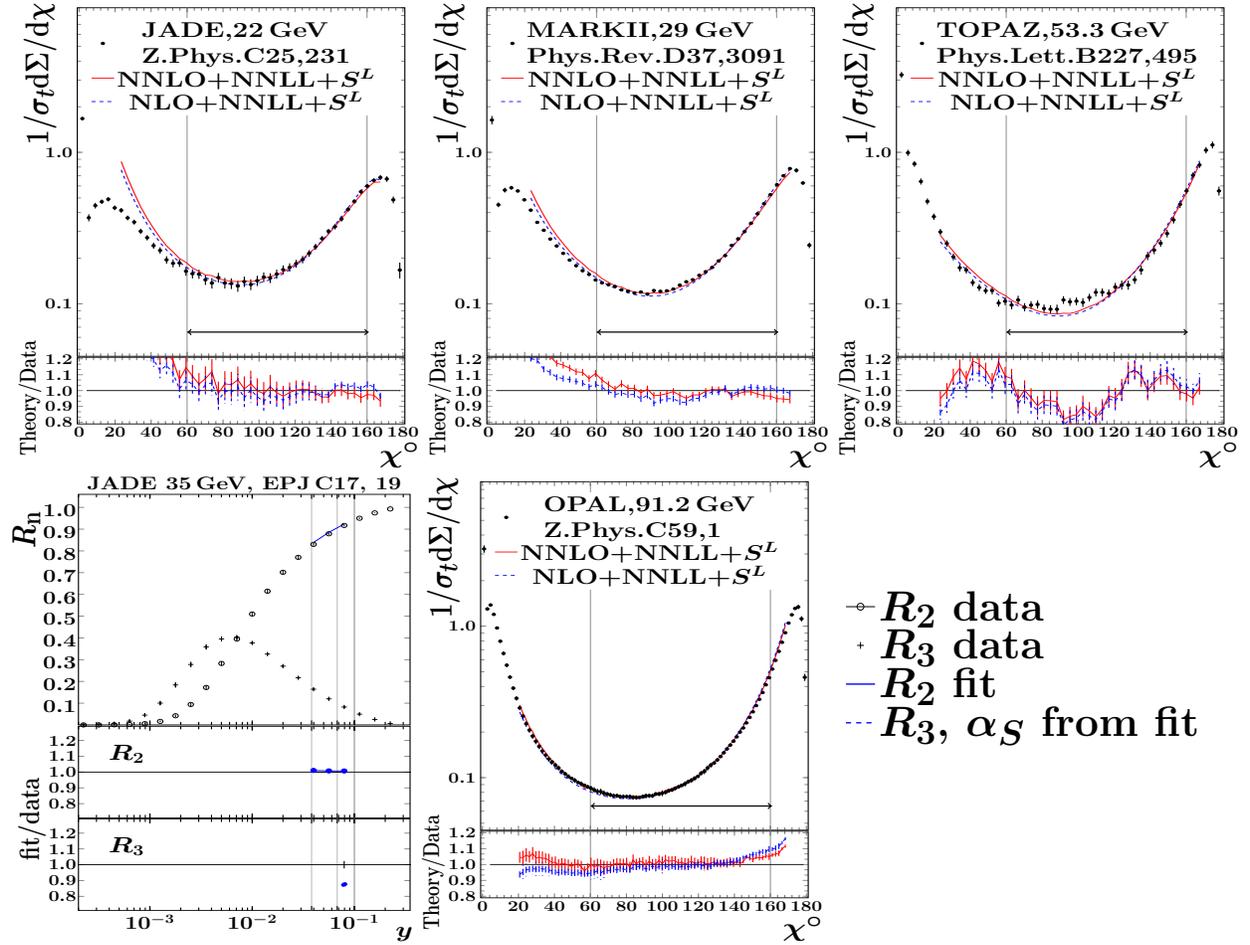

Figure 2: Selected data and fit results in the EEC (top) and jet rates (bottom) analyses.

## Systematic uncertainties and validity checks of the results

The full description of validity checks performed in both analyses is given in Refs. [12,10]. Below we describe briefly the way of estimation of the main systematic uncertainties. The systematic uncertainties were estimated with procedures used in previous studies [15]. To estimate the effects caused by the absence of higher-order terms in the perturbative predictions, the scale variation procedures were performed. The fits were repeated, with variation of the resummation $\mu_{\text{res.}} = x_R \times Q$ scale and renormalization $\mu_{\text{ren.}} = x_L \times Q$ scale by a factor $2^{\pm 1}$, see results in Fig. 3. The corresponding estimations are labelled below as ('ren.') for renormalization and as ('res.') for resummation scale variation.

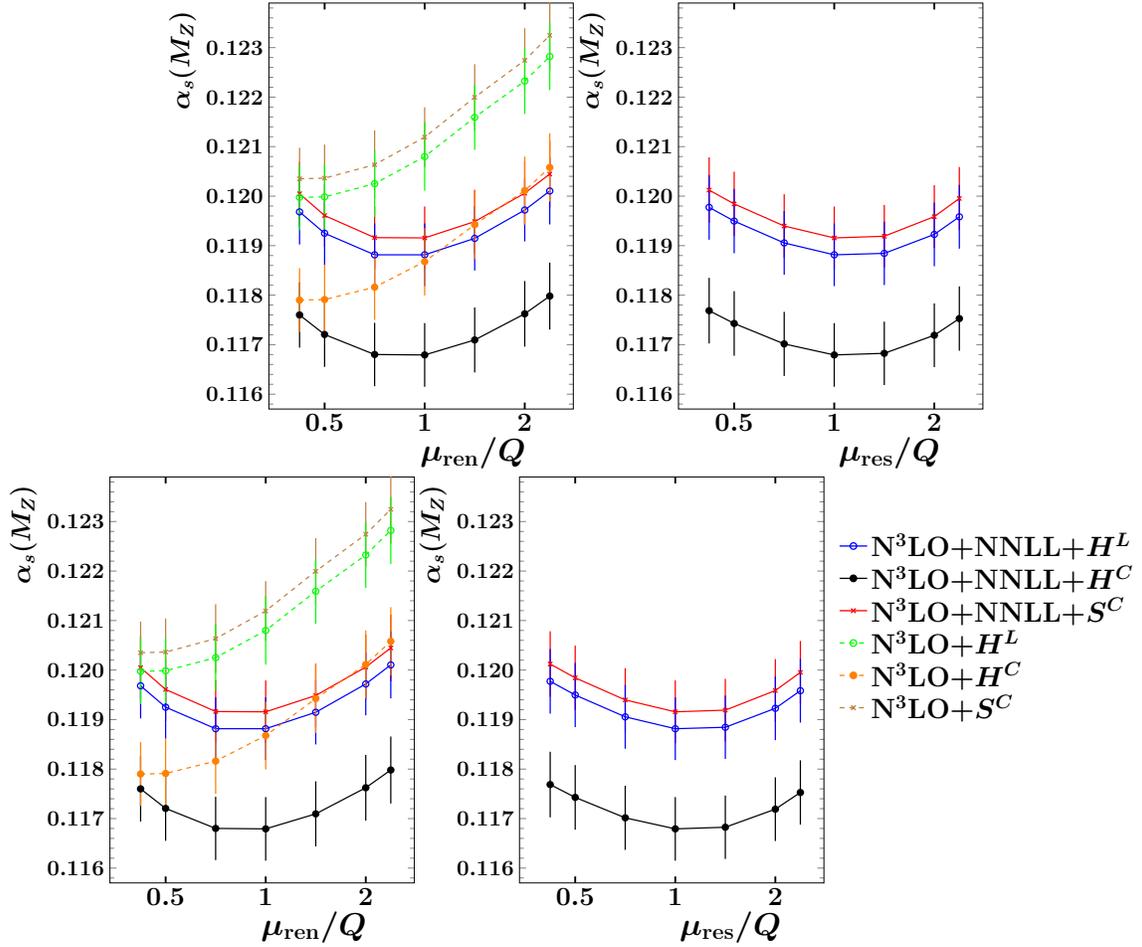

Figure 3: Dependence of fit results in the EEC (top) and jet rates (bottom) analyses on the renormalization and resummation scales.

The bias of hadronization model selection ('hadr.') is studied with alternative setups for MC hadronization corrections described above, i.e. from $S^C$ for the EEC analysis and $H^C$ for the jet rates analysis.

## Summary

For the central value of the final result for EEC (jet rates) analysis, we quote the results obtained from the fits with $S^L$ ($H^L$) hadronization model with uncertainties and estimations of biases obtained as described above. The final result of the EEC analysis is

$$\alpha_{\mathrm{s}}(m_{\mathrm{z}}) = 0.11750 \pm 0.00018\,(exp.) \pm 0.00102\,(hadr.) \pm 0.00257\,(ren.) \pm 0.00078\,(res.),$$

and for the jet rates analysis it is

$$\alpha_{\mathrm{s}}(m_{\mathrm{z}}) = 0.11881 \pm 0.00063\,(exp.) \pm 0.00101\,(hadr.) \pm 0.00045\,(ren.) \pm 0.00034\,(res.).$$

Both results are in agreement with the latest world average $\alpha_\text{s}(m_\text{z}) = 0.1181 \pm 0.0011$ [16].

Both analyses provide determinations of $\alpha_\text{s}(m_\text{z})$ determination which have one of the highest numerical and theory precisions ever obtained from the corresponding observables. In addition to that, in the case of the jet rates analysis, for the first time the hadronization-related uncertainty is much larger than other uncertainties.

# The strong coupling from $e^+e^- \to$ hadrons

Diogo Boito[1], **Maarten Golterman**[2], Alexander Keshavarzi[3], Kim Maltman[4,5],
Daisuke Nomura[6], Santiago Peris[7], and Thomas Teubner[8]

[1] Instituto de Física de São Carlos, Univ. São Paulo CP 369, 13570-970, São Carlos, SP, Brazil
[2] Department of Physics & Astronomy, San Francisco State University, San Francisco, CA 94132, USA
[3] The University of Mississippi - Dept. of Physics & Astronomy, Oxford, Mississippi 38677-1848, USA
[4] Department of Mathematics and Statistics, York University, Toronto, ON Canada M3J 1P3
[5] CSSM, University of Adelaide, Adelaide, SA 5005 Australia
[6] KEK Theory Center, Tsukuba, Ibaraki 305-0801, Japan
[7] Dept. of Physics and IFAE-BIST, Universitat Autònoma de Barcelona, 08193 Bellaterra, Barcelona
[8] Department of Mathematical Sciences, University of Liverpool, Liverpool L69 3BX, UK

**Abstract:** We present a brief summary of our recent determination of $\alpha_s$ from $e^+e^- \to$ hadrons, in the region $3 \leq s \leq 4$ GeV$^2$, with $s$ the square of the center-of-mass energy.

In a recent paper [1], we used a new compilation of data for the $R$-ratio $R(s)$, measured in the process $e^+e^- \to$ hadrons, to extract a value for the strong coupling, $\alpha_s$, using finite energy sum rules (FESRs). This determination can directly be compared with the determination from hadronic $\tau$ decays. Here we present a brief summary of this determination. A more extensive informal overview can be found in Ref. [2]; full details can be found in Ref. [1].

The data set we employed for our work is that of Ref. [3], and it is shown in the left panel of Fig. 1. This plot shows the $R$-ratio as a function of the square of the center-of-mass energy $s$, in GeV$^2$, below the threshold for charm production. In the right panel of Fig. 1 we show a blow-up of these same data, for 2 GeV$^2 \leq s \leq 6$ GeV$^2$. This plot shows more clearly that there are a lot more data in the region $s \leq 4$ GeV$^2$, where $R(s)$ was compiled from summing exclusive-channel experiments, than in the region $s \geq 4$ GeV$^2$, where $R(s)$ was compiled from inclusive experiments. A detailed analysis shows that an extraction of $\alpha_s$ employing FESRs using all data below 4 GeV$^2$ will yield a value with a smaller error than an extraction of $\alpha_s$ from $R(s)$ by direct comparison with QCD perturbation theory.

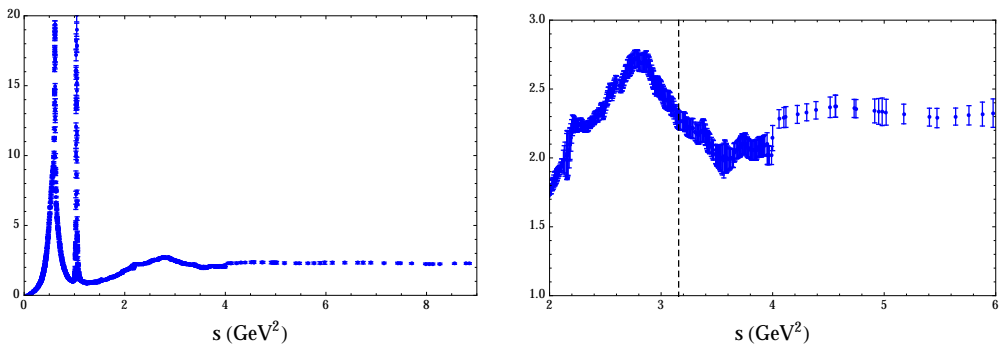

Figure 1: Left: $R$-ratio data from Ref. [3], as a function of $s$, the hadronic invariant squared mass. Right: A blow-up of the region $2 \leq s \leq 6$ GeV$^2$.

The sum rules we employ take on the form [1]

$$I^{(w)}(s_0) \equiv \frac{1}{s_0} \int_{m_\pi^2}^{s_0} ds\, w(s/s_0) \frac{1}{12\pi^2} R(s) = -\frac{1}{2\pi i s_0} \oint_{z=|s_0|} dz\, w(z/s_0)\, \Pi(z) \,, \qquad (1)$$

with $\Pi(z)$ the usual scalar electromagnetic polarization function, and $w(y)$ one of the following analytical weight functions

$$\begin{align}
w_0(y) &= 1 \,, \\
w_2(y) &= 1 - y^2 \,, \\
w_3(y) &= (1-y)^2(1+2y) \,, \\
w_4(y) &= (1-y^2)^2 \,.
\end{align} \qquad (2)$$

In Eq. (1), the left-hand side represents the "data" side, and it incorporates all data between threshold and $s = s_0$. The right-hand side represents the "theory" side, and, if $s_0$ is large enough, we can use the theory representation

$$\Pi(z) = \Pi_{\text{pert}}(z) + \Pi_{\text{OPE}}^{D>0}(z) + \Pi_{\text{DV}}(z) \,, \qquad (3)$$

where the first term, $\Pi_{\text{pert}}(z)$, represents massless perturbation theory, and is known to order $\alpha_s^4$ [5,6],[*] the second term represents mass-dependent perturbative and non-perturbative condensate contributions to the operator product expansion (OPE), while the "duality-violation" part $\Pi_{\text{DV}}(z)$ represents contributions to $\Pi(z)$ manifested by the presence of resonance peaks, which are not captured by perturbation theory or the OPE. In our analysis, we also included electromagnetic (EM) corrections to perturbation theory. For details, we refer to Ref. [1]. We just point out that duality violations, represented by the term $\Pi_{\text{DV}}(z)$, are expected to give a contribution which decreases exponentially with increasing $s_0$. In addition, their largest contribution to the integral on the right-hand side of Eq. (1) is expected to come from the part of the circle closest to the real axis, i.e., $z \approx s_0$ [8]. Their contribution is thus suppressed for $w = w_2$, which has a single zero at $z = s_0$ ($w_2$ is "singly pinched"), and more suppressed for $w = w_{3,4}$, which both have a double zero at $z = s_0$ ($w_{3,4}$ are "doubly pinched"). Note that the integral on the right-hand side of Eq. (1) with a polynomial weight containing $y^N$ receives a contribution from the effective condensate $C_D$ for $D = 2N + 2$ in the OPE.

Our fits of the FESRs (1) to the data were carried out on a window $s_0 \in [s_0^{\text{min}}, s_0^{\text{max}}]$, with $3.25\ \text{GeV}^2 \leq s_0^{\text{min}} \leq 3.80\ \text{GeV}^2$ and $s_0^{\text{max}} = 4\ \text{GeV}^2$, finding good stability for these values of $s_0^{\text{min}}$. In Fig. 2 we show typical fits for all four weights (2), with $s_0^{\text{min}} = 3.25\ \text{GeV}^2$. Fits were carried out neglecting the duality-violating term $\Pi_{\text{DV}}$ in Eq. (3). All fits take into account all the correlations in the data set, and have $p$-values varying from 0.09 to 0.42.

We note that the values of $s_0$ used in our fits are all larger than the square of the $\tau$ mass $m_\tau^2$, the kinematic end point for a similar analysis of spectral functions measured in hadronic $\tau$ decays. In particular, we notice that in the $e^+e^-$ case good fits are obtained neglecting duality violations, in contrast to the $\tau$-decay case (see below). For $w = w_0$, a remnant of integrated duality violations (the small oscillation in the upper left panel of Fig. 2) is visible, but the fit is consistent with the data. For the higher-degree weights (which are all pinched) no effect from integrated duality violations is visible.

As usual, two different resummations of the perturbative series are employed in our sum-rule analysis, FOPT (fixed-order perturbation theory) and CIPT (contour-improved perturbation theory [9]), leading to two different values for $\alpha_s$. For a more detailed discussion, we refer to Refs. [1,4,7] and references therein, as well as Ref. [10].

---

[*]We use an educated guess for the 5th order [7].

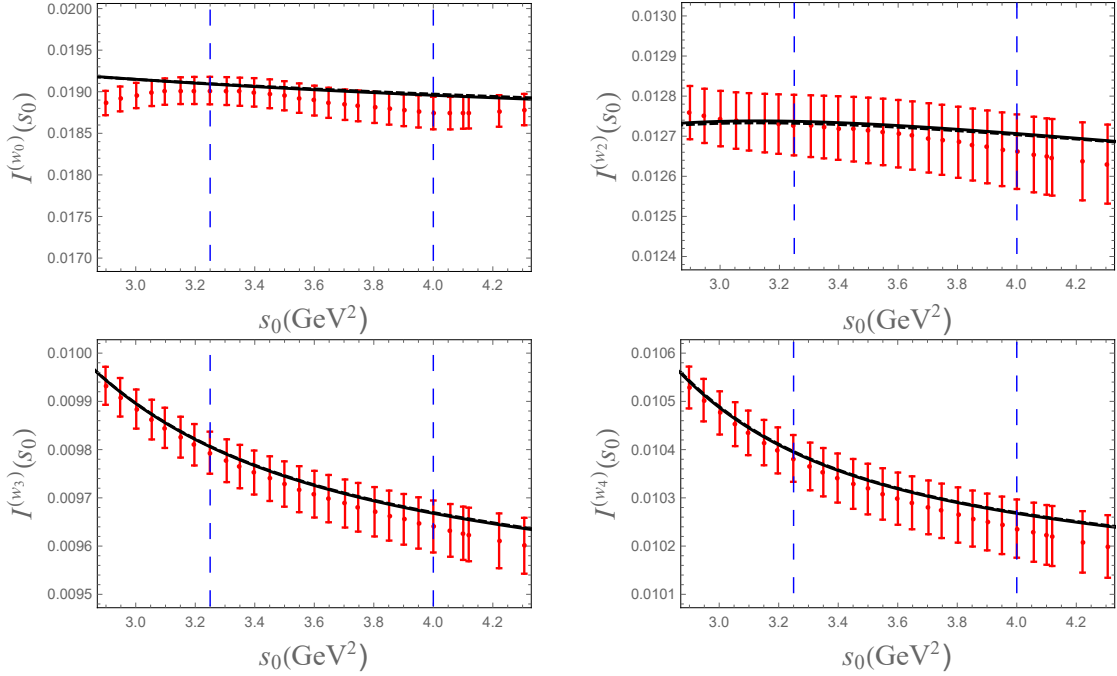

Figure 2: Comparison of the data for $I^{(w)}(s_0)$ with the fits on the interval $s_0^{\min} = 3.25$ to $4$ GeV$^2$, for $w = w_0$ (upper left panel), $w = w_2$ (upper right panel), $w = w_3$ (lower left panel), and $w = w_4$ (lower right panel). Solid black curves indicate FOPT fits, dashed curves CIPT. The fit window is indicated by the dashed vertical lines.

In Table 1 below, we show our results for the values of $\alpha_s(m_\tau^2)$ obtained from these fits, where we quote $\alpha_s$ at the $\tau$ mass in order to facilitate comparison with values obtained from hadronic $\tau$ decays. Clearly, there is excellent agreement between the values obtained from different weights. This agreement is also found for the fit values for the condensate $C_6$, between the weights $w_2$, $w_3$ and $w_4$ [1]. The errors shown are a combination of the fit error and the error due to the variation of $s_0^{\min}$; the first error dominates the total error.

We carried out a number of additional tests. First, we did a number of fits with $s_0^{\max}$ or both $s_0^{\min}$ and $s_0^{\max}$ in the inclusive region $s > 4$ GeV$^2$. We found results consistent with those reported in the table above but including data in the inclusive region does not lead to a reduction of the errors shown in the table.

Second, while fits without duality violations lead to good $p$-values, we tested the stability of the fits with weight $w_0$ against the inclusion of a model for duality violations. For a detailed discussion of this test, we refer to Ref. [1]. The upshot is that our fits are stable with respect to the inclusion of duality violations, and that duality violations can be ignored within current errors. The basic reason is that the analysis based on the $R$-ratio allows us to restrict our attention to values of $s_0$ large enough compared to $m_\tau^2$ that the exponentially decreasing duality violations are sufficiently suppressed.

Before coming to our final results, we present a brief comparison between FESR fits of moments of the non-strange $I=1$ vector spectral function obtained from hadronic $\tau$ decays [11], and FESR fits of the EM spectral function proportional to $R(s)$. Figure 3 shows fits of the moments $I^{(w_0)}(s_0)$ (upper panels) and $I^{(w_2)}(s_0)$ (lower panels), comparing fits based on the $\tau$ data (left panels) with fits based on the $e^+e^-$ data (right panels). The $\tau$-based fits have $s_0^{\max} = m_\tau^2$ and $s_0^{\min} = 1.55$ GeV$^2$; the $e^+e^-$-based fits have $s_0^{\max} = 4$ GeV$^2$ and $s_0^{\min} = 3.25$ GeV$^2$. In the $\tau$ panels, the blue curve represents FOPT fits with duality violations and the red dashed curve CIPT fits with duality violations. The black curves represent the perturbation theory plus OPE parts of these fits, omitting the duality-violating part. In the $e^+e^-$ panels, which just reproduce the top panels already shown in Fig. 2, the black curves represent FOPT (solid) and CIPT (dashed) fits, with no duality violations. Duality violations show up in the data points as oscillations around the perturbation theory plus OPE curves (black solid and dashed curves in all panels). Clearly, duality violations are very visible in the left panels. In contrast, they are barely visible in the upper right panel, and not visible in the lower right panel. These comparisons of theory with data show that duality violations cannot be ignored in the $\tau$-based results, while fits of moments of $R(s)$ at sufficiently higher $s_0$ are consistent with integrated duality violations being small enough at these higher values to be neglected, within current errors. This is consistent with the expected exponential decay of the duality-violating part of the spectral function with increasing $s$, as discussed in more detail in Refs. [12,13].

Table 1: Values for $\alpha_s(m_\tau^2)$ obtained from the various weights, with FOPT values in the second column, and CIPT values in the third.

| weight | $\alpha_s(m_\tau^2)$ (FOPT) | $\alpha_s(m_\tau^2)$ (CIPT) |
|--------|------------------------------|------------------------------|
| $w_0$  | 0.299(16)                    | 0.308(19)                    |
| $w_2$  | 0.298(17)                    | 0.305(19)                    |
| $w_3$  | 0.298(18)                    | 0.303(20)                    |
| $w_4$  | 0.297(18)                    | 0.303(20)                    |

Our final results for $\alpha_s(m_\tau^2)$ from the FESR-analysis of $R(s)$ are

$$\begin{aligned}
\alpha_s(m_\tau^2) &= 0.298(17) \quad \text{(FOPT)}, \\
&= 0.304(19) \quad \text{(CIPT)}.
\end{aligned} \quad (4)$$

We note that the error is dominated by the fit errors, obtained by propagating the errors on the data compilation of Ref. [3]. These results can be directly compared with the values obtained from the $\tau$-based analysis [11]:

$$\begin{aligned}
\alpha_s(m_\tau^2) &= 0.303(9) \quad \text{(FOPT)}, \\
&= 0.319(12) \quad \text{(CIPT)}.
\end{aligned} \quad (5)$$

There is excellent agreement between the results obtained from $e^+e^-$, and those obtained from $\tau$ decays. We note the much reduced difference between the FOPT and CIPT central values in the $e^+e^-$ analysis, which we believe can be partially ascribed to the fact that these values are extracted from spectral-weight moments at larger $s_0$, where the convergence properties of perturbation theory are expected to be better.

We also quote the $e^+e^-$-based values after running the values of Eq. (4) to the Z-mass, converting from three to five flavors:

$$\begin{aligned}
\alpha_s(m_Z^2) &= 0.1158(22) \quad \text{(FOPT)}, \\
&= 0.1166(25) \quad \text{(CIPT)}.
\end{aligned} \quad (6)$$

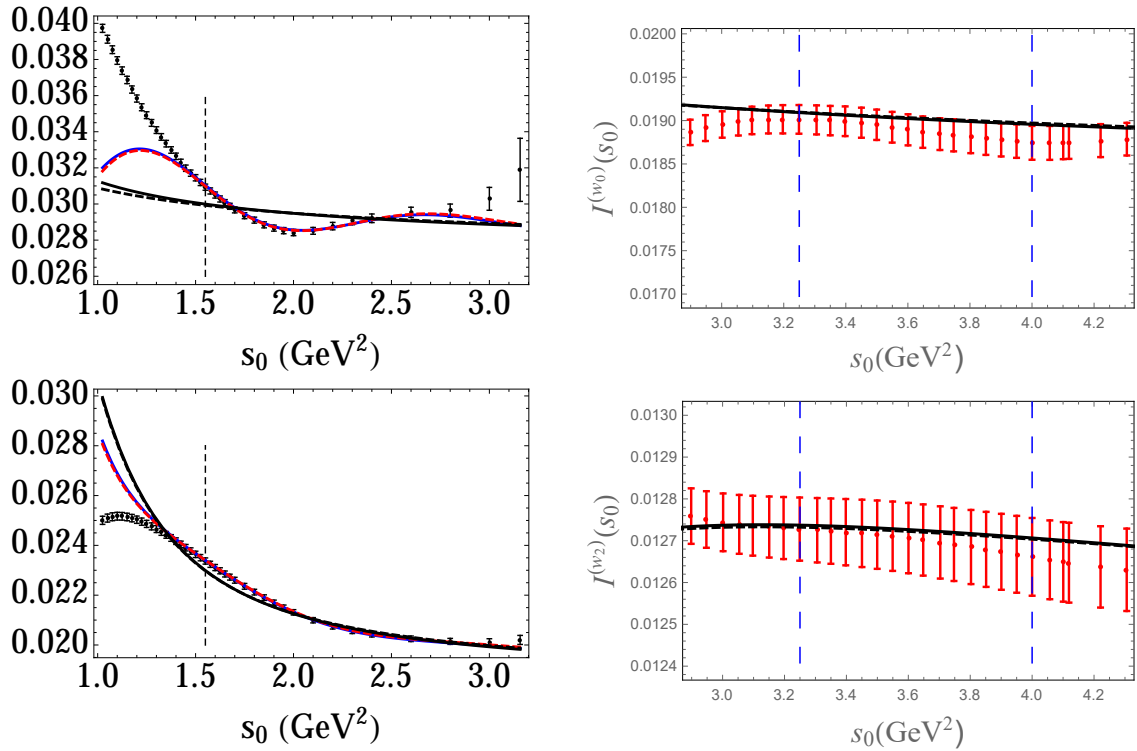

Figure 3: Comparison of FESR fits extracting $\alpha_s$ from hadronic $\tau$-decay data (left panels) *vs.* $e^+e^- \to$ hadrons($\gamma$) (right panels). Top panels show fits with weight $w_0$, bottom panels show fits with weight $w_2$. Because of the comparison between $\tau$-based moments and $e^+e^-$-based moments, we show those obtained from the vector channel in the plots on the left.

These values are both consistent, within errors, with the world average as reported in Ref. [14], confirming the running predicted by QCD between the scale of the $e^+e^-$ analysis and $m_Z$ [15].

Finally, we point out that the $R$-ratio data can be used to test results obtained in the $\tau$-based approach, as explained in the contribution by Peris to these proceedings [16].

**Acknowledgments.** We like to thank Claude Bernard and Matthias Jamin for helpful discussions. DB, AK and KM would like to thank the IFAE at the Universitat Autònoma de Barcelona for hospitality. The work of DB is supported by the São Paulo Research Foundation (FAPESP) Grant No. 2015/20689-9 and by CNPq Grant No. 305431/2015-3. The work of MG is supported by the U.S. Department of Energy, Office of Science, Office of High Energy Physics, under Award Number DE-FG03-92ER40711. The work of AK is supported by STFC under the consolidated grant ST/N504130/1. KM is supported by a grant from the Natural Sciences and Engineering Research Council of Canada. The work of DN is supported by JSPS KAKENHI grant numbers JP16K05323 and JP17H01133. SP is supported by CICYTFEDER-FPA2014-55613-P, 2014-SGR-1450. The work of TT is supported by STFC under the consolidated grant ST/P000290/1.

# $\alpha_{\rm s}$ from soft QCD jet fragmentation functions

**Redamy Pérez-Ramos**[1,2*] and David d'Enterria[3,†]

[1] *DRII-IPSA, Bis, 63 Boulevard de Brandebourg, 94200 Ivry-sur-Seine, France*
[2] *Sorbonne Universités, UPMC Univ Paris 06, UMR 7589, LPTHE, F-75005, Paris, France*
[3] *CERN, EP Department, CH-1211 Geneva 23, Switzerland*

**Abstract:** We present an extraction of the QCD coupling $\alpha_{\rm s}$ from the energy evolution of the first two moments (multiplicity and peak position) of the parton-to-hadron fragmentation functions at low fractional hadron momentum $z$. A fit of the experimental jet data, from $e^+e^-$ and deep-inelastic $e^\pm, \nu$-p collisions, to NNLO$^\star$+NNLL predictions yields $\alpha_{\rm s}(m_{\rm z}) = 0.1205 \pm 0.0010\,({\rm exp})^{+0.0022}_{-0.0000}\,({\rm th})$, in good agreement with the current $\alpha_{\rm s}$ world average.

# Introduction

In the chiral limit of zero quark masses and for fixed number of colours $N_c$, the $\alpha_{\rm s}$ coupling that determines the strength of the interaction among quarks and gluons is the only parameter of quantum chromodynamics (QCD), the theory of the strong interaction. Starting at an energy scale of order $\Lambda_{\rm QCD} \approx 0.2$ GeV, where the perturbatively-defined coupling diverges, $\alpha_{\rm s}$ decreases with energy $Q$ following a $1/\ln(Q^2/\Lambda^2_{\rm QCD})$ dependence at leading order. The current $\pm 0.9\%$ uncertainty of $\alpha_{\rm s}$ at the Z mass pole, $\alpha_{\rm s}(m_{\rm z}) = 0.1181 \pm 0.011$ [1], makes of the QCD coupling the least precisely known of all fundamental constants in nature. Improving our knowledge of $\alpha_{\rm s}$ is crucial in order to reduce the uncertainties in perturbative-QCD calculations of higher-order corrections of all hadronic cross sections and decays at colliders [2], as well as for precision electroweak fits of the Standard Model in indirect searches for new physics at future $e^+e^-$ machines [3]. The parametric dependence on $\alpha_{\rm s}$ accounts for a significant fraction of the theoretical uncertainties in e.g. the calculations of the Higgs boson H $\to b\bar{b}, c\bar{c}, gg$ partial widths [4]. The value of $\alpha_{\rm s}(m_{\rm z})$ and its evolution have also far-reaching implications including the stability of the electroweak vacuum [5], the existence of new coloured sectors at high energies [6], and our understanding of physics approaching the Planck scales, such as e.g. on the precise energy at which the interaction couplings may unify.

Having at hand new independent approaches to determine $\alpha_{\rm s}$, with experimental and theoretical uncertainties comparable to (or, even better, smaller than) those of the methods currently used, is crucial to reduce the overall uncertainty in the combined $\alpha_{\rm s}$ world-average value [2]. In Refs. [7], we presented a novel technique to extract $\alpha_{\rm s}$ from the energy evolution of the moments of the parton-to-hadron fragmentation functions (FFs) computed at increasingly higher degree of theoretical accuracy, including up to approximate next-next-to-leading-order (NNLO$^\star$) fixed-order and next-to-next-to-leading-log (NNLL) resummation corrections. We review here the latest NNLO$^\star$+NNLL theoretical calculations for the jet-energy dependence of the hadron multiplicity and the FF peak position. A fit of the analytical predictions to experimental jet measurements from $e^+e^-$ and deep-inelastic $e^\pm, \nu$-p collisions over $Q \approx 2$–200 GeV provides a new high-precision extraction of $\alpha_{\rm s}(m_{\rm z})$.

---

*e-mail: redamy.perez-ramos@ipsa.fr
†e-mail: dde@cern.ch

# DGLAP+MLLA evolution of the fragmentation functions

The conversion of a quark and gluon (collectively called partons) into a final jet of hadrons is driven by perturbative dynamics dominated by soft and collinear gluon bremsstrahlung [8] followed by the final transformation into hadrons of the last partons produced in the QCD shower at non-perturbative scales approaching $\Lambda_{\text{QCD}}$. The distribution of hadrons inside a jet is encoded in its fragmentation function, $D_{a \to h}(z, Q)$, describing the probability that an initial parton $a$ eventually fragments into a hadron $h$ carrying a fraction $z = p_{\text{hadron}}/p_{\text{parton}}$ of the parent parton's momentum. Starting with a parton at a given $\delta$-function energy $Q$, its evolution to any other energy scale $Q'$ is driven by a branching process of parton radiation and splitting, $a \to b\,c$, that can be perturbatively computed. At large $z \gtrsim 0.1$ one uses the DGLAP evolution equations [9], whereas the Modified Leading Logarithmic Approximation (MLLA) [10], resumming soft and collinear singularities, provides the proper theoretical framework at small $z$. In the latter approach, describing the region of low hadron momenta that dominates the jet fragments, one writes the FF as a function of the log of the inverse of $z$, $\xi = \ln(1/z)$. Due to colour coherence and interference in gluon radiation (known as "angular ordering"), not the softest partons but those with intermediate energies multiply most effectively in QCD cascades, leading to a final FF with a typical "hump-backed plateau" (HBP) shape as a function of $\xi$. Such an HBP can be described, without any loss of generality, in terms of a distorted Gaussian (DG, Fig. 2):

$$D(\xi, Y, \lambda) = \mathcal{N}/(\sigma\sqrt{2\pi}) \cdot e^{\left[\frac{1}{8}k - \frac{1}{2}s\delta - \frac{1}{4}(2+k)\delta^2 + \frac{1}{6}s\delta^3 + \frac{1}{24}k\delta^4\right]}, \text{ with } \delta = (\xi - \bar{\xi})/\sigma, \qquad (1)$$

where $\mathcal{N}$ is the hadron multiplicity inside a jet, and $\bar{\xi}$, $\sigma$, $s$, and $k$ are respectively the mean peak, dispersion, skewness, and kurtosis of the distribution. In Refs. [7], we described a new approach that solves the set of integro-differential equations for the FF evolution combining both DGLAP and MLLA corrections. This is done by expressing the Mellin-transformed hadron distribution in terms of the anomalous dimension $\gamma$: $D \simeq C(\alpha_s(t)) \exp\left[\int^t \gamma(\alpha_s(t'))dt\right]$ where $t = \ln Q$. Such an expression leads to a perturbative expansion in half powers of $\alpha_s$: $\gamma \sim \mathcal{O}(\alpha_s^{1/2}) + \mathcal{O}(\alpha_s) + \mathcal{O}(\alpha_s^{3/2}) + \mathcal{O}(\alpha_s^2) + \mathcal{O}(\alpha_s^{5/2}) + \cdots$, where integer powers of $\alpha_s$ correspond to fixed-order corrections, and half-integer terms can be identified with increasingly accurate resummations of soft and collinear logarithms, as schematically indicated in the following table:

| Order | LL (DLA) | NLL | NNLL | N³LL | N⁴LL |
|---|---|---|---|---|---|
| LO $P_{ac}^{(0)}$ | $\mathcal{O}(\sqrt{\alpha_s})$ | $\mathcal{O}(\alpha_s)$ | $\mathcal{O}(\alpha_s^{3/2})$ | $\mathcal{O}(\alpha_s^2)$ | $\mathcal{O}(\alpha_s^{5/2})$ |
| NLO $P_{ac}^{(1)}$ | ⋯ | ⋯ | $\mathcal{O}(\alpha_s^{3/2})$ | $\mathcal{O}(\alpha_s^2)$ | $\mathcal{O}(\alpha_s^{5/2})$ |
| NNLO $P_{ac}^{(2)}$ | ⋯ | ⋯ | ⋯ | ⋯ | $\mathcal{O}(\alpha_s^{5/2})$ |
| LO $\alpha_s$ | ⋯ | $\mathcal{O}(\alpha_s)$ | $\mathcal{O}(\alpha_s^{3/2})$ | $\mathcal{O}(\alpha_s^2)$ | $\mathcal{O}(\alpha_s^{5/2})$ |
| NLO $\alpha_s$ | ⋯ | ⋯ | $\mathcal{O}(\alpha_s^{3/2})$ | $\mathcal{O}(\alpha_s^2)$ | $\mathcal{O}(\alpha_s^{5/2})$ |
| NNLO $\alpha_s$ | ⋯ | ⋯ | ⋯ | ⋯ | $\mathcal{O}(\alpha_s^{5/2})$ |

The full set of NLO $\mathcal{O}(\alpha_s^2)$ terms for the anomalous dimension, including the two-loop splitting functions $P_{ac}^{(1)}$ and the two-loop running of $\alpha_s$, plus a fraction of the $\mathcal{O}(\alpha_s^{5/2})$ terms, coming from the NNLO expression for the $\alpha_s$ running, have been computed [12]. Upon inverse-Mellin transformation, one can derive the analytical expressions for the energy evolution of the FF, and its associated moments, as a function of $Y = \ln(E/\Lambda_{\text{QCD}})$, for an initial parton energy $E$, down to a shower cut-off scale $\lambda = \ln(Q_0/\Lambda_{\text{QCD}})$ for $N_f = 3, 4, 5$ quark flavors. The resulting formulas for the energy evolution of the moments depend on $\Lambda_{\text{QCD}}$ as *single* free parameter. Simpler expressions are obtained in the limiting-spectrum case obtained for $\lambda = 0$, i.e. evolving the FF down to $Q_0 = \Lambda_{\text{QCD}}$, motivated by the "local parton hadron duality" hypothesis for infrared-safe observables that states that the distribution of partons in jets are simply renormalized in the hadronization process without changing their shape. Thus, by fitting to Eq. (1) the measured HBP at various energies, one can determine $\alpha_s$ from the corresponding jet energy-dependence of the FF moments $\mathcal{N}$, $\bar{\xi}$, $\sigma$, $s$, and $k$.

## NNLO*+NNLL evolution of the FFs moments

As for the Schrödinger equation in quantum mechanics, the system of equations for the $D_{a\to h}(z, Q)$ FFs can be written as an evolution Hamiltonian that mixes gluon and (anti)quark states expressed in terms of DGLAP splitting functions [9] for the branchings $g \to gg$, $q(\bar{q}) \to gq(\bar{q})$ and $g \to q\bar{q}$, where $g$, $q$ and $\bar{q}$ are a gluon, quark, and anti-quark respectively. The evolution Hamiltonian is diagonalized into two eigenvalues $\gamma_{\pm\pm}$ in the new $\mathcal{D}^\pm$ basis. The relevant one for the calculation of the FF moments is $\gamma_{++}$. The analytical solution obtained at NLO+NNLL from the Mellin transform of the expressions, including the full-resummed NNLL splitting functions [11], reads (as a function of the energy of the radiated gluon $\omega$ and the variables $Y$ and $\lambda$):

$$\gamma_\omega^{\text{NLO+NNLL}} = \frac{1}{2}\omega(\mathsf{s}-1) + \frac{\gamma_0^2}{4N_c}\left[-\frac{1}{2}a_1(1+\mathsf{s}^{-1}) + \frac{\beta_0}{4}(1-\mathsf{s}^{-2})\right]$$
$$+ \frac{\gamma_0^4}{256N_c^2}(\omega\mathsf{s})^{-1}\bigg[4a_1^2(1-\mathsf{s}^{-2}) + 8a_1\beta_0(1-\mathsf{s}^{-3}) + \beta_0^2(1-\mathsf{s}^{-2})(3+5\mathsf{s}^{-2})$$
$$- 64N_c\frac{\beta_1}{\beta_0}\ln 2(Y+\lambda)\bigg]$$
$$+ \frac{1}{4}\gamma_0^2\omega\bigg[a_2(2+\mathsf{s}^{-1}+\mathsf{s}) + a_3(\mathsf{s}-1) - a_4(1-\mathsf{s}^{-1}) - a_5(1-\mathsf{s}^{-3}) - a_6\bigg], \quad (2)$$

where $\gamma_0 = \sqrt{4N_c\alpha_s/(2\pi)}$ is the LL anomalous dimension, $\mathsf{s} = \sqrt{1+4\gamma_0^2/\omega^2}$, $a_1$ and $a_2$ are hard constants obtained in [7], and $a_3$, $a_4$, $a_5$ and $a_6$ are new constants resulting from incorporating the full-resummed NNLL splitting functions. The different moments of the DG can be finally derived from the anomalous dimension via:

$$\mathcal{N} = K_0, \quad \bar{\xi} = K_1, \quad \sigma = \sqrt{K_2}, \quad s = \frac{K_3}{\sigma^3}, \quad k = \frac{K_4}{\sigma^4}; \quad (3)$$

where

$$K_{n\geq 0}(Y,\lambda) = \int_0^Y dy \left(-\frac{\partial}{\partial\omega}\right)^n \gamma_\omega(Y+\lambda)\bigg|_{\omega=0}, \quad (4)$$

Currently, beyond the analytical result given by Eq. (2), we have incorporated all $\mathcal{O}(\alpha_s^{3/2})$ contributions and also added a few of the $\mathcal{O}(\alpha_s^2)$ and $\mathcal{O}(\alpha_s^{5/2})$ ones, reaching NNLO*+NNLL accuracy. The expressions are too long to be provided here but will be given in [12]. The full inclusion of all $\mathcal{O}(\alpha_s^{5/2})$ terms is work in progress. Figure 1 shows the energy evolution of the zeroth (multiplicity) and first (peak position) moments of the FF, computed at an increasingly higher level of accuracy (from LO up to NNLO*). The FF hadron multiplicity and peak increase exponentially and logarithmically with energy respectively, and the theoretical convergence of their evolutions appears robust as indicated by the small changes introduced by adding higher-order terms.

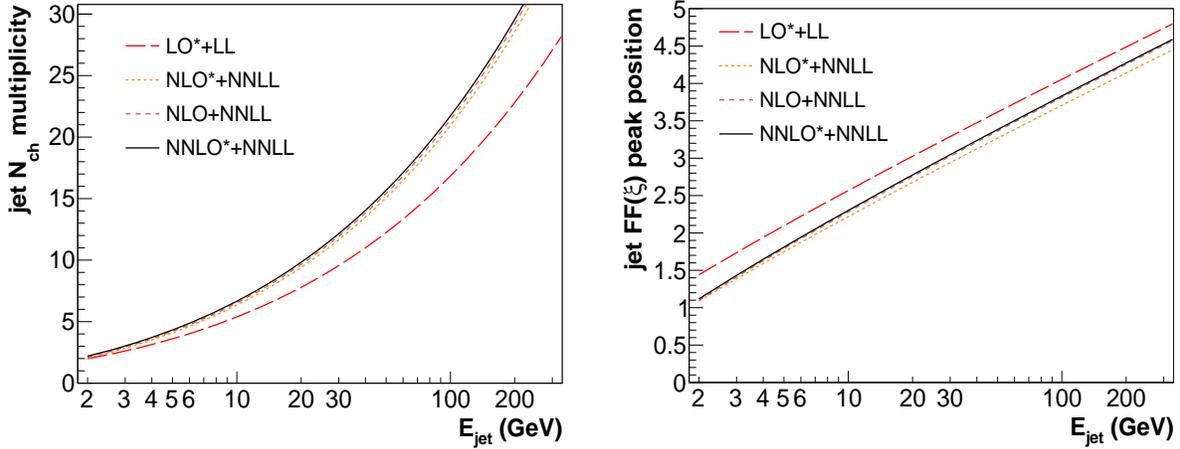

Figure 1: Theoretical energy evolution of the jet charged-hadron multiplicity (left) and FF peak position (right) at four levels of accuracy, from LO+LL up to NNLO$^\star$+NNLL.

## Data-theory comparison and $\alpha_\mathrm{s}$ extraction

In the phenomenological analysis, we first start by fitting to Eq. (1) all existing jet FF data measured in $e^+e^-$ and $e^\pm, \nu$-p collisions over $\sqrt{s} \approx 2$–200 GeV (Fig. 2), and thereby derive the corresponding FF moments at each jet energy. The overall normalization of the HBP spectrum ($\mathcal{K}_\mathrm{ch}$), which determines the average charged-hadron multiplicity of the jet (i.e. the zeroth moment of the FF), is an extra free parameter in the DG fit that, nonetheless, plays no role in the finally derived $\Lambda_\mathrm{QCD}$ value that is *solely* dependent on the evolution of the multiplicity, and not on its absolute value at any given jet energy. The impact of finite hadron-mass effects in the DG fit are taken into account through a rescaling of the theoretical (massless) parton momenta with an effective mass $m_\mathrm{eff} \approx m_\pi$. Varying such effective mass from zero to a few hundred MeV, results in small propagated uncertainties into the final extracted $\Lambda_\mathrm{QCD}$ value, as discussed in Refs. [7].

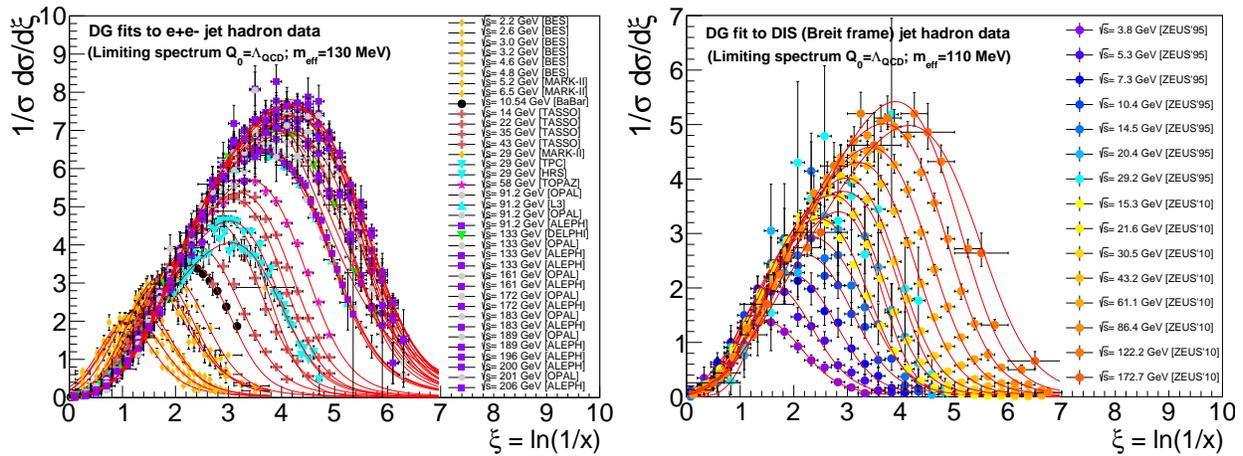

Figure 2: "Hump-backed plateau" charged-hadron distributions in jets as a function of $\xi = \ln(1/z)$ measured in $e^+e^-$ at $\sqrt{s} \approx 2$–200 GeV (left) and $e^\pm, \nu$-p (Breit frame, scaled up by $\times 2$ to account for the full hemisphere) at $\sqrt{s} \approx 4$–180 GeV (right), fitted to the DG given by Eq. (1).

Once the FF moments have been obtained, one can perform a combined fit of them as a function of the original parton energy. In the case of $e^+e^-$ collisions, the latter corresponds to half the centre-of-mass energy $\sqrt{s}/2$ whereas, for DIS, the invariant four-momentum transfer $Q_{\rm DIS}$ is used. The experimental and theoretical evolutions of the hadron multiplicity and FF peak position as a function of jet energy are shown in Fig. 3. The hadron multiplicities measured in DIS jets appear somewhat smaller (especially at high energy) than those from $e^+e^-$ collisions, due to limitations in the FF measurement only in half (current Breit) $e^\pm$p hemisphere and/or in the determination of the relevant $Q$ scale [7]. The NNLO$^\star$+NNLL limiting-spectrum ($\lambda=0$) predictions for $N_f=5$ active quark flavours[‡], leaving $\Lambda_{\rm QCD}$ as a free parameter, reproduce very well the data. Fit results for the rest of the FF moments can be found in [7]. Among FF moments, the peak position $\xi_{\rm max}$ appears as the most "robust" for the determination of $\Lambda_{\rm QCD}$, being relatively insensitive to most of the uncertainties associated with the extraction method (DG fits, energy evolution fits, finite-mass corrections, ...) as well as to higher-order corrections (Fig. 3 right).

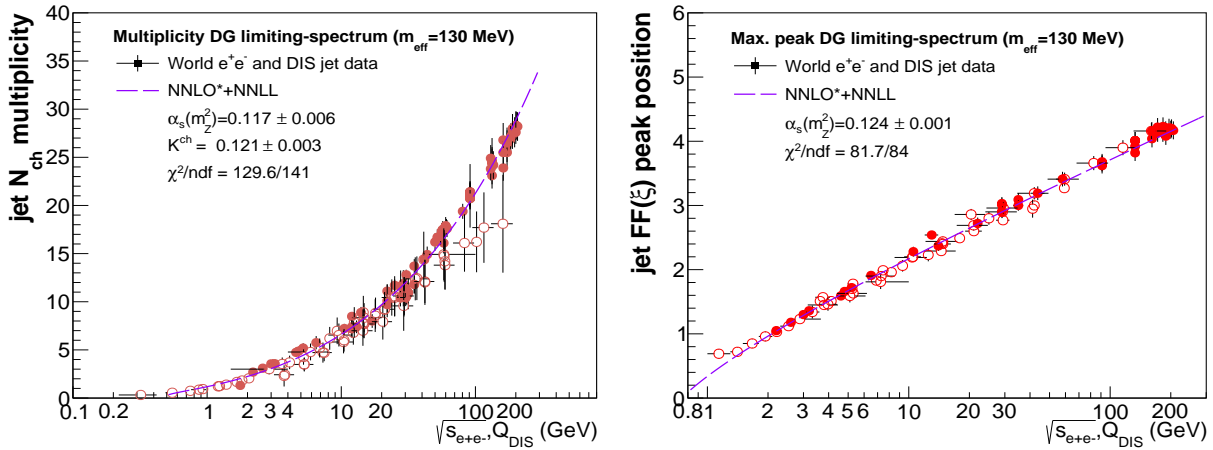

Figure 3: Energy evolution of the charged-hadron multiplicity (left) and of the FF peak position (right) measured in $e^+e^-$ and DIS data fitted to the NNLO$^\star$+NNLL predictions. The obtained $\mathcal{K}_{\rm ch}$ normalization constant, individual NNLO$^\star$ $\alpha_{\rm s}(m_{\rm z})$ values, and the goodness-of-fit per degree-of-freedom $\chi^2$/ndf, are quoted.

The QCD coupling obtained from the combined fit of the multiplicity and peak position is $\alpha_{\rm s}(m_{\rm z}) = 0.1205 \pm 0.0010^{+0.0022}_{-0.0000}$, where the first uncertainty includes all experimentally-related sources discussed in Refs. [7], and the second one is a theoretical scale uncertainty derived at NLO by stopping the parton evolution of the FFs at $Q_0 = 1$ GeV rather than at the limiting spectrum value $Q_0 = \Lambda_{\rm QCD}$. As shown in Fig. 4, our extracted $\alpha_{\rm s}(m_{\rm z})$ value is consistent with all other NNLO results from the latest PDG compilation [1], as well as with other determinations with a lower degree of theoretical accuracy [13]. The precision of our result $(+2\%, -1\%)$ is competitive with the other extractions, with a totally different set of experimental and theoretical uncertainties.

# References

[1] C. Patrignani *et al.* [Particle Data Group], Chin. Phys. C **40** (2016) 100001.

[2] D. d'Enterria, P. Z. Skands (eds.), arXiv:1512.05194 [hep-ph].

---

[‡]The moments of the lowest-$\sqrt{s}$ data have a few-percent correction applied to account for the slightly different ($N_f = 3,4$) evolutions below the charm and bottom production thresholds

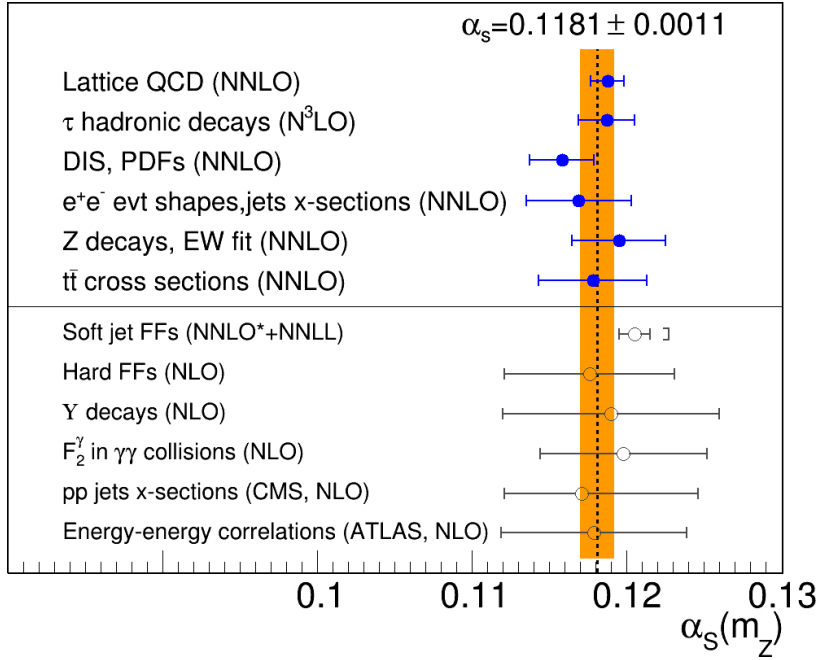

Figure 4: Summary of $\alpha_s$ determinations using different methods. The top points show $N^{2,3}LO$ extractions currently included in the PDG [1], the bottom ones shown those obtained with other approaches at lower degree of accuracy today [13], including the result of our work. The dashed line and shaded (orange) band indicate the current PDG world-average and its uncertainty.

# $\alpha_\text{s}$ from jets in $pp$ collisions

João Pires[1,2]

[1]*CFTP, Instituto Superior Técnico, Universidade de Lisboa, P-1049-001 Lisboa, Portugal*
[2]*LIP, Avenida Professor Gama Pinto 2, P-1649-003 Lisboa, Portugal*

**Abstract:** A determination of the strong coupling constant $\alpha_\text{s}$ from the single jet inclusive cross section measurements at the LHC is envisaged, using theoretical predictions at next-to-next-to-leading-order (NNLO) in QCD.

The observation of jet production at hadron colliders directly probes the basic parton-parton scattering process in QCD. As such, a number of fundamental quantities can be inferred from these measurements, such as, for example, the QCD coupling constant. As will be shown below, the interplay between the sensitivity of the inclusive jet $p_T$-spectrum to $\alpha_\text{s}$, and the experimental precision of the measurement is very favourable for a strong coupling extraction. This stems from the fact that for sufficiently high-$p_T$ jets, the jet cross section has a large rate at the LHC and a clean and simple definition, allowing jet measurements to become very precise.

Single jet inclusive measurements at the LHC have been performed by the ATLAS and CMS experiments at $\sqrt{s} = 7$ TeV [1,2,3], $\sqrt{s} = 8$ TeV [4,5] and $\sqrt{s} = 13$ TeV [6,7]. At present, the systematic uncertainty in the measurement is dominated by the jet energy scale (JES) at the 1–2% level, which, due to the steeply falling jet-$p_T$ spectrum, translates to a $< 10\%$ uncertainty on the cross section as shown in Fig. 1. Over a wide range in jet-$p_T$, it is similarly observed in ATLAS and CMS a 5% systematic uncertainty in the measurement and a statistical uncertainty at the subpercent level, which paves the way towards jet precision physics studies at the LHC. To this end, the availability of multiple single jet inclusive datasets allows for an investigation of the consistency of the data and a possible simultaneous inclusion in a combined $\alpha_\text{s}$ determination.

In reference [4], an $\alpha_\text{s}$ extraction from the single jet inclusive observable was performed using 19.7 fb$^{-1}$ of data recorded by the CMS detector from $pp$ collisions at $\sqrt{s} = 8$ TeV. In this study, 185 data points of the double differential inclusive jet cross section in the $p_T$ range 74 GeV to 2116 GeV and rapidity bins with $|y| < 3.0$, together with their statistical and systematic uncertainties and their correlation were used. The double differential inclusive jet cross section measurement [4] is shown in Fig. 2 (left) together with the NLO QCD prediction given by,

$$\frac{d\sigma}{dp_T} = \alpha_s^2(\mu_R)\hat{X}^{(0)}(\mu_F, p_T)\left[1 + \alpha_s(\mu_R)\mathrm{K1}(\mu_R, \mu_F, p_T)\right] \quad (1)$$

where $\alpha_s$ is the strong coupling, $\hat{X}^{(0)}(\mu_F, p_T)$ represents the LO contribution to the cross section and $\mathrm{K1}(\mu_R, \mu_F, p_T)$ is the NLO correction. A good agreement between the measured cross section and the theoretical prediction can be seen in Fig. 2 (right), within the errors of the NLO calculation, estimated by varying the renormalization and factorization scales in the following six combinations of scale factors: $(\mu_R/\mu, \mu_F/\mu) = (0.5, 0.5), (2, 2), (1, 0.5), (1, 2), (0.5, 1), (2, 1)$, with $\mu$ the default choice equal to the jet $p_T$. The sensitivity of the theory prediction to the $\alpha_s$ choice in the PDF is illustrated in Fig. 3, where predictions corresponding to 16 different $\alpha_s(m_Z)$ values in the range 0.112 to 0.127 in steps of 0.001 are plotted [4].

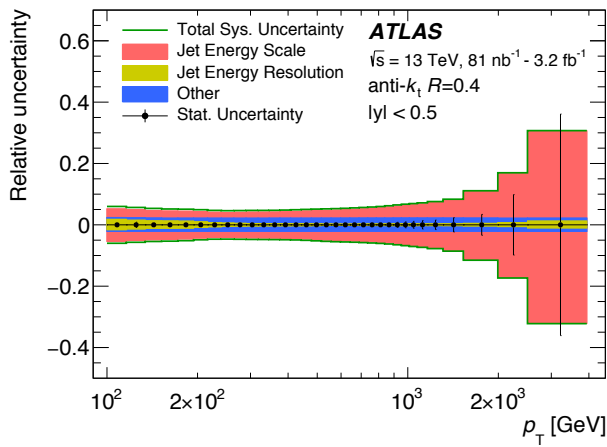

Figure 1: Relative systematic uncertainty for the inclusive jet cross-section as a function of the jet $p_T$ [7].

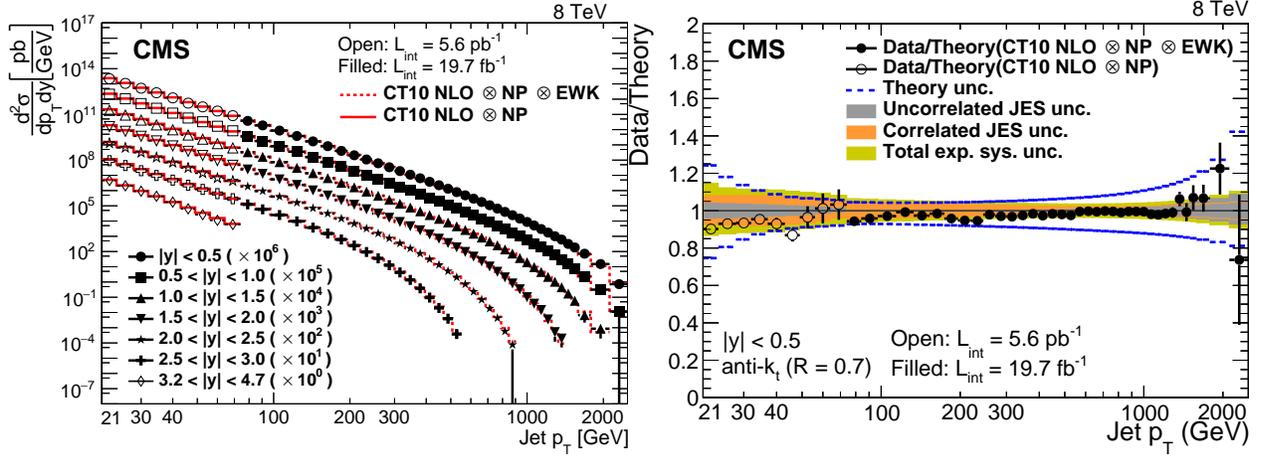

Figure 2: Left: Double-differential inclusive jet cross sections as function of jet $p_T$. Data and NLO predictions based on the CT10 PDF set [4]. Right: Ratios of data to the theory prediction using the CT10 PDF set. For comparison, the total theoretical (band enclosed by dashed lines) and the total experimental systematic uncertainties (band enclosed by full lines) are shown as well [4].

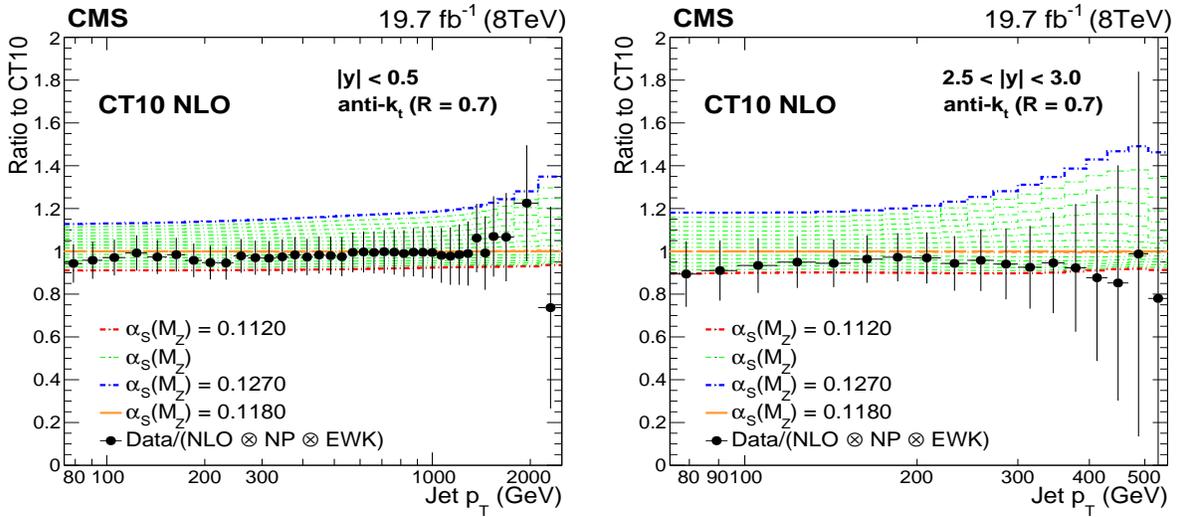

Figure 3: Ratio of data over theory prediction (closed circles) using the CT10 NLO PDF set, with the default $\alpha_s(m_Z)$ value of 0.118. Dashed lines represent the ratios of the predictions evaluated with different $\alpha_s(m_Z)$ values, to the central one. The error bars correspond to the total uncertainty of the data [4].

A comparison with the measured spectrum using the CT10 NLO PDF set, gives the best fitted $\alpha_s(m_Z)$ value [4] of,

$$\alpha_s(m_Z)(\text{NLO}) = 0.1164^{+0.0025}_{-0.0029}(\text{PDF})^{+0.0053}_{-0.0028}(\text{scale}) \pm 0.0001(\text{NP})^{+0.0014}_{-0.0015}(\text{exp}) = 0.1164^{+0.0060}_{-0.0043},$$

where the largest source of uncertainty in the $\alpha_s$ determination comes from the scale uncertainty of the NLO theory prediction, a strong indicator to the need of including higher-order corrections in the theoretical calculation.

To this end, NNLO corrections to the single jet inclusive observable including the dominant leading colour contribution from all partonic subprocesses in all channels, have been computed recently in Refs. [8,9]. These recent results provide new opportunities for QCD studies at hadron colliders, enabling precise theoretical predictions for jet observables to be used in $\alpha_s$ extractions from LHC jet data. In particular, the scale uncertainty of the jet cross section at NNLO has been thoroughly investigated in Ref. [9], leading to the observation that for this observable, the central scale choices $\mu = 2p_T$ and $\mu = \hat{H}_T{}^*$ are clearly found to be favoured in terms of stability and convergence of the predictions for single jet inclusive production.

In the remainder of this contribution we present our numerical predictions for the $\sqrt{s} = 8$ TeV single jet inclusive measurement from CMS [4], which are relevant for an $\alpha_s$ determination at NNLO. The high-precision differential QCD jet calculations to NNLO accuracy are performed by using the newly developed parton-level generator NNLOJET [10] with the MMHT2014 PDF set [11].

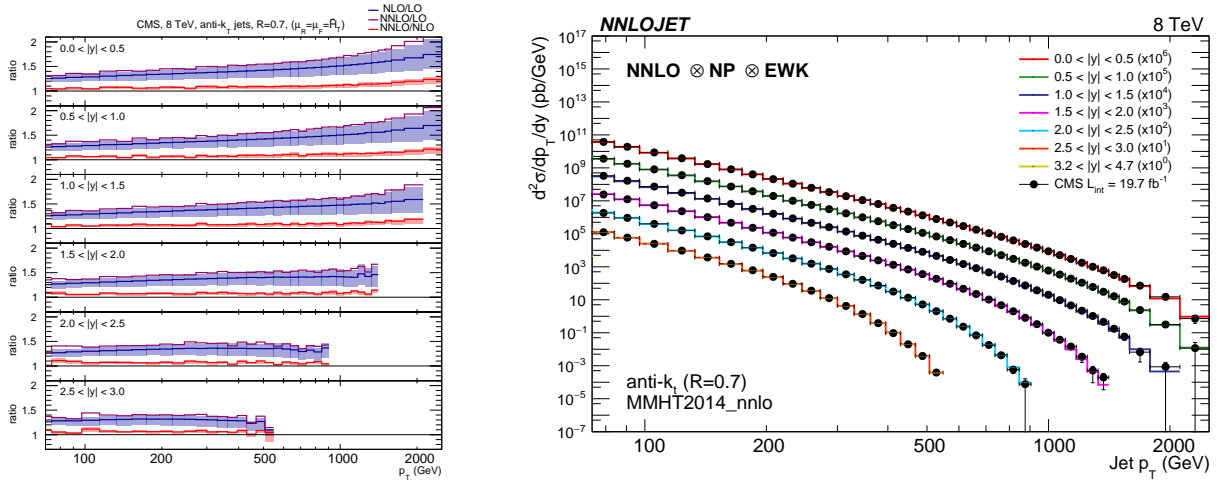

Figure 4: Left: Double-differential inclusive jet cross sections $k$-factors NNLO/NLO (red), NLO/LO (blue) and NNLO/LO (purple). The shaded bands represent the scale uncertainty of the theory predictions by varying $\mu_R$ and $\mu_F$ as described in the text. Right: The entire range of the CMS measurement compared to NNLO predictions corrected by non-perturbative (NP) and electroweak corrections (EWK) as estimated in the CMS publication [4].

---

*Where $\hat{H}_T = \sum_{i \in \text{partons}} p_{T,i}$ is defined as the transverse momentum sum of all partons in the event.

In Fig. 4 (left) we show the impact of the newly computed NNLO contribution by plotting explicitly the ratio between the NNLO prediction and the NLO result (in red), together with the ratio between the NLO cross section and the LO result (in blue) for the central scale choice $\mu = \hat{H}_T$. The NNLO correction ranges from 10% at low-$p_T$ to 20% at the highest-$p_T$ of 2.5 TeV, with scale uncertainties at the 5% level, a significant reduction with respect to NLO. In the same figure, on the right, the entire range of the CMS measurement is compared to NNLO predictions corrected by non-perturbative (NP) and electroweak corrections (EWK) as estimated by the CMS collaboration [4].

The remarkable improvement in the description of the CMS data at NNLO can be seen more clearly in Fig. 5 that explicitly shows the CMS data (black data points) and the NNLO prediction (in red), normalised to the NLO result (in blue). Over a wide range in jet-$p_T$ and rapidity $|y|$ we observe an excellent description of the jet spectrum at NNLO. A full $\alpha_{\rm s}$ fit at NNLO to the CMS dataset requires now a fast evaluation of the NNLO cross section for different input PDF and $\alpha_{\rm s}$ values.

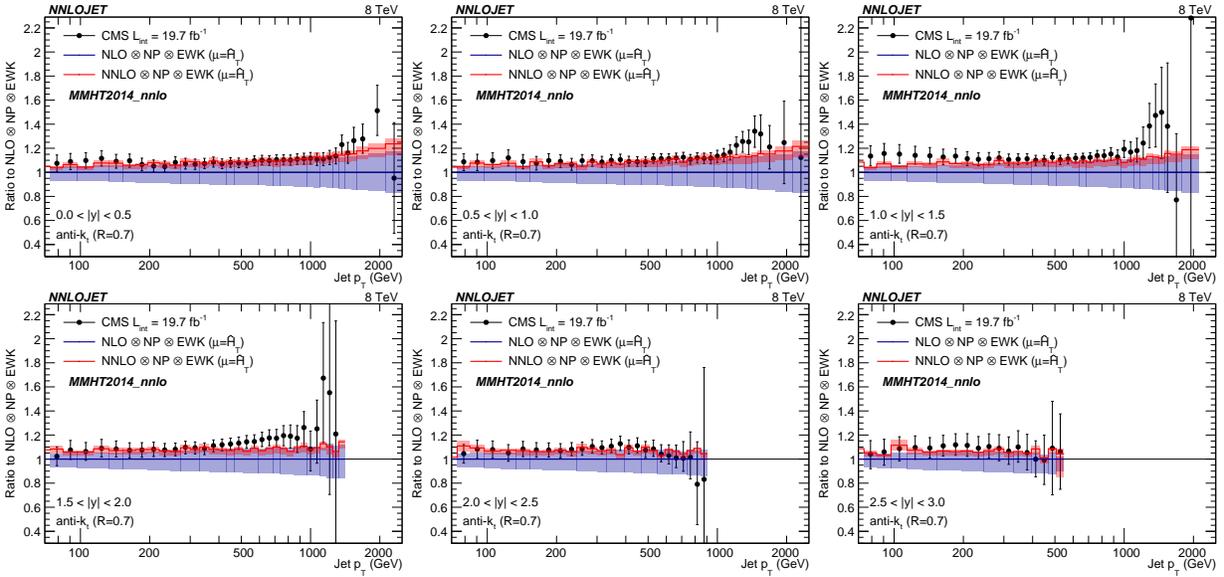

Figure 5: CMS data [4] (black data points) and NNLO prediction (in red) normalised to the NLO result (in blue) for the rapidity $|y|$ bins of the CMS $\sqrt{s} = 8$ TeV single jet inclusive measurement. Non-perturbative (NP) and electroweak corrections (EWK) as estimated in the CMS publication [4] are applied multiplicatively on top of the perturbative QCD results.

To this end, in close collaboration with experts from the FASTNLO [12] and APPLGRID [13] collaborations, an APPLFAST interface to the program NNLOJET [10] is in development to provide a fast and flexible way to reproduce the results of full jet cross sections at NLO and NNLO for the first time, in both FASTNLO and APPLGRID formats, which are suitable for state-of-the-art fits of PDF and $\alpha_{\rm s}$ with LHC jet data at NNLO.

It is anticipated that these results can substantially reduce the uncertainties on current determinations of $\alpha_{\rm s}$ from jet production at the LHC, which are largely dominated by the scale uncertainty of the NLO prediction. The scale uncertainties at NNLO are below the 5% level, which, depending on the jet-$p_T$, correspond to a reduction by more than a factor of 2 with respect to NLO. The incorporation of NNLO corrections in $\alpha_{\rm s}$ fits to jet cross sections at hadron colliders and a consistent combination of the multiple datasets available at the LHC, can in this way, open up a new contribution to the $\alpha_{\rm s}$ PDG average [14] with a target precision at the ∼1–2% level in the upcoming years.

**Acknowledgments.** It is a pleasure to thank J. Currie, A. Gehrmann, T. Gehrmann, N. Glover, A. Huss for collaboration on the work reported here. Financial support from the Fundação para a Ciência e Tecnologia (FCT-Portugal) under projects UID/FIS/00777/2019, CERN/FIS-PAR/0022/2017 and the hospitality of the CERN theory group are acknowledged.

# Jet substructure and a possible determination of the QCD coupling

Felix Ringer

*Nuclear Science Division, Lawrence Berkeley National Laboratory, Berkeley, CA 94720, USA*

**Abstract:** We review possible avenues toward an extraction of the QCD strong coupling $\alpha_\text{s}$ using jet substructure techniques. A range of jet substructure observables have been measured recently at the LHC with unprecedented precision. In addition, theoretical advances make it possible to directly compare LHC data and first principles calculations in QCD. LHC jet substructure observables may provide an independent handle on extracting $\alpha_\text{s}$ and they are particularly well suited to complement current extractions from electron-positron annihilation data. However, further theoretical and experimental efforts are needed in order to obtain a competitive extraction.

## Introduction

Jets and jet substructure are important tools for many analyses carried out at the LHC. For example, jet substructure techniques are used extensively for tagging boosted $W, Z$, discriminating between quark/gluon jets and for searches of particles beyond the standard model. The development of jet grooming techniques now also facilitate QCD precision studies using jet substructure. Jet grooming techniques are designed to remove soft wide-angle radiation from the jet. These techniques reduce the hadronization correction and remove the soft contamination from the underlying event (mostly from multiparton interactions) leaving behind only the hard core of the jet. In particular, the soft drop grooming algorithm [1] allows for first principles calculations within perturbative QCD. Despite the fact that most jet substructure observables are sensitive to very soft scales, it is possible to make direct one-to-one comparisons between theory and data. For example, jet substructure observables can be used for the tuning of parton showers, improving our understanding of the fragmentation/hadronization mechanism, and jets can be used as a well calibrated probe of the quark-gluon plasma in heavy-ion collisions. See [2] for a recent review of jet substructure techniques. Recently, it has been proposed to measure the strong coupling constant $\alpha_\text{s}$ using jet substructure observables [3]. Such an extraction is particularly challenging due to the required precision both from the experimental and the theoretical side. Here, we first review the soft drop grooming procedure and the status of theoretical calculations. We then discuss possible avenues toward an extraction of the QCD strong coupling constant.

## Soft drop groomed jet substructure observables

We consider inclusive jet production $pp \to \text{jet} + X$ at the LHC where jets are identified in a given transverse momentum $p_T$ and rapidity $\eta$ interval and no further constraints are imposed on the final state configuration. The soft drop grooming algorithm was introduced in [1] which we briefly review here. First, the jet constituents are reclustered with the Cambridge/Aachen algorithm, where the distance metric between particles only depends on their geometric distance in the $\eta$–$\phi$ plane. The obtained clustering tree is then declustered recursively, where at each step, the soft drop criterion is tested

$$\frac{\min[p_{T1}, p_{T2}]}{p_{T1} + p_{T2}} > z_\text{cut} \left(\frac{\Delta R_{12}}{R}\right)^\beta . \tag{1}$$

Here $p_{Ti}$ denote the transverse momenta of the two branches obtained at each declustering step and $\Delta R_{12}/R$ denotes their distance in the $\eta$–$\phi$ plane divided by the jet radius $R$. The soft threshold $z_{\text{cut}}$ and the angular exponent $\beta$ are free parameters of the grooming procedure. If the softer branch fails the criterion, it is removed from the jet and otherwise, the algorithm terminates and returns the groomed jet. An important feature of the soft drop grooming algorithm is that non-global logarithms are absent ($\beta = 0$) or power suppressed ($\beta > 0$) for the type of observables discussed here, except for logarithms in the grooming parameter $z_{\text{cut}}$.

Various jet substructure observables in proton-proton collisions have been calculated within perturbative QCD taking into account the effect of soft drop grooming. A class of observables which is particularly well suited for the extraction of the QCD coupling constant are jet shape observables. Similar to event shape observables in $e^+e^-$ collisions, a single number is determined in order to characterize the radiation pattern of the observed jet. The close analogy to $e^+e^-$ event shape observables makes this class of observables particularly well suited for the extraction of the QCD strong coupling constant (see for example [4,5,6]). In particular, it may be possible to learn about universality aspects of the relevant nonperturbative physics [7,8]. Examples of jet shape observables considered in the literature are jet angularities $\tau_a$ [9] and two-point correlation functions $e_2^{(\alpha)}$ [10] which are defined as

$$\tau_a = \frac{1}{p_T} \sum_{i \in J} p_{Ti} \Delta R_{iJ}^{2-a}, \qquad e_2^{(\alpha)} = \frac{1}{p_T^2} \sum_{i<j \in J} p_{Ti} p_{Tj} \Delta R_{ij}^{\alpha}. \qquad (2)$$

Here the $p_{Ti}$ denote the transverse momenta (relative to the beam) of the particles inside the jet, $\Delta R_{iJ}$ is their distance to the jet axis, and $\Delta R_{ij}$ denotes their pairwise distance in the $\eta$–$\phi$ plane. The parameters $a$ and $\alpha$ in Eq. (2) are free parameters as long as the resulting observable is infrared-collinear safe. For example, the jet mass (jet broadening) case is obtained for $a = 0$ ($a = 1$). The soft drop groomed jet mass distribution in proton-proton collisions was calculated in [11,12,13]. The more general two-point functions can be found in [11] and jet angularities in [14]. The groomed jet mass distribution was measured recently by both ATLAS [15] and CMS [16].

Here we briefly outline the QCD factorization structure using jet angularities as an example [14]. For sufficiently collimated jets, parametrically $R \ll 1$, we may separate the hard-scattering functions $H_{ab}^c$ from the formation of the jet taken into account by a jet function $\mathcal{G}_c$. We have [17,18,19,20]

$$\frac{d\sigma}{d\eta dp_T d\tau_a} = \sum_{abc} f_a(x_a, \mu) \otimes f_b(x_b, \mu) \otimes H_{ab}^c(x_a, x_b, \eta, p_T/z, \mu) \otimes \mathcal{G}_c(z, p_T R, \tau_a, \mu, z_{\text{cut}}, \beta), \qquad (3)$$

where $f_{a,b}$ are the parton distribution functions and the $\otimes$ denote integrals over the longitudinal momentum fractions $x_{a,b}$ and $z$. Here $z$ is the fraction of momentum contained in the observed jet relative to the initiating parton. For the phenomenologically relevant kinematic regime of $\tau_a^{1/(2-a)}/R \ll z_{\text{cut}} \ll 1$, we can further refactorize the jet function as

$$\begin{aligned}\mathcal{G}_c(z, p_T R, \tau_a, \mu, z_{\text{cut}}, \beta) &= \sum_i \mathcal{H}_{c \to i}(z, p_T R, \mu) S_i^{\notin \text{gr}}(z_{\text{cut}} p_T R, \beta, \mu) \\ &\times C_i(\tau_a, p_T, \mu) \otimes S_i^{\text{gr}}(\tau_a, p_T, R, \mu, z_{\text{cut}}, \beta).\end{aligned} \qquad (4)$$

The factorization and the associated renormalization group evolution equations allow for the simultaneous resummation of three classes of potentially large logarithms $\alpha_s^n \ln^n R$, $\alpha_s^n \ln^{2n}(\tau_a^{1/(2-a)}/R)$ and $\alpha_s^n \ln^{2n} z_{\text{cut}}$. Note that in Eqs. (3) and (4) only the collinear $C_i$ and soft function $S_i^{\text{gr}}$ depend on the jet angularity $\tau_a$. All other functions can be thought of as perturbatively calculable quark/gluon fractions. In Fig. 1, we show the jet mass distribution for LHC kinematics at next-to-leading logarithmic (NLL) (left) and next-to-next-to-leading logarithmic (NNLL) accuracy. The purely perturbative calculation is shown (dashed black, yellow band) and the result including non-perturbative effects (red) using the shape function of [7] with $\Omega = 1$ GeV. A more rigorous treatment of nonperturbative effects for groomed observables can be found in [8]. In addition, we show the PYTHIA 8 result [21] for comparison. We observe very good agreement and the nonperturbative effects are only important at very small jet mass values.

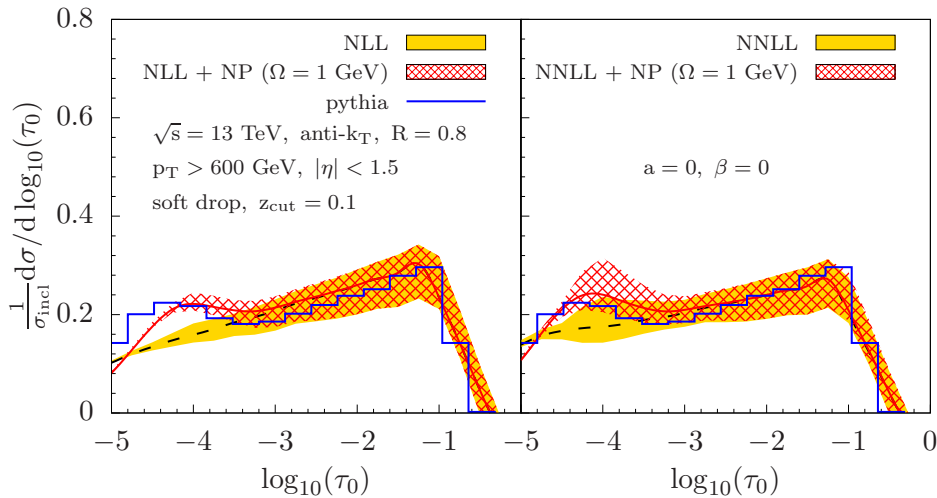

Figure 1: The soft drop groomed jet mass distribution at NLL (left) and NNLL (right) compared to PYTHIA 8 results for $\beta = 0$ and $z_{\mathrm{cut}} = 0.1$ [14]. Exemplary LHC kinematics are chosen as shown in the figure.

## A possible determination of the QCD strong coupling constant

Observables that are well suited for an extraction of the QCD strong coupling constant should be rather insensitive to nonperturbative effects and, at the same time, retain the sensitivity to $\alpha_s$. Following the arguments in [3], soft drop groomed observables have a significantly reduced sensitivity to nonperturbative effects compared to their ungroomed counterparts which can be seen as follows. The smallest scale that appears in the calculation of the groomed jet mass is $\mu_S^{\mathrm{gr}} = p_T \tau / R (z_{\mathrm{cut}} R^2 / \tau^2)^{1/(2+\beta)}$. Instead without grooming it would be $\mu_S = p_T \tau / R$. If we set $\mu_S = \Lambda_{\mathrm{QCD}} \sim 1$ GeV, we find that the onset of nonperturbative physics is shifted to significantly lower values when the grooming procedure is included

$$\tau_{\mathrm{gr}} = \tau_{\mathrm{ungr}} \left( \frac{\Lambda_{\mathrm{QCD}}}{z_{\mathrm{cut}} p_T R} \right)^{\frac{1}{1+\beta}} . \qquad (5)$$

For typical values of jet kinematics at the LHC, one finds that the onset of nonperturbative effects is pushed down by two orders of magnitude in $\tau$. The fact that the nonperturbative physics start to dominate only at very small values of $\tau$ can be seen also in Fig. 1. We note that grooming can also be considered in $e^+e^-$ collisions [22]. In proton-proton collisions it is necessary in particular to also remove from the jet soft particles resulting from the underlying event.

In Ref. [3] a range of observables and grooming parameters $z_{\mathrm{cut}}, \beta$ were explored. The criteria to determine which observables and parameters are most suitable for an extraction of $\alpha_s$ were as follows. First, the robustness to nonperturbative physics was assessed by turning hadronization effects on/off in a parton shower event generator. Second, the sensitivity to $\alpha_\mathrm{s}$ as a function of $\tau$ was determined by varying $\alpha_\mathrm{s}$ by $\pm 10\%$. The main takeaways are that gluon dominated regions exhibit a larger sensitivity to $\alpha_\mathrm{s}$ and observables such as jet broadening, $a = 1$ in Eq. (2), may be better suited than the jet mass. Given the current status of experimental results and theoretical calculations, the overall uncertainty of an extraction of $\alpha_\mathrm{s}$ was estimated to be of the order of 10%. However, this estimate is expected to improve significantly in the future.

On the theoretical side, the accuracy of the existing calculations will have to be extended to full NNLO+NNLL′ or even N$^3$LL in order to achieve a competitive determination of $\alpha_\mathrm{s}$. See for example [23,24] for recent precision calculations at fixed order. While the quark/gluon fractions may be calculated perturbatively following the factorization in Eqs. (3) and (4) above, a fit of $\alpha_\mathrm{s}$ might have to be combined with a determination of the PDFs [3]. In addition, it will be interesting to explore universality aspects of nonperturbative physics using either shape functions or Monte Carlo techniques [8]. Possible extensions are also multi-observable fits, see for example [25]. On the experimental side, the current precision may be increased by considering track based observables. However, this would require further theoretical efforts to achieve the required precision.

## Conclusions

Jet substructure techniques may help in the future to precisely constrain the QCD strong coupling constant $\alpha_\mathrm{s}$. Soft drop grooming allows for first principles calculations in perturbative QCD which may be compared directly to data taken at the LHC. The grooming procedure largely removes soft wide-angle radiation making jet substructure observables robust in the complicated LHC environment while retaining the sensitivity to $\alpha_\mathrm{s}$. In addition, new insights into universality aspects of the relevant nonperturbative physics may be obtained. In the future, it will be important to make further progress and achieve an improved precision both from the experimental and the theoretical side.

**Acknowledgements.** I would like to thank Z.-B. Kang, K. Lee, X. Liu, B. Nachman and J. Talbert for helpful discussions. This work was supported by the Department of Energy under Contract No. DE-AC0205CH11231, and the LDRD Program at LBNL.

# Extractions of the QCD coupling in ATLAS

Fernando Barreiro
On behalf of the ATLAS Collaboration

*Facultad de Ciencias, Universidad Autónoma de Madrid, 28049 Madrid*

**Abstract:** I discuss two recent measurements by the ATLAS collaboration [1] at the LHC on transverse energy-energy correlations and on the transverse momentum and rapidity dependence of dijet azimuthal decorrelations at $\sqrt{s} = 8$ TeV. They are based on the 2012 sample with an integrated luminosity of 20.2 fb$^{-1}$. They are used to determine the strong coupling constant and to probe its running up to scales of order 2 TeV.

## Measurement of transverse energy-energy correlations

Transverse energy-energy correlations (TEEC) are defined as the weighted average of azimuthal differences between jet pairs [2] i.e.:

$$\frac{1}{\sigma}\frac{d\Sigma}{d(\cos\phi)} = \frac{1}{\sigma}\sum_{ij}\int \frac{d\sigma}{dx_{\mathrm{T}i}dx_{\mathrm{T}j}d(\cos\phi)} x_{\mathrm{T}i} x_{\mathrm{T}j} dx_{\mathrm{T}i} dx_{\mathrm{T}j}, \qquad (1)$$

where the sum runs over all pairs of jets in the final state with azimuthal angular difference $\phi$ and $x_{\mathrm{T}i} = \frac{E_{\mathrm{T}i}}{E_{\mathrm{T}}}$ is the transverse energy carried by jet $i$ in units of the sum of jet transverse energies $E_{\mathrm{T}} = \sum_i E_{\mathrm{T}i}$.

In order to cancel uncertainties which are constant over $\cos\phi \in [-1, 1]$, it is useful to define the azimuthal asymmetry of the TEEC (ATEEC) as

$$\frac{1}{\sigma}\frac{d\Sigma^{\mathrm{asym}}}{d(\cos\phi)} \equiv \frac{1}{\sigma}\frac{d\Sigma}{d(\cos\phi)}\bigg|_{\phi} - \frac{1}{\sigma}\frac{d\Sigma}{d(\cos\phi)}\bigg|_{\pi-\phi}. \qquad (2)$$

Next to leading order (NLO) corrections have been recently calculated [3] using the NLOJET++ code [4]. They have been found to be moderate, with PDF uncertainties also well under control, thus making this observable suitable for a precise test of pQCD and for a determination of the strong coupling constant. Recent measurements of the TEEC have been published by the ATLAS collaboration at 8 TeV [5]. The analysis represents an extension of previous measurements at 7 TeV [6]. The selection criteria require at least two anti-$k_{\mathrm{T}}$ jets (R = 0.4) such that $H_{\mathrm{T}2} = p_{\mathrm{T}1} + p_{\mathrm{T}2} \geq 800$ GeV with the transverse momenta of any additional jet above 100 GeV. To ensure that jet reconstruction is optimal, jets are required to be central with $|\eta_{\mathrm{jet}}| \leq 2.5$. In the 2012 data sample used in this analysis the total number of selected events is $6.2 \times 10^6$. The data is further binned in six intervals in $H_{\mathrm{T}2}$, the first one between 800–850 GeV and the last one above 1400 GeV. The data are corrected for detector effects using either a bin-by-bin correction or a Bayesian unfolding. The systematic uncertainties are dominated by the choice of the MC model used in the unfolding, modelling, and by the jet energy scale, JES, which include 67 independent sources. The total systematic uncertainty for the TEEC measurements are always below the 4–5% level.

The unfolded data is fitted to NLOJET++ predictions which are dependent on $\alpha_s(m_Z)$. The choice of renormalization and factorization scales is $\mu_{\mathrm{R}} = H_{\mathrm{T}2}/2$ with $\mu_{\mathrm{F}} = \mu_{\mathrm{R}}/2$. The theoretical uncertainties are dominated by the scale uncertainties, while those due to the PDF eigenvectors are subdominant. The comparison between unfolded data for the TEEC (ATEEC) and theoretical predictions is fair as illustrated in Fig. 1 for the highest bin in $H_{T2}$ namely $H_{T2} \geq 1400\,GeV$.

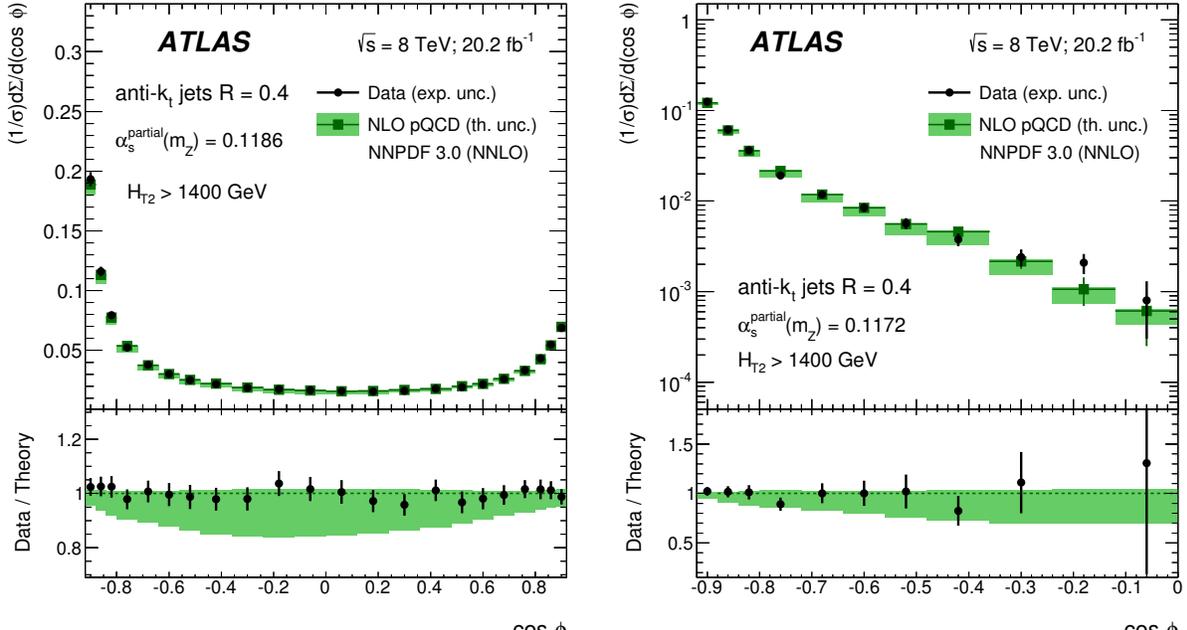

Figure 1: Results of fitting the unfolded data on the TEEC (left) and its asymmetry (right) to NLOJET++ predictions [5].

The values for $\alpha_s$ obtained at scales $H_T/2$ from the TEEC and its asymmetry are shown in Fig. 2. They are in very good agreement with the renormalization group equation (RGE) dependence predicted in QCD. In fact, the goodness of this agreement has been used recently to put limits on new coloured fermions in a way which is independent of assumptions about their decay modes [7].

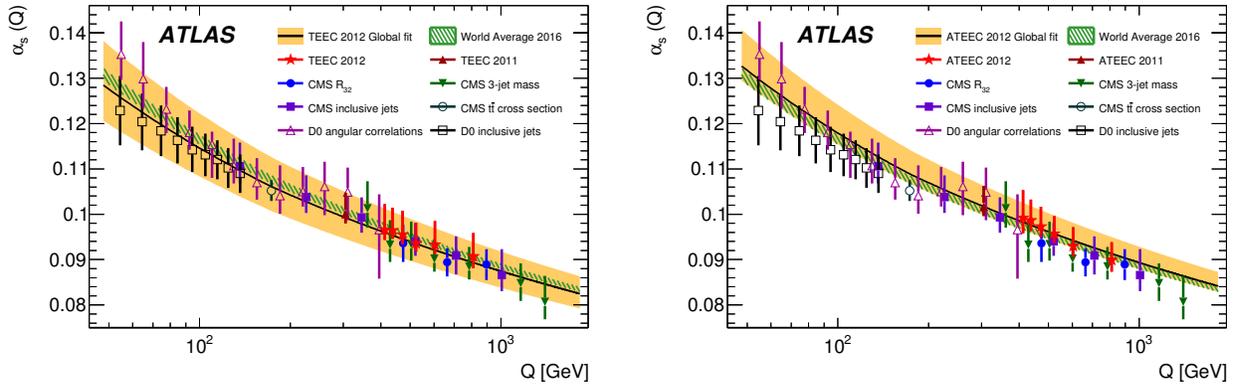

Figure 2: Scale dependence of $\alpha_s$ values obtained from TEEC (left) and ATEEC (right) measurements [5].

From a global fit to the complete data sample the following values for the strong coupling constant at the $Z$ boson mass are obtained:

$$\alpha_s^{\text{TEEC}}(m_Z) = 0.1162 \pm 0.0011\,(\text{exp})\,^{+0.0076}_{-0.0061}\,(\text{scale}) \pm 0.0018\,(\text{PDF}) \pm 0.0003\,(\text{NP})$$

$$\alpha_s^{\text{ATEEC}}(m_Z) = 0.1196 \pm 0.0013\,(\text{exp})\,^{+0.0061}_{-0.0013}\,(\text{scale}) \pm 0.0017\,(\text{PDF}) \pm 0.0004\,(\text{NP})$$

## Measurement of azimuthal decorrelations

The azimuthal decorrelations are defined as the fraction of the inclusive dijet cross-section for which the azimuthal difference between the two leading jets is smaller than a given value, $\Delta\phi_{\max}$, [8]:

$$R_{\Delta\phi}(H_{\mathrm{T}}, y^*, \Delta\phi_{\max}) = \frac{d^2\sigma_{\mathrm{dijet}}(\Delta\phi_{\mathrm{dijet}} < \Delta\phi_{\max})/dH_{\mathrm{T}}dy^*}{d^2\sigma_{\mathrm{dijet}}(\mathrm{inclusive})/dH_{\mathrm{T}}dy^*} . \quad (3)$$

ATLAS has recently presented measurements on azimuthal decorrelations [9] as an alternative method to determine the strong coupling constant and to probe pQCD at high scales. Anti-$k_{\mathrm{T}}$ jets (R=0.6) are selected with $p_{\mathrm{Tmin}} = 100$ GeV and $|y| \leq 2.5$. The selection criteria require $H_{\mathrm{T}} = \sum_i p_{\mathrm{T}i} \geq 450$ GeV, $p_{\mathrm{T}1} > H_{\mathrm{T}}/3$, $y^* = |y_1 - y_2|/2 < 2$ and $|y_i - y_{\mathrm{boost}}| < 0.5$ with $y_{\mathrm{boost}} = |y_1 + y_2|/2$. The data are presented as a function of $H_{\mathrm{T}}$ in three $y^*$ bins and four values of $\Delta\phi_{\max}$. The data are further corrected for detector effects using a bin-by-bin reweighting procedure. The systematic uncertainties are well under control, typically around the few percent level. The theoretical calculations have been performed as in the previous section using NLOJET++. The renormalization and factorization scales are set at $\mu_{\mathrm{R}} = \mu_{\mathrm{F}} = H_{\mathrm{T}}/2$. The corrected data are shown in the left hand side of Fig. 3 along with a comparison to the NLO predictions. The agreement is fair. A closer look at the data/theory ratios indicates that the predictions do best for $\Delta\phi_{\max} = 7\pi/8$ as expected. Therefore for a determination of $\alpha_{\mathrm{s}}$ the data for this particular value of $\Delta\phi_{\max}$ is integrated over $y^*$ and its $H_{\mathrm{T}}$ dependence fitted to pQCD predictions at NLO accuracy. The results of this fit are shown on the right hand side of Fig. 3. They yield the following value for the strong coupling constant at the $Z$ mass: $\alpha_{\mathrm{s}}^{\mathrm{decorr}}(m_{\mathrm{Z}}) = 0.1127^{+0.0063}_{-0.0027}$ (total). The total uncertainty is dominated again by the theoretical scales dependence.

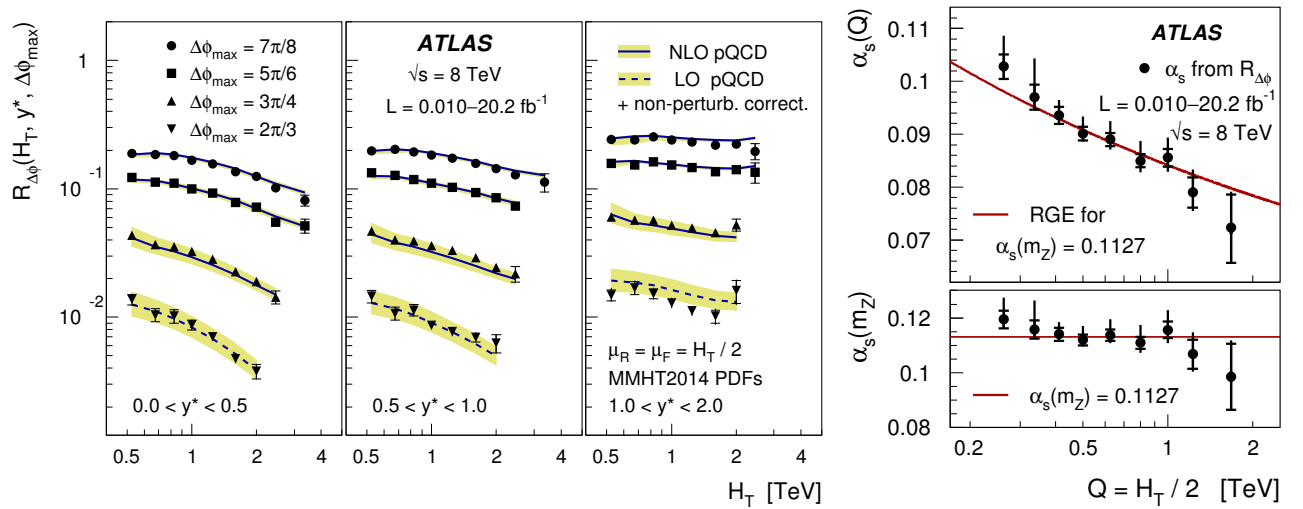

Figure 3: Left: Azimuthal decorrelations as a function of $H_{\mathrm{T}}$ compared to NLO pQCD predictions [9]. Right: Values of $\alpha_{\mathrm{s}}$ obtained from fits to $R_{\Delta\phi}$ for $\Delta\phi_{\max} = 7\pi/8$.

## Summary and conclusions

To summarize, two recent results on jet physics from the ATLAS collaboration at the LHC have been discussed as ways to probe pQCD at high scales. The main quantitative result is that the coupling constant has been measured with good precision as illustrated in Fig. 4 and its running tested to unprecedented scales of the order of 2 TeV. As a matter of fact, the level of experimental precision is similar to that of the LEP experiments. This calls for improved calculations beyond the NLO accuracy for three jet cross sections in pp collisions [10].

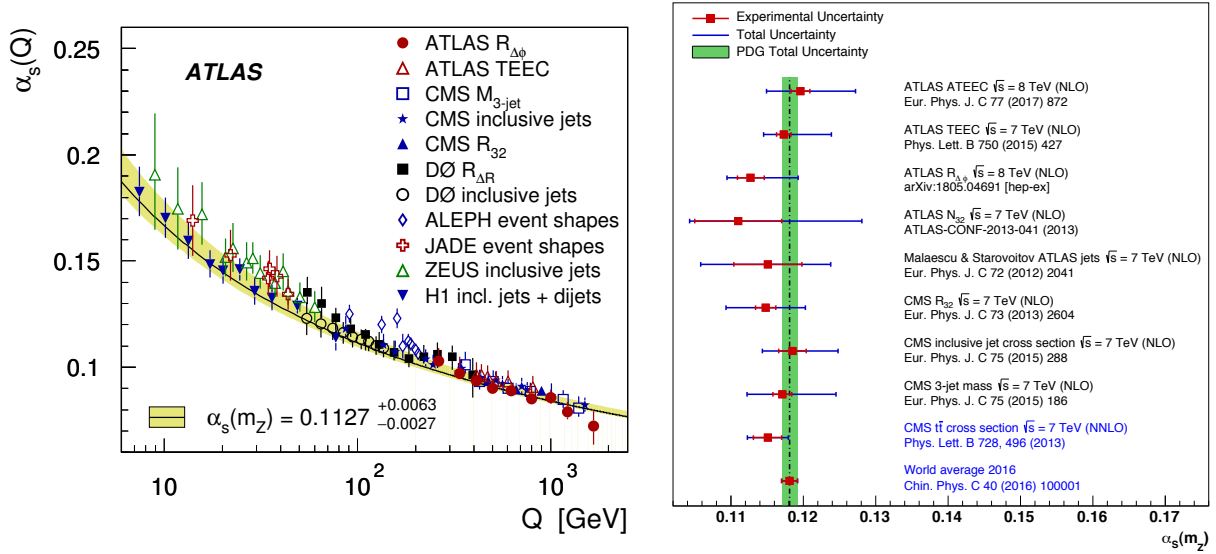

Figure 4: Left: Dependence of $\alpha_s$ on the scale from [9]. Right: Summary of $\alpha_s(m_Z)$ values obtained at colliders. In blue, those based on NNLO calculations. In green is the PDG average value [11].

# $\alpha_\text{s}$ determinations from CMS

K. Rabbertz (on behalf of the CMS Collaboration)

*KIT, Karlsruhe, Germany*

**Abstract:** Significant progress in experimental and theoretical techniques allow a determination of the strong coupling constant $\alpha_\text{s}$ from proton-proton collisions with much improved precision. Results of the CMS experiment [1] at the LHC are reviewed, which are based on measurements of jet and of top-quark pair production.

In hadron-initiated collisions jets are produced abundantly and offer the opportunity to determine $\alpha_\text{s}(m_\text{Z})$, where $m_\text{Z}$ is the mass of the Z boson. Moreover, the dependence of $\alpha_\text{s}(\mu_R)$ on the renormalisation scale $\mu_R$ can be studied up to the TeV range by identifying $\mu_R$ with the jet momenta as relevant momentum or energy scale $Q$ of the scattering process.

The CMS Collaboration has compared their measurements of the inclusive jet production cross section at 7 and 8 TeV centre-of-mass energy [2,3], which reach up to 2.5 TeV of jet transverse momentum, to predictions of perturbative quantum chromodynamics (QCD) at next-to-leading order (NLO) accuracy from NLOJet++ [4]. Including corrections for nonperturbative and electroweak effects, $\alpha_\text{s}(m_\text{Z})$ has been determined to lie in the range 0.1164–0.1192 with uncertainties of 1.3 to 3.0% from all sources other than the truncation of the perturbative expansion, cf. Table 1. The latter effect of missing higher orders is conventionally estimated by varying the renormalisation and factorisation scales $\mu_R$ and $\mu_F$ independently by factors of two avoiding the extreme cases of $\mu_R/\mu_F = 1/4$ and $\mu_R/\mu_F = 4$. At NLO this scale uncertainty amounts to 2–5% and clearly dominates the uncertainty of the extracted values of $\alpha_\text{s}(m_\text{Z})$. Since next-to-next-to-leading order (NNLO) calculations for inclusive jet and dijet production have recently become available [5,6], this uncertainty can be significantly reduced in the future.

The determination of $\alpha_\text{s}(m_\text{Z})$ from jet cross sections in hadron-initiated collisions cannot be independent of assumptions on the hadron structure. For proton-proton collisions in particular the parton distribution function (PDF) of the gluon inside protons is correlated with the strong coupling constant. This effect has been considered by CMS through either an additional uncertainty (included in the column "other" in Table 1), or by performing a simultaneous fit of $\alpha_\text{s}(m_\text{Z})$ and the proton PDFs using supplementary data on deep-inelastic scattering (DIS) from the H1 and ZEUS experiments. All four results from inclusive jet measurements are consistent among each other and with a simultaneous fit to triple-differential dijet production as reported by CMS in Ref. [7]. The "running" of the strong coupling constant, i.e. its scale dependence $\alpha_\text{s}(Q)$ as predicted by perturbative QCD, is found to be consistent with the jet measurements ranging up to 2 TeV.

Requiring additional partons, respectively jets, in the final state leads to 3-jet cross sections, which are sensitive to $\alpha_\text{s}^3$ instead of $\alpha_\text{s}^2$ as for the previous jet cross sections. Here, perturbative QCD is available only up to NLO and electroweak corrections, important at momentum scales in the TeV range, have been calculated only after the corresponding CMS publications on the 3-jet mass cross section [8] and the 3- to 2-jet ratio $R_{3/2}$ at 7 TeV [9] and 8 TeV [10] (preliminary). Compared to the 3-jet cross section the ratio $R_{3/2}$ has the advantage that numerous uncertainties cancel at least partially in the ratio. However, this comes at the price of an additional scale in the process, the $p_\text{T}$ of the third jet, and a reduced sensitivity to $\alpha_\text{s}$. The latter can be overcome by looking into multi-jet production ratios.

The results on $\alpha_s(m_Z)$ reported by CMS, cf. Table 1, suffer from somewhat enlarged scale uncertainties at NLO, but are compatible among each other and with the previous extractions, although the ratio $R_{3/2}$ exhibits a slight tendency towards smaller values of $\alpha_s(m_Z)$.

Table 1: Summary of $\alpha_s(m_Z)$ determinations from CMS. For each process the power in $\alpha_s$ of the leading order (LO), the centre-of-mass energy, the integrated luminosity, the accessed range of scale $Q$, and the number of fitted data points is given. $H'$ signifies the sum of transverse masses of all final state partons. Theory is employed at NLO accuracy for all jet related observables and the $t\bar{t}$ differential distributions. In case of the $t\bar{t}$ production cross section theory is used at NNLO+NNLL precision at 7 TeV and at NNLO precision at 13 TeV centre-of-mass energy. The last two columns compare the scale uncertainty to the quadratic sum of all other uncertainties affecting the $\alpha_s$ determinations.

| Observable | PDF fit? | LO $\alpha_s^n$ | $\sqrt{s}$ [TeV] | $L_{\text{int}}$ [fb$^{-1}$] | Q [GeV] | $N_p$ | $\alpha_s(m_Z)$ | $\delta\alpha_s(m_Z) \cdot 1000$ other | scale | Ref. |
|---|---|---|---|---|---|---|---|---|---|---|
| incl. jets | − | 2 | 7 | 5.0 | 114–2116 | 133 | 0.1185 | 35 | $^{+53}_{-24}$ | [2] |
|  | ✓ |  |  |  |  |  | 0.1192 | $^{+23}_{-19}$ | $^{+24}_{-39}$ |  |
| incl. jets | − | 2 | 8 | 19.7 | 74–2500 | 185 | 0.1164 | $^{+29}_{-33}$ | $^{+53}_{-28}$ | [3] |
|  | ✓ |  |  |  |  |  | 0.1185 | $^{+19}_{-26}$ | $^{+22}_{-18}$ |  |
| Dijet $p_{\text{T,avg}}$ | ✓ | 2 | 8 | 19.7 | 133-1784 | 122 | 0.1199 | $^{+15}_{-16}$ | $^{+31}_{-19}$ | [7] |
| 3-jet mass | − | 3 | 7 | 5.0 | 332–1635 | 46 | 0.1171 | 28 | $^{+69}_{-40}$ | [8] |
| $R_{3/2}$ | − | 1 | 7 | 5.0 | 420–1390 | 21 | 0.1148 | 23 | 50 | [9] |
| $R_{3/2}$ | − | 1 | 8 | 19.7 | 300–1680 | 29 | 0.1150 | 22 | +50 | [10] |
| $\sigma(t\bar{t})$ | − | 2 | 7 | 2.3 | $m_t$ | 1 | 0.1151 | $^{+27}_{-26}$ | $^{+9}_{-8}$ | [12] |
| $\sigma(t\bar{t})$ | − | 2 | 13 | 35.9 | $m_t$ | 1 | 0.1139 | 23 | $^{+14}_{-1}$ | [13] |
| $N_{\text{jet}}^{0,1+}, M(t\bar{t}), y(t\bar{t})$ | − | 2–3 | 13 | 35.9 | $H'/2$ | 24 | 0.1144 | 25 | $^{+16}_{-20}$ | [14] |
|  | ✓ |  |  |  |  |  | 0.1135 | $^{+18}_{-17}$ | $^{+11}_{-5}$ |  |

Exploiting the large centre-of-mass energy of 7 or even 13 TeV as compared to the top-quark mass $m_t$ of around 172 GeV, the top-quark pair production has become a very good candidate for precision studies of QCD processes. Moreover, CMS could extract a value of $\alpha_s(m_Z)$ for the first time at NNLO from hadron-hadron collisions using the theory prediction from Ref. [11]. This first result [12] as well as a new one at 13 TeV [13] with many more data are reported in rows nine and ten of Table 1. The most obvious difference to the previous results with jets is a much reduced scale uncertainty at NNLO. Also, the general tendency of smaller $\alpha_s(m_Z)$ values at NNLO than at NLO is respected here.

One complication of the $\sigma(t\bar{t})$ observable is posed by its dependency on the top-quark mass $m_t$. Since there is only one measurement point, one can either assume $m_t$ (and a PDF set) and extract $\alpha_s(m_Z)$ or do it the other way round. Both has been performed by CMS as reported in the quoted publications.

A strategy to remedy the $m_t$ dependence in $t\bar{t}$ production consists in the exploitation of many data points of a multi-differential cross section. Concretely, CMS studied in total 24 data points of the normalised $t\bar{t}$ cross section as a function of the mass $M(t\bar{t})$ and rapidity $y(t\bar{t})$ of the top-quark pair, and of the additional jet multiplicity $N_{\text{jet}}$ [14]. From these three quantities it could be shown that $M(t\bar{t})$ and $y(t\bar{t})$ are particularly sensitive to PDFs, $M(t\bar{t})$ to $m_t$, and $N_{\text{jet}}$ to $\alpha_s(m_Z)$. As a consequence, $\alpha_s(m_z)$ and $m_t$ can be determined simultaneously, even together with PDFs provided the H1 and ZEUS DIS data are added to the fit as before with jet measurements. Unfortunately though, the theory for the multi-differential distributions was available only at NLO. The values for $\alpha_s(m_z)$ with and without PDF fit are given in the last two rows of Table 1. The preliminary results [14] reported here have in the meantime been finalised and submitted to a journal [15].

To summarise, CMS has determined the strong coupling constant $\alpha_s(m_z)$ at NLO from jet cross sections and for the first time at NNLO from $t\bar{t}$ production cross sections with a significantly reduced scale uncertainty as compared to around 3–5% at NLO. Including further experimental uncertainties of 1–2%, PDF uncertainties around 1–2% as well as nonperturbative effects, the extracted values of $\alpha_s(m_z)$ are compatible among each other and with the last update of the world average as reported in Ref. [16], although small tensions are visible. Figure 1 presents an overview of the CMS results. The advent of theory predictions at NNLO and corresponding tools for fast fits promises a multitude of new results in the near future to be included in the $\alpha_s(m_z)$ combination of the review of particle physics.

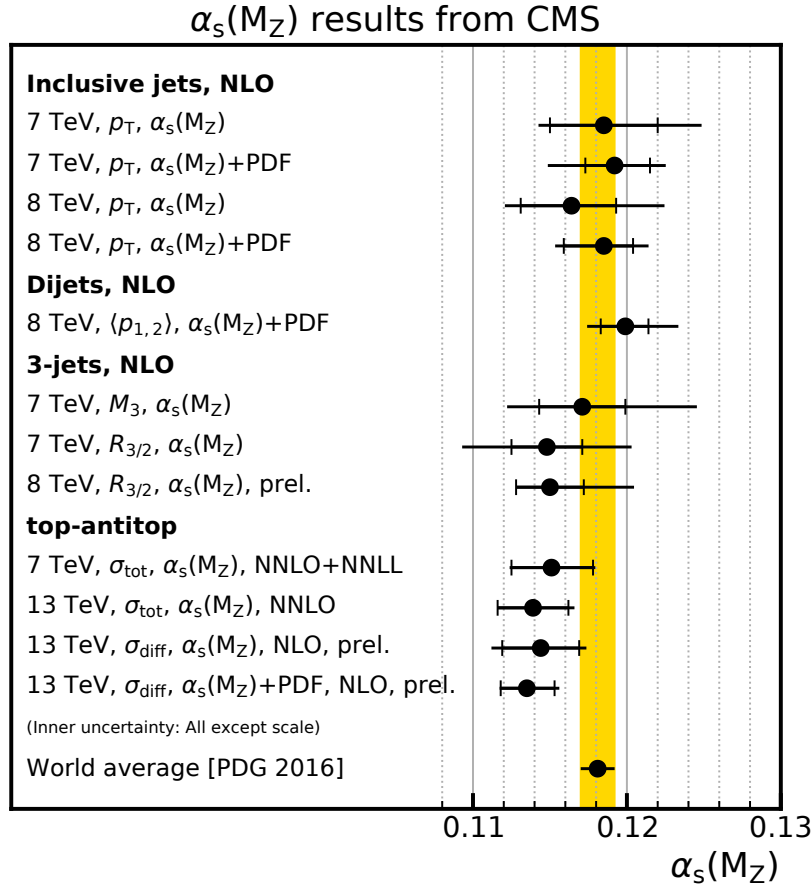

Figure 1: Summary of $\alpha_s(m_z)$ determinations from CMS. The data points show the values of $\alpha_s(m_z)$ for the various determinations as listed in Table 1 together with all uncertainties except the scale dependence (inner error bars) and the total uncertainty.

# $\alpha_\mathrm{s}$ from inclusive $W^\pm$ and Z cross sections in pp collisions at the LHC

David d'Enterria[1], **Andres Põldaru**[2], and Xiao Weichen[2]

[1] CERN, EP Department, CH-1211 Geneva 23, Switzerland
[2] LMU-University of Munich, Munich, D-85748 Garching

**Abstract:** Twenty-eight different measurements of inclusive $W^\pm$ and Z cross sections in pp collisions at the LHC are compared to the corresponding theoretical predictions, at NNLO accuracy in perturbative QCD including NLO electroweak corrections, in order to extract the QCD coupling at the Z pole, $\alpha_\mathrm{s}(m_\mathrm{z})$. The theoretical cross sections are computed for four different parton distribution functions (PDFs): CT14, HERAPDF 2.0, MMHT14, and NNPDF 3.0. The calculated cross sections reproduce well the data within experimental and theoretical uncertainties. A linear fit of the $\alpha_\mathrm{s}$ dependence of the theoretical cross sections is used to extract the $\alpha_\mathrm{s}(m_\mathrm{z})$ value that best reproduces the measured cross sections. The 28 $\alpha_\mathrm{s}(m_\mathrm{z})$ values extracted from each measurement are combined into a single result by properly taking into account their uncertainties and correlations. The following NNLO values of the QCD coupling for each one of the PDF sets are obtained: $\alpha_\mathrm{s}(m_\mathrm{z}) = 0.1181 \pm 0.0016$ (CT14), $0.1209 \pm 0.0015$ (MMHT14), and $0.1163 \pm 0.0019$ (NNPDF 3.0), with a final uncertainty at the 1.5% level, in good agreement with the $\alpha_\mathrm{s}(m_\mathrm{z})$ world average.

## Introduction

The QCD coupling is the least accurately known of all fundamental interaction couplings: the current world average at the Z boson mass, $\alpha_\mathrm{s}(m_\mathrm{z}) = 0.1181 \pm 0.0011$, has a 0.9% uncertainty [1]. New extraction methods with different types of experimental and theoretical uncertainties than those of the current determinations are needed in order to eventually improve the precision on $\alpha_\mathrm{s}(m_\mathrm{z})$ through a combined analysis of all existing results [2]. In this context, we propose a novel approach to extract $\alpha_\mathrm{s}(m_\mathrm{z})$ based on the comparison of inclusive $W^\pm$ and Z production cross sections measured at the LHC [3,4,5,6,7,8,9,10,11] to next-to-next-to-leading-order (NNLO) perturbative QCD (pQCD) calculations [12]. The method is similar to the one used to extract $\alpha_\mathrm{s}$ from the inclusive $t\bar{t}$ cross sections at hadron colliders [13,14], except that the underlying physical process is quite different: whereas $\sigma(t\bar{t})$ depends at leading order (LO) on $\alpha_\mathrm{s}$ albeit with $\sim$5% theoretical and experimental uncertainties, $\sigma(\mathrm{W},\mathrm{Z})$ is precisely known (down to $\sim$1% experimental and theoretical uncertainties) but at the Born level is a pure electroweak process with a dependence on $\alpha_\mathrm{s}(m_\mathrm{z})$ that comes only through higher-order pQCD corrections. Implementing the typical fiducial cuts of the $\mathrm{pp}, \mathrm{p\bar{p}} \to W^\pm, Z + X$ measurements performed at the Tevatron and LHC in MCFM v8.0 [12] at LO and NNLO accuracy, one can see that the higher-order QCD terms increase the Born W, Z cross sections by around 30%:

| Fiducial W, Z cross sections: | CDF | D0 | ATLAS | CMS | LHCb |
|---|---|---|---|---|---|
| NNLO/LO ratio | 1.35 | 1.35 | 1.22 | 1.33 | 1.29 |

thereby confirming their significant dependence on $\alpha_\mathrm{s}(m_\mathrm{z})$.

# Experimental data

Table 1 collects all the fiducial cross sections for W+, W−, and Z production measured in pp collisions at various center-of-mass energies ($\sqrt{s}$ = 7, 8, 13 TeV) by ATLAS (7 results), CMS (12 results), and LHCb (9 results). The experimental selection criteria on lepton transverse momentum $p_T$ and pseudorapidity $\eta$ are listed along with the measured experimental cross sections and their uncertainties. The $\ell$, $\mu$, $e$ labels refer to different measurements performed in the fully leptonic, muonic, or electronic final-states respectively. In terms of experimental uncertainties, the integrated luminosity is the largest source (1–5%, fully correlated for a given experiment at a given $\sqrt{s}$), the systematics one amounts to a 1–3% effect (partially correlated among measurements, see later), and the statistical one (0–2%, fully uncorrelated among measurements) is the smallest one.

Table 1: Summary of the experimental W$^+$, W$^-$, and Z cross sections measured at the LHC. The lepton acceptance cuts ($\ell$ for inclusive leptons, and $\mu$, $e$ for individual muon or electron final-states; $p_T^\nu$ for the missing $p_T$) are listed, along with the statistical, systematic, and integrated luminosity uncertainties.

| Selection | Cross section |
|---|---|
| **ATLAS (pp, $\sqrt{s}$ = 7 TeV)** | |
| $p_T^\ell > 25\,\text{GeV}$, $p_T^\nu > 25\,\text{GeV}$, $|\eta^\ell| < 2.5$, $m_T > 40\,\text{GeV}$ | $\sigma(\text{W}^+) = 2947 \pm 1_{(\text{stat})} \pm 15_{(\text{syst})} \pm 53_{(\text{lum})}$ pb $= 2947 \pm 55$ pb |
| $p_T^\ell > 25\,\text{GeV}$, $p_T^\nu > 25\,\text{GeV}$, $|\eta^\ell| < 2.5$, $m_T > 40\,\text{GeV}$ | $\sigma(\text{W}^-) = 1964 \pm 1_{(\text{stat})} \pm 11_{(\text{syst})} \pm 35_{(\text{lum})}$ pb $= 1964 \pm 37$ pb |
| $p_T^\ell > 20\,\text{GeV}$, $|\eta^\ell| < 2.5$, $66 < m_Z < 116\,\text{GeV}$ | $\sigma(\text{Z}) = 502.2 \pm 0.3_{(\text{stat})} \pm 1.7_{(\text{syst})} \pm 9.0_{(\text{lum})}$ pb $= 502.2 \pm 9.2$ pb |
| **ATLAS (pp, $\sqrt{s}$ = 8 TeV)** | |
| $p_T^\ell > 20\,\text{GeV}$, $|\eta^\ell| < 2.4$, $66 < m_Z < 116\,\text{GeV}$ | $\sigma(\text{Z}) = 537.10$ pb $\pm 0.45\%_{(\text{syst})} \pm 2.8\%_{(\text{lum})} = 537.10 \pm 15.23$ pb |
| **ATLAS (pp, $\sqrt{s}$ = 13 TeV)** | |
| $p_T^\ell > 25\,\text{GeV}$, $p_T^\nu > 25\,\text{GeV}$, $|\eta^\ell| < 2.5$, $m_T > 50\,\text{GeV}$ | $\sigma(\text{W}^+) = 4.53 \pm 0.01_{(\text{stat})} \pm 0.09_{(\text{syst})} \pm 0.10_{(\text{lum})}$ nb $= 4.53 \pm 0.13$ nb |
| $p_T^\ell > 25\,\text{GeV}$, $p_T^\nu > 25\,\text{GeV}$, $|\eta^\ell| < 2.5$, $m_T > 50\,\text{GeV}$ | $\sigma(\text{W}^-) = 3.50 \pm 0.01_{(\text{stat})} \pm 0.07_{(\text{syst})} \pm 0.07_{(\text{lum})}$ nb $= 3.50 \pm 0.10$ nb |
| $p_T^\ell > 25\,\text{GeV}$, $|\eta^\ell| < 2.5$, $66 < m_Z < 116\,\text{GeV}$ | $\sigma(\text{Z}) = 0.779 \pm 0.003_{(\text{stat})} \pm 0.006_{(\text{syst})} \pm 0.016_{(\text{lum})}$ nb $= 0.779 \pm 0.017$ nb |
| **CMS (pp, $\sqrt{s}$ = 7 TeV)** | |
| $p_T^e > 25\,\text{GeV}$, $|\eta^e| < 2.5$ | $\sigma(\text{W}_e^+) = 3.404 \pm 0.012_{(\text{stat})} \pm 0.067_{(\text{syst})} \pm 0.136_{(\text{lum})}$ nb $= 3.404 \pm 0.152$ nb |
| $p_T^e > 25\,\text{GeV}$, $|\eta^e| < 2.5$ | $\sigma(\text{W}_e^-) = 2.284 \pm 0.010_{(\text{stat})} \pm 0.043_{(\text{syst})} \pm 0.091_{(\text{lum})}$ nb $= 2.284 \pm 0.101$ nb |
| $p_T^e > 25\,\text{GeV}$, $|\eta^e| < 2.5$, $60 < m_Z < 120\,\text{GeV}$ | $\sigma(\text{Z}_e) = 0.452 \pm 0.005_{(\text{stat})} \pm 0.010_{(\text{syst})} \pm 0.018_{(\text{lum})}$ nb $= 0.452 \pm 0.021$ nb |
| $p_T^\mu > 25\,\text{GeV}$, $|\eta^\mu| < 2.1$ | $\sigma(\text{W}_\mu^+) = 2.815 \pm 0.009_{(\text{stat})} \pm 0.042_{(\text{syst})} \pm 0.113_{(\text{lum})}$ nb $= 2.815 \pm 0.121$ nb |
| $p_T^\mu > 25\,\text{GeV}$, $|\eta^\mu| < 2.1$ | $\sigma(\text{W}_\mu^-) = 1.921 \pm 0.008_{(\text{stat})} \pm 0.027_{(\text{syst})} \pm 0.077_{(\text{lum})}$ nb $= 1.921 \pm 0.082$ nb |
| $p_T^\mu > 20\,\text{GeV}$, $|\eta^\mu| < 2.1$, $60 < m_Z < 120\,\text{GeV}$ | $\sigma(\text{Z}_\mu) = 0.396 \pm 0.003_{(\text{stat})} \pm 0.007_{(\text{syst})} \pm 0.016_{(\text{lum})}$ nb $= 0.396 \pm 0.018$ nb |
| **CMS (pp, $\sqrt{s}$ = 8 TeV)** | |
| $p_T^e > 25\,\text{GeV}$, $|\eta^e| < 1.44$, $1.57 < |\eta^e| < 2.5$ | $\sigma(\text{W}_e^+) = 3.54 \pm 0.02_{(\text{stat})} \pm 0.11_{(\text{syst})} \pm 0.09_{(\text{lum})}$ nb $= 3.54 \pm 0.14$ nb |
| $p_T^e > 25\,\text{GeV}$, $|\eta^e| < 1.44$, $1.57 < |\eta^e| < 2.5$ | $\sigma(\text{W}_e^-) = 2.39 \pm 0.01_{(\text{stat})} \pm 0.06_{(\text{syst})} \pm 0.06_{(\text{lum})}$ nb $= 2.39 \pm 0.09$ nb |
| $p_T^e > 25\,\text{GeV}$, $|\eta^e| < 1.44$, $1.57 < |\eta^e| < 2.5$, $60 < m_Z < 120\,\text{GeV}$ | $\sigma(\text{Z}_e) = 0.45 \pm 0.01_{(\text{stat})} \pm 0.01_{(\text{syst})} \pm 0.01_{(\text{lum})}$ nb $= 0.45 \pm 0.02$ nb |
| $p_T^\mu > 25\,\text{GeV}$, $|\eta^\mu| < 2.1$ | $\sigma(\text{W}_\mu^+) = 3.10 \pm 0.01_{(\text{stat})} \pm 0.04_{(\text{syst})} \pm 0.08_{(\text{lum})}$ nb $= 3.10 \pm 0.09$ nb |
| $p_T^\mu > 25\,\text{GeV}$, $|\eta^\mu| < 2.1$ | $\sigma(\text{W}_\mu^-) = 2.24 \pm 0.01_{(\text{stat})} \pm 0.02_{(\text{syst})} \pm 0.06_{(\text{lum})}$ nb $= 2.24 \pm 0.06$ nb |
| $p_T^\mu > 25\,\text{GeV}$, $|\eta^\mu| < 2.1$, $60 < m_Z < 120\,\text{GeV}$ | $\sigma(\text{Z}_\mu) = 0.40 \pm 0.01_{(\text{stat})} \pm 0.01_{(\text{syst})} \pm 0.01_{(\text{lum})}$ nb $= 0.40 \pm 0.02$ nb |
| **LHCb (pp, $\sqrt{s}$ = 7 TeV)** | |
| $p_T^\ell > 20\,\text{GeV}$, $2.0 < \eta^\ell < 4.5$ | $\sigma(\text{W}^+) = 878.0 \pm 2.1_{(\text{stat})} \pm 6.7_{(\text{syst})} \pm 9.3_{(\text{en})} \pm 15.0_{(\text{lum})}$ pb $= 878.0 \pm 19.0$ pb |
| $p_T^\ell > 20\,\text{GeV}$, $2.0 < \eta^\ell < 4.5$ | $\sigma(\text{W}^-) = 689.5 \pm 2.0_{(\text{stat})} \pm 5.3_{(\text{syst})} \pm 6.3_{(\text{en})} \pm 11.8_{(\text{lum})}$ pb $= 689.5 \pm 14.5$ pb |
| $p_T^\ell > 20\,\text{GeV}$, $2.0 < \eta^\ell < 4.5$, $60 < m_Z < 120\,\text{GeV}$ | $\sigma(\text{Z}) = 76.0 \pm 0.3_{(\text{stat})} \pm 0.5_{(\text{syst})} \pm 1.0_{(\text{en})} \pm 1.3_{(\text{lum})}$ pb $= 76.0 \pm 1.7$ pb |
| **LHCb (pp, $\sqrt{s}$ = 8 TeV)** | |
| $p_T^e > 20\,\text{GeV}$, $2.0 < \eta^e < 4.25$ | $\sigma(\text{W}_e^+) = 1124.4 \pm 2.1_{(\text{stat})} \pm 21.5_{(\text{syst})} \pm 11.2_{(\text{en})} \pm 13.0_{(\text{lum})}$ pb $= 1124.4 \pm 27.6$ pb |
| $p_T^e > 20\,\text{GeV}$, $2.0 < \eta^e < 4.25$ | $\sigma(\text{W}_e^-) = 809.0 \pm 1.9_{(\text{stat})} \pm 18.1_{(\text{syst})} \pm 7.0_{(\text{en})} \pm 9.4_{(\text{lum})}$ pb $= 809.0 \pm 21.6$ pb |
| $p_T^\mu > 20\,\text{GeV}$, $2.0 < \eta^\mu < 4.5$ | $\sigma(\text{W}_\mu^+) = 1093.6 \pm 2.1_{(\text{stat})} \pm 7.2_{(\text{syst})} \pm 10.9_{(\text{en})} \pm 12.7_{(\text{lum})}$ pb $= 1093.6 \pm 18.3$ pb |
| $p_T^\mu > 20\,\text{GeV}$, $2.0 < \eta^\mu < 4.5$ | $\sigma(\text{W}_\mu^-) = 818.4 \pm 1.9_{(\text{stat})} \pm 5.0_{(\text{syst})} \pm 7.0_{(\text{en})} \pm 9.5_{(\text{lum})}$ pb $= 818.4 \pm 13.0$ pb |
| $p_T^\mu > 20\,\text{GeV}$, $2.0 < \eta^\mu < 4.5$, $60 < m_Z < 120\,\text{GeV}$ | $\sigma(\text{Z}_\mu) = 95.0 \pm 0.3_{(\text{stat})} \pm 0.7_{(\text{syst})} \pm 1.1_{(\text{en})} \pm 1.1_{(\text{lum})}$ pb $= 95.0 \pm 1.7$ pb |
| **LHCb (pp, $\sqrt{s}$ = 13 TeV)** | |
| $p_T^\ell > 20\,\text{GeV}$, $2.0 < \eta^\ell < 4.5$, $60 < m_Z < 120\,\text{GeV}$ | $\sigma(\text{Z}) = 194.3 \pm 0.9_{(\text{stat})} \pm 3.3_{(\text{syst})} \pm 7.6_{(\text{lum})}$ pb $= 194.3 \pm 8.3$ pb |

## Cross sections: data vs. NNLO

We use MCFM v8.0 [12] to calculate the cross sections at NNLO pQCD accuracy with four PDF sets: CT14 [15], HERAPDF 2.0 [16], MMHT14 [17], and NNPDF 3.0 [18], interfaced through LHAPDF v6.1.6 [19]; and for 5 or 7 different input QCD coupling values over the range $\alpha_s(m_z) = 0.111$–$0.121$. The experimental kinematical cuts on the final state lepton(s) for each system, listed in Table 1, were implemented in the code. For all the 28 systems shown in Table 1, this resulted in about 20 000 computing jobs. Since MCFM does not include electroweak corrections, we used MCSANC v1.01 [20] to compute them: For each system, we calculated the NLO pQCD cross section with electroweak corrections on and off, and applied this ratio to the MCFM results at NNLO accuracy. We calculated the PDF uncertainties using the procedures corresponding to each PDF set (i.e. using their corresponding symmetric or asymmetric eigenvalues or replicas). To assess the uncertainty from missing higher-order corrections, we varied the renormalization and factorization scales from their default values ($\mu_{R,F} = m_{W,Z}$) by factors of 2 and 1/2, and took the maximum variation in the cross section from the central value. All in all, in terms of theoretical uncertainty sources, the largest one is associated with the PDF uncertainty ($\sim$2–3%), followed by the theoretical scale uncertainty (about 1%), and the numerical uncertainty (around 0.7%).

Figure 1 shows a comparison of a subset of representative $W^+$, $W^-$, and Z cross sections measured at the LHC (horizontal black line with grey bands indicating the experimental uncertainties) compared to the NNLO theoretical predictions as a function of $\alpha_s(m_z)$. The ellipsoids show the convolution of theoretical and systematic uncertainties. All theoretical predictions agree with the experimental data within uncertainties, although not always for the same fixed value of $\alpha_s(m_z)$, in particular for the calculations obtained with HERAPDF 2.0, a fact that indicates some underlying tensions. Usually, a hierarchy of NNLO cross section predictions as a function of $\alpha_s(m_z)$ is apparent with, for $\alpha_s(m_z) = 0.1181$ fixed at the world average, the results obtained with HERAPDF 2.0 (NNPDF 3.0) overestimating (underestimating) the experimental data.

## Extraction of $\alpha_s(m_z)$

To extract the value of $\alpha_s(m_z)$ preferred by each experimental measurement, we proceed as follows. First, we fit to a first-order polynomial the observed dependence of the theoretical cross section on $\alpha_s(m_z)$ for each individual system (such a linear fit goes through the ellipsoids plotted in Fig. 1). The slope of this curve indicates the sensitivity of the theoretical cross section to the underlying $\alpha_s(m_z)$ value. The plots of Fig. 1 indicate that the predictions obtained with HERAPDF 2.0 (MMHT14) have the smallest (largest) slope, i.e. have the least (most) sensitivity to $\alpha_s$ variations. The crossing point of each theoretical $\sigma_{W,Z}^{th}$-versus-$\alpha_s(m_z)$ curve with the experimental cross section (straight flat line in Fig. 1) gives the preferred $\alpha_s(m_z)$ value for each system. It can be easily shown that the theoretical and experimental uncertainties of the cross sections can be properly propagated to the derived $\alpha_s(m_z)$ value by dividing each cross section uncertainty by the slope of the $\sigma_{W,Z}^{th}$-versus-$\alpha_s(m_z)$ curve.

The procedure described above yields 28 extractions of $\alpha_s(m_z)$ for each one of the 4 PDF sets. Those are combined properly into a single $\alpha_s(m_z)$ value per PDF by taking into account their correlations and uncertainties using the CONVINO tool (with the Neyman $\chi^2$ prescription) [21]. For the correlations between different measurements, we make the following assumptions. The experimental statistical and theoretical numerical errors are fully uncorrelated. The integrated luminosity is taken to have a 0.5 correlation at the same $\sqrt{s}$ for different experiments, full correlation within the same experiment at the same $\sqrt{s}$, and zero for different $\sqrt{s}$. For the PDF and scales, we calculate the Pearson correlation coefficients for all data points and take it as the correlations of their corresponding uncertainties. For the experimental systematic uncertainties, we did a detailed study based on the CMS measurements [10,11] and, preliminarily, apply the same CMS correlations to both LHCb and ATLAS results. This results in relatively strong correlations of the experimental measurements performed with the same lepton. We then insert the 28 $\alpha_{\rm s}(m_{\rm Z})$ results per PDF and their correlation matrices into CONVINO, with a $\chi^2$ minimization taking into account asymmetric uncertainties, to determine the best $\alpha_{\rm s}(m_{\rm Z})$ value per PDF set. This combination gives the results shown in Fig. 2.

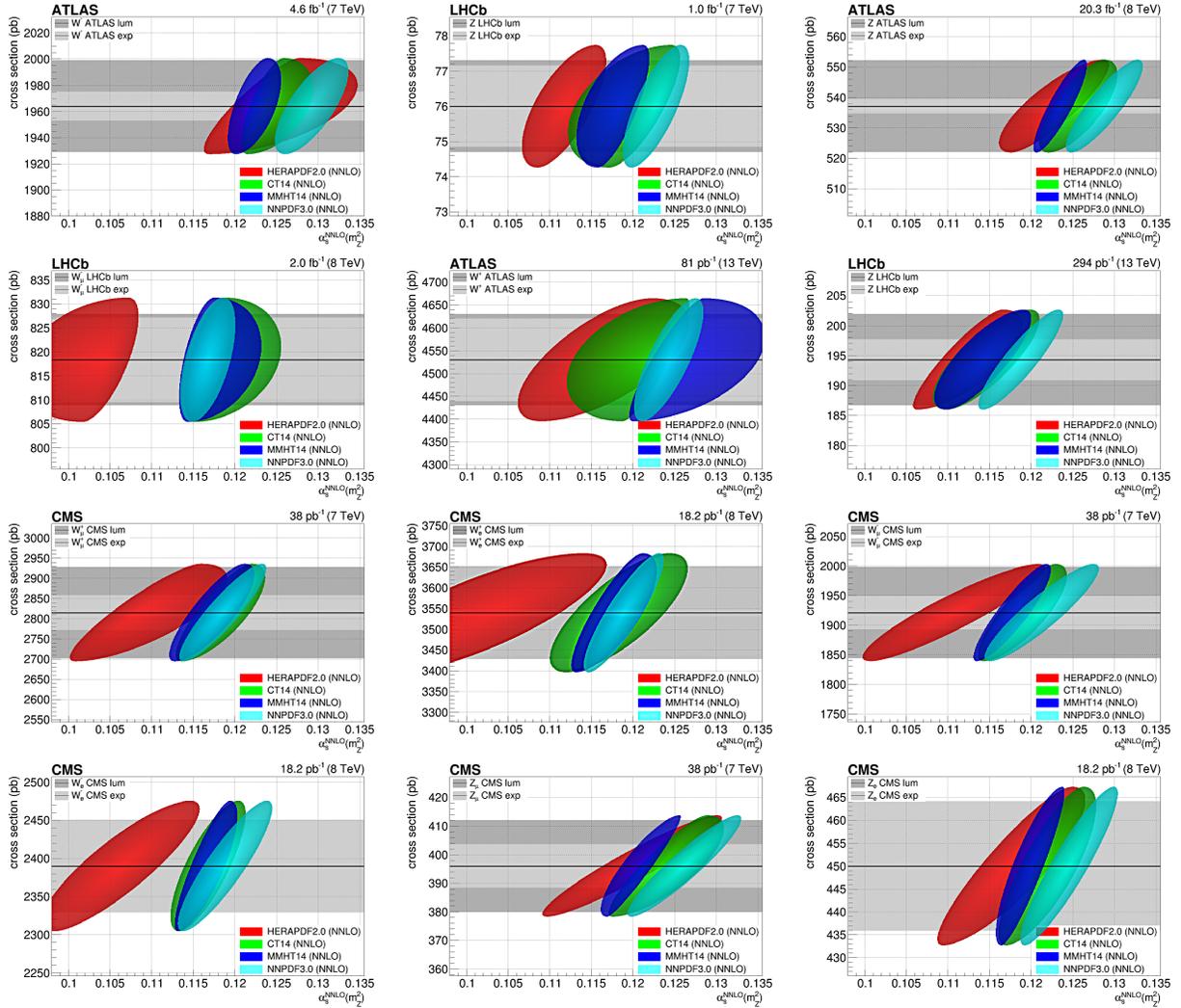

Figure 1: Examples of experimental W$^{\pm}$ and Z cross sections (lines with grey uncertainty bands) compared to theoretical NNLO predictions (ellipsoids, for each PDF set) as a function of $\alpha_{\rm s}(m_{\rm Z})$.

For CT14 we obtain $\alpha_{\rm s}(m_{\rm Z}) = 0.1181 \pm 0.0016$, for MMHT14: $\alpha_{\rm s}(m_{\rm Z}) = 0.1209 \pm 0.0015$, and for NNPDF 3.0: $\alpha_{\rm s}(m_{\rm Z}) = 0.1163 \pm 0.0019$. In this preliminary analysis, CONVINO did not converge on a stable result for HERAPDF 2.0. We see that despite the fact that PDF uncertainties are asymmetric for all sets except NNPDF 3.0, the final $\alpha_{\rm s}(m_{\rm Z})$ uncertainties turn out to be symmetric. All extractions are in reasonable agreement with each other and with the world average, considering that the uncertainty bars correspond to one standard deviation.

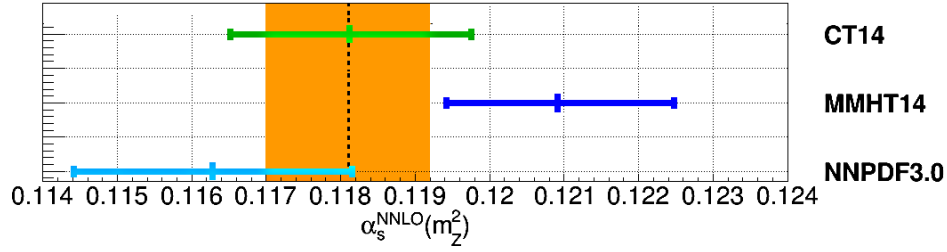

Figure 2: Final $\alpha_s(m_z)$ obtained by combining 28 individual extractions based on the $W^\pm, Z$ cross sections listed in Table 1 for the CT14, MMHT14, and NNPDF 3.0 parton densities, compared to the world-average (orange box).

In order to test the stability of our $\alpha_s$ extraction, we ran an analysis to determine the sensitivity of each final $\alpha_s(m_z)$ value on the data sets, their individual correlations and uncertainties. Figure 3 shows, for each PDF, the $\alpha_s(m_z)$ results obtained with the default assumptions (top point), with symmetrized PDF uncertainties (second point), when adding an extra 1% statistical uncertainty to all cross sections (third point), when using only the 7 or 8 TeV cross sections (fourth and fifth point), when assuming the integrated luminosity to be fully correlated at the same $\sqrt{s}$ between different experiments (sixth point), when dividing the PDF or experimental systematic correlations by a factor of two (seventh and eighth point), and when using only 7 TeV results and in addition also symmetrizing the PDF errors or adding an extra 1% uncorrelated uncertainty (last two points). This figure shows that the derived $\alpha_s(m_z)$ values mostly remain within one standard deviation of the default results plotted in Fig. 2.

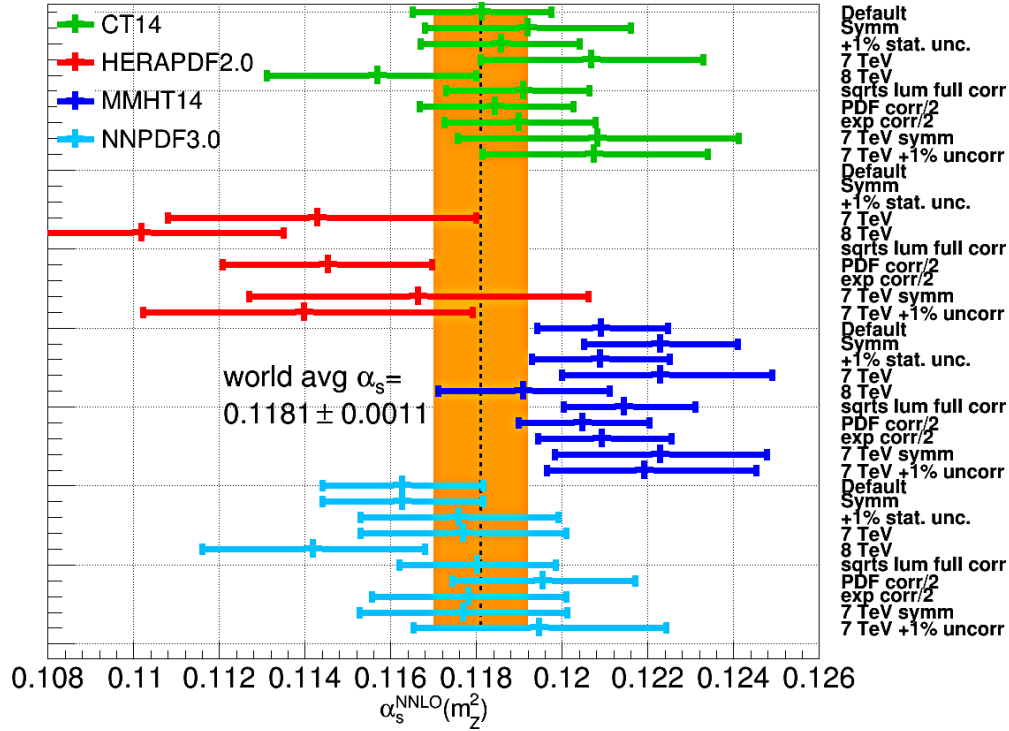

Figure 3: Overview of the sensitivity of the final $\alpha_s(m_z)$ value extracted per PDF to various assumptions on the data, the theoretical and experimental uncertainties and on their correlations.

# Summary and conclusions

We have used 28 measurements of the inclusive fiducial $W^{\pm}$ and Z production cross sections in proton-proton collisions at $\sqrt{s}$ = 7, 8, and 13 TeV, carried out in the electron and muon decay channels by the ATLAS, CMS, and LHCb experiments, to extract the QCD coupling at the Z mass pole $\alpha_{\rm s}(m_{\rm Z})$. The procedure is based on a detailed comparison of the measured weak boson cross sections to theoretical calculations computed at NNLO accuracy with the CT14, HERAPDF2.0, MMHT14, and NNPDF3.0 parton densities. The overall data–theory agreement is good within the experimental and theoretical uncertainties, but the CT14 and MMHT14 parton densities seem to provide the overall best description of all experimental data for the default value of the QCD coupling, $\alpha_{\rm s}(m_{\rm Z})$ = 0.118 in all PDF sets. A procedure has been employed to combine the 28 individual $\alpha_{\rm s}$ extractions per PDF into a single value by properly taking into account all individual sources of experimental and theoretical uncertainties and their correlations. The following QCD coupling values are extracted at NNLO accuracy: $\alpha_{\rm s}(m_{\rm Z})$ = 0.1181±0.0016 (CT14), 0.1209±0.0015 (MMHT14), and 0.1163 ± 0.0019 (NNPDF3.0). The largest propagated uncertainties, combined here in quadrature into a single uncertainty for each final $\alpha_{\rm s}(m_{\rm Z})$ value, are associated with the experimental integrated luminosity and theoretical intra-PDF uncertainties. In this preliminary analysis, using the correlation matrices derived from the CMS experiment alone, the combination procedure did not converge on a stable result for HERAPDF 2.0. All other three $\alpha_{\rm s}(m_{\rm Z})$ extractions appear robust and stable with respect to variations in the data and theory cross sections, their uncertainties, and correlations. The final values are fully compatible with the world average value, and have competitive ∼1.5% uncertainties that are similar to those obtained with other precise methods (e.g. hadronic $\tau$ lepton decays) [22].

# Determination of $\alpha_{\rm s}(m_{\rm Z})$ from the Z-boson transverse momentum distribution

Stefano Camarda

CERN, EP Department, CH-1211 Geneva 23, Switzerland

**Abstract:** The strong-coupling constant $\alpha_{\rm s}(m_{\rm Z})$ is measured from the transverse-momentum distribution of Z bosons measured at $\sqrt{s} = 1.96$ TeV with the CDF experiment, using predictions based on $q_{\rm T}$-resummation at NNLO+NNLL, as implemented in the `DYTurbo` program. The measurement is performed through a simultaneous fit of $\alpha_{\rm s}(m_{\rm Z})$ and the non-perturbative Sudakov form factor.

The strong-coupling constant has been measured at hadron colliders in final states with jets [1,2,3], and more recently from top-antitop production cross sections [4]. Such measurements allow probing the strong coupling at high values of momentum transfer. However, they generally suffer from large uncertainties, and do not provide a competitive determination of the strong coupling at the scale of the Z-boson mass, $\alpha_{\rm s}(m_{\rm Z})$. This contribution aims at discussing a new technique for precisely measuring $\alpha_{\rm s}(m_{\rm Z})$ at hadron colliders from a semi-inclusive (i.e. radiation inhibited) observable: the Z-boson transverse-momentum distribution. This measurement has all the desirable features for a precise determination of $\alpha_{\rm s}(m_{\rm Z})$ [5]: large observables sensitivity to $\alpha_{\rm s}(m_{\rm Z})$ compared to the experimental precision; high accuracy of the theoretical prediction; small size of non-perturbative QCD effects.

Measuring $\alpha_{\rm s}(m_{\rm Z})$, or equivalently $\Lambda_{\rm QCD}^{\overline{\rm MS}}$, from semi-inclusive Drell-Yan cross sections was first proposed in Ref. [6], by using Monte Carlo parton showers to determine $\Lambda_{\rm QCD}^{\rm MC}$ and later convert it to $\Lambda_{\rm QCD}^{\overline{\rm MS}}$. The conversion is based on resummation arguments showing that a set of universal QCD corrections can be absorbed in coherent parton showers by applying a simple rescaling, the so-called Catani-Marchesini-Webber (CMW) rescaling.

The Z-boson transverse-momentum distribution at small transverse momentum is one of such semi-inclusive observables. The recoil of Z bosons produced in hadron collisions is mainly due to QCD initial-state radiation, and the Sudakov form factor is responsible for the existence of a Sudakov peak in the distribution, at transverse-momentum values of approximately 4 GeV. The position of the peak is sensitive to the value of the strong-coupling constant. The arguments of Ref. [6] can be used to interpret the ATLAS result of a PYTHIA 8 Monte Carlo tuning to the Z-boson transverse-momentum distribution [7] as a measurement of $\alpha_{\rm s}(m_{\rm Z})$. Table 1 shows the results of the ATLAS tune of PYTHIA 8, named AZ, where the Monte Carlo parameter $\alpha_{\rm s}^{\rm ISR}(m_{\rm Z})$ was determined simultaneously with primordial $k_{\rm T}$ and the parton shower infrared cut-off from a fit to the ATLAS measurement of the Z-boson transverse-momentum distribution. The CMW conversion leads to: $\alpha_{\rm s}^{\rm ISR}(m_{\rm Z}) = 0.124 \rightarrow \alpha_{\rm s}^{\overline{\rm MS}}(m_{\rm Z}) = 0.116$.

Table 1: Results of the AZ tune of the PYTHIA 8 Monte Carlo to the ATLAS measurement of the Z-boson transverse-momentum distribution [7].

| | |
|---|---|
| ISR $\alpha_{\rm s}^{\rm ISR}(m_{\rm Z})$ | $0.1237 \pm 0.0002$ |
| primordial $k_{\rm T}$ [GeV] | $1.71 \pm 0.03$ |
| ISR cut-off [GeV] | $0.59 \pm 0.08$ |

The relative uncertainty on $\alpha_s(m_Z)$ of the ATLAS tune is 0.2%, which includes the experimental uncertainties, as well as the non-perturbative QCD uncertainties, since the non-perturbative QCD parameters are fitted simultaneously with $\alpha_s^{\text{ISR}}(m_Z)$. This naive result is missing important theory uncertainties as PDFs and missing higher order corrections. However, this simple exercise already shows a great experimental sensitivity and relatively small non-perturbative QCD uncertainties.

Turning this idea into an actual measurement poses several challenges. In order to achieve higher precision, it is highly desirable to employ for the measurement analytic predictions of the Z-boson transverse-momentum distribution including resummation of large logarithmic corrections of the form $\log(p_T/m)$. Such $q_T$-resummed predictions are available since long time, and they have recently reached N$^3$LL logarithmic accuracy [8]. The measurements of Z-boson transverse-momentum distribution have small experimental uncertainties, at the level of 2% at the Tevatron and 0.5% at the LHC. High numerical precision of the theory predictions is required to match such small uncertainties, which is a great challenge for these complicated high-order QCD calculations. Large correlations between $\alpha_s(m_Z)$ and non the perturbative QCD effects would spoil the measurement. Small correlations were observed with the PYTHIA 8 model, but they need to be studied also in the case of analytic predictions. At the LHC, significant heavy-flavour initiated production, at the level of 6% for $c\bar{c} \to Z$ and and 3% for $b\bar{b} \to Z$ introduce additional uncertainties.

For the measurement of $\alpha_s(m_Z)$ from the Z-boson transverse-momentum distribution it is necessary to rely on fast computing codes which allow the calculation of variations in the input parameters with small numerical uncertainties. To this end, the `DYTurbo` program has been created. It aims to provide fast and numerically precise predictions of fully-differential Drell–Yan production cross sections, for phenomenological applications such as QCD analyses and extraction of fundamental parameters of the Standard Model. The enhancement in performance over previous programs is achieved by overhauling pre-existing code, by factorising the fully-differential cross section into production and decay variables, and by introducing the usage of one-dimensional and multi-dimensional numerical integration based on interpolating functions. The `DYTurbo` program is a reimplementation of the `DYRes` [9] program for the small-$q_T$ resummed cross sections at up to next-to-next-to-leading-logarithmic (NNLL) accuracy. As an example of fast and numerically precise predictions, `DYTurbo` can compute the Z-boson transverse-momentum distribution at 13 TeV in full-lepton phase space, in 100 equally-spaced bins from zero to 25 GeV, with a target relative numerical uncertainty of $10^{-4}$, in 4 min. at NLO+NLL and 3.4 h at NNLO+NNLL, using simultaneously 20 parallel threads. The great majority of the computation time is spent in evaluating the LO or NLO $V$+jet term. However, it is possible to use `ApplGrid` [10] interfaced to `MCFM` for this term. Once this is done, the computation requires 6 s (10 s) at NLO+NLL (NNLO+NNLL).

The CDF measurement of Z-boson transverse-momentum distribution [11] at the Tevatron collider is ideal for testing the extraction of $\alpha_s(m_Z)$ with `DYTurbo` predictions. This measurement was performed with the angular coefficients technique, which allows extrapolating the cross section to full-lepton phase space with small theoretical uncertainties. The full-lepton phase space cross section allows fast predictions and avoid any theoretical uncertainties on the modelling of the Z-boson polarisation. Another advantage of this measurement with respect to similar measurements performed at the LHC is the fact that Tevatron is a proton-antiproton collider, and the Z-boson production has reduced contribution from heavy-flavour-initiated processes compared to proton-proton collisions at the LHC.

The CDF measurement is performed in the electron channel, with central ($|\eta^e| < 1.1$) and forward ($1.2 < |\eta^e| < 2.8$) electrons, allowing a coverage up to Z-boson rapidity of $|y| = 2.8$, and a small extrapolation to the full rapidity range $|y_{\max}| \approx 3.1$ of Z-boson production at $\sqrt{s} = 1.96$ TeV. The data sample is characterised by low values of the average number of interactions per bunch

crossing, and by good electron resolution, at the level of 0.9 GeV for central electrons, and 1.1 GeV for forward electrons. The good resolution allows fine transverse-momentum binning (0.5 GeV) while keeping the bin-to-bin correlations smaller than 20%.

The non-perturbative QCD corrections to the Z-boson transverse-momentum distribution are modelled by including a non-perturbative term in the Sudakov form factor: $S(b) \to S(b) \cdot S_{\rm NP}(b)$. The general form of $S_{\rm NP}(b)$ is mass and centre-of-mass energy dependent [12]. However, at fixed invariant mass $q = m_{\rm Z}$, and for one value of centre-of-mass energy, the form of $S_{\rm NP}(b)$ can be simplified to depend on a single parameter $g$: $S_{\rm NP}(b) = \exp(-g \cdot b^2)$. The non-perturbative parameter $g$ is generally determined from the data, and its value depends on the chosen prescription to avoid the Landau pole in the impact-parameter b-space, which corresponds to a divergence of the Sudakov form factor. The divergence is avoided by using the so-called $b_\star$ prescription, which freezes $b$ at a given value $b_{\rm lim}$: $b \to b_\star = \frac{b}{1+b^2/b_{\rm lim}^2}$. In this analysis $b_{\rm lim}$ is set to the value of the Landau pole, and a variation to half its value is considered as a systematic uncertainty.

The sensitivity of the Z-boson transverse-momentum distribution to $\alpha_{\rm s}(m_{\rm Z})$ mainly comes from the position of the Sudakov peak, and is related to the average recoil scale $\langle p_{\rm T} \rangle \approx 10$ GeV. The sensitivity of the Z-boson transverse-momentum distribution to $g$ also comes from the position of the Sudakov peak. However, the scale of the non-perturbative smearing governed by $g$ corresponds to the value of primordial $k_T$. Typical values of $g \approx 0.8$ GeV$^2$ corresponds to a primordial $k_T$ of approximately 1.8 GeV. It is possible to disentangle the perturbative contribution to the Sudakov form factor, governed by $\alpha_{\rm s}(m_{\rm Z})$, from the non-perturbative one, determined by $g$, thanks to their different scale, as shown in Figure 1.

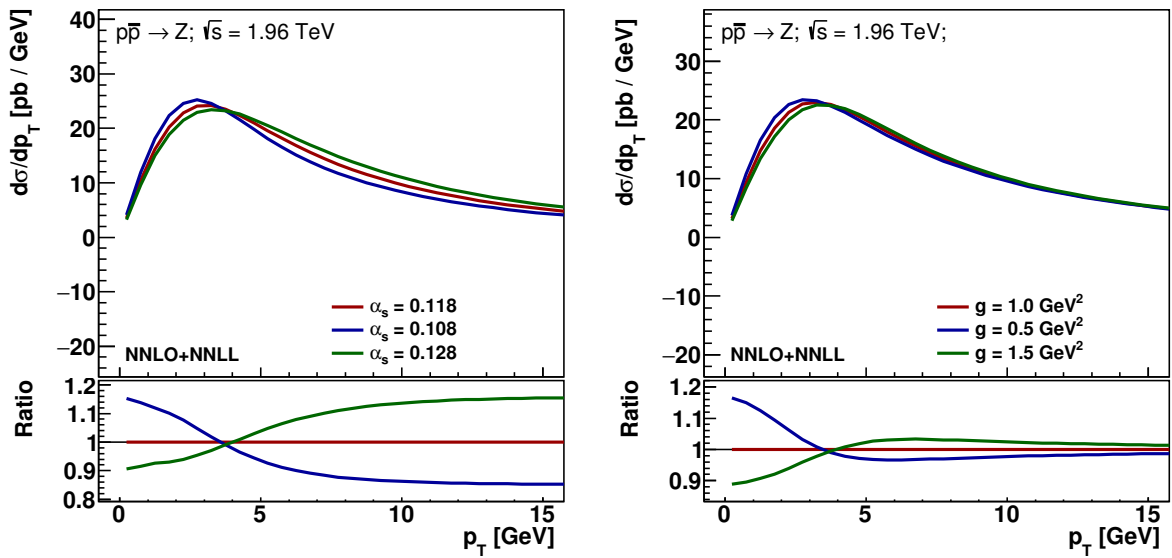

Figure 1: Sensitivity of the Z-boson transverse-momentum distribution to $\alpha_{\rm s}(m_{\rm Z})$ (left) and to the non-perturbative QCD parameter $g$ (right).

The statistical analysis leading to the determination of $\alpha_{\rm s}(m_{\rm Z})$ is performed by interfacing DYTurbo to xFitter [13]. The agreement between data and predictions is assessed by means of a $\chi^2$ function, which includes experimental and PDFs theoretical uncertainties. The non-perturbative form factor is added as unconstrained nuisance parameter in the $\chi^2$ definition, i.e. it is left free in the fit. The fit to the data is performed in the region of transverse momentum $p_{\rm T} < m_{\rm Z}$ by minimising the $\chi^2$ as a function of $\alpha_{\rm s}(m_{\rm Z})$, with $\alpha_{\rm s}$ variations as provided in LHAPDF. The corrections to the Z-boson transverse-momentum distribution due to QED initial-state radiation are estimated with PYTHIA 8, and applied as multiplicative corrections. They are the level of 1%, and are responsible for a shift in the measured value of $\alpha_{\rm s}(m_{\rm z})$ of $\Delta\alpha_{\rm s} = 0.0004$.

Table 2: Results of the fit of $\alpha_{\rm s}(m_{\rm z})$ to the CDF measurement of Z-boson transverse-momentum distribution.

|  | MMHT | CT14 | NNPDF3.1 |
| --- | --- | --- | --- |
| $\alpha_{\rm s}(m_{\rm z})$ | 0.1202 | 0.1193 | 0.1198 |
| Stat. unc. | 0.0008 | 0.0008 | 0.0007 |
| Syst. unc. | 0.0002 | 0.0003 | 0.0002 |
| PDF unc. | 0.0006 | 0.0007 | 0.0006 |
| $g$ [GeV$^2$] | $0.48 \pm 0.07$ | $0.51 \pm 0.07$ | $0.35 \pm 0.08$ |
| $\chi^2$/dof | 56/71 | 54/71 | 57/71 |

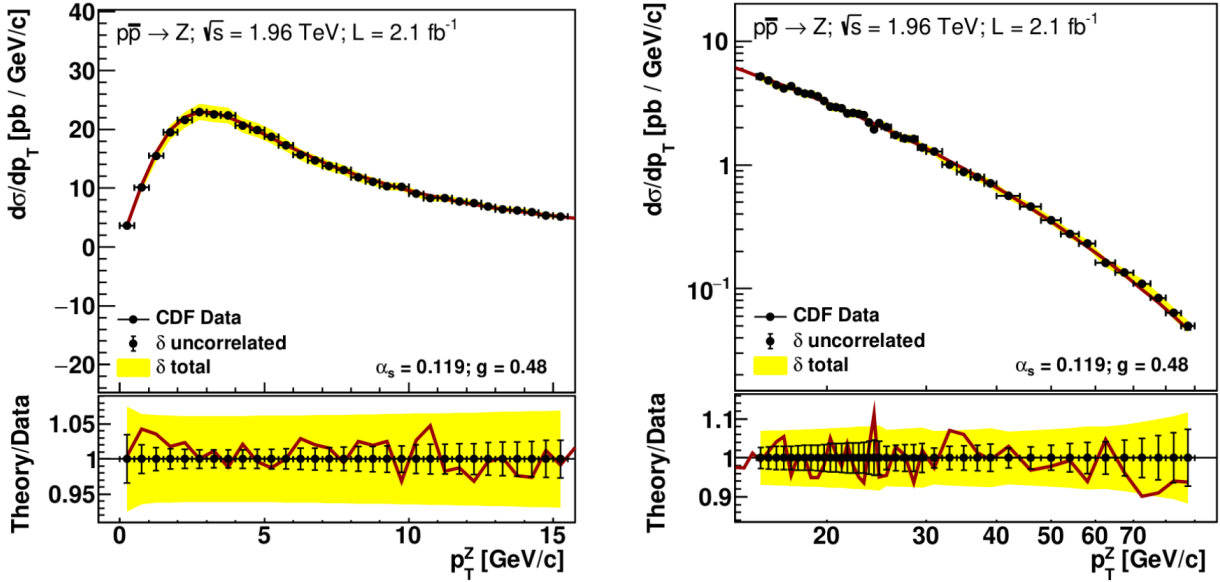

Figure 2: Results of the fit of $\alpha_{\rm s}(m_{\rm z})$ to the CDF measurement of Z-boson transverse-momentum distribution. Post-fit predictions are compared to the measured distributions.

The results of the fit of $\alpha_{\rm s}(m_{\rm z})$ to the CDF measurement of Z-boson transverse-momentum distribution are shown in Table 2 for three different PDF sets, MMHT2014, CT14, and NNPDF3.1. The fit with the CT14 PDF set has the smallest $\chi^2$ and is considered as central result. The post-fit predictions are compared with data in Figure 2. Additional sources of theoretical uncertainties are considered in the analysis. The predictions depend on the choice of the renormalisation, factorisation, and resummation scales. The central values of these scales are set to $\mu_{\rm R} = \mu_{\rm F} = \mu_{\rm res} = m_{\rm Z}/2$. Uncertainties arising from missing higher order corrections are estimated from the envelope of all possible combinations of factor of two variations, excluding variations where any pair of scales differ by a factor of four. The resulting uncertainty are $+0.0035, -0.0027$ for $\alpha_{\rm s}(m_{\rm z})$ and $+0.61, -0.28$ GeV$^2$ for $g$. An uncertainty related to the matching between resummation and fixed order prediction is estimated by switching off the resummation corrections above $p_{\rm T} = m_{\rm Z}/2$. The resulting shift of $\alpha_s(m_z)$ is $\Delta\alpha_s = +0.0001$. An uncertainty related to the particular prescription used to avoid the Landau pole is estimated by setting $b_{\text{lim}}$ to half the value of the Landau pole. The resulting shifts of $\alpha_s(m_z)$ and $g$ are $\Delta\alpha_s = -0.0008$ and $\Delta g = +0.18$ GeV$^2$. These values are considered as additional uncertainties.

The final result for the measurement of $\alpha_s(m_z)$ and the simultaneous determination of the non-perturbative parameter $g$ is:

$$\alpha_s(m_z) = 0.119^{+0.004}_{-0.003}$$
$$g = 0.51^{+0.64}_{-0.34} \text{ GeV}^2$$

The result is dominated by missing higher order uncertainties, estimated with scale variations, which are at the level of 3%. Predictions at higher logarithmic accuracy, namely N$^3$LL, are now available, as well as O($\alpha_s^3$) corrections to the Z-boson transverse-momentum distribution, which are expected to lead to a factor of 3–5 reduction in the uncertainty [8]. Measurements of Z-boson transverse-momentum at the LHC are significantly more precise than at the Tevatron. The ATLAS measurement at $\sqrt{s} = 7$ TeV yields 0.2% of relative experimental uncertainty on $\alpha_s(m_z)$. Three times smaller uncertainties are expected with ATLAS and CMS measurements at $\sqrt{s} = 8$ TeV, and it is likely to reach a few $10^{-4}$ with measurements based on the full Run 2 data sample. However, in order to perform this $\alpha_s(m_z)$ determination at the LHC, it is primordial to improve the modelling of heavy-flavour-initiated Z-boson production.

# $\alpha_\mathrm{s}$ from non-strange hadronic $\tau$ decays

Diogo Boito[1], Maarten Golterman[2], Kim Maltman[3,4], and **Santiago Peris**[5]

[1]*Instituto de Física de São Carlos, Univ. São Paulo CP 369, 13570-970, São Carlos, SP, Brazil*
[2]*Department of Physics and Astronomy, San Francisco State University, San Francisco, CA 94132, USA*
[3]*Department of Mathematics and Statistics, York University, Toronto, ON Canada M3J 1P3*
[4]*CSSM, University of Adelaide, Adelaide, SA 5005 Australia*
[5]*Dept of Physics and IFAE-BIST, Universitat Autònoma de Barcelona, 08193 Bellaterra, Barcelona*

**Abstract:** We review how the current precision attained in the extraction of $\alpha_\mathrm{s}(m_\tau)$ from hadronic $\tau$ decays requires the inclusion of Duality Violations (DVs) in the analysis, even though these decays are largely dominated by perturbation theory. A weighted average using the OPAL and ALEPH experimental data yields $\alpha_\mathrm{s}(m_\mathrm{Z}) = 0.1165 \pm 0.0012$ and $\alpha_\mathrm{s}(m_\mathrm{Z}) = 0.1185 \pm 0.0015$ in fixed-order and contour-improved perturbation theory, respectively.

The $\alpha_\mathrm{s}$ determination from hadronic $\tau$ decay usually relies on Finite Energy Sum Rules (FESRs). A FESR analysis takes advantage of the analyticity of the current-current correlator

$$\begin{aligned}
\Pi_{\mu\nu}(q) &= i\int d^4x\, e^{iqx}\langle 0|T\left\{J_\mu(x)J_\nu^\dagger(0)\right\}|0\rangle \\
&= \left(q_\mu q_\nu - q^2 g_{\mu\nu}\right)\Pi^{(1)}(q^2) + q_\mu q_\nu \Pi^{(0)}(q^2) \\
&= \left(q_\mu q_\nu - q^2 g_{\mu\nu}\right)\Pi^{(1+0)}(q^2) + q^2 g_{\mu\nu}\Pi^{(0)}(q^2)\ ,
\end{aligned} \qquad (1)$$

where $J_\mu$ stands for the non-strange $V$ or $A$ current, $\overline{u}\gamma_\mu d$ or $\overline{u}\gamma_\mu \gamma_5 d$, and the superscripts (0) and (1) label spin, to obtain the following identity [1]

$$\frac{1}{s_0}\int_0^{s_0} ds\, w(s)\, \rho_{V/A}^{(1+0)}(s) = -\frac{1}{2\pi i\, s_0}\oint_{|s|=s_0} ds\, w(s)\, \Pi_{V/A}^{(1+0)}(s)\ , \qquad (2)$$

which is valid for any $s_0 > 0$ and any weight $w(s)$ analytic inside and on the contour depicted in Fig. 1. The combinations $\Pi^{(1+0)}(q^2)$ and $q^2\Pi^{(0)}(q^2)$ are convenient because they are free of kinematic singularities. In Eq. (2), $\rho^{(1+0)}(s) = \frac{1}{\pi}\,\mathrm{Im}\,\Pi^{(1+0)}(s)$ designates the spectral function and $s = q^2$. From now on, we will suppress the index $(1+0)$.

As it stands, Eq. (2) is exact if for $\Pi(s)$ one is using the exact function. When $s_0$ is large enough, it begins to make sense to replace this function by its OPE representation from which it may be possible to extract the value of $\alpha_\mathrm{s}$*. As the OPE is expected to be asymptotic, and breaks down on the Minkowski axis, there will be a nonvanishing difference between the exact and the OPE representations. We will denote this difference by $\Pi_{DV}(s)$, where $DV$ stands for Duality Violations (DVs). Explicitly,

$$\Pi(s) = \Pi_{OPE}(s) + \Pi_{DV}(s)\ . \qquad (3)$$

In a hypothetical world in which the OPE converged, DVs would vanish by definition.

Using Eq. (3), one may rewrite Eq. (2) conveniently as [2]

$$\frac{1}{s_0}\int_0^{s_0} ds\, w(s)\, \rho^{exp}(s) = -\frac{1}{2\pi i s_0}\oint_{|z|=s_0} dz\, w(z)\, \Pi_{OPE}^{(\alpha_\mathrm{s})}(z) - \frac{1}{s_0}\int_{s_0}^\infty ds\, w(s)\, \frac{1}{\pi}\,\mathrm{Im}\,\Pi_{DV}(s)\ , \qquad (4)$$

---

*Here, we will consider the perturbative series as the contribution from the unit operator to the OPE.

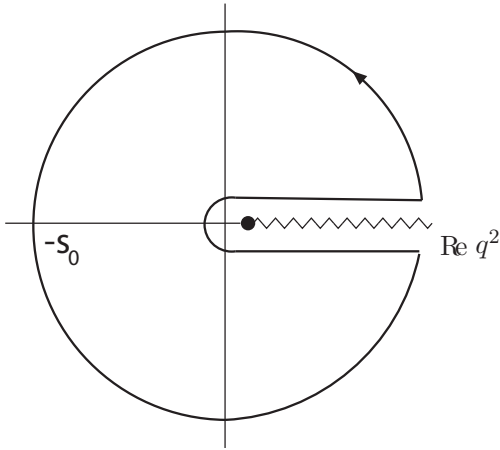

Figure 1: Contour used in the derivation of the FESRs, Eq. (2).

where we have indicated that $\rho^{exp}(s)$ is to be obtained from experimental data and $\Pi_{OPE}^{(\alpha_s)}(s)$ contains the value of $\alpha_s$ to be determined. In practice $w(s)$ will be taken to be a polynomial. At this point several important remarks are in order.

First, as a consequence of the residue theorem, a monomial $w(s) = (s/s_0)^N$ produces an OPE contribution $(-1)^N C_{2N+2}/s_0^{N+1}$ to the right-hand side of Eq. (4), where the coefficients $C_{2N+2}$ are related to the condensates of dimension $2N + 2$ (see also the discussion below). The $C_{2N+2}$ are typically not known *a priori* but can be determined using Eq. (4) if $\alpha_s$ and $\text{Im}\,\Pi_{DV}(s)$ have been previously determined. For a given set of experimental data $\rho^{exp}(s)$, the presence of DVs affects any $C_{2N+2}$ determined in this way, except in the case $N = 0$ ($w(s) = 1$) where $C_2$ vanishes for the $V$ and $A$ correlators[†]. Second, although $\text{Im}\,\Pi_{DV}(s)$ is certainly non-zero as a result of the non-convergence of the OPE on the Minkowski axis, its precise form is in principle unknown. In early $\tau$-decay analyses this problem was dealt with by assuming that the use of polynomials $w(s)$ with zeros at $s = s_0$ of sufficiently high order ("pinching") would provide sufficient suppression of contributions from the region near the Minkowski axis to allow DVs to be safely neglected. This assumption is predicated on the expectation that $\Pi_{DV}$ will be maximal in the vicinity of the Minkowski axis,[3] where, given that it represents a contribution missed by the asymptotic OPE, one expects $\text{Im}\,\Pi_{DV}(s) \sim e^{-\gamma s} \times (\text{oscillation})$, in analogy to the way the asymptotic renormalon series misses a non-perturbative term of order $e^{-b/\alpha_s}$. Third, the use of pinching, regrettably, poses a problem: a polynomial with a high-order zero necessarily also has a high degree, and a high-degree polynomial produces contributions from $C_{2N+2}$ with large $N$ to the right-hand side of the FESR (4). Such contributions are not known (unless DVs and $\alpha_s$ have somehow already been determined, as remarked above). This leads to a "no-go" theorem [4,5]:

"It is not possible to simultaneously suppress DV and high-dimension condensate contributions."

In order to avoid the contribution from $C_{2N+2}$ with large $N$, one could use a low-degree polynomial, but this could then fail to provide enough pinching to be able to safely neglect DVs. In summary, one way or the other, the inclusion of DVs in Eq. (2) is unavoidable. This requires a concrete parametrization of $\text{Im}\,\Pi_{DV}(s)$ which then allows its parameters to be determined with the help of Eq. (4), through a fit in an appropriate window of large-enough $s_0$.

Recently, such parametrization has been obtained [6]. The assumptions needed to derive it are rather mild: First, an asymptotic Regge spectrum for mesons at $N_c = \infty$ and, second, a constant width-over-mass ratio in the limit that the radial excitation number $\to \infty$, for $N_c = 3$. Both these assumptions are true in QCD in two dimensions (where all these properties can actually be computed), are supported by the string picture of hadrons [7], and are in agreement with phenomenology [8]. The resulting expression for $\mathrm{Im}\,\Pi_{DV}(s)$ then reads

$$\frac{1}{\pi}\mathrm{Im}\,\Pi_{DV}(q^2) \sim e^{-2\pi \frac{a}{N_c} \frac{q^2}{\Lambda_{\mathrm{QCD}}^2}} \sin\left[\frac{2\pi}{\Lambda_{\mathrm{QCD}}^2}\left(q^2 - c - b\log\frac{q^2}{\Lambda_{\mathrm{QCD}}^2}\right)\right]\left(1 + \mathcal{O}\left(\frac{1}{N_c}; \frac{1}{q^2}; \frac{1}{\log q^2}\right)\right) \,, \quad (5)$$

where $\Lambda_{\mathrm{QCD}} \sim 1\,\mathrm{GeV}$ is the characteristic QCD scale, related to the string tension. The result (5) is in accord with our expectations for an asymptotic OPE described above. Apart from a mild logarithmic dependence, modulated by the constant $b$ and subleading at large $q^2$, this form can be conveniently expressed as [9]

$$\frac{1}{\pi}\mathrm{Im}\,\Pi_{DV}(q^2) = e^{-\delta - \gamma q^2}\,\sin\left(\alpha + \beta\, q^2\right)\,, \qquad q^2 \gg \Lambda_{\mathrm{QCD}}^2 \,, \quad (6)$$

and this is, in fact, the parametrization we have used in our analyses. We emphasize that, in principle, a different set of parameters $\delta_{V,A}, \gamma_{V,A}, \alpha_{V,A}$ and $\beta_{V,A}$ should be used for the $V$ and $A$ channels since they are related to the resonance spectrum.

---

[†] The $u$ and $d$ quark masses are very small and, consequently, neglected.

This expression (6) was not available for use in the first $\tau$ decay determinations of $\alpha_s$ [10,11,12]. These pioneering analyses employed a strategy, which we will refer to as the truncated-OPE strategy (tOPE), in which pinched weights were used and both DVs and high-dimension OPE contributions were neglected. Recent examples of the continued use of this strategy may be found in Refs. [13,14]. The tOPE strategy proceeds as follows. A set of five polynomials,

$$w_{kl}(y) = (1 + 2y)(1 - y)^{2+k}\, y^l \,, \quad y = s/s_0, \quad s_0 = m_\tau^2 \,, \quad (7)$$

with $(k,l) \in \{(0,0), (1,0), (1,1), (1,2), (1,3)\}$, is chosen, and the corresponding set of weighted spectral integrals, evaluated at $s_0 = m_\tau^2$ only, used to extract four parameters: $\alpha_s$ and the coefficients $C_{D=4,6,8}$. Since these polynomials reach up to degree 7 in $s$, the FESR (4), in principle, receives contributions also from the $C_D$ with $D = 10, 12, 14$ and $16$, which, because they are unknown, are neglected. This neglect is predicated on an assumed $O\left(\Lambda_{\mathrm{QCD}}^D/m_\tau^D\right)$ suppression of dimension $D$ OPE contributions. In other words, the OPE is effectively treated as if it were convergent at the scale $s = m_\tau^2$. The term with DVs in Eq. (4) is also neglected. While this strategy may have been reasonable in the early work of Refs. [10,11,12], when the error in the extracted $\alpha_s$ was $\sim 10 - 15\%$, it is clear that, as time goes by, and errors decrease, the assumptions underpinning this approach need to be checked and, if necessary, the method needs to be revised.

Partly with this idea in mind, Ref. [14] has recently generalized the tOPE strategy by investigating a variety of alternate polynomial combinations, obtaining, in all cases, consistent results with good-quality fits. However, although the analysis of Ref. [14] showed no obvious sign that the results obtained might be unreliable and the value of $\alpha_s$ extracted might be polluted by a systematic error that the variations studied might not be capable of identifying, all these results, as shown in Ref. [15], *do* contain a hidden $\sim +6\%$ systematic error in the extracted value of $\alpha_s(m_\tau^2)$. For a full account of this systematic error, we refer to Ref. [15]. Here, we will just report on one particularly clean test that illustrates the point.

The test works as follows. We consider a model designed to closely match the actual experimental spectral data, but constructed to have an input value of $\alpha_s(m_\tau^2)$, $\alpha_s(m_\tau^2)^{fake} = 0.312$, and corresponding chosen values for the DV parameters[‡]. We then generate a set of fake data for the $V + A$ spectral function[§], using exactly the same binning and the same correlations as in the actual experiment, by letting the data points fluctuate according to a multivariant Gaussian distribution defined with the experimental covariance matrix.¶ An example of the resulting spectral distribution is shown in Fig. 2. The point is that, if the tOPE strategy were reliable, it should be able to reproduce the input value of $\alpha_s(m_\tau^2)^{fake}$ from the fake data set, within errors, and in spite of the neglect of higher-dimension OPE contributions and the absence of a representation of integrated DVs in the theoretical form it assumes.

---

[‡]See Ref.[15], for more details.
[§]This is the channel that Ref. [14] considers to be optimal for the reliability of the tOPE strategy.

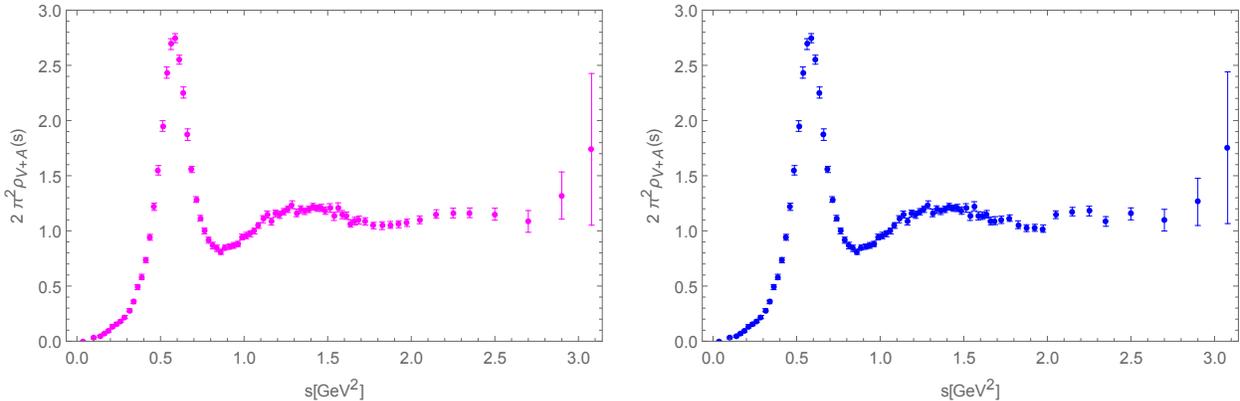

Figure 2: Left: $V+A$ fake data, generated as described in the text, as a function of $s$. Right: True ALEPH data [13] as a function of $s$. The fake data has been generated for $s \geq 1.55 \, \text{GeV}^2$; below this value the two sets of data are identical.

However, when we use the sets of polynomials suggested in Ref. [14] in fits to the fake data, we always find the value of $\alpha_s(m_\tau^2)$ to be overestimated by $\sim +0.02$, a systematic error of $+6\%$, which, in terms of the statistical errors of these fits, amounts to 5–7$\sigma$. Therefore, the tOPE strategy clearly fails. One might think that the tOPE could have also reproduced the right result, had the fake data set been generated without DVs. Such fake data, however, would not be able to reproduce the residual oscillations present even in the $V+A$ spectral distribution (see below). And the fact remains that the tOPE strategy, in ignoring higher-dimension terms in the OPE without justification, and failing to detect the presence of residual DVs in the fake data case, can produce a systematic shift in the extracted value of $\alpha_s$ whose presence cannot be exposed by looking at the variation in the output $\alpha_s$ produced when the tOPE analysis is performed using the various polynomial set choices considered in Ref. [14].

It has been argued [14], referring to the left panel of Fig. 3, that the spectral function in the $V+A$ channel is so flat at high $s$ as to be free from DVs. This argument, however, is rather misleading. The right panel in Fig. 3 shows the same spectral function, but now with the ($\alpha_s$-independent) parton-model contribution subtracted. The black dashed curve in this panel shows the corresponding result from perturbation theory. One sees that, even at $s = 2 \, \text{GeV}^2$, the data points agree, within errors, with the parton model. In other words, the $\alpha_s$-dependent part of the perturbative contribution cancels against the DV oscillation at that point. There is no sense in which the DVs are small *relative* to the $\alpha_s$-dependent perturbative contributions, from which the value of $\alpha_s$ is extracted, and, therefore, there is no sense in which the DVs may be reliably neglected. A similar effect is seen at $s \simeq 2.2 \, \text{GeV}^2$ but, this time, DVs and the $\alpha_s$-dependent part of perturbation theory add up, rather than cancel each other. Again, the size of DVs is comparable to that of the $\alpha_s$-dependent perturbative contributions. Notice that these data are very correlated, so the fact that a group of three data points, with central values very close together at $s \simeq 2.2\,\text{GeV}^2$, are above the perturbative curve while another group of three data points, again very close together at $s \simeq 2\,\text{GeV}^2$, are below the perturbative curve is difficult to explain as a fluctuation in the data, and not as the sign of a true residual oscillation. Above $s = 2.5\,\text{GeV}^2$ the errors are too large to tell. Furthermore, there is no doubt that both $V$ and $A$ separately contain DV oscillations, so the safest assumption is that $V + A$ also has them, even if they are smaller for $V + A$ than for the individual $V$ and $A$ cases. At any rate, smaller or not, we have seen that they can easily affect the extraction of $\alpha_s$, as illustrated in the fake-data test discussed above. Reference [15] contains more details of the different tests one may carry out, all of them leading to the conclusion that the tOPE strategy is unreliable, with an associated systematic error of $\sim +0.02$ in the value of $\alpha_s(m_\tau^2)$.

---

¶The fake data is generated only for $s_0 \geq 1.55 \, \text{GeV}^2$, which is the value we obtained in our true-data fits for the onset of the asymptotic DV expression (6). Below this $s_0$, the two data sets, fake and true, are identical.

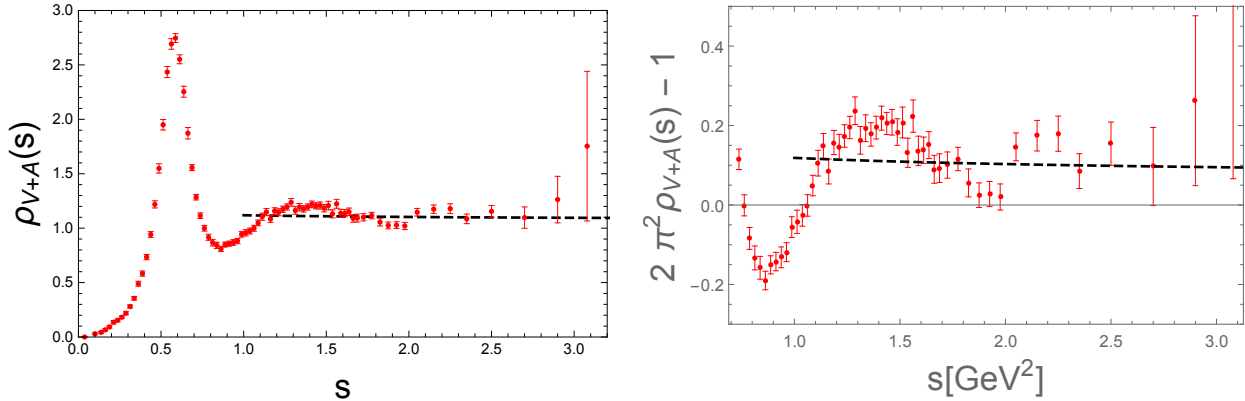

Figure 3: Left: $V + A$ spectral function. Right: $V + A$ spectral function, after the parton model contribution has been subtracted. The black dashed curve is the result of perturbation theory.

Given this state of affairs, we have recently proposed [5,16] a different strategy that takes DVs into account explicitly, parametrized as in Eq. (6), and employs the 3 polynomials (to be considered together or separately)

$$w_0 = 1\,, \quad w_2 = 1 - y^2 \quad \text{and} \quad w_3 = (1-y)^2(1+2y)\,. \tag{8}$$

The $\alpha_s$ value and the two coefficients $C_{6,8}$ (which are the only OPE coefficients contributing to the $w_2$ and $w_3$ FESRs) are then fit using the integrated spectral data in a window of $s_0$ extending from $m_\tau^2$ down to a lower value determined by the fit itself. The choice of polynomials is dictated by a desire to avoid, first, contributions from high-dimension terms in the OPE and, second, the use polynomials with a term linear in $y$ (which receive a contribution from $C_4$, associated with the gluon condensate) since model studies suggest that perturbation theory behaves poorly for such weights, whether with the FOPT or the CIPT choice for the scale $\mu$ [17][||].

A large variety of different fits using Eq. (4) and the three polynomials above, and employing the $V$ channel alone, or the $V$ and $A$ channels combined, were carried out in Ref. [16], to which we refer for more details. The results obtained were consistent in all cases, not only for $\alpha_s$ but also for the $C_{6,8}$ coefficients.

In Fig. (4) we show how the associated $V$ and $A$ spectral functions are described by our parametrization in Eq. (6) at high $s$, where the asymptotic DV form is expected to apply. A number of additional

---

[||]Fixed-order perturbation theory (FOPT) refers to the choice of the scale $\mu^2 = s_0$, where $s_0$ is the radius of the contour in Eq. (4). Contour-improved perturbation theory (CIPT) refers to the choice $\mu^2 = z$, where $z$ is the complex integration variable along the contour in Eq. (4).

consistency checks were also carried out; for example, the first Weinberg sum rule. In Fig. 5 we show the result of this sum rule, i.e.,

$$\int_0^\infty ds\ (\rho_V(s) - \rho_A(s)) - 2f_\pi^2 = 0\ , \tag{9}$$

as a function of the point $s_{sw}$ at which one switches from the experimental data to the corresponding theoretical description. The left panel shows the case without DVs in the theoretical description; in this case, the experimental data switches to zero since the perturbative contribution cancels in the $V - A$ difference. The right panel shows the case where the DV parametrization (6) is employed in the theoretical description, which, through the second term on the righthand side of Eq. (4), allows us to extend the upper limit in the sum rule to infinity. It is clear that taking DVs into account constitutes an improvement.

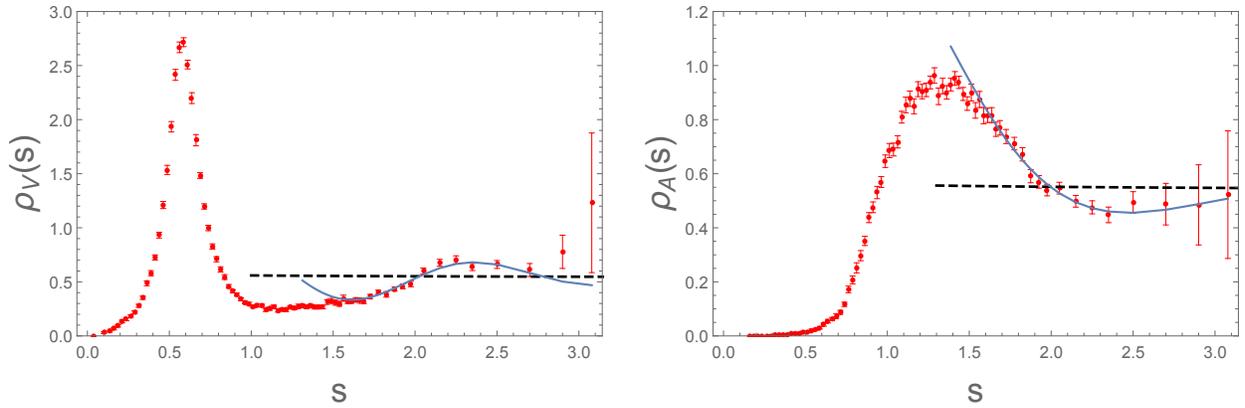

Figure 4: Left: $V$ spectral function, together with the parton model result (dashed black curve) and the result from the DV parametrization (6) obtained from Eq. (4) (blue curve). Right: The same for the $A$ spectral function.

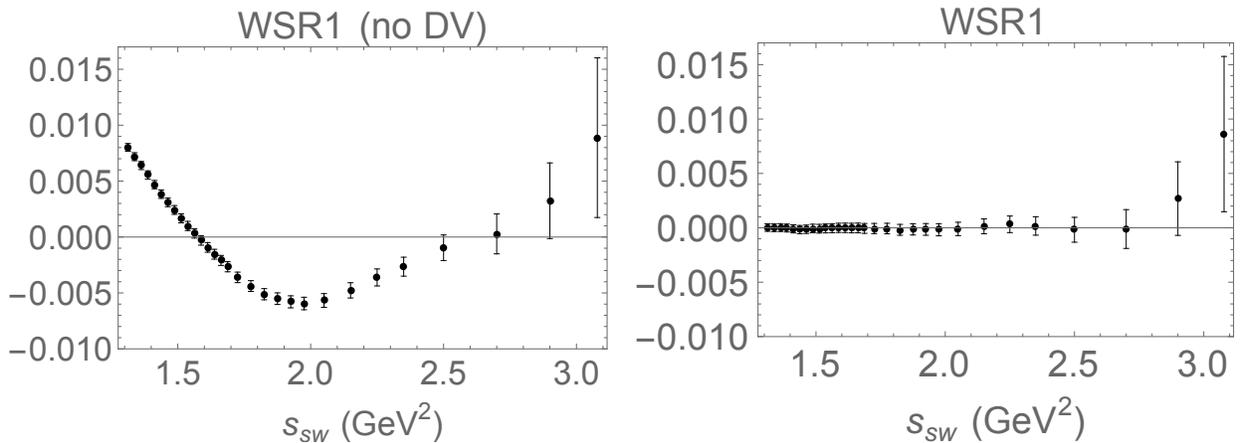

Figure 5: Left: First Weinberg sum rule without DVs. Right: First Weinberg sum rule with DVs taken into account.

Other tests were also considered. Two tests we find particularly interesting probe the idea of truncating the OPE. Using again the polynomials $w_{kl}$ in Eq. (7), the left panels of Fig. 6 show the example of $w_{11}$ and $w_{13}$ as a function of $s_0$, as obtained with the tOPE strategy. We emphasize that, within this strategy, fits are being done solely at $s_0 = m_\tau^2$. Therefore, it is not surprising that the data agree rather well with the theory curve at this $s_0$. However, the theory description quickly departs from the data as soon as $s_0$ is lowered, which is a clear sign that the $s_0$ scaling on the theory side of the corresponding FESR is not correct. This is a consequence of neglecting the higher-dimension terms in the OPE that contribute to these sum rules. For comparison, we also show the same result once DVs are taken into account, and the corresponding OPE coefficients have been determined with the help of Eqs. (7.3) of Ref. [16] (which are versions of the FESR (4)). We emphasize that what this figure shows is that, as a result of the absence of higher-dimension terms assumed negligible in the tOPE strategy, the tOPE version of the theory side of the FESRs (4) fails to scale correctly with $s_0$ as $s_0$ decreases below $m_\tau^2$. In other words, the argument that the scale $m_\tau$ is large enough to effectively suppress the contributions from the higher-dimension $C_D$ to the FESR (4), based on an assumed naive $\Lambda_{\rm QCD}^D/m_\tau^D$ scaling, turns out to be incorrect. This is compatible with the known asymptotic character of the OPE, which implies that these coefficients must eventually become significantly larger than implied by this naive scaling for sufficiently large dimension $D$.

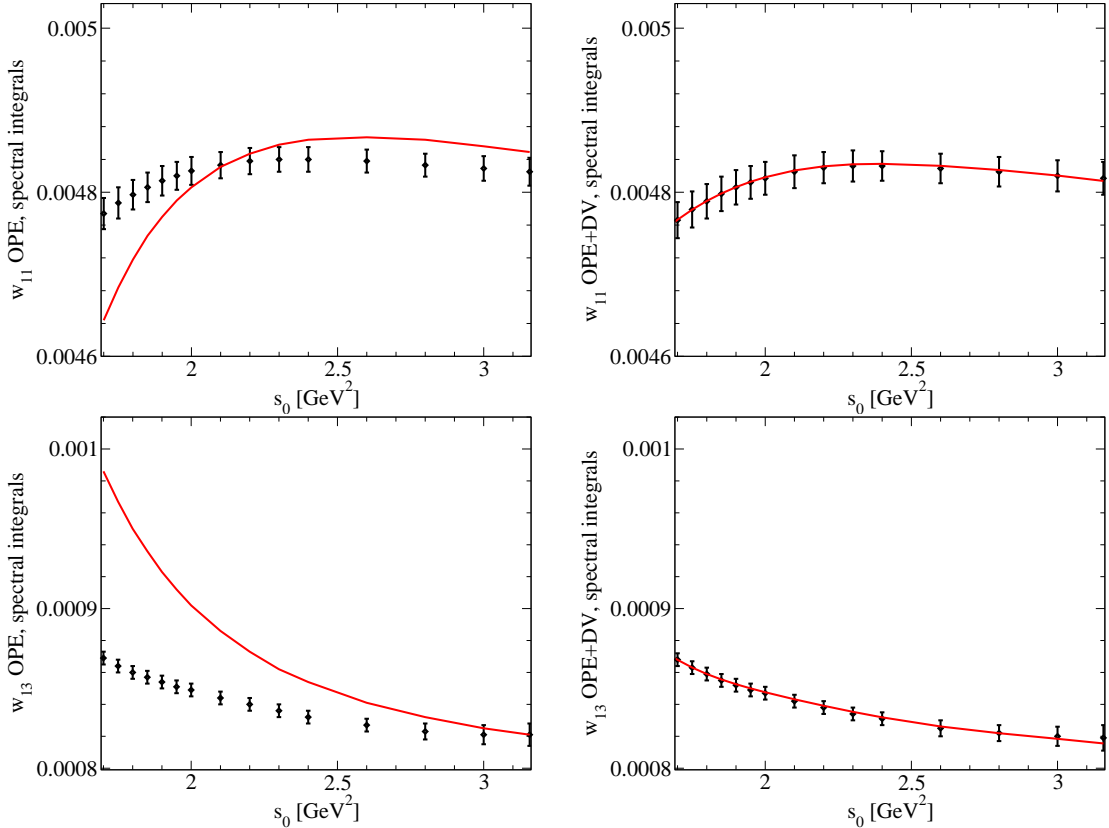

Figure 6: Comparison of the agreement between the lefthand and righthand sides of the FESR (4) for the weights $w_{11}$ and $w_{13}$ within the tOPE strategy, with DVs and high-dimension condensates neglected (left), and with DVs taken into account and condensates determined via Eqs. (7.3) of Ref. [16] (right).

It would be very instructive to be able to test this assumption about the simple $C_D/s_0^{D/2}$ suppression for scales $s_0 \geq m_\tau^2$. Clearly, if the higher-dimension terms in the OPE are suppressed at the scale $m_\tau^2$, they should be even more suppressed at scales larger than $m_\tau^2$. Although, regrettably, it is not possible to test this with the $\tau$ data, with some mild assumptions it is possible to do so using data for $e^+e^- \to$ hadrons [18,19].

Using the so-called "optimal" weights proposed in Ref. [14]:

$$w^{(2,n)}(y) = 1 - (n+2)y^{n+1} + (n+1)y^{n+2} , \qquad (10)$$

with $n = 1, ..., 5$, which are doubly pinched, one may determine $\alpha_s$ and $C_{6,8,10}$ at $s_0 = m_\tau^2$, provided one neglects $C_{12,14,16}$ in the FESRs (4). In the SU(3) limit, one finds that the correlator of two electromagnetic currents is $2/3$ times the correlator of two isospin currents, as they would appear in the $V$ channel in $\tau$ decay. Consequently, the physics of these two situations cannot be very different. In Ref. [19] we presented a preliminary version of this type of analysis. The result is shown in Fig. 7. In this figure we plot the result for the difference between the contribution of the OPE to the FESR (i.e., the righthand side of Eq. (4) without the DV term) at a variable $s_0$ minus the same for $s_0 = m_\tau^2$, as a function of $s_0$**. The result is represented by the two black curves (dashed for CIPT and solid for FOPT). We also plot the same difference, but now computed with the $e^+e^-$ data as the red points. The fact that they both agree, and vanish at $s = s_0$, is nothing but a consequence of our definition. What is more interesting is that, not only for $s_0 < m_\tau^2$ but also for $s_0 > m_\tau^2$, the two descriptions clearly disagree. This is, again, a rather clear sign that the assumption that higher-dimension terms in the OPE are negligible is not supported by the data. Similar conclusions follow from using the weights of Eq. (7) instead of those of Eq. (10).

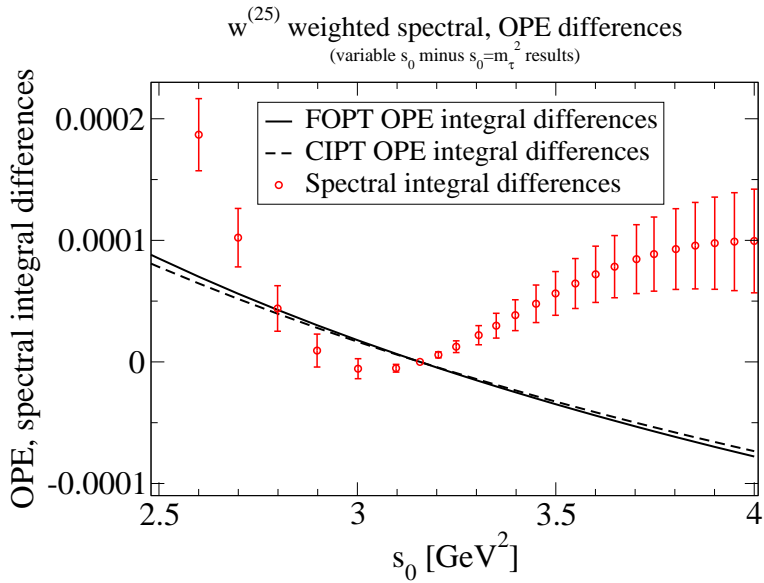

Figure 7: Electromagnetic FESR tests of the tOPE strategy using the set of weights of Eq. (10). Comparisons of differences between a variable $s_0$ and $s_0 = m_\tau^2$ versions of the OPE and spectral integrals. The OPE parameter values are obtained from the implementation of the tOPE strategy using the weights of Eq. (10) and $s_0 = m_\tau^2$ only in the fits.

In summary, we have presented conclusive evidence that the neglect of higher-dimension terms in the OPE and DVs at the core of the truncated OPE strategy leads to an irreducible systematic error of the order of $+0.02$ in the value of $\alpha_s(m_\tau^2)$. This method should therefore be considered unreliable and, consequently, no longer be used, at least not without adding a potential $\sim +0.02$ systematic error to the tOPE results. As an alternative, we have proposed a strategy that parametrizes DVs as in Eq. (6) and includes them in the analysis from the start, and which makes no *a priori* assumptions about the values of the relevant OPE coefficients, which are to be determined by the data through fits employing Eq. (4) in a window of values of $s_0$ ranging up to $m_\tau^2$. The result of these fits to ALEPH data leads to [16]

$$\alpha_{\rm s}(m_\tau) = 0.296 \pm 0.010 \quad \longrightarrow \quad \alpha_{\rm s}(m_Z) = 0.1155 \pm 0.0014 \quad ({\rm FOPT}) \;,$$
$$\alpha_{\rm s}(m_\tau) = 0.310 \pm 0.014 \quad \longrightarrow \quad \alpha_{\rm s}(m_Z) = 0.1174 \pm 0.0019 \quad ({\rm CIPT}) \;. \qquad (11)$$

---

**We do this to account for correlations.

Combining these results with those based on the OPAL data, we obtain as our final result

$$\alpha_{\rm s}(m_Z) = 0.1165 \pm 0.0012 \quad ({\rm FOPT}) \;, \qquad \alpha_{\rm s}(m_Z) = 0.1185 \pm 0.0015 \quad ({\rm CIPT}) \;. \qquad (12)$$

These results are in very good agreement with the value for $\alpha_{\rm s}$ obtained from the same type of FESRs using the $e^+e^-$ data below the charm threshold [20]. We emphasize that, in this case, the $s_0$ values being used are sufficiently larger than $m_\tau^2$ to make the contribution from DVs marginal, if not negligible. We should also recall that the $\tau$-based results rely on the assumption that our theory representation, which is expected to be valid for asymptotically large $s_0$, holds in a region of $s$ extending down to below the $\tau$ mass. The good agreement shown in Fig. 4 and the consistency with the value obtained from $e^+e^-$ is evidence for the validity of this assumption.

**Acknowledgments.** The work of D.B. is supported by the São Paulo Research Foundation (Fapesp) grant No. 2015/20689-9 and by the Brazilian National Council for Scientific and Technological Development (CNPq), grant No. 305431/2015-3. This material is based upon work supported by the U.S. Department of Energy, Office of Science, Office of High Energy Physics, under Award Number DE-SC0013682 (MG). KM is supported by a grant from the Natural Sciences and Engineering Research Council of Canada, and SP by CICYTFEDER-FPA2014-55613-P, 2014-SGR-1450 and the CERCA Program/Generalitat de Catalunya.

# QCD coupling: scheme variations and tau decays

Matthias Jamin[1,2] and **Ramon Miravitllas**[2]

[1]*ICREA, Pg. Lluís Companys 23, 08010 Barcelona, Spain*
[2]*IFAE, BIST, Campus UAB, 08193 Bellaterra (Barcelona) Spain*

**Abstract:** We introduce a QCD coupling redefinition which has a simple scheme transformation. As an application, we discuss possible improvements on perturbative predictions of QCD physical quantities. In particular, we focus our attention to the Adler function, which is relevant for the extraction of $\alpha_s$ from tau decays.

## The $C$-scheme coupling

We consider the Adler function [1], which governs theoretical predictions of the inclusive decay rate of tau leptons into hadronic final states [2]. After proper normalisation, a perturbative expansion for the Adler function is given by $D(a) = 1 + a + \mathcal{O}(a^2)$, where $a = \alpha_s(Q^2)/\pi$ and $Q$ is the momentum transfer of the related physical process.

In the large-$\beta_0$ approximation (see [3] for a review), this function has the following Borel integral representation [4,5]:*

$$D(a) = \frac{2}{\beta_1} \int_0^\infty \mathrm{d}u\, e^{-2u/(\beta_1 a)} B[D](u)\,, \tag{1}$$

where

$$B[D](u) = \frac{32}{3} \frac{e^{-Cu}}{2-u} \sum_{k \geq 0} \frac{(-1)^k k}{[k^2 - (1-u)^2]^2}\,. \tag{2}$$

$C$ parametrises the scheme dependence of $a$, arising from the renormalisation of the gluon chain $1/(1 + \Pi_0)$. A constant $C$ remains after subtracting the divergence from the fermion loop in dimensional regularisation:

$$\Pi_0(k^2) = -\beta_1 \alpha_s \left[\log\left(-\frac{k^2}{\mu^2}\right) + C\right]\,. \tag{3}$$

In the $\overline{\mathrm{MS}}$ scheme, $C = -5/3$.

Because the Adler function is a physical quantity, it is independent of theoretical conventions. In particular, the Borel integral in Eq. (1) is independent of $C$ and we conclude that the combination $2/(\beta_1 a) + C$ has to be $C$ invariant. Therefore, the $C$ dependence of the coupling $a$ is given by

$$\frac{1}{a(C)} = \frac{1}{a(C=0)} - \frac{\beta_1}{2} C\,. \tag{4}$$

Our goal is to define a new coupling in full QCD with similar scheme properties to those in Eq. (4). For that, we define the scale invariant $\Lambda_{\mathrm{QCD}}$ parameter

$$\Lambda_{\mathrm{QCD}} = Q e^{-1/(\beta_1 a)} a^{-\beta_2/\beta_1^2} \exp\left(\int_0^a \frac{\mathrm{d}a}{\tilde{\beta}(a)}\right), \tag{5}$$

---

*In our notation, the $\beta$ coefficients are defined through $\beta(a) = -\mu\, \mathrm{d}a/\mathrm{d}\mu = \beta_1 a^2 + \beta_2 a^3 + \ldots$

where
$$\frac{1}{\tilde{\beta}(a)} = \frac{1}{\beta(a)} - \frac{1}{\beta_1 a^2} + \frac{\beta_2}{\beta_1^2 a} \tag{6}$$

is free of singularities at $a = 0$.

Although $\Lambda_{\text{QCD}}$ by definition is scale independent, it does depend on the scheme. If the coupling is $a$ in one scheme and $a'$ in another so that both couplings are related by $a' = a + c_1 a^2 + \mathcal{O}(a^3)$, then the scheme transformation of $\Lambda_{\text{QCD}}$ is given by [6]

$$\Lambda' = \Lambda e^{c_1/\beta_1}, \tag{7}$$

where $\Lambda$ ($\Lambda'$) is the $\Lambda_{\text{QCD}}$ parameter in the $a$ ($a'$) coupling. The $\Lambda_{\text{QCD}}$ parameter only depends on $c_1$ and is insensitive to the rest of the expansion coefficients.

We fix the coupling $a$ in a specific scheme and define a new coupling $\hat{a}$ through

$$f(\hat{a}) = \beta_1 \log\left(\frac{Q}{\Lambda_{\text{QCD}}}\right) + \frac{\beta_1}{2} C, \tag{8}$$

where $f$ is some function to be specified later. The right hand side of the equation has a very simple scheme transformation originating from Eq. (7), which for convenience we reparametrised in terms of $C$ instead of $c_1$. Thus, the new coupling $\hat{a}$ has the same property regardless of the choice of $f$. Combining Eq. (5) and Eq. (8), we find

$$f(\hat{a}) - \frac{\beta_1}{2} C = \beta_1 \log\left(\frac{Q}{\Lambda_{\text{QCD}}}\right) = \frac{1}{a} + \frac{\beta_2}{\beta_1} \log(a) - \beta_1 \int_0^a \frac{\mathrm{d}a}{\tilde{\beta}(a)}, \tag{9}$$

so we choose

$$f(\hat{a}) = \frac{1}{\hat{a}} + \frac{\beta_2}{\beta_1} \log(\hat{a}) \tag{10}$$

in order to match both sides of Eq. (9). The coupling $\hat{a}$ is then implicitly defined through

$$\frac{1}{\hat{a}} + \frac{\beta_2}{\beta_1} \log(\hat{a}) = \beta_1 \log\left(\frac{Q}{\Lambda_{\text{QCD}}}\right) + \frac{\beta_1}{2} C. \tag{11}$$

The choice of $f$ is not arbitrary, but it is necessary so that the perturbative relation between $a$ and $\hat{a}$ remains a simple power expansion $\hat{a} = a + \sum_{n \geq 1} c_n a^{n+1}$. It is in this sense that $\hat{a}$ is a legitimate coupling redefinition. We call $\hat{a}$ the $C$-scheme coupling, which was first introduced in [7].

We note that in the large-$\beta_0$ approximation (where $\beta_n = 0$ for all $n \geq 2$), $\hat{a}$ has the same scheme dependence as in Eq. (4).

## $C$-scheme coupling evolution

The $C$-scheme coupling has simple properties regarding scale and $C$ transformations. Differentiating Eq. (11) with respect to either $C$ or $Q$, we find

$$-Q \frac{\mathrm{d}\hat{a}}{\mathrm{d}Q} \equiv \hat{\beta}(\hat{a}) = \frac{\beta_1 \hat{a}^2}{1 - \frac{\beta_2}{\beta_1} \hat{a}} = -2 \frac{\mathrm{d}\hat{a}}{\mathrm{d}C}. \tag{12}$$

So changes in the scheme $C$ are completely equivalent to changes in the scale $Q$. A shift in the scale from $Q_1$ to $Q_2$ can be compensated by a shift in the scheme from $C_1$ to $C_2$ so that $Q_1/Q_2 = e^{C_1 - C_2}$. In addition, because $\beta_1$ and $\beta_2$ are both scheme independent parameters, then the $\beta$ function of $\hat{a}$ is explicitly scheme independent.

## Cancellation of even $\zeta$'s in perturbative expansions

As an example for the discussion of this section, we consider the second derivative of the scalar correlator

$$\Psi''\left(Q^2\right) \sim K\frac{m^2}{Q^2} \sum_{n\geq 0} b_n a^n, \tag{13}$$

with the scale choice $\mu^2 = Q^2$ for both mass and coupling. The perturbative expansion in Eq. (13) is currently known up to fourth order. The coefficients $b_n$ contain different values of the Riemann $\zeta$ function [8,9,10]:

$$b_2 = -\frac{35}{2}\zeta_3 + \ldots \tag{14}$$

$$b_3 = \frac{715}{12}\zeta_5 - \frac{5}{4}\zeta_4 - \frac{65869}{216}\zeta_3 + \ldots \tag{15}$$

$$b_4 = \frac{52255}{256}\zeta_7 - \frac{625}{48}\zeta_6 + \frac{59875}{108}\zeta_5 - \frac{14575}{576}\zeta_4 + \ldots \tag{16}$$

We will rewrite the perturbative series of Eq. (13) in two steps. In the first step, we replace the mass $m$ by its scale-invariant version

$$\widehat{m} = m\,(\pi a)^{-\gamma_1/\beta_1} \exp\left(\int_0^a \mathrm{d}a\left[\frac{\gamma(a)}{\beta(a)} - \frac{\gamma_1}{\beta_1 a}\right]\right) \tag{17}$$

and obtain

$$\Psi''\left(Q^2\right) \sim K\frac{\widehat{m}^2}{Q^2}(\pi a)^{2\gamma_1/\beta_1} \sum_{n\geq 0} b'_n a^n, \tag{18}$$

where the coefficients $b'_n$ are combinations of the initial coefficients $b_n$ and contributions coming from the exponential factor in Eq. (17). The $\zeta_4$ present in $b_3$ and the $\zeta_6$ present in $b_4$ are cancelled by these contributions, but the $\zeta_4$ in $b_4$ still remains. The respective cancellations have also been observed in [11] for a related quantity.

In the second step, we replace in Eq. (18) the QCD coupling $a$ by the $C$-scheme coupling $\hat{a}$. The result is that the remaining $\zeta_4$ term in $b'_4$ also cancels against a corresponding $\zeta_4$ present in the $\beta_5$ coefficient that arises from this replacement. Thus, the coefficients of the perturbative expansion become free of any even $\zeta$ term at least up to order $\hat{a}^4$ (although we expect this to be true to all orders in perturbation theory). This result has also been demonstrated for the gluonium correlator [12] and for several more physical quantities [13,14].

## Borel models

It is well known that perturbative expansions are divergent [15]. Adding more terms to an expansion would in general give better theoretical predictions, but there is a turning point when the factorial divergence of the coefficients dominate over the suppression of the coupling. From there, the precision degrades as more terms are added.

Conventionally, a finite value is assigned to the divergent expansions by considering its Borel sum:

$$D(a) = \frac{2}{\beta_1} \int_0^\infty \mathrm{d}u\, e^{-2u/(\beta_1 a)} B[D](u) + \ldots \tag{19}$$

However, the corresponding Borel transform $B[D]$ has singularities on the positive real axis which interfere with the Borel integral, producing imaginary ambiguities that are exponentially suppressed like $e^{-S/a}$, where $S > 0$ is the position of the singularity. These ambiguities indicate the presence of opposing exponential corrections to the original perturbative expansion. We have

$$D(a) \sim \sum_{n \geq 0} b_n a^n \pm i\, b\, e^{-S/a} a^{-\lambda} \sum_{n \geq 0} b'_n a^n + \ldots, \quad a \to 0^+, \tag{20}$$

which is conventionally written as an operator product expansion (OPE) with the exponential factors written in terms of $\Lambda^2_{\text{QCD}}/Q^2$. Each exponential factor corresponds to a different singularity $S$ in the Borel transform.

Imposing ambiguity cancellation between the two sectors of the expansion yields a relation between the large $n$ behaviour of the $b_n$ and the low $n$ behaviour of the $b'_n$. This connection reads

$$b_n = b\, \frac{(-1)^{n+1}}{\pi} \frac{\Gamma(n+\lambda)}{(-S)^{n+\lambda}} \left[ 1 + \frac{-S\, b'_1}{n+\lambda-1} + \mathcal{O}\left(\frac{1}{n^2}\right) \right], \tag{21}$$

$$\pm i\, b\, e^{-S/a} a^{-\lambda} \left[ a + b'_1 a^2 + \mathcal{O}(a^3) \right]. \tag{22}$$

Even if it is impossible to compute the coefficients $b_n$ at large $n$ from Feynman diagrams, we can still determine its large order behaviour thanks to this connection with the OPE. Namely, from the known structure of the OPE, we can fix the parameters $\lambda$ and $S$, and from the computation of the Wilson coefficients we can determine the $b'_n$. But the residue $b$ can only be computed from the condensates and very little is known about them at the present time.

In [16], an ansatz is proposed for the Borel transform of the Adler function in full QCD. This ansatz incorporates the closest singularities to the origin ($u = -1, 2, 3$) and their respective residues $b$ are fit with the first few known coefficients of the Adler function. This strategy already presumes that the large order behaviour of the $b_n$ sets in fast enough, but it is possible that a far-away singularity has an artificially high residue in such a way that its contribution to low orders is not negligible.

Using the $C$-scheme coupling, we want to improve on the above strategy by making sure that already at low order the large order behaviour is a good approximation to the true result. For that, we compute the Borel transform of the Adler function with respect to $\hat{a}$ (instead of $a$). The residues of this Borel transform change with the scheme $C$ and it is then possible to choose an optimal value of $C$ for which the residues of the closest singularities are enhanced with respect to far-away singularities.

The large-$\beta_0$ approximation is a good play field to qualitatively investigate how these changes take place. Going back to Eq. (2), we see that the residue at $u = S$ goes like $b \approx e^{-CS}$. Thus, as a rough approximation, $C > 0$ enhances negative poles, while $C < 0$ enhances positive poles. Lessons learned from this model then can be extrapolated to full QCD.

We hope theoretical uncertainties in $\alpha_s$ extractions that arise from the truncation of perturbative expansions can be reduced by using the procedure described in this section.

# High-precision $\alpha_\text{s}$ from W and Z hadronic decays

David d'Enterria[1,*]

[1] CERN, EP Department, CH-1211 Geneva 23, Switzerland

**Abstract:** The extraction of the QCD coupling $\alpha_\text{s}$ from the comparison of experimental data on inclusive W and Z bosons hadronic decays to state-of-the-art perturbative QCD calculations is reviewed. The relatively small amount of W data from $e^+e^- \to W^+W^-$ collisions at LEP leads today to an non-competitive extraction of the strong coupling at the Z mass from the measured $R_W$ ratio of hadronic-to-leptonic branching fractions, $\alpha_\text{s}(m_\text{z}) = 0.117 \pm 0.042_\text{exp} \pm 0.004_\text{th} \pm 0.001_\text{par}$ with a ~35% propagated experimental uncertainty. Analysis of the much more abundant hadronic results at the Z pole leads to $\alpha_\text{s}(m_\text{z}) = 0.1203 \pm 0.0030$, with a 2.5% uncertainty by combining three different pseudo-observables (ratio of hadronic-to-leptonic widths $R_Z$, hadronic peak cross section $\sigma_\text{Z}^\text{had}$, and total width $\Gamma_\text{Z}^\text{tot}$). An $\alpha_\text{s}$ determination with per mille uncertainty requires high-statistics W and Z bosons data samples at future $e^+e^-$ colliders, such as the FCC-ee, combined with even higher precision (N$^4$LO) pQCD calculations.

## Introduction

The strong coupling $\alpha_\text{s}$ is one of the fundamental parameters of the Standard Model (SM), and its value not only directly affects the stability of the electroweak vacuum [1] but it chiefly impacts the theoretical calculations of all scattering and decay processes involving real and/or virtual quarks and gluons [2]. Known today with a 0.9% precision, $\alpha_\text{s}$ is the worst known of all fundamental interaction couplings in nature [3], and such an imprecision propagates as an input parametric uncertainty in the calculation of many important physics observables, in particular in the electroweak (EW), Higgs, and top-quark SM sectors [4]. The current world-average value, $\alpha_\text{s}(m_\text{z}) = 0.1181 \pm 0.0011$ [3], is derived from a combination of six subclasses of approximately-independent observables measured in $e^+e^-$ collisions (hadronic Z boson and $\tau$ decays, plus event shapes and jet rates), deep-inelastic scattering DIS (structure functions and global fits of parton distributions functions PDFs), and p-p collisions (inclusive top-pair cross sections), as well as from lattice QCD computations constrained by the empirical values of hadron masses and decay constants. In order to be combined into the $\alpha_\text{s}(m_\text{z})$ world-average, the experimental (or lattice) results need to have a counterpart perturbative QCD (pQCD) prediction at next-to-next-to-leading-order NNLO (or beyond) accuracy.

In principle, among the theoretically and experimentally "cleanest" $\alpha_\text{s}$ extractions are those based on the hadronic decays of electroweak bosons. This is so because (i) the inclusive hadronic W and Z decays can be very accurately measured in $e^+e^-$ collisions provided one has large enough data samples, (ii) the corresponding theoretical predictions can be computed with a very high theoretical accuracy, today up to $\mathcal{O}(\alpha_\text{s}^4)$, i.e. N$^3$LO, in pQCD [5], plus mixed $\mathcal{O}(\alpha\alpha_\text{s})$ pQCD-EW [6,7] and (in the Z case) the full two-loop $\mathcal{O}(\alpha^2)$ EW corrections [8], and (iii) non-pQCD effects are suppressed thanks to the large energy scale given by the electroweak masses ($m_\text{W,Z} \gg \Lambda_\text{QCD} \approx 0.2$ GeV). The common high-precision hadronic observables used to extract $\alpha_\text{s}$ in $e^+e^-$ annihilation at the W and Z boson masses can be schematically decomposed as follows:

---

[*] e-mail: dde@cern.ch

- total W and Z hadronic width:

$$\Gamma^{\text{had}}_{\text{W,Z}}(Q) = \frac{\sigma(e^+e^- \to (\text{W,Z}) \to \text{hadrons})}{\sigma(e^+e^- \to (\text{W,Z}) \to \text{X})}$$

$$= \Gamma^{\text{Born}}_{\text{W,Z}} \left(1 + \sum_{i=1}^{4} c_i(Q) \left(\frac{\alpha_s(Q)}{\pi}\right)^i + \mathcal{O}(\alpha_s^5) + \delta_{\text{EW}}(\alpha, \alpha^2) + \delta_{\text{m}}(\alpha\alpha_s) + \delta_{\text{np}}\right) \quad (1)$$

where the Born width $\Gamma^{\text{Born}}_{\text{W,Z}} = f(G_F, N_C, m^3_{\text{W,Z}}; \sum |V_{ij}|^2)$ depends on the Fermi constant $G_F$ and the number of colours $N_C$, and in the W case on the sum of CKM matrix elements $|V_{ij}|^2$, and

- ratio of inclusive hadronic-to-leptonic widths (that commonly includes also the $\tau$ lepton, which proceeds via offshell W decays):

$$R_{\tau,\text{W,Z}}(Q) = \frac{\sigma(e^+e^- \to (\tau,\text{W,Z}) \to \text{hadrons})}{\sigma(e^+e^- \to (\tau,\text{W,Z}) \to \ell^+\ell^-)}$$

$$= R^{\text{EW}}_{\tau,\text{W,Z}}(\alpha, \alpha^2; Q) \left(1 + \sum_{i=1}^{4} c_i(Q) \left(\frac{\alpha_s(Q)}{\pi}\right)^i + \mathcal{O}(\alpha_s^5) + \delta m(\alpha\alpha_s) + \delta_{\text{np}}\right) \quad (2)$$

where the $R^{\text{EW}}_{\tau,\text{W,Z}}$ prefactor accounts for the purely electroweak dependence of the ratio.

In both expressions (1) and (2), $Q = m_\tau, m_\text{W}, m_\text{Z}$ is the relevant momentum transfer in the process, $c_i$ are coefficients of the pQCD expansion calculated today up to a finite order $i = 4$, the $\mathcal{O}(\alpha_s^5)$ term indicates (sub-permille) corrections at N$^4$LO accuracy not yet computed, and $\delta_{\text{m}}(\alpha\alpha_s)$ and $\delta_{\text{np}}(\Lambda^p_{\text{QCD}}/Q^p)$ correspond to mixed pQCD-EW and power-suppressed non-perturbative corrections, respectively. It is important to note that the Born level term in the calculation of W and Z hadronic decays is completely independent of the QCD coupling, and that all $\alpha_s$ sensitivity comes through (small) higher-order loop corrections. Indeed, for $\alpha_s(m_\text{z}) \approx 0.118$, the size of the QCD sum in Eq. (2) amounts to a $\sim$3% effect in the calculation of $R_{\text{W,Z}}$, and thereby at least permille measurement accuracies in this ratio are required for a competitive $\alpha_s(m_\text{z})$ determination. Such an experimental precision has been achieved in $\tau$ and Z boson measurements, but not in the W boson case, and that is why the latter does not yet provide a precise $\alpha_s$ extraction [9] as discussed below. Reaching permille uncertainties in $\alpha_s$ determinations requires many orders of magnitude smaller uncertainties in the experimental $\tau$, W and Z measurements than today, a situation only reachable at a future $e^+e^-$ collider such as the FCC-ee [10] (or before, at B-factories, for the $\tau$ lepton).

It is instructive to consider the $\alpha_s$ extraction via $\tau$ lepton decays using Eq. (2), which proceeds via offshell W hadronic decays (involving only the kinematically allowed $u$, $d$, and $s$ quarks), before studying the (onshell) electroweak bosons case. In this case, the ratio of hadronic to leptonic decays, known experimentally to within $\pm$0.23%, $R_{\tau,\text{exp}} = 3.4697 \pm 0.0080$, yields $\alpha_s(m_\text{z}) = 0.1192 \pm 0.0018$ with a 1.5% uncertainty, through a combination of results from different N$^3$LO calculations (contour-improved CIPT, and fixed-order FOPT, perturbation theory) with different treatments of the non-pQCD corrections [11,12]. The non-perturbative power-suppressed $\delta_{\text{np}}$ term in Eq. (2) is $\mathcal{O}(\Lambda^2_{\text{QCD}}/m^2_\tau) \approx 10^{-2}$, and thereby not negligible at variance with the much heavier W and Z bosons case. Reducing the current $\alpha_s(m_\text{z})$ extraction uncertainties from the $\tau$ lepton requires controlling the non-pQCD uncertainties through better experimental data (in particular, $\tau$ spectral functions) than those from ALEPH and OPAL currently available (e.g., from B-factories now, and FCC-ee in the future) [12]. solving CIPT-FOPT discrepancies, and eventually extending the calculations to N$^4$LO accuracy.

## Extraction of $\alpha_{\rm s}(m_{\rm z})$ from hadronic W decays

The current state-of-the-art calculations of W boson hadronic decays include N³LO pQCD [5], one-loop $\mathcal{O}(\alpha)$ EW [13], and mixed two-loop $\mathcal{O}(\alpha\alpha_{\rm s})$ pQCD-EW [6] corrections. Numerically, the relative weights of the different terms appearing in Eqs. (1) and (2) amount to [9]: $\Gamma_{\rm W}^{\rm Born}$, $R_{\rm W}^{\rm EW} \approx 96.6\%$, $\mathcal{O}(\alpha_{\rm s}^1) \approx 3.7\%$, $\mathcal{O}(\alpha_{\rm s}^2) \approx 0.2\%$, $\mathcal{O}(\alpha_{\rm s}^3) \approx -0.1\%$, $\mathcal{O}(\alpha_{\rm s}^4) \approx -0.02\%$, $\mathcal{O}(\alpha) \approx -0.35\%$, $\delta_{\rm m} \approx -0.05\%$, with negligible $\delta_{\rm np}$ non-pQCD effects, suppressed by $\mathcal{O}(\Lambda_{\rm QCD}^4/m_{\rm W}^4)$ power corrections. However, the calculations suffer from a significant parametric uncertainty from the input CKM matrix elements. Indeed, the Born-level W hadronic decay width is directly proportional to the sum over the first two rows of the CKM matrix, $\Gamma_{\rm W}^{\rm Born} \propto \sum_{\rm u,c,d,s,b} |V_{\rm ij}|^2$ (the top quark is kinematically forbidden in W decays), whose uncertainty is dominated by the 1.6% imprecision of the measured charm-strange quark mixing element, $|V_{\rm cs,exp}| = 0.986 \pm 0.016$ [3]. Thus, using the experimental CKM elements, the prefactor $\sum_{\rm u,c,d,s,b} |V_{\rm ij}|^2 = 2.024 \pm 0.032$ propagates as a final 1.6% uncertainty into any hadronic W decay calculation today. In order to assess the impact of such a parametric uncertainty, one can impose CKM unitarity and take $\sum_{\rm u,c,d,s,b} |V_{\rm ij}|^2 \equiv 2$.

Unfortunately, on the experimental side the situation is even much less precise. The relevant LEP $W^+W^-$ data are statistically poor, based on about $5 \cdot 10^4$ W bosons alone, and the associated extraction of $\alpha_{\rm s}(m_{\rm z})$ is truly non-competitive today. From the current value of the W hadronic width, $\Gamma_{\rm W}^{\rm had,exp} = \Gamma_{\rm W}^{\rm tot,exp} \cdot \mathcal{B}_{\rm W}^{\rm had,exp} = 1405 \pm 29$ MeV with a 2% uncertainty [3], one can barely constraint the QCD coupling: $\alpha_{\rm s}(m_{\rm z}) = 0.069 \pm 0.065_{\rm exp} \pm 0.050_{\rm par}$, or assuming CKM unitarity, $\alpha_{\rm s}(m_{\rm z}) = 0.107 \pm 0.066_{\rm exp} \pm 0.002_{\rm par} \pm 0.001_{\rm th}$. If one uses, instead, the value of the hadronic/leptonic ratio experimentally known with a 1.2% precision ($R_{\rm W}^{\rm exp} = 2.068 \pm 0.025$), one obtains $\alpha_{\rm s}(m_{\rm z}) = 0.00 \pm 0.04_{\rm exp} \pm 0.16_{\rm par}$ (with the experimental CKM matrix) or $\alpha_{\rm s}(m_{\rm z}) = 0.117 \pm 0.042_{\rm exp} \pm 0.004_{\rm th} \pm 0.001_{\rm par}$ (assuming CKM unitarity) [9]. This last value shows that, in the best scenario, the derived $\alpha_{\rm s}(m_{\rm z})$ value has currently a huge $\pm 36\%$ propagated uncertainty (Fig. 1, left).

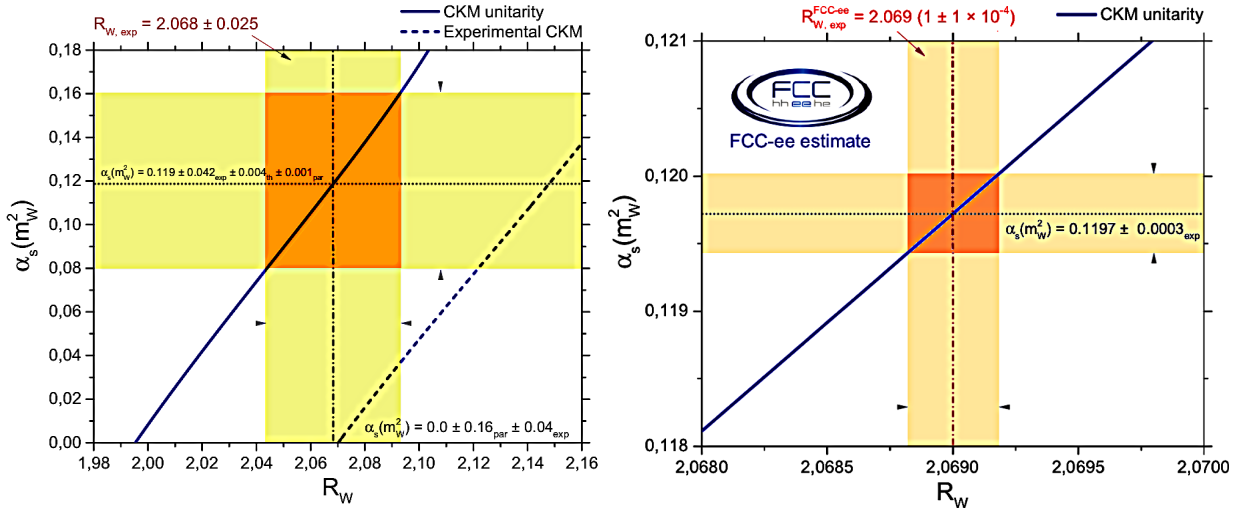

Figure 1: Extraction of $\alpha_{\rm s}$ from the hadronic/leptonic W decay ratio $R_{\rm W}$, using the current data (left) and expected at the FCC-ee with experimental uncertainties alone (right) [9]. Note the wildly different x- and y-axes scales. The diagonal blue line in both plots assumes CKM matrix unitarity.

At the FCC-ee, the total W width $\Gamma_{\rm W}^{\rm tot}$ can be accurately measured through a threshold $e^+e^- \to W^+W^-$ scan around $\sqrt{s} = 2m_{\rm W}$, and also the $R_{\rm W}$ ratio will profit from the huge sample of $5 \cdot 10^8$

W bosons (a thousand times more than those collected at LEP) thereby reducing the statistical uncertainty of $R_W$ to around 0.005%. Neglecting parametric uncertainties, the high-precision W decay measurements at the FCC-ee would significantly improve the extraction of $\alpha_s$ with propagated experimental uncertainties of order 0.4%. A value that could be further reduced to ∼0.2% through the measurement of the $R_W$ ratio in three $e^+e^- \to W^+W^-$ final states ($\ell\nu\,\ell\nu$, $\ell\nu\,qq$, $qq\,qq$), and/or combining it with the $\alpha_s$ value derived from the total width $\Gamma_W^{\rm tot}$. Indeed, the ratio of cross sections $\sigma(WW \to qq\,qq)/\sigma(WW \to \ell\nu\,\ell\nu)$ is proportional to $(R_W)^2$, thereby gaining a factor two in statistical sensitivity, and being totally independent of potential modifications of the weak coupling running as well as free from cross section normalization uncertainties [10]. Figure 1 (right) shows the estimated $\alpha_s$ extraction from the expected improved measurement of $R_W$ at FCC-ee, assuming that $V_{cs}$ has a negligible uncertainty (or, identically, assuming CKM matrix unitarity). A full determination of $\alpha_s$ with permille uncertainty including also parametric and theoretical uncertainties will require two more developments: (i) a significantly reduced uncertainty of the $V_{cs}$ CKM element, and (ii) computing the N$^4$LO pQCD term $\mathcal{O}(\alpha_s^5)$, as well as missing two-loop electroweak corrections (available now for the Z boson) of Eqs. (1) and (2).

## Extraction of $\alpha_s(m_Z)$ from hadronic Z decays

On the theory side, the current state-of-the-art Z boson hadronic decays calculations include N$^3$LO pQCD [5], plus full two-loop $\mathcal{O}(\alpha)$ EW, and mixed two-loop $\mathcal{O}(\alpha\alpha_s)$ pQCD-EW corrections (see Ref. [8] for a complete list of relevant references). Numerically, the size of the Born term appearing in Eqs. (1) and (2) is $\Gamma_Z^{\rm Born}$, $R_Z^{\rm EW} \approx 96.8\%$, and one can see again that the $\alpha_s$ dependence on these observables only enters through (small) higher-order corrections. However, as for the W boson case, the non-perturbative effects encoded in the $\delta_{\rm np}$ term are power-suppressed by $\mathcal{O}(\Lambda_{\rm QCD}^4/m_Z^4)$. The current QCD coupling extraction based on Z hadronic decays uses not just $\Gamma_Z^{\rm tot}$ ($0.1209 \pm 0.0049$) and $R_Z$ ($0.1237 \pm 0.0043$), but also the hadronic peak cross section $\sigma_Z^{\rm had} = 12\pi/m_Z \cdot \Gamma_Z^e \Gamma_Z^{\rm had}/(\Gamma_Z^{\rm tot})^2$ ($0.1078 \pm 0.0076$) measured at LEP (based on a data sample of $1.7 \cdot 10^7$ Z bosons) [14], to derive $\alpha_s(m_Z) = 0.1203 \pm 0.0030$ with a 2.5% uncertainty [15] (the extraction based on LEP-only data is $\alpha_\mathrm{s}(m_\mathrm{Z}) = 0.1221 \pm 0.0031$ as quoted in the Electroweak chapter of the PDG [3]). Alternatively, fixing all SM parameters to their measured values and letting free $\alpha_\mathrm{s}$ in the full SM electroweak fit results in $\alpha_\mathrm{s} = 0.1194 \pm 0.0029$ with a ~2.4% uncertainty (blue curve in Fig. 2) [15].

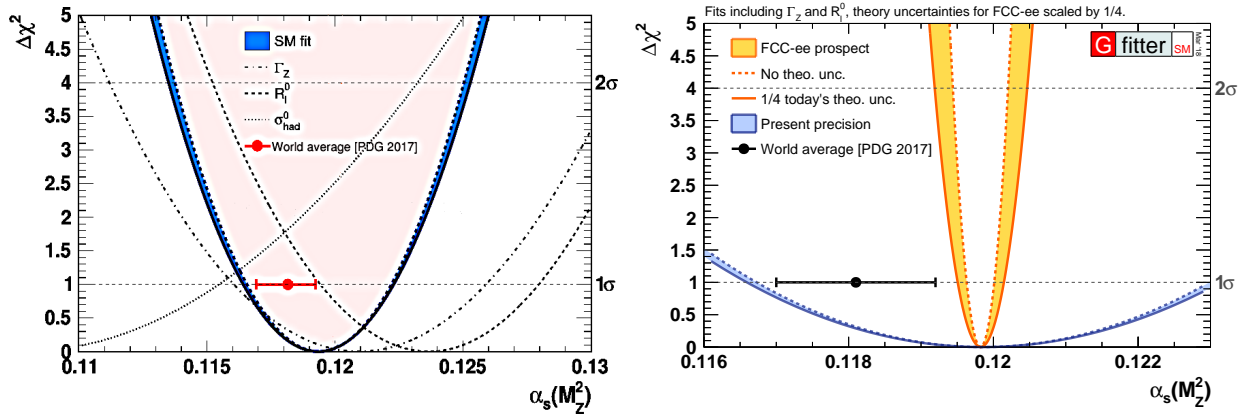

Figure 2: Extracted $\alpha_s$ values from hadronic Z decay data compared to the current world-average (circle). Left: Using the current experimental measurements of $\Gamma_Z^{\rm tot}$ (dashed-dotted), $R_Z$ (dashed), and $\sigma_Z^{\rm had}$ (dotted lines). Right: Expected at the FCC-ee from $\Gamma_Z^{\rm tot}$ and $R_Z$ (yellow band) without theoretical uncertainties (dotted curve) and with the current ones divided by a factor of four (solid curve). The blue band in both plots shows the result of the full SM electroweak fit today [15].

At the FCC-ee, the availability of $10^{12}$ Z bosons providing high-precision measurements with $\Delta m_\mathrm{Z} = 0.1$ MeV, $\Delta\Gamma_\mathrm{Z}^\mathrm{tot} = 0.1$ MeV, $\Delta\mathrm{R}_\mathrm{Z} = 10^{-3}$ (achievable thanks to the possibility to perform a threshold scan including energy self-calibration with resonant depolarization) will reduce the $\alpha_\mathrm{s}(m_\mathrm{Z})$ uncertainty to ~0.15%. Figure 2 (right) shows the expected $\alpha_\mathrm{s}$ extractions from $\mathrm{R}_\mathrm{Z}$ and $\Gamma_\mathrm{Z}^\mathrm{tot}$ at the FCC-ee (yellow band) without theoretical uncertainties (dotted red curve) and with the theoretical uncertainties reduced to one-fourth of their current values (solid red curve) [15], a result that is ~25 times more precise than that from the current full SM electroweak fit today (blue band). Of course, since the main FCC-ee goal is to carry out "stress precision tests" of the SM in searches for physics beyond the SM, one would need to carefully compare the results of both extractions in order to identify possible deviations due to new physics (which, would potentially affect differently the result derived from the Z-pole data alone, and from the full SM fit).

# Summary of the workshop discussions

All workshop participants listed in page 2

**Abstract:** A summary of the main points raised during the talks discussions and their follow-up questions, as well as in the round table of the last day of the workshop, is presented. The discussions not only focused on particular issues affecting each one of the individual $\alpha_\text{s}$ extractions, but also on the current PDG categorization of $\alpha_\text{s}$ measurements and on the methods used to average them into a single $\alpha_\text{s}(m_\text{z})$ value. Most of the listed points are open and sources of potential controversies, which we highlight here as one might expect that ongoing progress in the field will lead to their clarification and resolution.

The results of the discussions during the presentations and round-table session on the last day of the workshop, are summarized here ordered according to $\alpha_\text{s}$ extraction category. An important point of discussion was the organization of $\alpha_\text{s}$ determination categories, the incorporation of new developments, and the methodology used for the $\alpha_\text{s}(m_\text{z})$ averaging in the PDG review [1]. The last point of this contribution deals with those latter issues. We note that the topic of $\alpha_\text{s}$ determinations was also discussed in a 2018 workshop [2] where complementary details can be found.

- **Lattice QCD:**

  The lattice-QCD practitioners suggested the PDG $\alpha_\text{s}$ review to include an expert member of this community. Alternatively, the $\alpha_\text{s}(m_\text{z})$ average of the FLAG collaboration report [3] could be incorporated into the lattice-QCD PDG chapter and propagated as input into the world-average value. Despite the fact that it is difficult to reach full agreement on the averaging of a very broad range of observables, FLAG has gathered the expertise of a large fraction of the lattice collaborations in its team in order to reach a rough consensus. The fact that the most recent FLAG report provides subaverages for the various different $\alpha_\text{s}$ extractions, helps to carry out reanalyses of these results if desired. It was pointed out that several lattice results are now dominated by higher-order uncertainties in the pQCD counterpart observables mostly computed at NNLO accuracy, except for the static QCD energy that uses a N$^3$LO result [4]. Thus more efforts should be put on the perturbative side of the calculations in the coming years in order to understand and reduce these errors, in parallel to evaluating observables non-perturbatively at higher and higher scales. It was emphasized that the lattice community considers that their $\alpha_\text{s}$ subcategories are as different from each other as e.g. the category of $\tau$-decay is from DIS. Therefore it was suggested that the PDG average includes the results of the lattice QCD subcategories as categories parallel to $\tau$-decays, DIS, etc.

- **Hadronic $\tau$ decays:**

  It was proposed to collect the latest results from $\tau$ decays as well as the novel low-energy $e^+e^-$ annihilation (R$_{\text{e}^+\text{e}^-}$) extraction [5] under one single $\alpha_\text{s}$ group labelled "$\tau$ and $e^+e^-$ continuum below charm" (or similar) as they share many theoretical and experimental coincidences. The $\alpha_\text{s}$ extraction from hadronic $\tau$ decays is significantly affected by the spectral functions measured with limited precision by ALEPH and OPAL in $e^+e^-$ collisions at LEP. It was stressed the need to discuss with the BaBar/BELLE-II communities the use of large $\tau$ decays data samples from B-factory experiments to improve on the $\alpha_\text{s}(m_\tau)$ determination. Although, up until now, all recent extractions of $\alpha_\text{s}$ have been based on the same data, they differ mainly in the treatment of non-perturbative physics. In particular, there is the need to resolve the duality violation treatment and pinching strategy in connection with the properties of the Operator Product Expansion.

- **$e$-p scattering and fits of parton distribution functions:**

  First, it was pointed out that the current "Deep inelastic lepton-nucleon scattering (DIS)" label of this extraction category should be changed to "e-p scattering and global PDF fits" (or similar) to properly include all available $\alpha_s$ determinations in this domain. In particular, to take into account the fact that there are new $\alpha_s$ determinations from HERA data, e.g. based on NNLO jets in DIS and in (anticipated) photoproduction studies, which will be included into the world average. A discussion followed on how to properly merge the novel DIS jets $\alpha_s(m_Z)$ result together with the more inclusive structure function results [6]. The question was raised on what to do with novel NNLO hadron collider extractions, e.g. based on jet or electroweak boson production at the LHC, that have a strong explicit PDF dependence. It seemed that the inclusive W, Z cross sections should go under the "hadron collider" category, as those are *total* cross section like the $t\bar{t}$ ones that are not explicitly included into the global PDF fits, whereas any extraction based on *differential* jet cross sections at NNLO should be rather included as part of the $\alpha_s$ determinations derived in parallel with the future global PDF fits that include these jet spectra too.

  It was reminded that there is currently no consensus on how to reliably estimate theoretical uncertainties of $\alpha_s$ extractions, from missing higher-order corrections in NNLO PDF+$\alpha_s$ fits. This complicates not only the comparison with other categories, but also comparisons within this category to some extent.

- **Hadronic final states in $e^+e^-$ annihilation:**

  The novel results from $e^+e^-$ annihilation based on energy-energy correlations (EEC) and jet rates (R2) [7], further justify organizing the $\alpha_s$ subgroup extractions of this category based on the hadronization correction method employed, i.e. based on Monte Carlo event generators or on analytic models for the non-perturbative effects. New developments in jet substructure techniques [8] applied to $e^+e^-$ studies will reduce the hadronization corrections and, once they reach NNLO accuracy, will allow to reanalyze the LEP data with smaller non-perturbative uncertainties. These latest more precise results open up the potential substitution of older LEP analyses, with larger uncertainties, from the world average. New applications of the Principle of Maximum Conformality (PMC) for determining renormalization scales in $\alpha_s$ extractions via $e^+e^-$ event-shape variables were discussed during the workshop [9]. The PMC renormalization scales depend on the event-shape kinematics, reflecting the virtuality of the underlying QCD subprocess. Work is ongoing to provide a systematic evaluation of the theory uncertainties of the PMC predictions for pQCD at high orders.

- **Hadronic Z and W boson decays:**

  Why does the PDG world-average for this category prefer the $\alpha_s$ value derived from the global electroweak (Gfitter) SM fit, rather than the value derived from stand-alone analysis of the pseudo-observables directly measured at the Z boson pole? The result from the global SM fit, $\alpha_s(m_Z) = 0.1194 \pm 0.0029$ with a $\sim$2.4% uncertainty, is only slightly more precise than the latter, $\alpha_s(m_Z) = 0.1203 \pm 0.0028$ with $\sim$2.5% uncertainty, but the former is more prone to potential biases from new physics present in other sectors of the SM [10].

- **Hadronic final states at pp, p$\bar{\text{p}}$ colliders:**

  The breadth of LHC data and the associated recent NNLO pQCD theoretical developments have provided various new $\alpha_\text{s}$ extractions from pp collisions, including inclusive pp $\to t\bar{t}$, W, and Z [11] production cross sections, as well as differential jets cross sections. The question was raised whether adding more measurements could lead to not improving the world average with the currently used linear pre-averaging method. To be able to fully exploit all the experimental data, via e.g. a $\chi^2$-based BLUE-type average [12,13], the correlation matrices among measurements must be provided by the experimental collaborations. It has also been stressed that (future) systematically improved parton showers, if possible including corrections at NLL (or beyond) accuracy, are essential to fully gain control of the MC uncertainties involved in these analyses and also for $\alpha_\text{s}$ determinations based on $e^+e^-$ data.

  One of the strengths of the LHC data is the possibility to test asymptotic freedom at high energy scales, in the TeV regime, never explored before. In this context, it was pointed out that e.g. some ATLAS analyses of data from pp collisions covering scales at large total event energy ($H_T$) seem to run (evolve with scale) faster than expected. This effect can depend on the choice of renormalization scale setting in the extraction of $\alpha_\text{s}$, and this should be carefully checked for each chosen observable. The $\alpha_\text{s}$ running plots in the PDG summary should incorporate the pp $\to$ jets results at NNLO that extend the range up to about 2 TeV at the proper scale of each observable (leading jet $p_T$, sum of jet $p_T$'s,...). For the $\alpha_\text{s}$ running in the low energy range, the lattice results should be added. Last but not least, it was not clear, i.e. not explicitly documented in the publications, if all analyses at scales above the top quark production threshold used the proper number of active free flavours $n_f = 6$ in the prediction and the evolution calculations.

- **$\alpha_\text{s}$ categorization, combination and averaging of $\alpha_\text{s}(m_\text{z})$ results:**

  The current $\alpha_\text{s}(m_\text{z})$ PDG world-average [1] is derived from different measurements grouped, first, into subcategories that are subsequently combined into six overall categories. The individual subcategories are grouped following experimental measurements and theoretical methods (e.g. sharing a similar treatment of hadronization corrections), and the overall categories share basically the same underlying physical process. Suggestions were made to change some of the labels of the categories and/or to rearrange them to include newly available $\alpha_\text{s}$ extractions (see more detailed cases discussed above). In order to enter into the world average, the current conditions are that the $\alpha_\text{s}$ analysis has at least an NNLO theoretical accuracy, includes reliable estimates of experimental, systematic, and theoretical uncertainties, and the results are published in a peer-reviewed journal. It was discussed the possibility to drop relatively old analyses (e.g. from LEP $e^+e^-$ final states), because the same data have been reanalized in newer studies and/or because old hadronization corrections may have been superseded. Although the large $\alpha_\text{s}(m_\text{z})$ uncertainties of the oldest results likely have a small numerical impact on the final world-average, the results of more recent developments could be considered instead. In any case, it was emphasized, as done now in the PDG, that one should clearly study and define *beforehand* the rules for the selection of the analyses to be incorporated into the world average, and then follow them strictly to avoid any bias.

  It was highlighted that the individual lattice-QCD results have total 0.5 to 1% uncertainties, which are a factor of 2–4 smaller than all other $\alpha_\text{s}(m_\text{z})$ individual extractions (with 1.5–4% uncertainties). What does this imply for averaging, which will be driven by the most precise result? One way to control this (as already performed in recent $\alpha_\text{s}$ combinations) is to drop categories from the average and check the consistency of the results by explicitly quoting $\alpha_\text{s}(m_\text{z})$ averages without a subset of the measurements. The possibility to eventually use the lattice extraction as *the single* $\alpha_\text{s}(m_\text{z})$ PDG world-average, as it is the most precise value, it is based on experimental data (hadron masses and decays), and now contains also the running up to high scales, was considered. Some people expressed concerns on that proposal, given the need to always cross-check the lattice-QCD extraction with hadronic data in the explicitly perturbative regime. It was also pointed out that there are a number of dedicated high-precision determinations of the strong coupling from various methods that may eventually be inconsistent with the lattice-QCD results (if their derived central $\alpha_\text{s}(m_\text{z})$ values do not change, and the uncertainties shrink). In addition to the discussion on the averaging, were concrete decisions on the weight of these analysis have to be made, future average analyses should also point out these discrepancies factually to motivate further studies and progress.

The technical averaging procedure was also discussed. Currently it uses linear preaverages for the subcategories, then $\chi^2$-average with floating correlation with the PDF "$\chi^2$ reweighting" prescription (enlarged uncertainties, if needed, until $\chi^2/\text{dof} = 1$). Alternative averaging methods, e.g. a $\chi^2$ average in the groups with a correlation model following the BLUE or CONVINO approaches [12,13], were suggested, in particular to combine LHC measurements (see above). Values obtained with alternative methods should be provided together with the "default" world-average to check the overall robustness and stability of the final $\alpha_\text{s}(m_\text{z})$ averaging procedure.

All in all, the meeting featured lively and stimulating discussions among different experts on controversial issues, as well as on technical details, whose clarification will ultimately have an impact on more accurate and precise $\alpha_\text{s}$ determinations. Novel ideas to extract $\alpha_\text{s}$, estimation of expected reductions in the theoretical and experimental uncertainties of each method, as well as issues to be addressed in the coming future to improve the combination of all results, were discussed. There was a common agreement of the usefulness of organizing similar dedicated $\alpha_\text{s}$ workshops every $\sim 2$ years, following the 2011 [14] and 2015 [15] meetings. Whereas the strong force decreases with energy, the scientific interest in the QCD interaction clearly increases with time.